The Diverse Solar Phase Curves of Distant Icy Bodies. Part I: Photometric Observations of 18 Trans-Neptunian Objects, 7 Centaurs, and Nereid

Short Title: Opposition Surges of 26 Distant Icy Bodies

David L. Rabinowitz[1], Bradley E. Schaefer[2], Suzanne W. Tourtellotte[3]

[1]Center for Astronomy and Astrophysics, Yale University, P. O. Box 208121, New Haven CT 06520-8121 email: david.rabinowitz@yale.edu

[2]Department of Physics & Astronomy, Louisiana State University, 234 Nicholson, Baton Rouge LA 70803-0001

[3]Astronomy Department, Yale University, P. O. Box 208121, New Haven CT 06520-8121

ABSTRACT

We have measured the solar phase curves in B, V, and I for 18 Trans-Neptunian Objects, 7 Centaurs, and Nereid and determined the rotation curves for 10 of these targets. For each body, we have made ~100 observations uniformly spread over the entire visible range. We find that all the targets except Nereid have linear phase curves at small phase angles (< 2 deg) with widely varying phase coefficients (0.0 to 0.4 mag deg$^{-1}$). At phase angles > 3 deg, the Centaurs (54598) Bienor and (32532) Thereus have phase curves that flatten. The recently discovered Pluto-scale bodies (2003 UB313, 2005 FY9, and 2003 EL61), like Pluto, have neutral colors compared to most TNOs and small phase coefficients (< 0.1 mag deg$^{-1}$). Together these two properties are a likely indication for large TNOs of high-albedo, freshly coated icy surfaces. We find several bodies with significantly wavelength-dependent phase curves. The TNOs (50000) Quaoar, (120348) 2004 TY364 (47932), and 2000 GN171 have unusually high I-band phase coefficients (0.290±0.038, 0.413±0.064, 0.281±0.033 mag deg-1, respectively) and much lower coefficients in the B and V bands. Their phase coefficients increase in proportion to wavelength by 0.5 - 0.8 mag deg$^{-1}$ μm$^{-1}$. The phase curves for TNOs with small B-band phase coefficients (< 0.1 mag deg$^{-1}$) have a similar but weaker wavelength dependence. Coherent backscatter is the likely cause for the wavelength dependence for all these bodies. We see no such dependence for the Centaurs, which have visual albedos ~0.05.

## 1. INTRODUCTION

Most airless bodies in the solar system exhibit a brightness enhancement, known as an opposition surge, when they are observed at low solar phase angle, $\alpha$ (the Sun-object-Earth angle). The phenomenon has been known for many years and is well studied (Gehrels 1956, Hapke 1993, Nelson et al. 1998, Shkuratov et al. 2002). The effect is due to the granular structure of the material at the surface of these bodies. As the phase angle approaches zero, the shadows cast by one grain upon another disappear, increasing the total intensity of the scattered light. There is also an influence from coherent backscatter. At $\alpha$=0, the path length of a light ray from the sun hitting one particle, scattering to a second nearby particle, and then scattering back to the observer is identical to the path of another ray initially hitting the second particle, scattering to the first, and then scattering back. Provided the separation of the two particles is not much greater than wavelength of the light, then the interference is constructive between all such ray pairs regardless of the orientation of the scattering particles, thus leading to a total brightness enhancement. Both shadow hiding and coherent backscatter are important for explaining opposition surges, with coherent backscatter especially important for highly reflective surfaces.

In this paper we present our measurements of the opposition phase curves for 26 distant solar system bodies with widely varying sizes and orbit. These include 18 trans-Neptunian objects (defined here as bodies with perihelion, q > 19.2 AU and semimajor axis, a > 30.1 AU), 7 Centaurs (defined here as bodies with 19.2 > q > 5.2 AU), and Neptune's satellite Nereid. This is the most extensive survey to date dedicated to the long-term measurement of the light curves of distant bodies. We have already reported our observations for two of these bodies, (38628) Huya (Schaefer & Rabinowitz 2002) and 2003 EL61 (Rabinowitz et al. 2006). The target list includes one inner-Oort cloud object (90377 Sedna), all of the recently discovered Pluto-sized TNOs (2003 UB313, 2005 FY9, and 2003 EL61), and nearly all the known TNOs and Centaurs with apparent magnitude V < 20 and phase angle $\alpha < 2$ deg at the time of our survey. Table 1 lists the names and orbital elements for each of these targets, and also the period and amplitude of the rotational light curve (determined in this paper or elsewhere) and the visual albedo ($p_v$) where these values are known.

For each of our targets we have made numerous observations spread uniformly in time and covering the entire observable range of phase angle. This allows us to measure both the non-linearity of the phase curve and also the rotational light curve. In most cases, however, our targets are observable only at small phase angles ($\alpha < 2$ deg) where their phase curves are linear and the phase coefficient (slope of the curve) is the only measure of the magnitude of the opposition surge. Where the rotational modulation is significant, we subtract the modulation from our observations to properly measure the phase-dependence of the light curve. We have also measured most of the phase curves in Johnson-Cousins B, V, and I filters to look for a wavelength dependence to the surge. This is expected if coherent backscatter is an important influence (Hapke 1993, Schaefer & Rabinowitz 2002).

Our goal is to explore the range of shapes for the opposition surges (slope and amplitude) in order to constrain the surface structures and compositions of distant icy bodies. For the well-studied asteroids in the main belt, the shape of the phase curve is known to correlate with spectral type and albedo (Bowell & Lumme 1979, Belskaya & Shevchenko 2000). For many of the icy satellites in the outer solar systems, steep and narrow opposition surges have been observed (Buratti et al. 1992, Domingue et al. 1995, Schaefer & Tourtellotte 2001). These narrow surges likely result from coherent backscatter of the highly reflective surfaces (Nelson et al. 2000). Since it is known that the visible colors and albedos of TNOs and Centaurs are diverse (Tegler & Romanishin 1998, Jewitt & Luu 2001, Hainaut & Delsanti 2002, Grundy, Knoll, & Stephens 2005), it is reasonable to expect similar diversity for their opposition surges and to find correlations between the phase curve shape, albedo, and color. In a follow up to this paper, we will discuss the relationships we find between surge amplitude, color, albedo, orbit, and absolute magnitude in greater detail. Here, our main purpose is to present the observations, our method for reducing the data, and preliminary conclusions.

## 2. OBSERVATIONS

The observations we report here were made by on-site operators at Cerro Tololo using the 1.3-m telescope of Small and Moderate Aperture Research Telescope System (SMARTS) consortium. Images were recorded with the optical channel of the permanently mounted, dual infrared/optical CCD camera known as A Novel Dual-Imaging Camera (ANDICAM). The optical channel is a Fairchild 2Kx2K CCD which we binned in 2x2 mode to obtain 0.37" per pixel and a 6.3 'x 6.3' field of view. All exposures are auto-guided, and typical seeing is 1-2". Because the telescope is queue-scheduled for shared use by all members of the SMARTS consortium, we were able to obtain ~15 minutes of observing time per target every night or every other night for the entire duration of the apparition. We typically observed two targets per night, each with a sequence of three or four exposures (B-V-V-I or B-V-I), sometimes including the R band to better characterize the color. Users of the telescope share dome and sky flats, bias frames, and observations in B, V, R, and I of Landolt stars taken at a variety of air masses on all photometric nights. For each of our targets, Table 2 lists the average V magnitude (<V>) of our observations, the minimum and maximum phase angle observed ($\alpha_{min}$ and $\alpha_{max}$), the average geocentric and heliocentric distance of the target (<d> and <r>, respectively), the number of observations ($N_{obs}$), the range of observing dates, and the exposure times for each filter.

Our reduction procedure is identical for all our observations, and is described in detail elsewhere for the case of 2003 EL61 (Rabinowitz et al. 2006). Briefly, we correct all images using bias frames and flats recorded nightly. We use selected field stars appearing in each target image as secondary standards, and determine their apparent magnitudes from observations of the field stars and of Landolt standards made on the same photometric nights (for ~5% of the observations, the field star magnitudes were determined from photometric observations later made with the SMARTS 0.9-m and 1.0-m telescopes and the McDonald Observatory 0.8-m telescope). We use a large pixel aperture (14.8” diameter) to measure the fluxes of the field stars in each image and thereby determine the transformation from instrumental to apparent magnitude in each filter. To determine the apparent magnitude of the targets, we measure the flux within a small aperture (2.2” diameter), and apply a correction determined separately for each image from the ratio of the aperture fluxes of the field stars. We also make color-dependent corrections in the final determination of the phase curves (see discussion in Section 3.1).

To speed and standardize our reduction procedure we use a set of software scripts. Initially we use standard, interactive IRAF routines to measure the fluxes of Landolt stars, to select and measure the pixel coordinates of the field stars, and to identify and measure the pixel coordinates of the target. In the process, we make a visual inspection of each image. This is the basis upon which we decide to reject some measurements (discussed further below). We then run our scripts, which use a non-interactive IRAF routine (“phot”) to measure the small and large aperture fluxes for the target and selected field stars. The scripts determine the magnitude transformation and aperture correction for each image, and compute the resulting apparent magnitude and its uncertainty for each target image. The magnitude uncertainty is the quadratic sum of the noise (within the small aperture) from the readout, sky, and target and of the uncertainty in the magnitude calibration.

After we have reduced all the observations of a given target, we then reject measurements with magnitude uncertainties exceeding 0.3, or for which the target is near a bright star or a noise artifact (such as a cosmic ray hit or spurious noisy pixel), or for which another star or galaxy is visible within a few pixels of the aperture radius. We also require at least three field stars significantly brighter than the target (usually V < ~17). After these rejections, we iteratively compute the dispersion of all the measured magnitudes in each filter, throwing out observations with magnitudes exceeding the mean by three times the standard deviation. After three iterations, we use the final set of observations for analysis.

## 3. RESULTS AND ANALYSIS

### 3.1 Light-Curve Data

Table 3 lists the Julian date for the mid-time of the exposure, apparent magnitude, magnitude uncertainty, Julian date corrected for light-travel time, reduced magnitude, phase angle ($\alpha$), heliocentric distance (r), geocentric distance (d), and the filter for each observation we accepted for each target (3282 observations total). The reduced magnitude is the apparent magnitude minus 5log(rd) with r and d expressed in AU. Extrapolated to $\alpha$=0 deg, the reduced magnitude is the absolute magnitude in the respective filter band. The correction for light-travel time, subtracted from the UT date, is $(d-d_0)/c$ where $d_0$ is the geocentric distance at the time of the earliest observation for each target and c is the speed of light. Table 3 does not list our previously reported observations of Huya and 2003 EL61.

We note here that the magnitudes we report in Table 3 do not account for color-dependent terms in our transformations from instrumental to apparent magnitude. The resulting corrections, which are magnitude offsets no larger than a few percent, depend on the color of the target, the atmospheric conditions, and instrumental variations. Our long-term observations with the same telescope and detector show that the color-dependent terms do not vary significantly from night to night. We are therefore able to determine mean corrections and add these to our listed magnitudes after initially reducing all the observations for each target without color correction. The uncorrected data yield a mean color that yields a more accurate correction than we can determine on any individual night owing to relatively high measurement uncertainties. After we analyze the data, compute rotation and phase curves and determine the colors of each object, we then determine the color corrections and add these to the absolute magnitudes derived in Section 3.3, below. The corrections ($\Delta$B, $\Delta$V, and $\Delta$I added to the uncorrected magnitudes) are linear functions of the uncorrected B-V and V-I values, with $\Delta B = 0.063(B-V) - 0.027$ mag, $\Delta V = -0.030(B-V) + 0.018$ mag, and $\Delta I = -0.073(V-I) + 0.051$ mag. No correction is needed for R.

### 3.2 Rotational Light Curves

For some of our targets, the scatter in the brightness measurements clearly exceeds the measurement error, and this is likely the result of rotational modulation. For these, we have attempted to compute a rotational light curve and measure the rotation period using the procedure we describe elsewhere for 2003 EL61 (Rabinowitz et al. 2006). In this procedure, we initially make a linear fit to the reduced magnitude as a function of $\alpha$ for each filter and subtract this filter-dependent fit from the observations. We then combine the residuals for all filter observations into one data set and use phase-dispersion minimization (Stellingwerf 1978) to search for rotational periodicity. The final rotation curve is the combined set of residuals phased by the measured period and binned by rotational phase.

Note that the linear fits to the $\alpha$-dependence in each filter are preliminary at this stage. We use them only to create a combined data set suitable for a period search, with the $\alpha$ dependence and wavelength dependence largely removed. If we are able to find a rotation period and compute a reliable rotation curve, we then go back and subtract this rotational dependence from the original data and then re-determine the $\alpha$-dependence (see next section). In most cases, the preliminary fits are very close to the final fits after removing the modulation. This is because rotational modulation usually averages to zero over the longer times scale of the $\alpha$ variation, and our observations sample the light curve many times over the longer time scale. There are exceptions, and we discuss these further below (Section 4.5).

Figure 1 shows the resulting rotation curves of those objects for which we are able to determine the period unambiguously from our own observations, or for which a reliable period has been published elsewhere that we can use to compute a rotation curve. Gaps appear in these curves where we have no rotational-phase coverage. In Table 1 we list the respective periods and light curve amplitudes along with their uncertainties. The period uncertainty is the range of values around the best-fit period for which the computed dispersion is small and for which the computed rotation curve shows clear peaks and troughs. For 2002 UX25 and (8405) Asbolus, our observations are not sufficient to reveal the rotation period unambiguously. For these two cases we compute rotation curves assuming the periods published by Rousselot et al. (2005) and Kern et al. (2000), respectively (these periods are listed in Table 1 without error bars).

For the cases of 2002 GN171 and 1999 TD10, we are able to determine rotation periods that are consistent with Sheppard and Jewitt (2002) and Mueller et al. (2004), respectively, but additional periodicities appear in our observations owing to our 24-hour sampling bias. Here we again used the periods determined by the other authors to compute rotation curves. The periods listed in Table 1 are the values reported by Sheppard and Jewitt and by Mueller et al, but the period uncertainties are from our analysis of our own data. In all, we are able to determine unambiguous rotation periods and rotational light curves for 6 of these targets. We rely on published periods to determine the remaining 4 rotation curves.

We attempted to find a rotation curve for Sedna because there are variations in the time-averaged light curve at the 5% level that could be due to short-term (< 1 d) or long-term (1-100 d) periodicity. However, the uncertainties in our individual observations (much larger than 5%) and the 24-hour sampling bias of our observations preclude an unambiguous measurement of the period. We also searched for long rotation periods for our remaining targets with undetermined periods but did not find any significant or unambiguous periodicity on these timescales.

### 3.3 Solar Phase Curves

Figure 2 shows the final solar phase curves for all of our targets (except Huya which is published in Schaefer & Rabinowitz 2002). For each filter of each target for which we have covered a significant range in $\alpha$, we present a separate phase curve and a separate linear fit. Table 4 lists the resulting phase coefficients (B', V', R', and I'), the intercepts at $\alpha$=0 ($B_0$, $V_0$, $R_0$, and $I_0$), and the uncertainties of these measurements. The intercepts are corrected for color-dependent terms in the magnitude transformations, as discussed above. Figure 2 represents these fits by solid lines, using dashed lines to show the ranges of uncertainty. Note that to have the best visual comparison of the different phase curves, we have normalized the plotted data so that $B_0$=–0.3, $V_0$=0.0, $R_0$=0.3, and $I_0$=0.6 mags, respectively, for all targets.

To determine these phase curves, we first subtract the computed rotation curves from the reduced magnitudes of the respective targets as listed in Table 3. See Rabinowitz et al. (2006) for a detailed description of this procedure. For those targets with no rotation curves, we make no correction. We then sort the observations in each filter of each target into N equally spaced bins in $\alpha$, and compute the average solar phase angle, $\alpha_i$, the weighted average of the reduced magnitude, $y_i$, and the error of the weighted average, $\sigma_i$, for each bin. These weighted averages and their errors include a systematic error of 0.015 mag added in quadrature to each measurement error. This additional error accounts for night-to-night uncertainties in our magnitude calibrations determined from our measurements of bright field stars. We set the bin size for each average to a multiple of 0.05 deg, chosen so that N~10 for all targets. After normalization, these are the data points with error bars shown in figure 2.

Each fit to the binned data is a line, $F(\alpha)=M_0+M'\alpha$, with slope M' and intercept $M_0$ that minimizes the chi-square sum,

(1) $\chi^2 = \Sigma_i [(F(\alpha_i)-y_i)/\sigma_i]^2$

(see Press et al. 1986). In some cases there are a few outliers to the fit that we reject by iterating the fitting procedure with a 3-sigma cutoff. On each iteration we recompute the rms residual to the fit,

(2) $\sigma_{rms}^2 = \Sigma_i[(F(\alpha_i)-y_i)^2]/\Sigma_i$ ,

and throw out observations, i, for which $|F(a_i)-y(a_i)| > 3\sigma_{rms}$. After iterating the fit up to three times (or less if there are no more outliers), the final iteration yields the best fit. The errors listed by Table 4 for the intercept and phase coefficient of each fit are calculated in the usual way by propagating the uncertainties of each unrejected observation, $\sigma_i$, through to the solutions for the $M_0$ and M'.

Table 5 lists the resulting values for $\chi^2$, N, and the likelihood, P, for the measured $\chi^2$ assuming that all observations are independent and that their uncertainties have Gaussian dispersion. Note that N is the number of bins after rejecting outliers. Most our fits yield $\chi^2 \sim N$ and $P > 10\%$, indicating that they are consistent with the observations. However, a few yield $\chi^2/N > 2$ and $P < 1\%$. For these cases (highlighted in bold font in Table 5) the phase curve may not be linear, there may be uncorrected rotational modulation, or the measurement errors may be larger than we have calculated. We discuss these possibilities on a case-by-case basis in Section 4.5, below.

## 4. DISCUSSION

### 4.1 phase coefficients

As shown by Figure 2, nearly all the objects we observe have linear phase curves at low phase angles ($\alpha < 2$ deg), consistent with previously published phase curves for TNOs (Schaefer & Rabinowitz 2002, Sheppard & Jewitt 2002, Rousselot et al. 2003, Rabinowitz et al. 2006). The only significant exceptions are Nereid at small phase angles and the two Centaurs (54598) Bienor and (32532) Thereus for $\alpha > 2$ deg. We discuss these exceptions further, below.

Phase curves that are linear at small angles are not unusual. Most asteroidal bodies in the solar system have linear phase curves at very small phase angles where the opposition surge is strongest. There is normally an inflection at larger phase angles where the opposition surge weakens and the phase curve flattens (Bowell et al. 1989, Hapke 1993). For TNOs, which we can not observe at $\alpha > 2$ deg owing to their large distances, the lack of an inflection in the phase curves limits our knowledge of the width of the opposition surges. We can only say that for each target, the width is larger than the maximum phase angle of the observations (see Table 2).

Unlike the phase curves reported for TNOs by previous investigators, the phase curves we observe for TNOs and Centaurs have coefficients ranging widely from 0.0 to 0.4 mag deg$^{-1}$. In the largest previous study of TNO phase curves Sheppard & Jewitt (2002) measured R-band phase curves for seven TNOs for which the coefficients ranged only from 0.13 to 0.19 mag deg$^{-1}$. Phase coefficients in this range were also measured by us for Huya (Schaefer & Rabinowitz 2002) and by Rousselot et al. (2003) for 1999 TD10. Until recently, the only known TNO with a phase coefficient outside this range was Pluto. Pluto has a very flat phase curve with phase coefficient 0.041±0.003 mag deg$^{-1}$ (Tholen & Tedesco 1994). Our observations now show that Pluto is not unusual in this respect. We find four additional TNOs – 2003 UB313, 2005 FY9, 2003 EL61, and (55636) 2002 TX300 – with low phase coefficients ranging from 0.0 to 0.10 mag deg$^{-1}$.

Given that 2003 UB313, 2005 FY9, and 2003 EL61 are icy bodies comparable in size or larger than Pluto, it is perhaps natural they should have phase curves similar to Pluto's. Recent infrared observations show that 2003 UB313 and 2005 FY9, like Pluto, have reflectance spectra dominated by the presence of methance ice (Brown et al. 2005, Licandro et al. 2006). The reflectance spectrum of 2003 EL61, on the other hand, has the strong signature of crystalline water ice (Trujillo et al. 2006). All three bodies have neutral colors compared to most TNOs, as does Pluto (Rabinowitz et al. 2006 and discussion below). Furthermore, the albedos of 2003 UB313 and 2003 EL61 (see Table 1) are known to match or exceed Pluto's albedo of 60% (Brown et al. 2006, Rabinowitz et al. 2006). Since ultraviolet light and cosmic radiation will redden and darken methane-rich ice over time (Luu & Jewitt 1996) and turn crystalline ice to amorphous ice (Jewitt & Luu 2004), the icy surfaces of these bodies must be regularly recoated, similar to the resurfacing of Pluto when it approaches perihelion (Brown et al. 2005, Trujillo et al. 2006). It is thus possible that the flat phase curve for Pluto and for these other large TNOs is a property resulting from both from their high reflectivity and from the granular structure of their freshly-coated icy surfaces. We suspect that 2002 TX300 has a similar surface since it too has a neutral reflectance and flat phase curve. This interpretation is supported by the observations of Pinilla-Alonso et al. (2004), who report an infrared reflectance with strong water-ice absorption bands, and by Grundy et al. (2005), who establish a lower limit for the R-band albedo of 19%.

We also observe flat phase curves for some of our Centaur targets. The phase coefficients for 2002 GZ32, 20002 PN34, Asbolus, and Thereus are all below 0.1 mag deg$^{-1}$. Unlike the phase curves of the Pluto-scale TNOs, however, the flat phase curves for these much smaller bodies are not an indication of high-albedo, freshly-coated icy surfaces. The Centaurs we have observed all have spectral slopes, $V_0$–$I_0$, that are ~30% redder than solar (see discussion below). This is an indication that their surfaces are more heavily contaminated by organics (Cruikshank & Dalleore 2003) than the largest TNOs. The very low albedos of 0.059 ± 0.016 and 0.047 ± 0.015 that have been measured for Asbolus and Thereus (Stansberry et al. 2005) show that any ices on the surfaces of these two Centaurs are mixed with or covered by a much darker material. Furthermore, Centaurs approach the sun closer and more often than TNOs. Any methane ice on their surfaces would be rapidly outgassed and would not be retained owing to the low surface gravity of these relatively small bodies. Finally, the Centaurs we have observed have sizes in the range that would be heavily eroded by collisions assuming they originated as bodies in the Kuiper Belt and have only recently (within ~100 Myr) acquired Centaur orbits (Durda & Stern 2000). Any initially pure water-ice covering would have been eroded away.

That the dark Centaurs and the bright Pluto-sized TNOs have similarly flat phase curves is not unexpected. Laboratory measurements of materials with the lowest and highest albedos also show that both materials can have flat phase curves. Nelson et al. (2000) measured the phase curves at small angles of highly reflective (> 90%) aluminum oxide powders with varying particles sizes. For particles size six times smaller than the wavelength of the illumination and for $\alpha = 0.5 - 5.0$ deg, they observed a linear phase curve with coefficient ~ 0.01 mag $deg^{-1}$. They also observed a strong opposition surge in this sample, but only for $\alpha < 0.5$ deg. Shkuratov et al. (2002) measured flat, linear phase curves over the range $\alpha = 0.2 - 5.0$ deg for both freshly fallen snow and for coarse graphite (respective coefficients 0.01 and 0.05 mag $deg^{-1}$). The graphite had a very weak opposition surge for $\alpha < 0.5$, but the snow did not.

4.2 wavelength-dependent phase curves

Another new property revealed by our survey is strong wavelength dependence for some of the TNO phase curves. This is shown most clearly by Figure 3, where we plot I' versus B' for all the targets we observed in both these bands. While most of the points lie close to the dashed line marking I' = B', there are significant outliers. In particular, the TNOs Quaoar and 2004 TY364 have unusually steep I-band phase curves (I' = 0.290 ± 0.038 and 0.413 ± 0.064 mag $deg^{-1}$, respectively) while their B-band phase curves are relatively flat (B' = 0.081 ± 0.028 and 0.136 ± 0.047 mag $deg^{-1}$, respectively). The TNO (47932) 2000 GN171 (not represented in Fig. 3 because we did not observe it in the B-band) has a large I-band coefficient (0.281 ± 0.033 mag $deg^{-1}$) and a significantly lower coefficient in the V-band (0.143 ± 0.031 mag $deg^{-1}$). For these three TNOs the phase coefficients increase proportionally with wavelength, with a similar dependence for all three bodies (0.56±0.13 and 0.53±0.17 mag $deg^{-1}$ $\mu m^{-1}$ for Quaoar and 2000 GN171, and 0.75±0.13 mag $deg^{-1}$ $\mu m^{-1}$ for 2004 TY364).

Figure 3 also shows that the TNOs with the flattest B-band phase curves generally have phase coefficients that are steeper in the I band. For the TNOs we observe with B' < 0.1 mag $deg^{-1}$, all five have I'/B' ratios larger than unity. The overall trend is for The TNO phase curves to become steeper with wavelength. For each band pass, the average phase coefficient for the TNOs and the standard error of the mean are $\langle B' \rangle = 0.12 \pm 0.02$, $\langle V' \rangle = 0.15 \pm 0.02$, and $\langle I' \rangle = 0.17 \pm 0.02$ mag $deg^{-1}$. Interestingly, none of the above trends hold true for the Centaurs. Their distribution is symmetric about the line, I' = B', in Fig 3 and their average phase coefficients do not change significantly with wavelength ($\langle B' \rangle = 0.07 \pm 0.02$, $\langle V' \rangle = 0.07 \pm 0.02$, and $\langle I' \rangle = 0.08 \pm 0.020$ mag $deg^{-1}$).

Previous investigators have not reported TNO phase curves with a significant dependence on wavelength, as we have observed. Sheppard and Jewitt (2002) report only the R-band measurements of TNO phase curves. Buratti et al. (2003) observe a small wavelength dependence to Pluto's phase coefficient, ranging from 0.037 ± 0.001 mag $deg^{-1}$ in the B band to 0.032 ± 0.001 mag $deg^{-1}$ in the V and R bands. Rousselot et al. (2003) measure the phase curve for 1999 TD10 in the B, V, and R bands and see no wavelength dependence. However, Rousselot et al. sampled the B and V phase curves of 1999 TD10 at only two phase angles and their measurements uncertainties do not preclude a wavelength dependence at the level we see for 1999 TD10.

Voyager observations of Europa (Buratti & Veverka 1983) and of the Uranian satellites (Buratti et al. 1990) do show wavelength dependence to these phase curves at large phase angles ($\alpha$ = 5 – 50 deg), but the dependence is opposite to the trend we observe for TNOs at small phase angles. At these large phase angles the phase coefficient are small at all wavelengths (0.01 to 0.03 mag/deg). The spacecraft observations show, however, that the phase coefficients are generally larger by ~50% at ultraviolet wavelengths (~0.3 μm) compared to visible wavelengths (0.6 – 0.7 μm). This is believed to occur because of increased multiple scattering by the surface particles at longer wavelengths owing to an increase in albedo with wavelength. The multiple scattering fills in the shadows cast by the surface particles, and hence decreases the slope of the phase curve.

The reason we see an opposite trend for TNOs at small $\alpha$ could be that coherent backscatter rather than shadow hiding is the dominant cause of the opposition surge at small phase angles. For coherent backscatter, it is generally true that as the albedo and multiple scattering increases, the strength of the resultant opposition surge also increases. This is demonstrated in laboratory measurements by Shkuratov et al. (2002) who observe an increase in the slope of the phase curve at small phase angles for samples of increasing albedo. They also find that red pigments have steeper phase curves at small angles in red light, where they are most reflective, than in blue light where they are darker. Thus, for the distant bodies we observe at small angles, the phase coefficient may increase with wavelength because the albedo increases with wavelength.

This albedo dependence would also explain why the Centaurs we observe do not have significantly wavelength-dependent phase curves. These are the bodies with the lowest albedos in our target list. Of those with reported values (see Table 1), the average visual albedo for Centaurs is 0.056 ± 0.004, whereas the average for TNOs is 0.28 ± 0.10. With little or no multiple scattering, we should not expect coherent backscatter to dominate shadow hiding as the cause for the opposition surge. Hence we should not expect a significant dependence on wavelength.

We did observe one body, Nereid, with an apparent wavelength-dependence opposing the general trend for the other TNOs. As shown by Figure 3, we measure a phase coefficient for Nereid in the B-band (0.310 ± 0.019 mag $deg^{-1}$) significantly larger than in the I band (0.205 ± 0.037 mag $deg^{-1}$). However, in this case phase coefficients are not an appropriate measure of the opposition surge. As we discuss in Sec 4.3, below, Nereid has a significantly nonlinear phase curve at low phase angles (see Sec. 4.3, below). Because of the nonlinearity, the phase coefficients we measure are sensitive to the relative weighting of the observations in each filter as a function of $\alpha$ and to their range in $\alpha$. Also, the uncertainties we calculate for the coefficent are incorrect because they assume a linear fit is valid. Hence, we cannot conclude that Nereid has a significantly wavelength-dependent phase curve based on this analysis.

### 4.3 non-linear phase curves

As discussed above, the slopes of the TNO and Centaur phase curves should become nonlinear and flatten out at larger phase angles. This flattening is not possible to verify for most of our targets because they are too distant to observe at large phase angles. However, some of our Centaurs targets were close enough for us to observe them at $\alpha > 3$ deg. Figure 4 shows the extended phase curves for Thereus and Bienor. Here we have combined the separately determined B, V, and I curves for each object to determine an average curve with reduced error. We did this by shifting the B and V curves, respectively, by the values of $B_0$-$V_0$ and $V_0$- $I_0$ (from Table 4), combining with the I-band data, and then taking the median average of the measurements within the same phase angle bins that we use to separately determine the B, V, and I curves. The error bars are the standard error of the mean for each bin. Note that we have shifted the resulting curve for Bienor by +2.0 magnitudes to better compare with Thereus. Best-fit parabolas are superimposed on each curve.

Figure 4 shows that phase curves for both Thereus and Bienor can be fit by curves which flatten as $\alpha$ increases rather than a straight line. In both of these cases the residuals with respect to a parabolic fit are smaller than with respect to a linear fit. For Bienor, the linear fit yields reduced $\chi^2$ =2.18 with 9 degrees of freedom whereas the parabolic fit yields reduced $\chi^2$=1.98 with 8 degrees of freedom. For Thereus, the improvement decreases the reduced $\chi^2$ from 1.13 with 10 degrees of freedom to 0.96 with 9 degrees of freedom. An F-test for significance of the improved fit (see Bevington 1992, Eq. 10-10) yields respective likelihoods of 21 % and 13% that these improvements are due to chance. While these improvements are marginal, the apparent curvature for both cases is in the direction we would expect as the opposition surge weakens with $\alpha$. We also have observations of Centaurs 2002 PN34, 1999 TD10, and Asbolus at phase angles exceeding 3 deg, but we find no significant departure from linearity (i.e. no decrease in $\chi^2$ going to a parabolic fit) for these phase curves.

Other than Thereus and Bienor, the only other target we observe with a noticeably nonlinear phase curve is Nereid. Figure 2 shows that the V-band curve has an inflection at $\alpha$=1 deg, with the curve flattening at larger phase angles. This is similar to the inflection in the V-band phase curve we measured at an earlier epoch (Schaefer & Tourtellotte 2001). In our earlier analysis, we were able to fit the phase curve by two phase coefficients (0.38 mag deg$^{-1}$ for $\alpha$<1 deg and 0.03 mag deg$^{-1}$ for $1 < \alpha < 2$ deg). We obtain similar coefficients if we split our current V-band data the same way (0.337±0.025 mag deg$^{-1}$ for $\alpha$<1 deg, –0.049±0.041 mag deg$^{-1}$ for $\alpha$>1 deg). Neither the B-band nor the I-band phase curves show this inflection, but they are less well resolved owing to poorer sampling at $\alpha$>1 deg. Fitting the entire V-band phase curve with a parabola yields a reduced $\chi^2$ of 1.38 with10 degrees of freedom whereas the reduced $\chi^2$ for the linear fit is 8.47 with 11 degrees of freedom. Because an F test yields a likelihood of less than 0.1% that this improvement is due to chance and because we have independent observations from an earlier epoch showing the same curvature, we believe the curvature is significant.

## 4.4 color distribution versus phase angle

Figs 5 shows the B, V, I color distribution we observe for our targets, where the colors are the values $B_0$-$V_0$ and $V_0$- $I_0$ at $\alpha$ = 0 deg (see Table 4). The figure also shows the sun's color and the mean B-V, V-I values listed by Hainaut & Delsanti (2002) for 100 TNOs and 24 Centaurs in their Minor Bodies in the Outer Solar System (MBOSS) database (www.sc.eso.org/~ohainaut/MBOSS). It is apparent that our targets have a color distribution similar to the larger MBOSS sample. A Kolmogorov-Smirnoff test (Press et al. 1986) yields respective probabilities 0.95 and 0.27 that the B-V and V-I distributions are drawn from the same distributions as the MBOSS sample. We note, however, that the color distribution we see for the TNOs depends on the phase angle that we choose to represent the colors. This is because some of the TNOs have phase coefficients, and hence colors, that depend on wavelength (see Sec. 4.2). This may help explaining the disparate results reported by previous observers for the bimodality of the TNO color distribution (Tegler & Romanishin 1998, Jewitt & Luu 2001, Hainaut & Delsanti 2002). We will explore this effect in detail in a future analysis.

4.5 phase curve uncertainties

As we discuss above, there are a few cases where our linear fits to the measured phase curves yield $\chi^2$ values with very low probabilities (see Table 5). We already addressed the problems for Nereid above (Sec 4.3). We address the remaining cases here.

Ixion. Here we had trouble fitting the B-band phase curve (the $\chi^2$ likelihood is only 0.008). Examination of the plotted curve (Fig. 2) shows that two of the B-band data points (at $\alpha$ = 0.3 and 0.4 deg) have very large error bars (~0.25 mags). Throwing these two points out does not change the slope significantly, but raises the likelihood of the fit to 0.02 ($\chi^2$ =19.8 for 9 degrees of freedom). The most likely explanation for the poor fit is that we have underestimated the measurement uncertainty.

1999 TC36. For this target the fit to the V-band curve is poor ($\chi^2$ likelihood of 0.006). The measurement uncertainties are all about the same magnitude (~0.1). The most likely explanation for the poor fit is that we have underestimated the measurement error. There could also be rotational modulation that we have not subtracted. Ortiz et al. (2003) report variability of ~0.06 mags with indeterminate periodicity.

Bienor. Both the B and V phase curves are poorly fit (likelihoods of 0.001 and 0.002, respectively). As discussed above, however, the phase curve has a marginally significant inflection at $\alpha \sim 3$ deg. Fitting these curves with a parabolic function yields lower values for $\chi^2$. Also, Bienor has significant rotational modulation. There is an uncertainty introduced by subtracting the light curve, and we do not account for this in our determination of $\chi^2$.

2002 PN34. Both the V and I bands are poorly fit (reduced $\chi^2$ of 3.01 and 2.23 with 9 and 11 degrees of freedom, respectively). Here it is likely that there is rotational modulation that we have not subtracted. We are able to find various rotation curves that reduce the scatter in the phase curves considerably after subtraction. For example, subtracting a rotation curve with period of 0.34473 d yields reduced $\chi^2$ values of 0.93 for the V-band phase curve (7 degrees of freedom) and 1.33 for the I band (10 degrees of freedom). However, there several other periods that work as well to reduce the scatter.

## 5. CONCLUSIONS

This is paper presents the results of the first survey dedicated to the measurement of solar phase curves of distant solar system bodies. The target list is diverse, including TNOs and Centaurs with widely varying sizes, orbits, and surface properties. The measurement we present are numerous, uniformly sampling the entire observable phase curve of each body at several visible wavelengths. For bodies showing significant variability on short timescales, the measurements are sufficient to determine and subtract the rotational modulation from the phase curves. Our preliminary conclusions are as follows.

(1) Small phase coefficients (< 0.10 mag $deg^{-1}$) are a salient feature of the phase curves for Pluto-scale TNOs with neutral colors, high albedos, and icy surfaces. The low amplitude of the opposition surge may be related to frequent resurfacing of these bodies with fresh ices. It is likely that this constellation of properties extends to other large (> 100 km) bodies. If so, measuring the color and phase coefficient may be sufficient to recognize other high-albedo members of the TNO population that are otherwise too faint for direct albedo measurements.

(2) Nearly all of the distant bodies we observe have linear phase curves at phase angles < 2 deg. The phase curves must flatten at larger phase angles where the opposition surge diminishes. Our observations of two Centaurs show the flattening may begin at phase angles as small as ~3 deg. Only the phase curve for Nereid shows recognizable deflection at phase angles < 2 deg.

(3) Three TNOs have phase curves that are significantly wavelength dependent, with the I-band phase coefficient exceeding the B-band coefficient by more than a factor of 2. Their phase coefficients increase linearly with wavelength with proportionality 0.5 to 0.8 mag $deg^{-1}$ $\mu m^{-1}$. There is a similar but much weaker trend for the TNO phase curves with flat B-band phase curves (B-band phase coefficients < 0.1 mag $deg^{-1}$). For these cases, coherent backscatter may be the dominant cause for the opposition surge at low phase angle. None of the Centaur phase curves are significantly wavelength dependent, consistent with recent observations that these bodies have very low albedos (~ 0. 05) for which the effect of coherent backscatter should be small.

(4) The color distribution we observe for the TNOs and Centaurs extrapolated to zero phase angle is generally consistent with the distributions determined from other surveys. However, the colors for some TNOs depend on the phase angle of the observations. It may be important to include this influence in the analyses of color distributions.

The above conclusions are the starting point for a succeeding paper examining in more detail the relation between the orbits, sizes, colors, phase coefficients, and albedos of the distant solar system bodies. Other investigators have found that the colors of the TNOs are related to their orbits, perhaps serving as markers for their place of origin within the solar system (Gomes 2003, Morbidelli & Brown 2002, Tegler & Romanishin 2003). A similar investigation of the orbital dependence of phase curves may further unravel the origins for compositional diversity in the Kuiper Belt.

This work was supported by the Planetary Astronomy program of the National Aeronautics and Space Administration under Grants NAG5-13533 and NAG5-13369. We specially thank our SMARTS queue manager, Rebeccah Winnick, for help scheduling the observations.

Figure Captions

Figure 1. Rotational light curves we observe for the TNOs and Centaurs we with measured rotation periods. For each light curve, the combined B, V, and I observations of the respective target have been phased by the rotation period and averaged together within equally spaced phase bins. Gaps appear where there is no phase coverage. The respective rotation periods are listed in Table 1, with order in the table corresponding to top-to-bottom, left-to-right order in the figure.

Figure 2. Normalized reduced magnitude versus solar phase angle in Johnson filters B (hexagons), V (diamonds), R (circles), and I (squares) for the observed targets. For those objects with measured rotation curves (indicated by asterisks after the name), we have subtracted the rotational phase curve from the observations. Solid lines show linear fits to the phase curves. Dashed lines show the range of uncertainty for each fit. All curves are normalized so that the linear fits intercept zero phase angle at –0.3, 0.0, 0.3, and 0.6 mags for B, V, R, and I, respectively. As in Fig. 1, the top-to-bottom, left-to-right order corresponds to the order in Table 1.

Figure 3. Phase coefficients, I' versus B', for the observed targets, with TNOs and Centaurs represented by filled squares and unfilled triangles, respectively. The dashed line shows I' = B'.

Figure 4. Normalized reduced magnitude versus solar phase angle showing the full phase-angle coverage for (32532) Thereus (filled squares) and (54598) Bienor (unfilled triangles). The solid lines are best-fit parabolas. The curve for Bienor is shifted by +2.0 mag.

Figure 5. V-I versus B-V extrapolated to zero phase angle for the observed targets (with TNOs and Centaurs represented by filled squares and unfilled triangles, respectively). A circle represents the sun's values. Small squares show the mean B-V, V-I values listed by Hainaut & Delsanti (2002) for 100 TNOs (unfilled circles) and 24 Centaurs (filled circles) in their Minor Bodies in the Outer Solar System (MBOSS) database (www.sc.eso.org/~ohainaut/MBOSS).

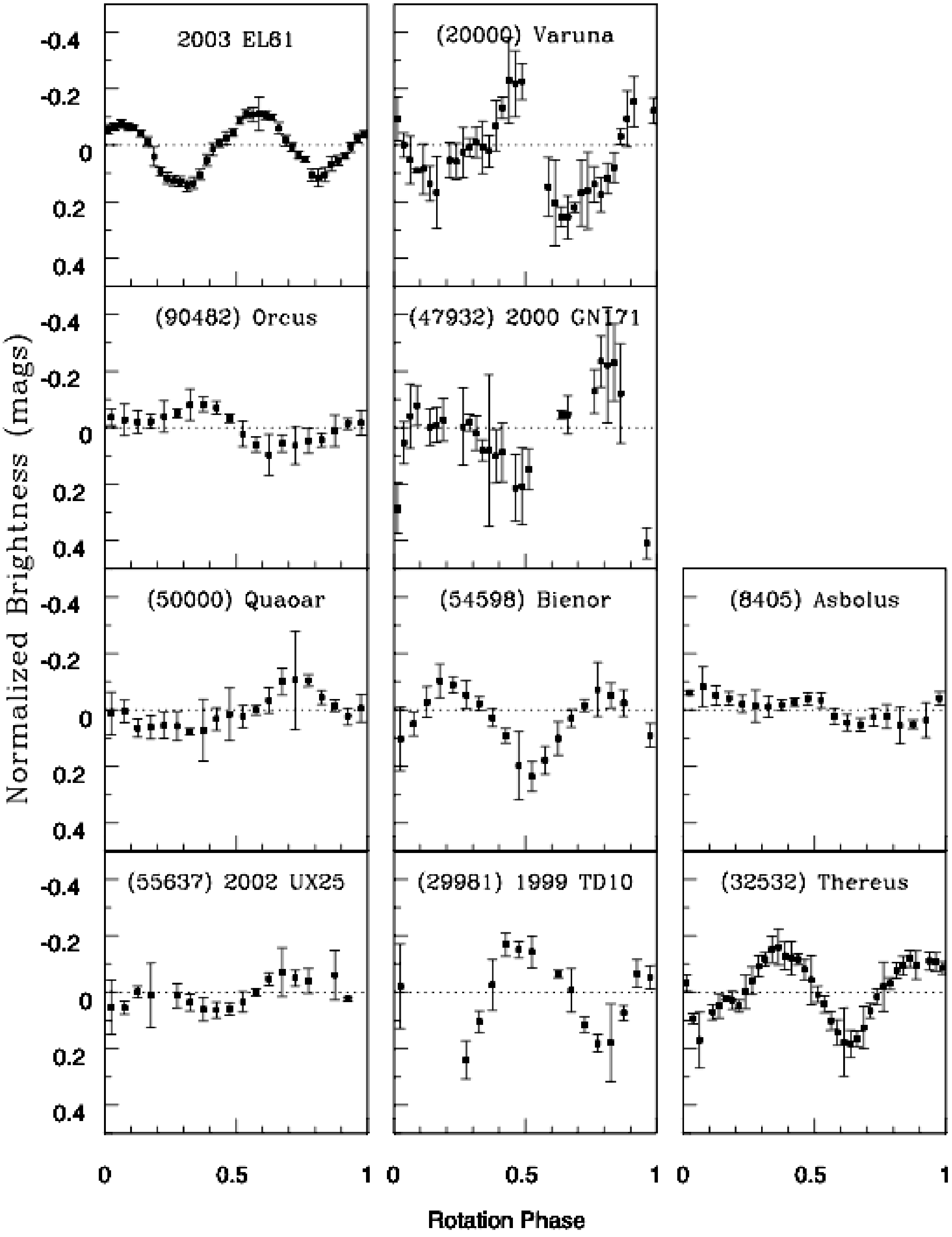

Figure 1

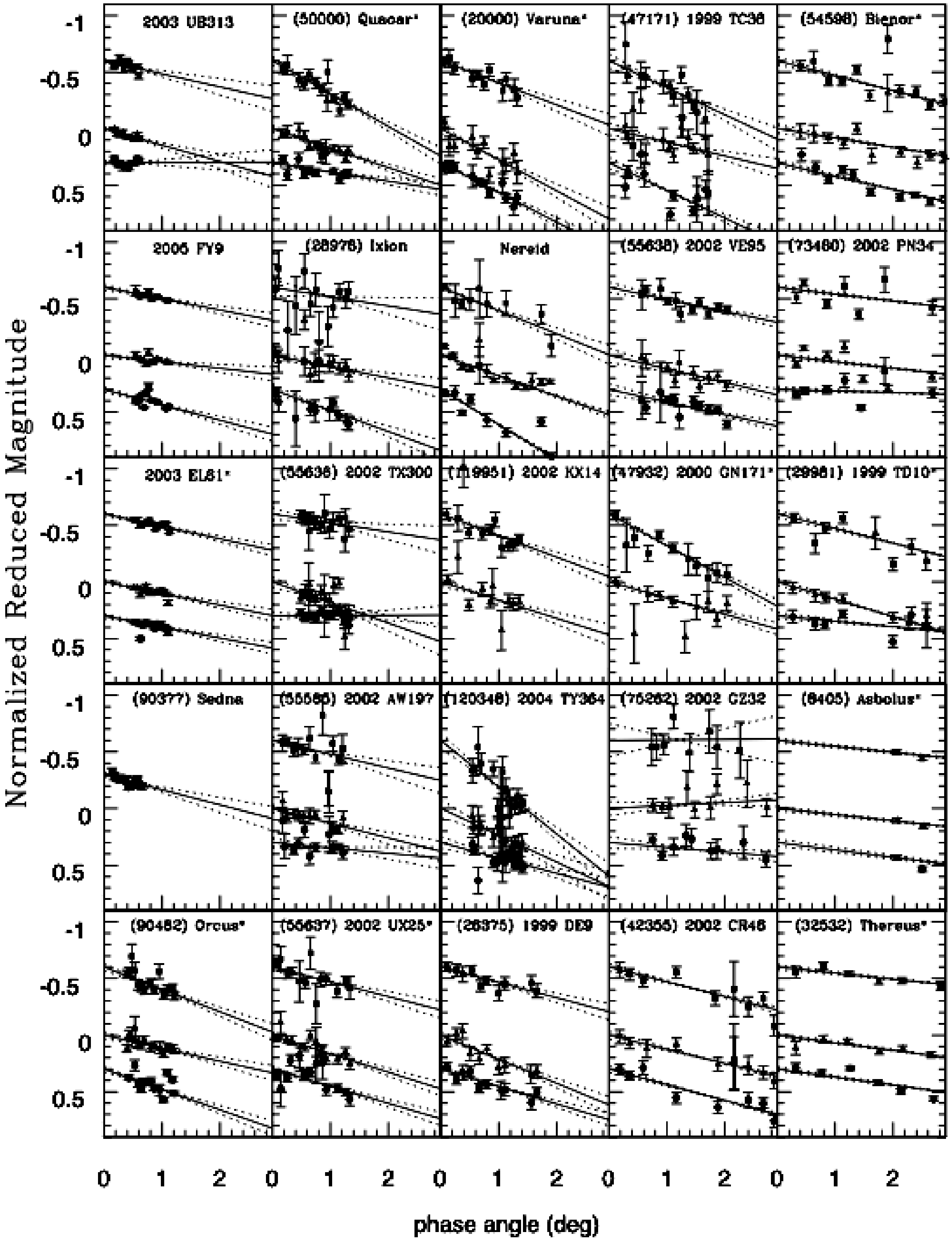

Figure 2

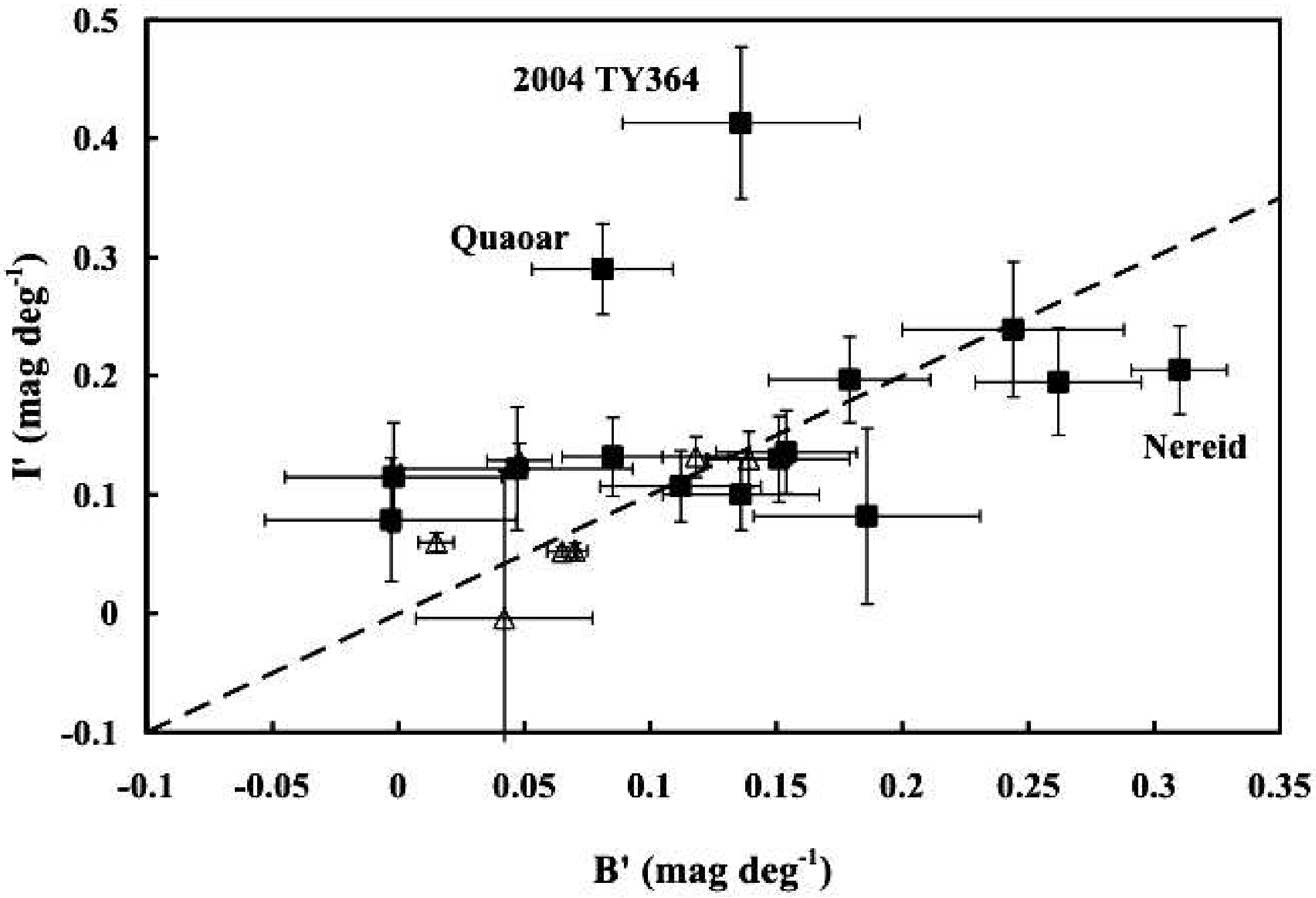

Figure 3

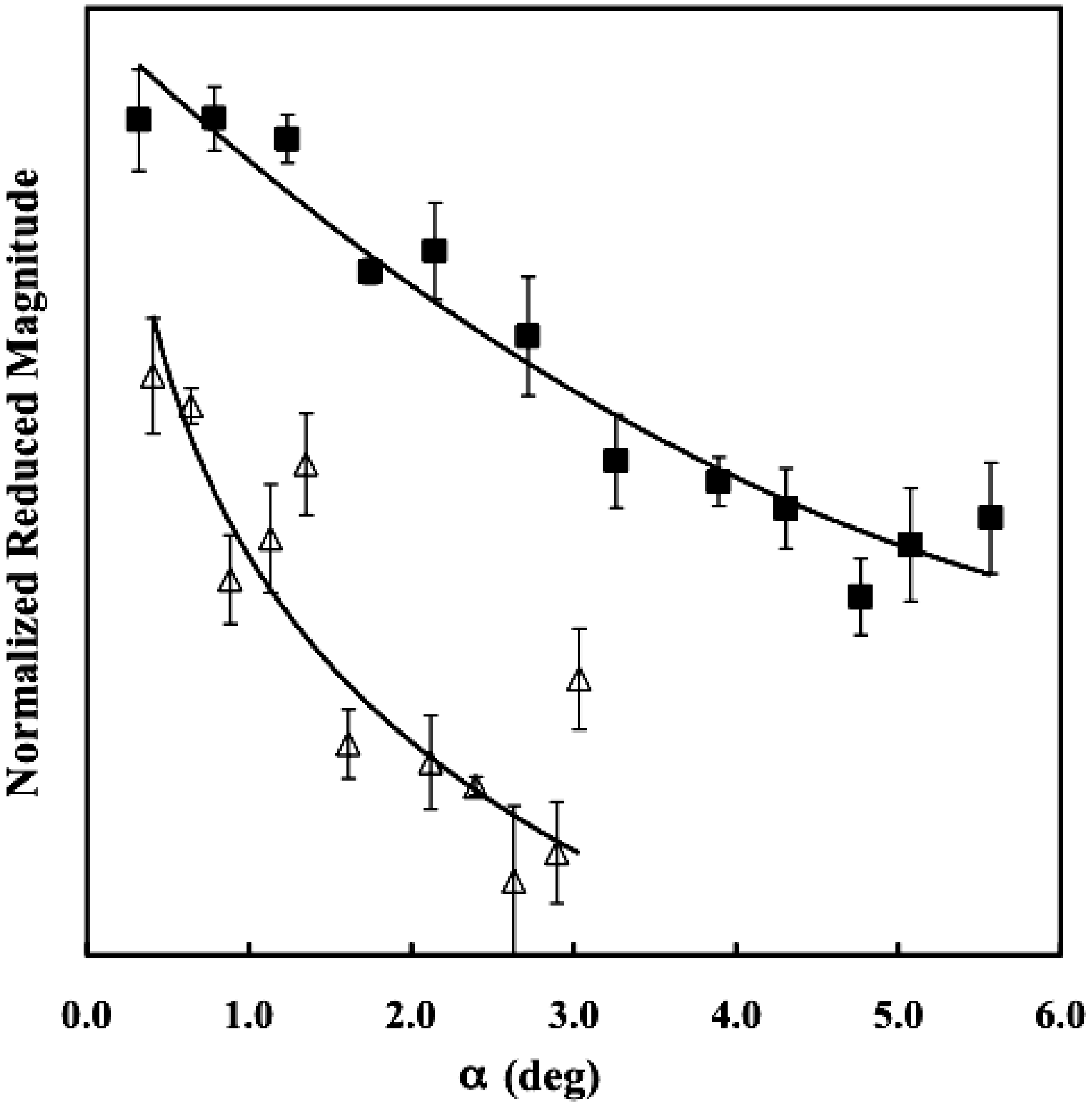

Figure 4

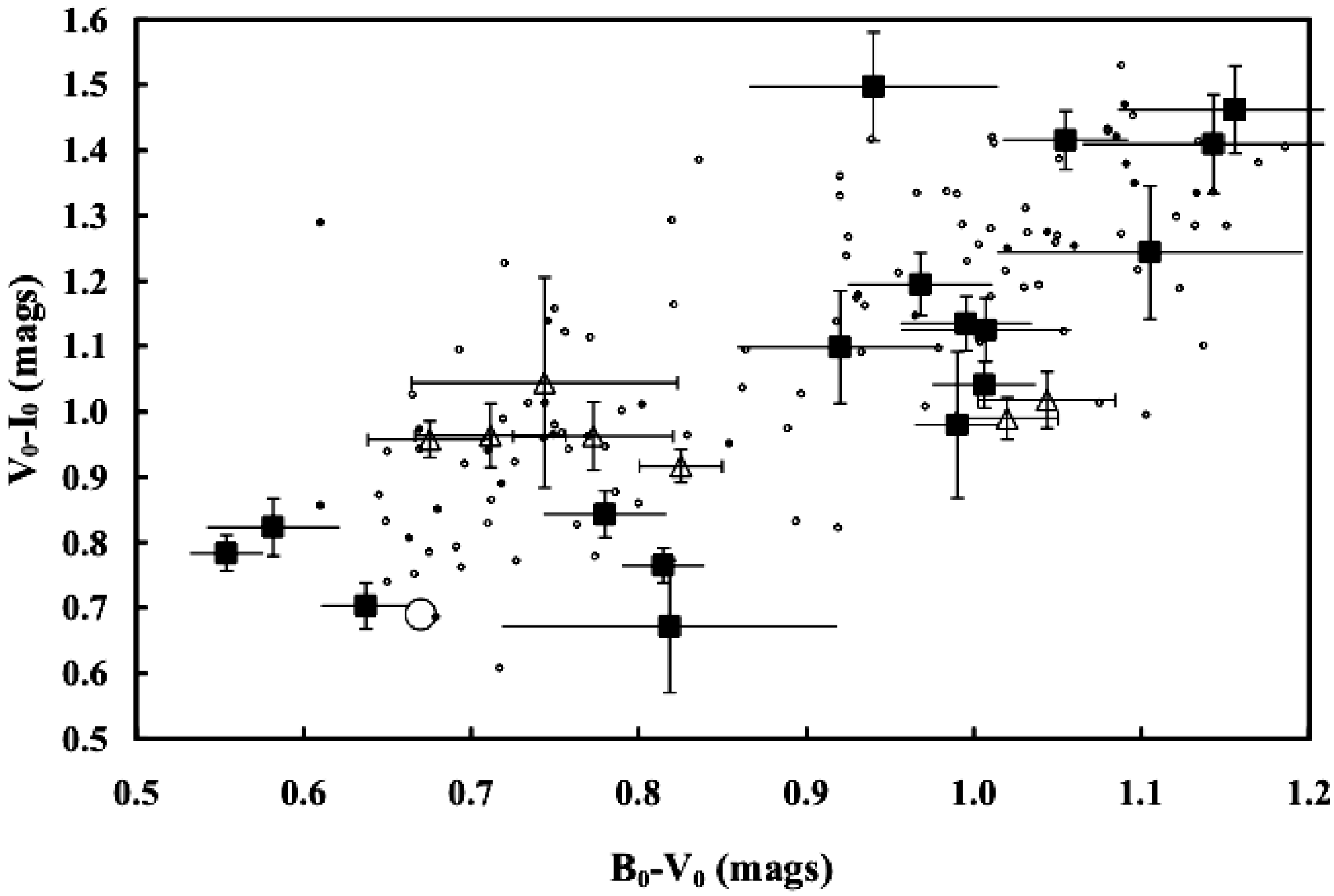

Figure 5

TABLE 1
Target Properties

| Target | q (AU) | Q (AU) | H (mag) | i (deg) | e | a (AU) | Amp. (mag) | Amp. error | Period (d) | Period error (s) | $p_v$ |
|---|---|---|---|---|---|---|---|---|---|---|---|
| TNOs: | | | | | | | | | | | |
| 2003 UB313 | 37.8 | 97.6 | -1.2 | 44.2 | 0.442 | 67.7 | | | | | 0.86±0.07[a] |
| 2005 FY9 | 38.6 | 52.8 | -0.4 | 29.0 | 0.155 | 45.7 | | | | | |
| 2003 EL61 | 35.1 | 51.5 | 0.1 | 28.2 | 0.189 | 43.3 | 0.28 | 0.04 | 0.163145 | 1 | 0.7±0.1[b] |
| (90377) Sedna | 76.1 | 902.0 | 1.6 | 11.9 | 0.844 | 489.0 | | | | | |
| (90482) Orcus | 30.7 | 48.1 | 2.3 | 20.6 | 0.220 | 39.4 | 0.18 | 0.08 | 0.549517 | 4 | |
| (50000) Quaoar | 42.0 | 45.1 | 2.6 | 8.0 | 0.035 | 43.5 | 0.18 | 0.10 | 0.368333 | 30 | 0.10±0.03[c] |
| (28978) Ixion | 30.1 | 49.2 | 3.2 | 19.6 | 0.241 | 39.6 | | | | | 0.24±0.13[d] |
| (55636) 2002 TX300 | 37.8 | 48.4 | 3.3 | 25.9 | 0.123 | 43.1 | | | | | |
| (55565) 2002 AW197 | 41.2 | 53.6 | 3.3 | 24.4 | 0.131 | 47.4 | | | | | 0.134±0.046[d] |
| (55637) 2002 UX25 | 36.5 | 48.6 | 3.6 | 19.5 | 0.142 | 42.5 | 0.13 | 0.09 | 0.699250[e] | … | |
| (20000) Varuna | 40.7 | 45.2 | 3.7 | 17.2 | 0.052 | 43.0 | 0.49 | 0.17 | 0.264342 | 9 | 0.155±0.075[d] |
| Nereid | 29.8 | 30.3 | 4.4 | 1.77 | 0.009 | 30.1 | | | | | 0.18±0.02[f] |
| (119951) 2002 KX14 | 37.4 | 40.6 | 4.4 | 0.4 | 0.041 | 39.0 | | | | | |
| (120348) 2004 TY364 | 36.1 | 41.3 | 4.5 | 24.9 | 0.067 | 38.7 | | | | | |
| (38628) Huya | 28.5 | 51.0 | 4.7 | 15.5 | 0.282 | 39.8 | | | | | 0.07±0.02[d] |
| (26375) 1999 DE9 | 32.3 | 79.4 | 4.7 | 7.6 | 0.421 | 55.9 | | | | | |
| (47171) 1999 TC36 | 30.6 | 47.9 | 4.9 | 8.4 | 0.221 | 39.2 | | | | | 0.083±0.028[j] |
| (55638) 2002 VE95 | 28.0 | 50.3 | 5.3 | 16.4 | 0.285 | 39.2 | | | | | |
| (47932) 2000 GN171 | 28.3 | 51.2 | 6.0 | 10.8 | 0.288 | 39.7 | 0.64 | 0.11 | 0.347046[g] | 9 | |
| Centaurs: | | | | | | | | | | | |
| (95626) 2002 GZ32 | 18.1 | 28.4 | 6.8 | 15.0 | 0.222 | 23.2 | | | | | |
| (42355) 2002 CR46 | 17.5 | 58.8 | 7.2 | 2.4 | 0.541 | 38.2 | | | | | 0.068±0.023[d] |
| (54598) Bienor | 13.2 | 19.8 | 7.6 | 20.8 | 0.201 | 16.5 | 0.34 | 0.08 | 0.382230 | 4 | 0.049±0.016[d] |
| (73480) 2002 PN34 | 13.3 | 48.5 | 8.2 | 16.6 | 0.569 | 30.9 | | | | | |
| (29981) 1999 TD10 | 12.3 | 178.0 | 8.8 | 6.0 | 0.871 | 95.1 | 0.41 | 0.08 | 0.640917[h] | 30 | 0.058±0.020[d] |
| (8405) Asbolus | 6.8 | 29.1 | 9.0 | 17.6 | 0.620 | 18.0 | 0.14 | 0.10 | 0.186148[i] | … | 0.059±0.016[d] |

| | | | | | | | | | | |
|---|---|---|---|---|---|---|---|---|---|---|
| (32532) Thereus | 8.5 | 12.7 | 9.0 | 20.4 | 0.198 | 10.6 | 0.34 | 0.08 | 0.347441 | 2 | 0.047±0.015[d] |

Notes: Orbital elements and H values are from the Minor Planet Center (http://cfa-www.harvard.edu/cfa/ps/mpc.html). Nereid is given Neptune's orbital elements. Where referenced, periods are the values determined by others (2002 UX25 and (8405) Asbolus) or else they are one of several multiple solutions from our own analysis that matches values independently determined by others. Unreferenced periods were determined solely from the data we report in this paper or in Rabinowitz et al. (2006). References: [a]Brown et al. (2006), [b]Rabinowitz et al. (2006), [c] Brown & Trujillo (2004), [d]Stansberry et al. (2005), [e] Rousselot et al. (2005), [f]Brown et al. (1998), [g]Sheppard & Jewitt (2002), [h]Mueller et al (2004), [i]Kern et al (2000), [j]Stansberry et al. (2006).

TABLE 2

Observing Circumstances

| Target | <V> (mag) | $\alpha_{min}$ (deg) | $\alpha_{max}$ (deg) | <d> (AU) | <r> (AU) | $N_{obs}$ | Observing Dates | Exposure Times (s) B | V | R | I |
|---|---|---|---|---|---|---|---|---|---|---|---|
| TNOs | | | | | | | | | | | |
| 2003 UB313 | 18.8 | 0.16 | 0.60 | 96.4 | 96.9 | 275 | 2005 Jan 6-9 | 300 | 180 | 180 | 180 |
| | | | | | | | 2005 Jul 10- 2006 Jan 26 | 300 | 180 | 180 | 180 |
| 2005 FY9 | 17.3 | 0.69 | 1.14 | 50.7 | 51.0 | 118 | 2005 Apr 4 – 2006 Apr 23 | 300 | 120 | 120 | 120 |
| 2003 EL61 | 17.6 | 0.51 | 1.11 | 50.7 | 51.2 | 382 | 2005 Jan 25 – 2006 Apr 23 | 300 | 120 | 120 | 120 |
| (90377) Sedna | 21.4 | 0.27 | 0.46 | 88.8 | 89.6 | 293 | 2003 Nov 26- Nov 27 | - | - | 300 | - |
| | | | | | | | 2003 Dec 2 – Dec 25 | 900 | 200 | 200 | 200 |
| | | | | | | | 2003 Dec 3 - 2006 Jan 29 | - | - | 600 | - |
| (90482) Orcus | 19.2 | 0.39 | 1.22 | 47.3 | 47.6 | 143 | 2004 Feb 21 – Jul 6 | 300 | 60 | 60 | 60 |
| (50000) Quaoar | 19.2 | 0.17 | 1.31 | 42.8 | 43.4 | 166 | 2003 Feb 25 – Jul 16 | 300 | 70 | - | 50 |
| (28978) Ixion | 20.2 | 0.03 | 1.32 | 42.3 | 42.9 | 144 | 2003 Feb 26 – Jul 25 | 400 | 100 | - | 60 |
| (55636) 2002 TX300 | 19.6 | 0.49 | 1.34 | 40.2 | 41.0 | 99 | 2004 Aug 6 – Dec 20 | 360 | 90 | - | 90 |
| (55565) 2002 AW197 | 20.4 | 0.17 | 1.22 | 46.7 | 47.2 | 130 | 2003 Feb 25 – June 3 | 600 | 140 | - | 100 |
| | | | | | | | 2003 Nov 29 – 2004 Feb 28 | 650 | 150 | - | 150 |
| (55637) 2002 UX25 | 20.2 | 0.02 | 1.36 | 41.8 | 42.6 | 130 | 2003 Jul 24 – 2003 Dec 22 | 600 | 150 | - | 150 |
| (20000) Varuna | 20.4 | 0.06 | 1.32 | 42.6 | 43.3 | 78 | 2004 Dec 30 –2005 Apr 17 | 600 | 120 | - | 120 |
| Nereid | 19.3 | 0.01 | 1.92 | 29.3 | 30.1 | 250 | 2003 Mar 30 – Aug 20 | 240 | 70 | - | 40 |
| | | | | | | | 2005 May 15 - Oct 26 | - | 900 | - | - |
| (119951) 2002 KX14 | 20.9 | 0.10 | 1.39 | 39.0 | 39.7 | 59 | 2003 Mar 2 – Jul 14 | - | 180 | - | 180 |
| (120348) 2004 TY364 | 20.6 | 0.55 | 1.41 | 39.3 | 39.8 | 97 | 2004 Oct 13 – 2005 Jan 24 | 480 | 120 | - | 120 |
| (38628) Huya | 19.9 | 0.28 | 1.96 | 29.5 | 29.8 | 214 | 2000 Dec 11 - 2001 Jul 24 | - | 300 | 300 | - |
| (26375) 1999 DE9 | 20.7 | 0.09 | 1.66 | 34.2 | 34.9 | 88 | 2005 Jan 30 – Jun 5 | 600 | 180 | - | 180 |
| (47171) 1999 TC36 | 20.3 | 0.28 | 1.71 | 30.4 | 31.2 | 101 | 2003 Jul 24 – Dec 6 | 360 | 90 | - | 40 |
| (55638) 2002 VE95 | 20.4 | 0.57 | 2.07 | 27.5 | 28.0 | 117 | 2004 Aug 6 – 2005 Jan 28 | 480 | 120 | - | 120 |
| (47932) 2000 GN171 | 21.1 | 0.02 | 2.04 | 27.8 | 28.5 | 67 | 2003 Mar 3 – 2004 Jul 16 | - | 180 | - | 180 |
| Centaurs: | | | | | | | | | | | |
| (95626) 2002 GZ32 | 20.5 | 0.73 | 2.80 | 20.1 | 20.9 | 88 | 2003 Feb 25 – Jul 26 | 600 | 140 | - | 100 |
| (42355) 2002 CR46 | 20.3 | 0.19 | 2.90 | 16.8 | 17.6 | 52 | 2005 Jan 31 – Apr 13 | 300 | 120 | - | 120 |
| (54598) Bienor | 20.5 | 0.33 | 3.05 | 18.2 | 18.8 | 180 | 2003 Jul 24 – Nov 19<br>2005 Sep 20 – Dec 16 | 650<br>300 | 190<br>180 | -<br>- | 70<br>180 |
| (73480) 2002 PN34 | 20.0 | 0.33 | 4.30 | 13.1 | 13.5 | 206 | 2003 Apr 3 – Jul 19<br>2003 Jul 22 - Nov 13 | 400<br>900 | 100<br>200 | -<br>- | 60<br>200 |
| (29981) 1999T D10 | 20.4 | 0.26 | 3.99 | 13.1 | 13.9 | 71 | 2003 Aug 14 – Dec 2 | 720 | 180 | - | 180 |
| (8405) Asbolus | 17.6 | 1.97 | 7.81 | 6.4 | 7.2 | 117 | 2003 Jul 24 – Nov 7 | 30 | 30 | - | 30 |
| (32532) Thereus | 19.3 | 0.22 | 5.60 | 8.9 | 9.7 | 204 | 2003 Aug 2 – Nov 30<br>2005 Sep 24 – Dec 23 | 600<br>300 | 100<br>120 | -<br>- | 100<br>120 |

Notes: Data for 2003 EL61 and Huya are reported in Rabinowitz et al. (2006) and Schaefer & Rabinowitz (2002), respectively.

TABLE 3
Measured Magnitudes for TNOs, Centaurs, and Nereid

| Target | JD-2450000 | Apparent Mag | Mag error | JD-245000 (corrected) | Reduced Mag | α (Deg) | r (AU) | d (AU) | Filter |
|---|---|---|---|---|---|---|---|---|---|
| 2003 UB313 | 3377.56375 | 18.344 | 0.016 | 3377.56375 | -1.520 | 0.580 | 96.937 | 96.884 | R |
| 2003 UB313 | 3377.56657 | 18.031 | 0.025 | 3377.56657 | -1.833 | 0.580 | 96.937 | 96.884 | I |
| 2003 UB313 | 3377.57628 | 18.360 | 0.018 | 3377.57628 | -1.504 | 0.580 | 96.937 | 96.884 | R |
| 2003 UB313 | 3377.57909 | 17.894 | 0.024 | 3377.57909 | -1.970 | 0.580 | 96.937 | 96.884 | I |
| 2003 UB313 | 3378.55277 | 18.271 | 0.045 | 3378.55267 | -1.593 | 0.580 | 96.937 | 96.901 | R |
| 2003 UB313 | 3378.55559 | 18.044 | 0.028 | 3378.55549 | -1.820 | 0.580 | 96.937 | 96.901 | I |
| 2003 UB313 | 3378.55907 | 19.294 | 0.022 | 3378.55897 | -0.570 | 0.580 | 96.937 | 96.901 | B |
| 2003 UB313 | 3378.56529 | 18.421 | 0.026 | 3378.56519 | -1.443 | 0.580 | 96.937 | 96.901 | R |
| 2003 UB313 | 3378.56810 | 17.989 | 0.024 | 3378.56800 | -1.875 | 0.580 | 96.937 | 96.901 | I |
| 2003 UB313 | 3379.55241 | 19.559 | 0.034 | 3379.55222 | -0.305 | 0.581 | 96.937 | 96.917 | B |
| 2003 UB313 | 3379.55864 | 18.367 | 0.034 | 3379.55845 | -1.497 | 0.581 | 96.937 | 96.917 | R |
| 2003 UB313 | 3379.56145 | 18.097 | 0.023 | 3379.56126 | -1.767 | 0.581 | 96.937 | 96.917 | I |
| 2003 UB313 | 3379.56492 | 19.694 | 0.045 | 3379.56473 | -0.170 | 0.581 | 96.937 | 96.918 | B |
| 2003 UB313 | 3379.57112 | 18.171 | 0.037 | 3379.57093 | -1.693 | 0.581 | 96.937 | 96.918 | R |
| 2003 UB313 | 3379.57393 | 17.972 | 0.027 | 3379.57374 | -1.892 | 0.581 | 96.937 | 96.918 | I |
| 2003 UB313 | 3380.57359 | 18.652 | 0.031 | 3380.57330 | -1.213 | 0.581 | 96.937 | 96.934 | V |
| 2003 UB313 | 3380.57635 | 18.342 | 0.033 | 3380.57606 | -1.523 | 0.581 | 96.937 | 96.934 | R |
| 2003 UB313 | 3380.57917 | 17.998 | 0.026 | 3380.57888 | -1.867 | 0.581 | 96.937 | 96.935 | I |
| 2003 UB313 | 3380.58267 | 19.522 | 0.046 | 3380.58238 | -0.343 | 0.581 | 96.937 | 96.935 | B |
| 2003 UB313 | 3380.58613 | 18.748 | 0.052 | 3380.58584 | -1.117 | 0.581 | 96.937 | 96.935 | V |
| 2003 UB313 | 3380.58889 | 18.362 | 0.018 | 3380.58860 | -1.503 | 0.581 | 96.937 | 96.935 | R |
| 2003 UB313 | 3380.59171 | 18.048 | 0.028 | 3380.59142 | -1.817 | 0.581 | 96.937 | 96.935 | I |
| 2003 UB313 | 3562.91503 | 19.538 | 0.022 | 3562.91471 | -0.326 | 0.601 | 96.912 | 96.939 | B |
| 2003 UB313 | 3562.91843 | 18.825 | 0.021 | 3562.91811 | -1.039 | 0.601 | 96.912 | 96.939 | V |
| 2003 UB313 | 3562.92381 | 18.111 | 0.024 | 3562.92349 | -1.753 | 0.601 | 96.912 | 96.939 | I |
| 2003 UB313 | 3563.87390 | 19.503 | 0.029 | 3563.87367 | -0.361 | 0.601 | 96.912 | 96.923 | B |
| 2003 UB313 | 3563.87730 | 18.805 | 0.037 | 3563.87707 | -1.059 | 0.601 | 96.912 | 96.923 | V |
| 2003 UB313 | 3563.87997 | 18.845 | 0.028 | 3563.87974 | -1.019 | 0.601 | 96.912 | 96.923 | V |
| 2003 UB313 | 3563.88269 | 18.121 | 0.035 | 3563.88246 | -1.743 | 0.601 | 96.912 | 96.923 | I |
| 2003 UB313 | 3568.87152 | 19.514 | 0.020 | 3568.87177 | -0.348 | 0.600 | 96.911 | 96.841 | B |
| 2003 UB313 | 3568.87493 | 18.773 | 0.023 | 3568.87518 | -1.089 | 0.600 | 96.911 | 96.841 | V |
| 2003 UB313 | 3568.87759 | 18.778 | 0.023 | 3568.87784 | -1.084 | 0.600 | 96.911 | 96.841 | V |
| 2003 UB313 | 3568.88032 | 18.102 | 0.029 | 3568.88057 | -1.760 | 0.600 | 96.911 | 96.841 | I |
| 2003 UB313 | 3570.89190 | 19.503 | 0.063 | 3570.89234 | -0.358 | 0.599 | 96.911 | 96.808 | B |
| 2003 UB313 | 3570.89530 | 18.823 | 0.029 | 3570.89574 | -1.038 | 0.599 | 96.911 | 96.808 | V |
| 2003 UB313 | 3570.89796 | 18.764 | 0.027 | 3570.89840 | -1.097 | 0.599 | 96.911 | 96.808 | V |
| 2003 UB313 | 3570.90069 | 18.024 | 0.029 | 3570.90113 | -1.837 | 0.599 | 96.911 | 96.808 | I |
| 2003 UB313 | 3573.80593 | 19.577 | 0.095 | 3573.80665 | -0.283 | 0.596 | 96.911 | 96.760 | B |
| 2003 UB313 | 3573.80934 | 18.781 | 0.054 | 3573.81006 | -1.079 | 0.596 | 96.911 | 96.760 | V |
| 2003 UB313 | 3573.81201 | 18.795 | 0.054 | 3573.81273 | -1.065 | 0.596 | 96.911 | 96.760 | V |
| 2003 UB313 | 3573.81474 | 18.106 | 0.041 | 3573.81546 | -1.754 | 0.596 | 96.911 | 96.760 | I |
| 2003 UB313 | 3575.80834 | 19.503 | 0.070 | 3575.80924 | -0.357 | 0.593 | 96.910 | 96.727 | B |
| 2003 UB313 | 3575.81174 | 18.766 | 0.054 | 3575.81264 | -1.094 | 0.593 | 96.910 | 96.727 | V |
| 2003 UB313 | 3575.81440 | 18.846 | 0.059 | 3575.81530 | -1.014 | 0.592 | 96.910 | 96.727 | V |
| 2003 UB313 | 3575.81712 | 18.113 | 0.048 | 3575.81803 | -1.747 | 0.592 | 96.910 | 96.727 | I |
| 2003 UB313 | 3575.87227 | 18.890 | 0.086 | 3575.87318 | -0.970 | 0.592 | 96.910 | 96.726 | V |
| 2003 UB313 | 3575.87227 | 18.867 | 0.065 | 3575.87318 | -0.993 | 0.592 | 96.910 | 96.726 | V |
| 2003 UB313 | 3575.87297 | 19.525 | 0.059 | 3575.87388 | -0.335 | 0.592 | 96.910 | 96.726 | B |
| 2003 UB313 | 3577.80743 | 19.508 | 0.061 | 3577.80852 | -0.351 | 0.589 | 96.910 | 96.695 | B |
| 2003 UB313 | 3577.81083 | 18.781 | 0.053 | 3577.81192 | -1.078 | 0.589 | 96.910 | 96.695 | V |
| 2003 UB313 | 3577.81349 | 18.879 | 0.056 | 3577.81458 | -0.980 | 0.589 | 96.910 | 96.695 | V |
| 2003 UB313 | 3577.81622 | 18.015 | 0.054 | 3577.81731 | -1.844 | 0.589 | 96.910 | 96.695 | I |
| 2003 UB313 | 3583.81072 | 19.494 | 0.047 | 3583.81236 | -0.363 | 0.574 | 96.909 | 96.600 | B |
| 2003 UB313 | 3583.81412 | 18.838 | 0.104 | 3583.81576 | -1.019 | 0.574 | 96.909 | 96.599 | V |
| 2003 UB313 | 3583.81678 | 18.726 | 0.059 | 3583.81842 | -1.131 | 0.574 | 96.909 | 96.599 | V |
| 2003 UB313 | 3585.85930 | 19.512 | 0.018 | 3585.86113 | -0.344 | 0.568 | 96.909 | 96.568 | B |
| 2003 UB313 | 3585.86270 | 18.770 | 0.019 | 3585.86453 | -1.086 | 0.568 | 96.909 | 96.568 | V |
| 2003 UB313 | 3585.86536 | 18.733 | 0.020 | 3585.86719 | -1.123 | 0.568 | 96.909 | 96.568 | V |
| 2003 UB313 | 3585.86809 | 18.019 | 0.021 | 3585.86992 | -1.837 | 0.568 | 96.909 | 96.567 | I |
| 2003 UB313 | 3586.80778 | 19.502 | 0.020 | 3586.80969 | -0.354 | 0.565 | 96.909 | 96.553 | B |
| 2003 UB313 | 3586.81118 | 18.780 | 0.020 | 3586.81309 | -1.076 | 0.565 | 96.909 | 96.553 | V |
| 2003 UB313 | 3586.81385 | 18.792 | 0.021 | 3586.81576 | -1.064 | 0.565 | 96.909 | 96.553 | V |
| 2003 UB313 | 3586.81657 | 18.004 | 0.023 | 3586.81848 | -1.852 | 0.565 | 96.909 | 96.553 | I |
| 2003 UB313 | 3587.84987 | 19.468 | 0.043 | 3587.85187 | -0.387 | 0.561 | 96.909 | 96.537 | B |
| 2003 UB313 | 3587.85327 | 18.806 | 0.043 | 3587.85527 | -1.049 | 0.561 | 96.909 | 96.537 | V |
| 2003 UB313 | 3587.85593 | 18.761 | 0.046 | 3587.85793 | -1.094 | 0.561 | 96.909 | 96.537 | V |
| 2003 UB313 | 3587.85866 | 18.017 | 0.054 | 3587.86067 | -1.838 | 0.561 | 96.909 | 96.537 | I |
| 2003 UB313 | 3588.83796 | 19.495 | 0.035 | 3588.84005 | -0.360 | 0.558 | 96.908 | 96.522 | B |
| 2003 UB313 | 3588.84136 | 18.727 | 0.024 | 3588.84345 | -1.128 | 0.558 | 96.908 | 96.522 | V |
| 2003 UB313 | 3588.84402 | 18.793 | 0.022 | 3588.84611 | -1.062 | 0.558 | 96.908 | 96.522 | V |
| 2003 UB313 | 3588.84674 | 18.015 | 0.045 | 3588.84883 | -1.840 | 0.558 | 96.908 | 96.522 | I |
| 2003 UB313 | 3590.85949 | 19.523 | 0.016 | 3590.86176 | -0.331 | 0.550 | 96.908 | 96.491 | B |
| 2003 UB313 | 3590.86290 | 18.799 | 0.017 | 3590.86517 | -1.055 | 0.550 | 96.908 | 96.491 | V |
| 2003 UB313 | 3590.86556 | 18.767 | 0.018 | 3590.86783 | -1.087 | 0.550 | 96.908 | 96.491 | V |
| 2003 UB313 | 3590.86829 | 18.013 | 0.021 | 3590.87056 | -1.841 | 0.550 | 96.908 | 96.491 | I |
| 2003 UB313 | 3591.80832 | 19.495 | 0.019 | 3591.81067 | -0.359 | 0.546 | 96.908 | 96.477 | B |
| 2003 UB313 | 3591.81172 | 18.789 | 0.020 | 3591.81407 | -1.065 | 0.546 | 96.908 | 96.477 | V |
| 2003 UB313 | 3591.81438 | 18.773 | 0.019 | 3591.81673 | -1.081 | 0.546 | 96.908 | 96.477 | V |
| 2003 UB313 | 3591.81710 | 18.001 | 0.022 | 3591.81945 | -1.853 | 0.546 | 96.908 | 96.477 | I |
| 2003 UB313 | 3592.87528 | 19.555 | 0.019 | 3592.87772 | -0.299 | 0.542 | 96.908 | 96.461 | B |
| 2003 UB313 | 3592.87868 | 18.853 | 0.020 | 3592.88112 | -1.001 | 0.542 | 96.908 | 96.461 | V |
| 2003 UB313 | 3592.88134 | 18.823 | 0.020 | 3592.88378 | -1.031 | 0.542 | 96.908 | 96.461 | V |
| 2003 UB313 | 3592.88407 | 18.027 | 0.024 | 3592.88651 | -1.827 | 0.542 | 96.908 | 96.461 | I |
| 2003 UB313 | 3594.79250 | 19.400 | 0.108 | 3594.79510 | -0.453 | 0.534 | 96.908 | 96.434 | B |
| 2003 UB313 | 3594.79591 | 18.756 | 0.129 | 3594.79851 | -1.097 | 0.534 | 96.908 | 96.433 | V |
| 2003 UB313 | 3594.79857 | 18.886 | 0.203 | 3594.80117 | -0.967 | 0.534 | 96.908 | 96.433 | V |
| 2003 UB313 | 3594.80132 | 18.091 | 0.179 | 3594.80392 | -1.762 | 0.534 | 96.908 | 96.433 | I |
| 2003 UB313 | 3596.81059 | 19.489 | 0.111 | 3596.81336 | -0.363 | 0.524 | 96.907 | 96.405 | B |
| 2003 UB313 | 3596.81454 | 18.779 | 0.130 | 3596.81731 | -1.073 | 0.524 | 96.907 | 96.405 | V |
| 2003 UB313 | 3596.81454 | 18.725 | 0.163 | 3596.81731 | -1.127 | 0.524 | 96.907 | 96.405 | V |
| 2003 UB313 | 3597.84243 | 19.430 | 0.099 | 3597.84528 | -0.422 | 0.519 | 96.907 | 96.390 | B |
| 2003 UB313 | 3600.82293 | 19.595 | 0.053 | 3600.82602 | -0.256 | 0.504 | 96.907 | 96.349 | B |
| 2003 UB313 | 3600.82632 | 18.818 | 0.043 | 3600.82941 | -1.033 | 0.504 | 96.907 | 96.349 | V |
| 2003 UB313 | 3600.82898 | 18.840 | 0.049 | 3600.83207 | -1.011 | 0.504 | 96.907 | 96.349 | V |
| 2003 UB313 | 3600.83171 | 17.963 | 0.044 | 3600.83480 | -1.888 | 0.504 | 96.907 | 96.349 | I |
| 2003 UB313 | 3603.86600 | 19.453 | 0.051 | 3603.86932 | -0.397 | 0.487 | 96.906 | 96.308 | B |
| 2003 UB313 | 3603.86940 | 18.772 | 0.045 | 3603.87272 | -1.078 | 0.487 | 96.906 | 96.308 | V |
| 2003 UB313 | 3603.87207 | 18.755 | 0.042 | 3603.87539 | -1.095 | 0.487 | 96.906 | 96.308 | V |
| 2003 UB313 | 3603.87479 | 17.962 | 0.029 | 3603.87812 | -1.888 | 0.487 | 96.906 | 96.308 | I |
| 2003 UB313 | 3606.82469 | 19.483 | 0.078 | 3606.82823 | -0.366 | 0.470 | 96.906 | 96.270 | B |
| 2003 UB313 | 3606.82810 | 18.696 | 0.061 | 3606.83164 | -1.153 | 0.470 | 96.906 | 96.270 | V |
| 2003 UB313 | 3606.83076 | 18.760 | 0.062 | 3606.83430 | -1.089 | 0.470 | 96.906 | 96.270 | V |
| 2003 UB313 | 3606.83348 | 17.969 | 0.061 | 3606.83702 | -1.880 | 0.470 | 96.906 | 96.270 | I |
| 2003 UB313 | 3608.80188 | 19.580 | 0.033 | 3608.80556 | -0.269 | 0.458 | 96.906 | 96.246 | B |
| 2003 UB313 | 3608.80529 | 18.772 | 0.039 | 3608.80898 | -1.077 | 0.458 | 96.906 | 96.246 | V |
| 2003 UB313 | 3608.80795 | 18.809 | 0.026 | 3608.81164 | -1.040 | 0.458 | 96.906 | 96.246 | V |
| 2003 UB313 | 3608.81068 | 18.037 | 0.030 | 3608.81437 | -1.812 | 0.458 | 96.906 | 96.246 | I |
| 2003 UB313 | 3619.75138 | 19.573 | 0.026 | 3619.75577 | -0.273 | 0.383 | 96.904 | 96.125 | B |
| 2003 UB313 | 3619.75479 | 18.802 | 0.033 | 3619.75918 | -1.044 | 0.383 | 96.904 | 96.125 | V |
| 2003 UB313 | 3619.75745 | 18.776 | 0.039 | 3619.76184 | -1.070 | 0.383 | 96.904 | 96.125 | V |
| 2003 UB313 | 3619.76018 | 17.912 | 0.029 | 3619.76457 | -1.934 | 0.383 | 96.904 | 96.124 | I |
| 2003 UB313 | 3621.72881 | 19.561 | 0.020 | 3621.73331 | -0.284 | 0.369 | 96.904 | 96.105 | B |
| 2003 UB313 | 3621.73221 | 18.797 | 0.020 | 3621.73671 | -1.048 | 0.369 | 96.904 | 96.105 | V |
| 2003 UB313 | 3621.73487 | 18.766 | 0.021 | 3621.73937 | -1.079 | 0.369 | 96.904 | 96.105 | V |
| 2003 UB313 | 3621.73760 | 17.987 | 0.025 | 3621.74210 | -1.858 | 0.369 | 96.904 | 96.105 | I |
| 2003 UB313 | 3624.70865 | 19.554 | 0.075 | 3624.71330 | -0.291 | 0.347 | 96.904 | 96.078 | B |
| 2003 UB313 | 3624.71205 | 18.743 | 0.055 | 3624.71670 | -1.102 | 0.347 | 96.904 | 96.078 | V |
| 2003 UB313 | 3624.71471 | 18.891 | 0.077 | 3624.71936 | -0.954 | 0.347 | 96.904 | 96.078 | V |
| 2003 UB313 | 3624.71744 | 17.957 | 0.056 | 3624.72209 | -1.888 | 0.347 | 96.904 | 96.078 | I |
| 2003 UB313 | 3626.70852 | 19.565 | 0.250 | 3626.71327 | -0.279 | 0.331 | 96.903 | 96.061 | B |
| 2003 UB313 | 3626.71192 | 18.607 | 0.133 | 3626.71667 | -1.237 | 0.331 | 96.903 | 96.061 | V |
| 2003 UB313 | 3635.69096 | 19.493 | 0.087 | 3635.69608 | -0.350 | 0.262 | 96.902 | 95.997 | B |
| 2003 UB313 | 3635.69437 | 18.750 | 0.094 | 3635.69949 | -1.093 | 0.262 | 96.902 | 95.997 | V |
| 2003 UB313 | 3635.69704 | 18.726 | 0.090 | 3635.70216 | -1.117 | 0.262 | 96.902 | 95.997 | V |
| 2003 UB313 | 3635.69977 | 17.955 | 0.129 | 3635.70489 | -1.888 | 0.262 | 96.902 | 95.997 | I |
| 2003 UB313 | 3638.68977 | 19.523 | 0.020 | 3638.69499 | -0.320 | 0.240 | 96.902 | 95.981 | B |
| 2003 UB313 | 3638.69316 | 18.787 | 0.021 | 3638.69838 | -1.056 | 0.240 | 96.902 | 95.981 | V |
| 2003 UB313 | 3638.69583 | 18.744 | 0.021 | 3638.70105 | -1.099 | 0.240 | 96.902 | 95.981 | V |
| 2003 UB313 | 3638.69856 | 17.967 | 0.025 | 3638.70378 | -1.876 | 0.240 | 96.902 | 95.981 | I |
| 2003 UB313 | 3640.69036 | 18.777 | 0.020 | 3640.69563 | -1.065 | 0.225 | 96.901 | 95.971 | V |
| 2003 UB313 | 3640.69036 | 18.742 | 0.021 | 3640.69563 | -1.100 | 0.225 | 96.901 | 95.971 | V |
| 2003 UB313 | 3640.69036 | 18.001 | 0.021 | 3640.69563 | -1.841 | 0.225 | 96.901 | 95.971 | I |
| 2003 UB313 | 3640.69106 | 19.528 | 0.019 | 3640.69633 | -0.314 | 0.225 | 96.901 | 95.971 | B |
| 2003 UB313 | 3644.76809 | 19.508 | 0.017 | 3644.77346 | -0.334 | 0.198 | 96.901 | 95.955 | B |
| 2003 UB313 | 3644.77150 | 18.707 | 0.019 | 3644.77687 | -1.135 | 0.198 | 96.901 | 95.955 | V |
| 2003 UB313 | 3644.77416 | 18.724 | 0.019 | 3644.77953 | -1.118 | 0.198 | 96.901 | 95.955 | V |
| 2003 UB313 | 3644.77688 | 17.955 | 0.026 | 3644.78225 | -1.887 | 0.198 | 96.901 | 95.955 | I |
| 2003 UB313 | 3646.72410 | 19.446 | 0.053 | 3646.72950 | -0.396 | 0.186 | 96.901 | 95.948 | B |
| 2003 UB313 | 3646.72751 | 18.733 | 0.053 | 3646.73291 | -1.109 | 0.186 | 96.901 | 95.948 | V |
| 2003 UB313 | 3646.73017 | 18.768 | 0.042 | 3646.73557 | -1.074 | 0.186 | 96.901 | 95.948 | V |
| 2003 UB313 | 3646.73290 | 17.940 | 0.042 | 3646.73830 | -1.902 | 0.186 | 96.901 | 95.948 | I |
| 2003 UB313 | 3650.66963 | 19.451 | 0.023 | 3650.67509 | -0.391 | 0.165 | 96.900 | 95.939 | B |
| 2003 UB313 | 3650.67304 | 18.741 | 0.021 | 3650.67850 | -1.101 | 0.165 | 96.900 | 95.939 | V |
| 2003 UB313 | 3650.67570 | 18.705 | 0.021 | 3650.68116 | -1.137 | 0.165 | 96.900 | 95.939 | V |
| 2003 UB313 | 3650.67843 | 17.960 | 0.026 | 3650.68389 | -1.882 | 0.165 | 96.900 | 95.939 | I |
| 2003 UB313 | 3652.65040 | 19.520 | 0.020 | 3652.65587 | -0.322 | 0.158 | 96.900 | 95.936 | B |
| 2003 UB313 | 3652.65381 | 18.721 | 0.020 | 3652.65928 | -1.121 | 0.158 | 96.900 | 95.936 | V |
| 2003 UB313 | 3652.65647 | 18.771 | 0.021 | 3652.66194 | -1.071 | 0.158 | 96.900 | 95.936 | V |
| 2003 UB313 | 3652.65920 | 18.071 | 0.023 | 3652.66467 | -1.771 | 0.158 | 96.900 | 95.936 | I |
| 2003 UB313 | 3664.64581 | 19.391 | 0.026 | 3664.65125 | -0.451 | 0.163 | 96.898 | 95.943 | B |
| 2003 UB313 | 3664.65150 | 18.626 | 0.024 | 3664.65694 | -1.216 | 0.163 | 96.898 | 95.943 | V |
| 2003 UB313 | 3666.56716 | 19.473 | 0.021 | 3666.57257 | -0.369 | 0.172 | 96.898 | 95.947 | B |
| 2003 UB313 | 3666.57056 | 18.681 | 0.022 | 3666.57597 | -1.161 | 0.172 | 96.898 | 95.948 | V |
| 2003 UB313 | 3666.57323 | 18.670 | 0.022 | 3666.57864 | -1.172 | 0.172 | 96.898 | 95.948 | V |
| 2003 UB313 | 3669.66987 | 19.363 | 0.055 | 3669.67522 | -0.479 | 0.189 | 96.897 | 95.958 | B |
| 2003 UB313 | 3669.67328 | 18.783 | 0.029 | 3669.67863 | -1.059 | 0.189 | 96.897 | 95.958 | V |
| 2003 UB313 | 3669.67593 | 18.781 | 0.031 | 3669.68128 | -1.061 | 0.189 | 96.897 | 95.958 | V |
| 2003 UB313 | 3669.67866 | 18.084 | 0.047 | 3669.68401 | -1.758 | 0.189 | 96.897 | 95.958 | I |
| 2003 UB313 | 3672.62461 | 19.477 | 0.025 | 3672.62989 | -0.365 | 0.207 | 96.897 | 95.970 | B |
| 2003 UB313 | 3672.62801 | 18.793 | 0.030 | 3672.63329 | -1.049 | 0.207 | 96.897 | 95.970 | V |
| 2003 UB313 | 3672.63068 | 18.781 | 0.026 | 3672.63596 | -1.061 | 0.207 | 96.897 | 95.970 | V |
| 2003 UB313 | 3672.63341 | 17.937 | 0.029 | 3672.63869 | -1.905 | 0.208 | 96.897 | 95.970 | I |
| 2003 UB313 | 3674.60098 | 19.531 | 0.021 | 3674.60620 | -0.311 | 0.221 | 96.897 | 95.979 | B |
| 2003 UB313 | 3674.60437 | 18.752 | 0.018 | 3674.60959 | -1.090 | 0.221 | 96.897 | 95.979 | V |
| 2003 UB313 | 3674.60703 | 18.761 | 0.026 | 3674.61225 | -1.081 | 0.221 | 96.897 | 95.979 | V |
| 2003 UB313 | 3674.60975 | 17.971 | 0.111 | 3674.61497 | -1.871 | 0.221 | 96.897 | 95.979 | I |
| 2003 UB313 | 3676.59101 | 19.524 | 0.025 | 3676.59617 | -0.319 | 0.235 | 96.897 | 95.990 | B |
| 2003 UB313 | 3676.59441 | 18.775 | 0.023 | 3676.59957 | -1.068 | 0.235 | 96.897 | 95.990 | V |
| 2003 UB313 | 3676.59706 | 18.782 | 0.025 | 3676.60222 | -1.061 | 0.235 | 96.897 | 95.990 | V |
| 2003 UB313 | 3676.59978 | 17.968 | 0.030 | 3676.60494 | -1.875 | 0.235 | 96.897 | 95.990 | I |
| 2003 UB313 | 3678.62086 | 19.507 | 0.017 | 3678.62595 | -0.336 | 0.250 | 96.896 | 96.002 | B |
| 2003 UB313 | 3678.62425 | 18.753 | 0.019 | 3678.62934 | -1.090 | 0.250 | 96.896 | 96.002 | V |
| 2003 UB313 | 3678.62691 | 18.732 | 0.024 | 3678.63200 | -1.111 | 0.250 | 96.896 | 96.002 | V |
| 2003 UB313 | 3678.62964 | 17.989 | 0.024 | 3678.63473 | -1.854 | 0.250 | 96.896 | 96.002 | I |
| 2003 UB313 | 3681.61878 | 19.532 | 0.026 | 3681.62376 | -0.311 | 0.273 | 96.896 | 96.022 | B |
| 2003 UB313 | 3681.62216 | 18.780 | 0.035 | 3681.62714 | -1.063 | 0.273 | 96.896 | 96.022 | V |
| 2003 UB313 | 3681.62483 | 18.755 | 0.030 | 3681.62981 | -1.088 | 0.273 | 96.896 | 96.022 | V |
| 2003 UB313 | 3681.62755 | 17.944 | 0.039 | 3681.63253 | -1.899 | 0.273 | 96.896 | 96.022 | I |
| 2003 UB313 | 3686.55509 | 19.566 | 0.053 | 3686.55985 | -0.278 | 0.311 | 96.895 | 96.060 | B |
| 2003 UB313 | 3686.55848 | 18.760 | 0.040 | 3686.56324 | -1.084 | 0.311 | 96.895 | 96.060 | V |
| 2003 UB313 | 3686.56114 | 18.760 | 0.050 | 3686.56590 | -1.084 | 0.311 | 96.895 | 96.060 | V |
| 2003 UB313 | 3686.56386 | 18.030 | 0.040 | 3686.56862 | -1.814 | 0.311 | 96.895 | 96.060 | I |
| 2003 UB313 | 3688.58497 | 19.517 | 0.082 | 3688.58963 | -0.328 | 0.326 | 96.895 | 96.078 | B |
| 2003 UB313 | 3688.58838 | 18.887 | 0.076 | 3688.59304 | -0.958 | 0.326 | 96.895 | 96.078 | V |
| 2003 UB313 | 3688.59107 | 18.745 | 0.081 | 3688.59573 | -1.100 | 0.326 | 96.895 | 96.078 | V |
| 2003 UB313 | 3688.59381 | 18.118 | 0.078 | 3688.59847 | -1.727 | 0.326 | 96.895 | 96.078 | I |
| 2003 UB313 | 3690.63730 | 19.535 | 0.066 | 3690.64185 | -0.310 | 0.342 | 96.895 | 96.096 | B |
| 2003 UB313 | 3690.64070 | 18.741 | 0.048 | 3690.64525 | -1.104 | 0.342 | 96.895 | 96.096 | V |
| 2003 UB313 | 3690.64336 | 18.670 | 0.045 | 3690.64791 | -1.175 | 0.342 | 96.895 | 96.096 | V |
| 2003 UB313 | 3690.64609 | 18.050 | 0.038 | 3690.65064 | -1.795 | 0.342 | 96.895 | 96.096 | I |
| 2003 UB313 | 3695.57196 | 19.504 | 0.022 | 3695.57623 | -0.342 | 0.378 | 96.894 | 96.145 | B |
| 2003 UB313 | 3695.57536 | 18.808 | 0.019 | 3695.57963 | -1.038 | 0.378 | 96.894 | 96.145 | V |
| 2003 UB313 | 3695.57802 | 18.761 | 0.022 | 3695.58229 | -1.085 | 0.378 | 96.894 | 96.145 | V |
| 2003 UB313 | 3695.58075 | 17.965 | 0.024 | 3695.58502 | -1.881 | 0.378 | 96.894 | 96.145 | I |
| 2003 UB313 | 3699.59219 | 19.575 | 0.017 | 3699.59620 | -0.272 | 0.406 | 96.893 | 96.189 | B |
| 2003 UB313 | 3699.59559 | 18.784 | 0.018 | 3699.59960 | -1.063 | 0.406 | 96.893 | 96.190 | V |
| 2003 UB313 | 3699.59825 | 18.761 | 0.018 | 3699.60226 | -1.086 | 0.406 | 96.893 | 96.190 | V |
| 2003 UB313 | 3699.60097 | 17.976 | 0.021 | 3699.60498 | -1.871 | 0.406 | 96.893 | 96.190 | I |
| 2003 UB313 | 3701.58713 | 19.563 | 0.020 | 3701.59101 | -0.285 | 0.420 | 96.893 | 96.213 | B |
| 2003 UB313 | 3701.59052 | 18.786 | 0.020 | 3701.59440 | -1.062 | 0.420 | 96.893 | 96.213 | V |
| 2003 UB313 | 3701.59318 | 18.798 | 0.020 | 3701.59706 | -1.050 | 0.420 | 96.893 | 96.213 | V |
| 2003 UB313 | 3701.59589 | 18.014 | 0.027 | 3701.59977 | -1.834 | 0.420 | 96.893 | 96.213 | I |
| 2003 UB313 | 3703.61274 | 19.553 | 0.017 | 3703.61648 | -0.295 | 0.433 | 96.893 | 96.237 | B |
| 2003 UB313 | 3703.61614 | 18.818 | 0.018 | 3703.61988 | -1.030 | 0.433 | 96.893 | 96.237 | V |
| 2003 UB313 | 3703.61883 | 18.798 | 0.018 | 3703.62257 | -1.050 | 0.433 | 96.893 | 96.237 | V |
| 2003 UB313 | 3705.60398 | 19.581 | 0.026 | 3705.60757 | -0.268 | 0.446 | 96.893 | 96.262 | B |
| 2003 UB313 | 3705.60738 | 18.820 | 0.026 | 3705.61097 | -1.029 | 0.446 | 96.893 | 96.262 | V |
| 2003 UB313 | 3705.61006 | 18.808 | 0.021 | 3705.61365 | -1.041 | 0.446 | 96.893 | 96.262 | V |
| 2003 UB313 | 3705.61281 | 17.962 | 0.045 | 3705.61640 | -1.887 | 0.446 | 96.893 | 96.262 | I |
| 2003 UB313 | 3708.56808 | 19.530 | 0.022 | 3708.57145 | -0.320 | 0.464 | 96.892 | 96.300 | B |
| 2003 UB313 | 3708.57151 | 18.817 | 0.021 | 3708.57488 | -1.033 | 0.464 | 96.892 | 96.300 | V |
| 2003 UB313 | 3708.57419 | 18.810 | 0.021 | 3708.57756 | -1.040 | 0.464 | 96.892 | 96.301 | V |
| 2003 UB313 | 3708.57694 | 18.094 | 0.023 | 3708.58031 | -1.756 | 0.464 | 96.892 | 96.301 | I |
| 2003 UB313 | 3710.58299 | 19.523 | 0.019 | 3710.58620 | -0.327 | 0.475 | 96.892 | 96.328 | B |
| 2003 UB313 | 3710.58640 | 18.868 | 0.020 | 3710.58961 | -0.982 | 0.475 | 96.892 | 96.328 | V |
| 2003 UB313 | 3710.58909 | 18.790 | 0.019 | 3710.59230 | -1.060 | 0.475 | 96.892 | 96.328 | V |
| 2003 UB313 | 3710.59184 | 18.085 | 0.022 | 3710.59505 | -1.765 | 0.476 | 96.892 | 96.328 | I |
| 2003 UB313 | 3712.56924 | 19.534 | 0.027 | 3712.57230 | -0.317 | 0.486 | 96.892 | 96.355 | B |
| 2003 UB313 | 3712.57265 | 18.812 | 0.021 | 3712.57571 | -1.039 | 0.486 | 96.892 | 96.355 | V |
| 2003 UB313 | 3712.57534 | 18.785 | 0.020 | 3712.57840 | -1.066 | 0.486 | 96.892 | 96.355 | V |
| 2003 UB313 | 3712.57809 | 18.029 | 0.022 | 3712.58115 | -1.822 | 0.486 | 96.892 | 96.355 | I |
| 2003 UB313 | 3714.56097 | 19.485 | 0.042 | 3714.56386 | -0.366 | 0.497 | 96.891 | 96.383 | B |
| 2003 UB313 | 3714.56438 | 18.777 | 0.039 | 3714.56727 | -1.074 | 0.497 | 96.891 | 96.383 | V |
| 2003 UB313 | 3714.56707 | 18.783 | 0.038 | 3714.56996 | -1.068 | 0.497 | 96.891 | 96.383 | V |
| 2003 UB313 | 3714.56982 | 17.947 | 0.037 | 3714.57271 | -1.904 | 0.497 | 96.891 | 96.383 | I |
| 2003 UB313 | 3716.56756 | 19.607 | 0.084 | 3716.57029 | -0.245 | 0.507 | 96.891 | 96.412 | B |
| 2003 UB313 | 3716.57097 | 18.886 | 0.072 | 3716.57370 | -0.966 | 0.507 | 96.891 | 96.412 | V |
| 2003 UB313 | 3716.57366 | 18.867 | 0.075 | 3716.57639 | -0.985 | 0.507 | 96.891 | 96.412 | V |
| 2003 UB313 | 3716.57641 | 18.036 | 0.064 | 3716.57914 | -1.816 | 0.507 | 96.891 | 96.412 | I |
| 2003 UB313 | 3718.58219 | 19.554 | 0.060 | 3718.58475 | -0.299 | 0.516 | 96.891 | 96.441 | B |
| 2003 UB313 | 3718.58564 | 18.791 | 0.045 | 3718.58820 | -1.062 | 0.516 | 96.891 | 96.442 | V |
| 2003 UB313 | 3718.58836 | 18.794 | 0.049 | 3718.59092 | -1.059 | 0.516 | 96.891 | 96.442 | V |
| 2003 UB313 | 3718.59114 | 18.051 | 0.032 | 3718.59370 | -1.802 | 0.516 | 96.891 | 96.442 | I |
| 2003 UB313 | 3720.57853 | 19.496 | 0.045 | 3720.58091 | -0.357 | 0.525 | 96.891 | 96.471 | B |
| 2003 UB313 | 3720.58198 | 18.841 | 0.039 | 3720.58436 | -1.012 | 0.525 | 96.891 | 96.471 | V |
| 2003 UB313 | 3720.58470 | 18.823 | 0.040 | 3720.58708 | -1.030 | 0.525 | 96.891 | 96.471 | V |
| 2003 UB313 | 3720.58748 | 17.975 | 0.027 | 3720.58986 | -1.878 | 0.525 | 96.891 | 96.472 | I |
| 2003 UB313 | 3722.57932 | 19.540 | 0.029 | 3722.58153 | -0.314 | 0.533 | 96.890 | 96.502 | B |
| 2003 UB313 | 3722.58276 | 18.864 | 0.027 | 3722.58497 | -0.990 | 0.533 | 96.890 | 96.502 | V |
| 2003 UB313 | 3722.58548 | 18.840 | 0.031 | 3722.58769 | -1.014 | 0.533 | 96.890 | 96.502 | V |
| 2003 UB313 | 3722.58826 | 18.113 | 0.033 | 3722.59047 | -1.741 | 0.533 | 96.890 | 96.502 | I |
| 2003 UB313 | 3724.55837 | 19.533 | 0.214 | 3724.56040 | -0.322 | 0.541 | 96.890 | 96.533 | B |
| 2003 UB313 | 3724.56182 | 18.823 | 0.240 | 3724.56385 | -1.032 | 0.541 | 96.890 | 96.533 | V |
| 2003 UB313 | 3724.56454 | 18.807 | 0.243 | 3724.56657 | -1.048 | 0.541 | 96.890 | 96.533 | V |
| 2003 UB313 | 3724.56733 | 18.081 | 0.291 | 3724.56936 | -1.774 | 0.541 | 96.890 | 96.533 | I |
| 2003 UB313 | 3726.54889 | 19.521 | 0.023 | 3726.55074 | -0.334 | 0.548 | 96.890 | 96.564 | B |
| 2003 UB313 | 3726.55234 | 18.772 | 0.180 | 3726.55419 | -1.083 | 0.548 | 96.890 | 96.564 | V |
| 2003 UB313 | 3726.55506 | 18.810 | 0.186 | 3726.55691 | -1.045 | 0.548 | 96.890 | 96.564 | V |
| 2003 UB313 | 3726.55785 | 18.059 | 0.220 | 3726.55970 | -1.796 | 0.548 | 96.890 | 96.564 | I |
| 2003 UB313 | 3728.55294 | 19.487 | 0.020 | 3728.55460 | -0.369 | 0.554 | 96.889 | 96.596 | B |
| 2003 UB313 | 3728.55638 | 18.806 | 0.019 | 3728.55804 | -1.050 | 0.554 | 96.889 | 96.596 | V |
| 2003 UB313 | 3728.55910 | 18.804 | 0.021 | 3728.56076 | -1.052 | 0.554 | 96.889 | 96.596 | V |
| 2003 UB313 | 3728.56189 | 18.012 | 0.025 | 3728.56355 | -1.844 | 0.554 | 96.889 | 96.596 | I |
| 2003 UB313 | 3746.52683 | 19.346 | 0.105 | 3746.52678 | -0.517 | 0.581 | 96.887 | 96.893 | B |
| 2003 UB313 | 3746.53028 | 18.879 | 0.070 | 3746.53023 | -0.984 | 0.581 | 96.887 | 96.893 | V |
| 2003 UB313 | 3746.53300 | 18.769 | 0.057 | 3746.53295 | -1.094 | 0.581 | 96.887 | 96.893 | V |
| 2003 UB313 | 3746.53579 | 18.070 | 0.051 | 3746.53574 | -1.793 | 0.581 | 96.887 | 96.893 | I |
| 2003 UB313 | 3748.52496 | 19.515 | 0.226 | 3748.52471 | -0.349 | 0.581 | 96.887 | 96.926 | B |
| 2003 UB313 | 3748.52841 | 18.610 | 0.077 | 3748.52816 | -1.254 | 0.581 | 96.887 | 96.926 | V |
| 2003 UB313 | 3748.53391 | 17.961 | 0.057 | 3748.53366 | -1.903 | 0.581 | 96.887 | 96.927 | I |
| 2003 UB313 | 3750.53462 | 18.814 | 0.064 | 3750.53418 | -1.050 | 0.580 | 96.886 | 96.960 | V |
| 2003 UB313 | 3750.53733 | 18.894 | 0.055 | 3750.53689 | -0.970 | 0.580 | 96.886 | 96.960 | V |
| 2003 UB313 | 3752.52881 | 18.755 | 0.081 | 3752.52818 | -1.110 | 0.578 | 96.886 | 96.993 | V |
| 2003 UB313 | 3752.53153 | 18.794 | 0.060 | 3752.53090 | -1.071 | 0.578 | 96.886 | 96.993 | V |
| 2003 UB313 | 3752.53431 | 18.034 | 0.049 | 3752.53368 | -1.831 | 0.578 | 96.886 | 96.993 | I |
| 2003 UB313 | 3754.52600 | 18.790 | 0.138 | 3754.52518 | -1.076 | 0.576 | 96.886 | 97.027 | V |
| 2003 UB313 | 3754.52872 | 18.939 | 0.107 | 3754.52790 | -0.927 | 0.576 | 96.886 | 97.027 | V |
| 2003 UB313 | 3754.53150 | 17.997 | 0.083 | 3754.53068 | -1.869 | 0.576 | 96.886 | 97.027 | I |
| 2003 UB313 | 3756.52653 | 19.534 | 0.173 | 3756.52551 | -0.332 | 0.573 | 96.886 | 97.060 | B |
| 2003 UB313 | 3756.52998 | 18.928 | 0.063 | 3756.52896 | -0.938 | 0.573 | 96.886 | 97.060 | V |
| 2003 UB313 | 3756.53269 | 18.763 | 0.043 | 3756.53167 | -1.103 | 0.573 | 96.886 | 97.060 | V |
| 2003 UB313 | 3756.53548 | 18.038 | 0.041 | 3756.53446 | -1.828 | 0.573 | 96.886 | 97.060 | I |
| 2003 UB313 | 3758.52490 | 19.587 | 0.135 | 3758.52370 | -0.280 | 0.569 | 96.885 | 97.093 | B |
| 2003 UB313 | 3758.52835 | 18.703 | 0.052 | 3758.52715 | -1.164 | 0.569 | 96.885 | 97.093 | V |
| 2003 UB313 | 3758.53107 | 18.885 | 0.050 | 3758.52986 | -0.982 | 0.569 | 96.885 | 97.093 | V |
| 2003 UB313 | 3758.53386 | 18.037 | 0.046 | 3758.53266 | -1.830 | 0.569 | 96.885 | 97.093 | I |
| 2003 UB313 | 3760.52732 | 19.375 | 0.101 | 3760.52593 | -0.493 | 0.565 | 96.885 | 97.125 | B |
| 2003 UB313 | 3760.53078 | 18.845 | 0.048 | 3760.52939 | -1.023 | 0.565 | 96.885 | 97.125 | V |
| 2003 UB313 | 3760.53349 | 18.763 | 0.034 | 3760.53210 | -1.105 | 0.565 | 96.885 | 97.125 | V |
| 2003 UB313 | 3760.53628 | 18.013 | 0.035 | 3760.53489 | -1.855 | 0.565 | 96.885 | 97.125 | I |
| 2003 UB313 | 3762.52369 | 18.931 | 0.086 | 3762.52211 | -0.938 | 0.560 | 96.885 | 97.157 | V |
| 2003 UB313 | 3762.52641 | 18.793 | 0.055 | 3762.52483 | -1.076 | 0.560 | 96.885 | 97.158 | V |
| 2003 UB313 | 3762.53071 | 18.146 | 0.056 | 3762.52913 | -1.723 | 0.560 | 96.885 | 97.158 | I |
| | | | | # | | | | | |
| 2005 FY9 | 3465.66844 | 18.144 | 0.013 | 3465.66844 | 1.101 | 0.691 | 51.008 | 50.221 | B |
| 2005 FY9 | 3465.67148 | 17.293 | 0.015 | 3465.67148 | 0.250 | 0.691 | 51.008 | 50.221 | V |
| 2005 FY9 | 3465.67417 | 16.802 | 0.013 | 3465.67417 | -0.241 | 0.691 | 51.008 | 50.221 | R |
| 2005 FY9 | 3465.67552 | 16.458 | 0.014 | 3465.67552 | -0.585 | 0.691 | 51.008 | 50.221 | I |
| 2005 FY9 | 3465.67859 | 18.158 | 0.011 | 3465.67859 | 1.115 | 0.691 | 51.008 | 50.221 | B |
| 2005 FY9 | 3465.68162 | 17.265 | 0.016 | 3465.68162 | 0.222 | 0.691 | 51.008 | 50.221 | V |
| 2005 FY9 | 3465.68432 | 16.738 | 0.012 | 3465.68432 | -0.305 | 0.691 | 51.008 | 50.221 | R |
| 2005 FY9 | 3465.68566 | 16.449 | 0.014 | 3465.68566 | -0.594 | 0.691 | 51.008 | 50.221 | I |
| 2005 FY9 | 3466.67503 | 18.163 | 0.015 | 3466.67500 | 1.120 | 0.701 | 51.008 | 50.227 | B |
| 2005 FY9 | 3466.67807 | 17.296 | 0.020 | 3466.67804 | 0.253 | 0.701 | 51.008 | 50.227 | V |
| 2005 FY9 | 3466.68077 | 16.813 | 0.012 | 3466.68074 | -0.230 | 0.701 | 51.008 | 50.227 | R |
| 2005 FY9 | 3466.68212 | 16.510 | 0.017 | 3466.68209 | -0.533 | 0.701 | 51.008 | 50.227 | I |
| 2005 FY9 | 3466.68520 | 18.141 | 0.019 | 3466.68517 | 1.098 | 0.701 | 51.008 | 50.227 | B |
| 2005 FY9 | 3466.68823 | 17.357 | 0.017 | 3466.68820 | 0.314 | 0.701 | 51.008 | 50.227 | V |
| 2005 FY9 | 3466.69093 | 16.866 | 0.013 | 3466.69090 | -0.177 | 0.701 | 51.008 | 50.227 | R |
| 2005 FY9 | 3466.69228 | 16.475 | 0.016 | 3466.69225 | -0.568 | 0.701 | 51.008 | 50.227 | I |
| 2005 FY9 | 3467.68781 | 18.103 | 0.014 | 3467.68774 | 1.060 | 0.711 | 51.008 | 50.233 | B |
| 2005 FY9 | 3467.69084 | 17.258 | 0.021 | 3467.69077 | 0.215 | 0.711 | 51.008 | 50.233 | V |
| 2005 FY9 | 3467.69353 | 16.788 | 0.011 | 3467.69346 | -0.255 | 0.711 | 51.008 | 50.233 | R |
| 2005 FY9 | 3467.69488 | 16.359 | 0.018 | 3467.69481 | -0.684 | 0.711 | 51.008 | 50.233 | I |
| 2005 FY9 | 3467.69796 | 18.076 | 0.013 | 3467.69789 | 1.033 | 0.711 | 51.008 | 50.233 | B |
| 2005 FY9 | 3467.70101 | 17.259 | 0.018 | 3467.70094 | 0.216 | 0.711 | 51.008 | 50.233 | V |
| 2005 FY9 | 3467.70370 | 16.802 | 0.011 | 3467.70363 | -0.241 | 0.711 | 51.008 | 50.233 | R |
| 2005 FY9 | 3467.70505 | 16.442 | 0.022 | 3467.70498 | -0.601 | 0.711 | 51.008 | 50.233 | I |
| 2005 FY9 | 3489.58981 | 18.123 | 0.030 | 3489.58860 | 1.071 | 0.926 | 51.008 | 50.430 | B |
| 2005 FY9 | 3489.59285 | 17.254 | 0.022 | 3489.59164 | 0.202 | 0.926 | 51.008 | 50.430 | V |
| 2005 FY9 | 3489.59488 | 16.479 | 0.019 | 3489.59367 | -0.573 | 0.926 | 51.008 | 50.430 | I |
| 2005 FY9 | 3489.59684 | 16.482 | 0.017 | 3489.59563 | -0.570 | 0.926 | 51.008 | 50.430 | I |
| 2005 FY9 | 3490.60990 | 18.072 | 0.011 | 3490.60863 | 1.020 | 0.936 | 51.008 | 50.441 | B |
| 2005 FY9 | 3490.61295 | 17.264 | 0.012 | 3490.61168 | 0.212 | 0.936 | 51.008 | 50.441 | V |
| 2005 FY9 | 3490.61499 | 16.446 | 0.013 | 3490.61372 | -0.606 | 0.936 | 51.008 | 50.441 | I |
| 2005 FY9 | 3490.61695 | 16.451 | 0.012 | 3490.61568 | -0.601 | 0.936 | 51.008 | 50.441 | I |
| 2005 FY9 | 3491.61327 | 18.085 | 0.011 | 3491.61193 | 1.032 | 0.944 | 51.008 | 50.452 | B |
| 2005 FY9 | 3491.61632 | 17.237 | 0.014 | 3491.61498 | 0.184 | 0.944 | 51.008 | 50.452 | V |
| 2005 FY9 | 3491.61835 | 16.485 | 0.020 | 3491.61701 | -0.568 | 0.944 | 51.008 | 50.452 | I |
| 2005 FY9 | 3491.62032 | 16.459 | 0.018 | 3491.61898 | -0.594 | 0.944 | 51.008 | 50.452 | I |
| 2005 FY9 | 3508.55447 | 18.217 | 0.031 | 3508.55190 | 1.155 | 1.069 | 51.009 | 50.667 | B |
| 2005 FY9 | 3508.55753 | 17.123 | 0.037 | 3508.55496 | 0.061 | 1.069 | 51.009 | 50.667 | V |
| 2005 FY9 | 3508.55956 | 16.452 | 0.055 | 3508.55699 | -0.610 | 1.069 | 51.009 | 50.667 | I |
| 2005 FY9 | 3508.56153 | 16.438 | 0.054 | 3508.55896 | -0.624 | 1.069 | 51.009 | 50.667 | I |
| 2005 FY9 | 3509.55564 | 18.130 | 0.049 | 3509.55299 | 1.068 | 1.075 | 51.009 | 50.680 | B |
| 2005 FY9 | 3509.55870 | 17.301 | 0.029 | 3509.55605 | 0.239 | 1.075 | 51.009 | 50.680 | V |
| 2005 FY9 | 3509.56073 | 16.450 | 0.025 | 3509.55808 | -0.612 | 1.075 | 51.009 | 50.680 | I |
| 2005 FY9 | 3509.56270 | 16.511 | 0.044 | 3509.56005 | -0.551 | 1.075 | 51.009 | 50.680 | I |
| 2005 FY9 | 3511.56645 | 18.098 | 0.023 | 3511.56364 | 1.034 | 1.085 | 51.009 | 50.708 | B |
| 2005 FY9 | 3511.56949 | 17.225 | 0.025 | 3511.56668 | 0.161 | 1.085 | 51.009 | 50.708 | V |
| 2005 FY9 | 3511.57153 | 16.425 | 0.025 | 3511.56872 | -0.639 | 1.085 | 51.009 | 50.708 | I |
| 2005 FY9 | 3511.57349 | 16.398 | 0.028 | 3511.57068 | -0.666 | 1.085 | 51.009 | 50.708 | I |
| 2005 FY9 | 3512.52967 | 18.116 | 0.026 | 3512.52678 | 1.052 | 1.090 | 51.009 | 50.721 | B |
| 2005 FY9 | 3512.53272 | 17.298 | 0.021 | 3512.52983 | 0.234 | 1.090 | 51.009 | 50.721 | V |
| 2005 FY9 | 3512.53474 | 16.486 | 0.017 | 3512.53185 | -0.578 | 1.090 | 51.009 | 50.721 | I |
| 2005 FY9 | 3512.53671 | 16.538 | 0.020 | 3512.53382 | -0.526 | 1.090 | 51.009 | 50.722 | I |
| 2005 FY9 | 3515.51660 | 18.138 | 0.038 | 3515.51347 | 1.072 | 1.103 | 51.009 | 50.763 | B |
| 2005 FY9 | 3515.51966 | 17.233 | 0.026 | 3515.51653 | 0.167 | 1.103 | 51.009 | 50.763 | V |
| 2005 FY9 | 3515.52169 | 16.492 | 0.024 | 3515.51856 | -0.574 | 1.103 | 51.009 | 50.763 | I |
| 2005 FY9 | 3515.52365 | 16.484 | 0.027 | 3515.52052 | -0.582 | 1.103 | 51.009 | 50.763 | I |
| 2005 FY9 | 3517.53260 | 18.109 | 0.016 | 3517.52930 | 1.042 | 1.111 | 51.009 | 50.792 | B |
| 2005 FY9 | 3517.53563 | 17.292 | 0.024 | 3517.53233 | 0.225 | 1.111 | 51.009 | 50.792 | V |
| 2005 FY9 | 3517.53766 | 16.451 | 0.032 | 3517.53436 | -0.616 | 1.111 | 51.009 | 50.793 | I |
| 2005 FY9 | 3517.53963 | 16.497 | 0.025 | 3517.53633 | -0.570 | 1.111 | 51.009 | 50.793 | I |
| 2005 FY9 | 3520.52747 | 18.153 | 0.011 | 3520.52392 | 1.084 | 1.121 | 51.009 | 50.835 | B |
| 2005 FY9 | 3520.53050 | 17.303 | 0.013 | 3520.52695 | 0.234 | 1.121 | 51.009 | 50.835 | V |
| 2005 FY9 | 3520.53253 | 16.482 | 0.015 | 3520.52898 | -0.587 | 1.121 | 51.009 | 50.835 | I |
| 2005 FY9 | 3520.53450 | 16.483 | 0.015 | 3520.53095 | -0.586 | 1.121 | 51.009 | 50.835 | I |
| 2005 FY9 | 3521.51534 | 18.116 | 0.012 | 3521.51171 | 1.046 | 1.124 | 51.009 | 50.849 | B |
| 2005 FY9 | 3521.51840 | 17.284 | 0.013 | 3521.51477 | 0.214 | 1.124 | 51.009 | 50.849 | V |
| 2005 FY9 | 3521.52043 | 16.471 | 0.019 | 3521.51680 | -0.599 | 1.124 | 51.009 | 50.849 | I |
| 2005 FY9 | 3521.52239 | 16.479 | 0.015 | 3521.51876 | -0.591 | 1.124 | 51.009 | 50.849 | I |
| 2005 FY9 | 3522.49780 | 18.097 | 0.030 | 3522.49409 | 1.027 | 1.127 | 51.009 | 50.864 | B |
| 2005 FY9 | 3522.50084 | 17.295 | 0.020 | 3522.49713 | 0.225 | 1.127 | 51.009 | 50.864 | V |
| 2005 FY9 | 3522.50287 | 16.496 | 0.016 | 3522.49916 | -0.574 | 1.127 | 51.009 | 50.864 | I |
| 2005 FY9 | 3522.50483 | 16.501 | 0.028 | 3522.50112 | -0.569 | 1.127 | 51.009 | 50.864 | I |
| 2005 FY9 | 3523.50681 | 18.067 | 0.032 | 3523.50301 | 0.996 | 1.129 | 51.009 | 50.879 | B |
| 2005 FY9 | 3523.50986 | 17.234 | 0.029 | 3523.50606 | 0.163 | 1.129 | 51.009 | 50.879 | V |
| 2005 FY9 | 3523.51189 | 16.473 | 0.020 | 3523.50809 | -0.598 | 1.129 | 51.009 | 50.879 | I |
| 2005 FY9 | 3523.51386 | 16.478 | 0.015 | 3523.51006 | -0.593 | 1.129 | 51.009 | 50.879 | I |
| 2005 FY9 | 3524.50054 | 18.023 | 0.617 | 3524.49666 | 0.951 | 1.131 | 51.009 | 50.893 | B |
| 2005 FY9 | 3524.50360 | 17.412 | 0.331 | 3524.49972 | 0.340 | 1.131 | 51.009 | 50.893 | V |
| 2005 FY9 | 3524.50565 | 16.301 | 0.214 | 3524.50177 | -0.771 | 1.131 | 51.009 | 50.893 | I |
| 2005 FY9 | 3524.50761 | 16.379 | 0.240 | 3524.50373 | -0.693 | 1.131 | 51.009 | 50.893 | I |
| 2005 FY9 | 3525.53698 | 17.172 | 0.447 | 3525.53301 | 0.100 | 1.133 | 51.009 | 50.909 | V |
| 2005 FY9 | 3525.53901 | 16.334 | 0.296 | 3525.53504 | -0.738 | 1.133 | 51.009 | 50.909 | I |
| 2005 FY9 | 3525.54098 | 16.336 | 0.289 | 3525.53701 | -0.736 | 1.133 | 51.009 | 50.909 | I |
| 2005 FY9 | 3526.50802 | 18.094 | 0.581 | 3526.50397 | 1.021 | 1.134 | 51.009 | 50.923 | B |
| 2005 FY9 | 3526.51108 | 17.405 | 0.524 | 3526.50703 | 0.332 | 1.134 | 51.009 | 50.923 | V |
| 2005 FY9 | 3526.51311 | 16.354 | 0.267 | 3526.50906 | -0.719 | 1.134 | 51.009 | 50.923 | I |
| 2005 FY9 | 3527.50132 | 18.088 | 0.572 | 3527.49719 | 1.015 | 1.136 | 51.009 | 50.937 | B |
| 2005 FY9 | 3527.50436 | 17.320 | 0.421 | 3527.50023 | 0.247 | 1.136 | 51.009 | 50.937 | V |
| 2005 FY9 | 3527.50639 | 16.324 | 0.309 | 3527.50225 | -0.749 | 1.136 | 51.009 | 50.937 | I |
| 2005 FY9 | 3527.50836 | 16.432 | 0.304 | 3527.50422 | -0.641 | 1.136 | 51.009 | 50.937 | I |
| 2005 FY9 | 3528.51698 | 18.120 | 0.583 | 3528.51276 | 1.046 | 1.137 | 51.009 | 50.952 | B |
| 2005 FY9 | 3528.52003 | 17.327 | 0.429 | 3528.51581 | 0.253 | 1.137 | 51.009 | 50.952 | V |
| 2005 FY9 | 3528.52206 | 16.394 | 0.267 | 3528.51784 | -0.680 | 1.137 | 51.009 | 50.952 | I |
| 2005 FY9 | 3542.47099 | 18.255 | 0.024 | 3542.46558 | 1.172 | 1.126 | 51.009 | 51.158 | B |
| 2005 FY9 | 3542.47404 | 17.410 | 0.019 | 3542.46863 | 0.327 | 1.126 | 51.009 | 51.158 | V |
| 2005 FY9 | 3542.47607 | 16.665 | 0.017 | 3542.47066 | -0.418 | 1.126 | 51.009 | 51.158 | I |
| 2005 FY9 | 3542.47804 | 16.612 | 0.015 | 3542.47263 | -0.471 | 1.126 | 51.009 | 51.158 | I |
| 2005 FY9 | 3543.46910 | 18.062 | 0.070 | 3543.46360 | 0.979 | 1.124 | 51.009 | 51.173 | B |
| 2005 FY9 | 3543.47213 | 17.446 | 0.082 | 3543.46663 | 0.363 | 1.124 | 51.009 | 51.173 | V |
| 2005 FY9 | 3543.47416 | 16.606 | 0.079 | 3543.46866 | -0.477 | 1.124 | 51.009 | 51.173 | I |
| 2005 FY9 | 3543.47612 | 16.598 | 0.105 | 3543.47062 | -0.485 | 1.124 | 51.009 | 51.173 | I |
| 2005 FY9 | 3544.45362 | 18.181 | 0.040 | 3544.44805 | 1.097 | 1.121 | 51.009 | 51.186 | B |
| 2005 FY9 | 3545.47094 | 18.140 | 0.045 | 3545.46528 | 1.055 | 1.118 | 51.009 | 51.202 | B |
| 2005 FY9 | 3545.47398 | 17.428 | 0.047 | 3545.46832 | 0.343 | 1.118 | 51.009 | 51.202 | V |
| 2005 FY9 | 3545.47601 | 16.564 | 0.044 | 3545.47035 | -0.521 | 1.118 | 51.009 | 51.202 | I |
| 2005 FY9 | 3545.47798 | 16.464 | 0.057 | 3545.47232 | -0.621 | 1.118 | 51.009 | 51.202 | I |
| 2005 FY9 | 3547.46816 | 18.212 | 0.013 | 3547.46234 | 1.126 | 1.111 | 51.009 | 51.230 | B |
| 2005 FY9 | 3547.47120 | 17.385 | 0.018 | 3547.46538 | 0.299 | 1.111 | 51.009 | 51.230 | V |
| 2005 FY9 | 3547.47323 | 16.599 | 0.025 | 3547.46741 | -0.487 | 1.111 | 51.009 | 51.230 | I |
| 2005 FY9 | 3547.47323 | 16.599 | 0.025 | 3547.46741 | -0.487 | 1.111 | 51.009 | 51.230 | I |
| 2005 FY9 | 3547.47738 | 16.618 | 0.022 | 3547.47156 | -0.468 | 1.111 | 51.009 | 51.230 | I |
| 2005 FY9 | 3774.82241 | 16.363 | 0.055 | 3774.82173 | -0.760 | 0.746 | 51.922 | 51.196 | I |
| 2005 FY9 | 3776.82497 | 16.494 | 0.021 | 3776.82438 | -0.628 | 0.726 | 51.922 | 51.180 | I |
| 2005 FY9 | 3776.82706 | 17.316 | 0.013 | 3776.82647 | 0.194 | 0.726 | 51.922 | 51.180 | V |
| 2005 FY9 | 3778.80455 | 18.021 | 0.042 | 3778.80405 | 0.899 | 0.707 | 51.923 | 51.165 | B |
| 2005 FY9 | 3778.80767 | 16.485 | 0.017 | 3778.80717 | -0.637 | 0.707 | 51.923 | 51.165 | I |
| 2005 FY9 | 3778.80976 | 17.289 | 0.024 | 3778.80926 | 0.167 | 0.707 | 51.923 | 51.165 | V |
| 2005 FY9 | 3785.82382 | 18.024 | 0.018 | 3785.82358 | 0.904 | 0.643 | 51.924 | 51.120 | B |
| 2005 FY9 | 3785.82694 | 16.451 | 0.015 | 3785.82670 | -0.669 | 0.643 | 51.924 | 51.120 | I |
| 2005 FY9 | 3785.82903 | 17.267 | 0.022 | 3785.82879 | 0.147 | 0.643 | 51.924 | 51.120 | V |
| 2005 FY9 | 3787.80048 | 18.036 | 0.011 | 3787.80030 | 0.917 | 0.627 | 51.924 | 51.110 | B |
| 2005 FY9 | 3787.80361 | 16.471 | 0.013 | 3787.80343 | -0.648 | 0.627 | 51.924 | 51.110 | I |
| 2005 FY9 | 3787.80570 | 17.222 | 0.018 | 3787.80552 | 0.103 | 0.627 | 51.924 | 51.110 | V |
| 2005 FY9 | 3789.79225 | 18.022 | 0.013 | 3789.79212 | 0.903 | 0.611 | 51.925 | 51.100 | B |
| 2005 FY9 | 3789.79536 | 16.451 | 0.015 | 3789.79523 | -0.668 | 0.611 | 51.925 | 51.100 | I |
| 2005 FY9 | 3789.79745 | 17.236 | 0.017 | 3789.79732 | 0.117 | 0.611 | 51.925 | 51.100 | V |
| 2005 FY9 | 3791.78812 | 18.036 | 0.011 | 3791.78804 | 0.917 | 0.597 | 51.925 | 51.092 | B |
| 2005 FY9 | 3791.79124 | 16.456 | 0.015 | 3791.79116 | -0.663 | 0.597 | 51.925 | 51.092 | I |
| 2005 FY9 | 3791.79334 | 17.234 | 0.012 | 3791.79326 | 0.115 | 0.597 | 51.925 | 51.092 | V |
| 2005 FY9 | 3793.78960 | 18.175 | 0.013 | 3793.78957 | 1.057 | 0.584 | 51.925 | 51.084 | B |
| 2005 FY9 | 3793.79272 | 16.487 | 0.021 | 3793.79269 | -0.631 | 0.584 | 51.925 | 51.084 | I |
| 2005 FY9 | 3793.79481 | 17.324 | 0.025 | 3793.79478 | 0.206 | 0.584 | 51.925 | 51.084 | V |
| 2005 FY9 | 3798.76558 | 18.026 | 0.024 | 3798.76563 | 0.908 | 0.558 | 51.926 | 51.070 | B |
| 2005 FY9 | 3798.76870 | 16.376 | 0.019 | 3798.76875 | -0.742 | 0.558 | 51.926 | 51.070 | I |
| 2005 FY9 | 3798.77079 | 17.194 | 0.016 | 3798.77084 | 0.076 | 0.558 | 51.926 | 51.070 | V |
| 2005 FY9 | 3800.77695 | 18.046 | 0.011 | 3800.77702 | 0.928 | 0.550 | 51.926 | 51.066 | B |
| 2005 FY9 | 3800.78007 | 16.438 | 0.011 | 3800.78014 | -0.680 | 0.550 | 51.926 | 51.066 | I |
| 2005 FY9 | 3800.78216 | 17.225 | 0.011 | 3800.78223 | 0.107 | 0.550 | 51.926 | 51.066 | V |
| 2005 FY9 | 3802.75348 | 16.372 | 0.062 | 3802.75357 | -0.745 | 0.545 | 51.927 | 51.063 | I |
| 2005 FY9 | 3802.75557 | 17.192 | 0.038 | 3802.75566 | 0.075 | 0.545 | 51.927 | 51.063 | V |
| 2005 FY9 | 3804.77045 | 18.058 | 0.029 | 3804.77055 | 0.941 | 0.542 | 51.927 | 51.061 | B |
| 2005 FY9 | 3804.77357 | 16.437 | 0.070 | 3804.77367 | -0.680 | 0.542 | 51.927 | 51.061 | I |
| 2005 FY9 | 3804.77566 | 17.249 | 0.049 | 3804.77576 | 0.132 | 0.542 | 51.927 | 51.061 | V |
| 2005 FY9 | 3813.70416 | 18.031 | 0.026 | 3813.70423 | 0.913 | 0.552 | 51.929 | 51.066 | B |
| 2005 FY9 | 3813.70728 | 16.436 | 0.016 | 3813.70735 | -0.682 | 0.552 | 51.929 | 51.066 | I |
| 2005 FY9 | 3813.70937 | 17.232 | 0.020 | 3813.70944 | 0.114 | 0.552 | 51.929 | 51.066 | V |
| 2005 FY9 | 3815.70726 | 18.080 | 0.013 | 3815.70731 | 0.962 | 0.560 | 51.929 | 51.070 | B |
| 2005 FY9 | 3815.71037 | 16.410 | 0.013 | 3815.71042 | -0.708 | 0.560 | 51.929 | 51.070 | I |
| 2005 FY9 | 3815.71246 | 17.258 | 0.016 | 3815.71251 | 0.140 | 0.560 | 51.929 | 51.070 | V |
| 2005 FY9 | 3817.67021 | 18.089 | 0.010 | 3817.67023 | 0.971 | 0.569 | 51.929 | 51.075 | B |
| 2005 FY9 | 3817.67332 | 16.425 | 0.018 | 3817.67334 | -0.693 | 0.569 | 51.929 | 51.075 | I |
| 2005 FY9 | 3817.67541 | 17.276 | 0.013 | 3817.67543 | 0.158 | 0.569 | 51.929 | 51.075 | V |
| 2005 FY9 | 3821.66415 | 18.085 | 0.014 | 3821.66409 | 0.966 | 0.594 | 51.930 | 51.088 | B |
| 2005 FY9 | 3821.66727 | 16.400 | 0.011 | 3821.66721 | -0.719 | 0.594 | 51.930 | 51.088 | I |
| 2005 FY9 | 3821.66936 | 17.201 | 0.013 | 3821.66930 | 0.082 | 0.594 | 51.930 | 51.088 | V |
| 2005 FY9 | 3823.70059 | 18.101 | 0.014 | 3823.70049 | 0.982 | 0.608 | 51.930 | 51.096 | B |
| 2005 FY9 | 3823.70371 | 16.398 | 0.030 | 3823.70361 | -0.721 | 0.608 | 51.930 | 51.096 | I |
| 2005 FY9 | 3823.70371 | 16.398 | 0.030 | 3823.70361 | -0.721 | 0.608 | 51.930 | 51.096 | I |
| 2005 FY9 | 3823.70581 | 17.284 | 0.023 | 3823.70571 | 0.165 | 0.608 | 51.930 | 51.096 | V |
| 2005 FY9 | 3828.68445 | 17.933 | 0.013 | 3828.68421 | 0.813 | 0.648 | 51.931 | 51.121 | B |
| 2005 FY9 | 3828.68966 | 17.240 | 0.015 | 3828.68941 | 0.120 | 0.648 | 51.931 | 51.121 | V |
| 2005 FY9 | 3831.61703 | 18.074 | 0.015 | 3831.61669 | 0.953 | 0.675 | 51.932 | 51.138 | B |
| 2005 FY9 | 3831.62015 | 16.414 | 0.017 | 3831.61981 | -0.707 | 0.675 | 51.932 | 51.138 | I |
| 2005 FY9 | 3831.62224 | 17.238 | 0.014 | 3831.62190 | 0.117 | 0.675 | 51.932 | 51.138 | V |
| 2005 FY9 | 3833.62948 | 18.054 | 0.024 | 3833.62906 | 0.933 | 0.693 | 51.932 | 51.151 | B |
| 2005 FY9 | 3833.63259 | 16.443 | 0.019 | 3833.63217 | -0.678 | 0.693 | 51.932 | 51.151 | I |
| 2005 FY9 | 3833.63468 | 17.201 | 0.017 | 3833.63426 | 0.080 | 0.693 | 51.932 | 51.151 | V |
| 2005 FY9 | 3837.63168 | 18.006 | 0.038 | 3837.63109 | 0.883 | 0.732 | 51.933 | 51.180 | B |
| 2005 FY9 | 3837.63933 | 17.975 | 0.056 | 3837.63874 | 0.852 | 0.732 | 51.933 | 51.180 | B |
| 2005 FY9 | 3837.64244 | 16.342 | 0.024 | 3837.64185 | -0.781 | 0.732 | 51.933 | 51.180 | I |
| 2005 FY9 | 3837.64453 | 17.235 | 0.029 | 3837.64394 | 0.112 | 0.732 | 51.933 | 51.180 | V |
| 2005 FY9 | 3840.63420 | 17.990 | 0.048 | 3840.63348 | 0.866 | 0.762 | 51.933 | 51.204 | B |
| 2005 FY9 | 3840.63732 | 16.419 | 0.024 | 3840.63660 | -0.705 | 0.762 | 51.933 | 51.204 | I |
| 2005 FY9 | 3840.63941 | 17.091 | 0.067 | 3840.63869 | -0.033 | 0.762 | 51.933 | 51.204 | V |
| 2005 FY9 | 3841.63142 | 17.914 | 0.037 | 3841.63065 | 0.790 | 0.772 | 51.933 | 51.212 | B |
| 2005 FY9 | 3841.63454 | 16.395 | 0.042 | 3841.63377 | -0.729 | 0.772 | 51.933 | 51.212 | I |
| 2005 FY9 | 3841.63663 | 17.213 | 0.029 | 3841.63586 | 0.089 | 0.772 | 51.933 | 51.212 | V |
| 2005 FY9 | 3844.63795 | 18.000 | 0.027 | 3844.63703 | 0.875 | 0.801 | 51.934 | 51.239 | B |
| 2005 FY9 | 3844.64107 | 16.376 | 0.067 | 3844.64015 | -0.749 | 0.801 | 51.934 | 51.239 | I |
| 2005 FY9 | 3844.64315 | 17.311 | 0.015 | 3844.64223 | 0.186 | 0.802 | 51.934 | 51.239 | V |
| 2005 FY9 | 3846.61797 | 17.936 | 0.037 | 3846.61694 | 0.810 | 0.821 | 51.934 | 51.257 | B |
| 2005 FY9 | 3846.62108 | 16.258 | 0.067 | 3846.62005 | -0.868 | 0.821 | 51.934 | 51.257 | I |
| 2005 FY9 | 3846.62316 | 17.323 | 0.015 | 3846.62213 | 0.197 | 0.821 | 51.934 | 51.257 | V |
| 2005 FY9 | 3847.61538 | 18.104 | 0.018 | 3847.61429 | 0.978 | 0.831 | 51.934 | 51.267 | B |
| 2005 FY9 | 3847.61851 | 16.498 | 0.025 | 3847.61742 | -0.628 | 0.831 | 51.934 | 51.267 | I |
| 2005 FY9 | 3847.62060 | 17.250 | 0.017 | 3847.61951 | 0.124 | 0.831 | 51.934 | 51.267 | V |
| 2005 FY9 | 3848.60758 | 18.085 | 0.015 | 3848.60644 | 0.958 | 0.840 | 51.934 | 51.276 | B |
| 2005 FY9 | 3848.61069 | 16.497 | 0.023 | 3848.60955 | -0.630 | 0.840 | 51.934 | 51.276 | I |
| 2005 FY9 | 3848.61277 | 17.302 | 0.017 | 3848.61163 | 0.175 | 0.840 | 51.934 | 51.276 | V |
| 2005 FY9 | 3848.61479 | 17.233 | 0.022 | 3848.61365 | 0.106 | 0.841 | 51.934 | 51.276 | V |
| 2005 FY9 | 3774.82241 | 16.363 | 0.055 | 3774.82173 | -0.760 | 0.746 | 51.922 | 51.196 | I |
| 2005 FY9 | 3776.82497 | 16.494 | 0.021 | 3776.82438 | -0.628 | 0.726 | 51.922 | 51.180 | I |
| | | | | # | | | | | |
| (90377) Sedna | 2969.82338 | 20.701 | 0.085 | 2969.82338 | 1.201 | 0.208 | 89.586 | 88.656 | R |
| (90377) Sedna | 2970.63344 | 20.571 | 0.068 | 2970.63342 | 1.071 | 0.215 | 89.585 | 88.659 | R |
| (90377) Sedna | 2971.53123 | 20.537 | 0.071 | 2971.53119 | 1.037 | 0.222 | 89.584 | 88.663 | R |
| (90377) Sedna | 2971.53123 | 20.537 | 0.071 | 2971.53119 | 1.037 | 0.222 | 89.584 | 88.663 | R |
| (90377) Sedna | 2976.60031 | 21.427 | 0.196 | 2976.60012 | 1.927 | 0.266 | 89.580 | 88.689 | V |
| (90377) Sedna | 2976.60620 | 20.659 | 0.107 | 2976.60601 | 1.159 | 0.267 | 89.580 | 88.689 | R |
| (90377) Sedna | 2976.61180 | 19.838 | 0.091 | 2976.61161 | 0.338 | 0.267 | 89.580 | 88.689 | I |
| (90377) Sedna | 2977.52379 | 20.436 | 0.127 | 2977.52357 | 0.935 | 0.275 | 89.579 | 88.694 | R |
| (90377) Sedna | 2977.55591 | 20.499 | 0.086 | 2977.55569 | 0.998 | 0.275 | 89.579 | 88.695 | R |
| (90377) Sedna | 2977.59738 | 20.689 | 0.084 | 2977.59716 | 1.188 | 0.275 | 89.579 | 88.695 | R |
| (90377) Sedna | 2977.76378 | 20.507 | 0.111 | 2977.76355 | 1.006 | 0.277 | 89.579 | 88.696 | R |
| (90377) Sedna | 2978.60546 | 20.528 | 0.237 | 2978.60520 | 1.027 | 0.284 | 89.578 | 88.701 | R |
| (90377) Sedna | 2978.60851 | 19.675 | 0.186 | 2978.60825 | 0.174 | 0.284 | 89.578 | 88.701 | I |
| (90377) Sedna | 2979.59920 | 19.810 | 0.289 | 2979.59890 | 0.309 | 0.293 | 89.577 | 88.708 | I |
| (90377) Sedna | 2980.61946 | 20.351 | 0.204 | 2980.61912 | 0.850 | 0.302 | 89.576 | 88.715 | R |
| (90377) Sedna | 2983.60151 | 21.300 | 0.284 | 2983.60105 | 1.799 | 0.328 | 89.573 | 88.736 | V |
| (90377) Sedna | 2983.73237 | 20.443 | 0.177 | 2983.73190 | 0.942 | 0.330 | 89.573 | 88.737 | R |
| (90377) Sedna | 2983.73542 | 20.010 | 0.200 | 2983.73495 | 0.509 | 0.330 | 89.573 | 88.737 | I |
| (90377) Sedna | 2984.53058 | 20.540 | 0.071 | 2984.53008 | 1.038 | 0.336 | 89.572 | 88.744 | R |
| (90377) Sedna | 2984.56577 | 20.649 | 0.053 | 2984.56526 | 1.147 | 0.337 | 89.572 | 88.744 | R |
| (90377) Sedna | 2984.60805 | 20.668 | 0.081 | 2984.60754 | 1.166 | 0.337 | 89.572 | 88.744 | R |
| (90377) Sedna | 2984.64972 | 20.688 | 0.094 | 2984.64921 | 1.186 | 0.338 | 89.572 | 88.744 | R |
| (90377) Sedna | 2984.68244 | 20.589 | 0.074 | 2984.68193 | 1.087 | 0.338 | 89.572 | 88.745 | R |
| (90377) Sedna | 2985.52615 | 20.483 | 0.143 | 2985.52560 | 0.981 | 0.345 | 89.571 | 88.752 | R |
| (90377) Sedna | 2985.57714 | 20.523 | 0.043 | 2985.57659 | 1.021 | 0.346 | 89.571 | 88.752 | R |
| (90377) Sedna | 2985.62181 | 20.668 | 0.076 | 2985.62125 | 1.166 | 0.346 | 89.571 | 88.752 | R |
| (90377) Sedna | 2985.66398 | 20.475 | 0.087 | 2985.66342 | 0.973 | 0.346 | 89.571 | 88.753 | R |
| (90377) Sedna | 2985.70870 | 20.656 | 0.077 | 2985.70814 | 1.154 | 0.347 | 89.571 | 88.753 | R |
| (90377) Sedna | 2986.58615 | 22.480 | 0.072 | 2986.58555 | 2.978 | 0.354 | 89.570 | 88.760 | B |
| (90377) Sedna | 2986.59321 | 21.410 | 0.100 | 2986.59261 | 1.908 | 0.354 | 89.570 | 88.760 | V |
| (90377) Sedna | 2986.59611 | 21.542 | 0.114 | 2986.59551 | 2.040 | 0.354 | 89.570 | 88.760 | V |
| (90377) Sedna | 2986.59910 | 20.610 | 0.065 | 2986.59850 | 1.108 | 0.354 | 89.570 | 88.760 | R |
| (90377) Sedna | 2986.60215 | 19.989 | 0.080 | 2986.60155 | 0.487 | 0.354 | 89.570 | 88.761 | I |
| (90377) Sedna | 2986.70132 | 22.499 | 0.244 | 2986.70071 | 2.997 | 0.355 | 89.570 | 88.761 | B |
| (90377) Sedna | 2986.70838 | 21.118 | 0.174 | 2986.70777 | 1.616 | 0.355 | 89.570 | 88.761 | V |
| (90377) Sedna | 2986.71427 | 20.769 | 0.140 | 2986.71366 | 1.267 | 0.355 | 89.570 | 88.761 | R |
| (90377) Sedna | 2986.71731 | 20.040 | 0.129 | 2986.71670 | 0.538 | 0.355 | 89.570 | 88.761 | I |
| (90377) Sedna | 2987.52317 | 20.456 | 0.186 | 2987.52252 | 0.954 | 0.362 | 89.570 | 88.768 | R |
| (90377) Sedna | 2987.60764 | 20.610 | 0.039 | 2987.60699 | 1.108 | 0.363 | 89.570 | 88.769 | R |
| (90377) Sedna | 2987.64978 | 20.591 | 0.041 | 2987.64913 | 1.089 | 0.363 | 89.570 | 88.770 | R |
| (90377) Sedna | 2987.68946 | 20.554 | 0.050 | 2987.68880 | 1.052 | 0.364 | 89.569 | 88.770 | R |
| (90377) Sedna | 2988.62048 | 20.544 | 0.036 | 2988.61978 | 1.042 | 0.372 | 89.569 | 88.778 | R |
| (90377) Sedna | 2988.66305 | 20.654 | 0.048 | 2988.66234 | 1.152 | 0.372 | 89.569 | 88.778 | R |
| (90377) Sedna | 2988.70606 | 20.645 | 0.059 | 2988.70535 | 1.143 | 0.372 | 89.569 | 88.779 | R |
| (90377) Sedna | 2989.56952 | 22.573 | 0.082 | 2989.56877 | 3.071 | 0.380 | 89.568 | 88.787 | B |
| (90377) Sedna | 2989.57657 | 21.389 | 0.100 | 2989.57582 | 1.887 | 0.380 | 89.568 | 88.787 | V |
| (90377) Sedna | 2989.57947 | 21.301 | 0.098 | 2989.57872 | 1.799 | 0.380 | 89.568 | 88.787 | V |
| (90377) Sedna | 2989.58246 | 20.664 | 0.070 | 2989.58171 | 1.161 | 0.380 | 89.568 | 88.787 | R |
| (90377) Sedna | 2989.58551 | 19.824 | 0.079 | 2989.58476 | 0.321 | 0.380 | 89.568 | 88.787 | I |
| (90377) Sedna | 2989.70041 | 22.574 | 0.113 | 2989.69965 | 3.071 | 0.381 | 89.568 | 88.788 | B |
| (90377) Sedna | 2989.70747 | 21.450 | 0.156 | 2989.70671 | 1.947 | 0.381 | 89.568 | 88.788 | V |
| (90377) Sedna | 2989.71036 | 21.378 | 0.184 | 2989.70960 | 1.875 | 0.381 | 89.568 | 88.788 | V |
| (90377) Sedna | 2989.71335 | 20.541 | 0.088 | 2989.71259 | 1.038 | 0.381 | 89.568 | 88.788 | R |
| (90377) Sedna | 2989.71640 | 20.133 | 0.123 | 2989.71564 | 0.630 | 0.381 | 89.568 | 88.788 | I |
| (90377) Sedna | 2990.56183 | 20.546 | 0.050 | 2990.56102 | 1.043 | 0.388 | 89.567 | 88.796 | R |
| (90377) Sedna | 2990.60846 | 20.492 | 0.051 | 2990.60765 | 0.989 | 0.388 | 89.567 | 88.796 | R |
| (90377) Sedna | 2990.65617 | 20.532 | 0.046 | 2990.65536 | 1.029 | 0.389 | 89.567 | 88.797 | R |
| (90377) Sedna | 2990.69417 | 20.551 | 0.055 | 2990.69336 | 1.048 | 0.389 | 89.567 | 88.797 | R |
| (90377) Sedna | 2991.57789 | 20.529 | 0.051 | 2991.57703 | 1.026 | 0.396 | 89.566 | 88.806 | R |
| (90377) Sedna | 2991.62048 | 20.616 | 0.050 | 2991.61961 | 1.113 | 0.397 | 89.566 | 88.806 | R |
| (90377) Sedna | 2991.66439 | 20.746 | 0.056 | 2991.66352 | 1.243 | 0.397 | 89.566 | 88.806 | R |
| (90377) Sedna | 2991.71414 | 20.667 | 0.062 | 2991.71327 | 1.164 | 0.397 | 89.566 | 88.807 | R |
| (90377) Sedna | 2992.66844 | 20.651 | 0.042 | 2992.66752 | 1.148 | 0.405 | 89.565 | 88.816 | R |
| (90377) Sedna | 2993.62540 | 22.567 | 0.098 | 2993.62442 | 3.064 | 0.413 | 89.564 | 88.826 | B |
| (90377) Sedna | 2993.63245 | 21.566 | 0.135 | 2993.63147 | 2.063 | 0.413 | 89.564 | 88.826 | V |
| (90377) Sedna | 2993.63535 | 21.339 | 0.116 | 2993.63437 | 1.836 | 0.413 | 89.564 | 88.826 | V |
| (90377) Sedna | 2993.63834 | 20.462 | 0.106 | 2993.63736 | 0.959 | 0.413 | 89.564 | 88.826 | R |
| (90377) Sedna | 2993.64139 | 19.900 | 0.082 | 2993.64041 | 0.397 | 0.413 | 89.564 | 88.826 | I |
| (90377) Sedna | 2993.68600 | 22.697 | 0.122 | 2993.68502 | 3.194 | 0.413 | 89.564 | 88.827 | B |
| (90377) Sedna | 2993.69307 | 21.291 | 0.115 | 2993.69209 | 1.788 | 0.414 | 89.564 | 88.827 | V |
| (90377) Sedna | 2993.69897 | 20.679 | 0.132 | 2993.69799 | 1.176 | 0.414 | 89.564 | 88.827 | R |
| (90377) Sedna | 2993.70202 | 19.931 | 0.100 | 2993.70104 | 0.428 | 0.414 | 89.564 | 88.827 | I |
| (90377) Sedna | 2995.57781 | 20.696 | 0.051 | 2995.57671 | 1.192 | 0.428 | 89.562 | 88.846 | R |
| (90377) Sedna | 2995.65052 | 20.760 | 0.055 | 2995.64942 | 1.256 | 0.429 | 89.562 | 88.847 | R |
| (90377) Sedna | 2996.54618 | 20.712 | 0.065 | 2996.54502 | 1.208 | 0.436 | 89.561 | 88.857 | R |
| (90377) Sedna | 2996.59052 | 20.639 | 0.056 | 2996.58936 | 1.135 | 0.436 | 89.561 | 88.857 | R |
| (90377) Sedna | 2996.64130 | 20.638 | 0.048 | 2996.64014 | 1.134 | 0.437 | 89.561 | 88.858 | R |
| (90377) Sedna | 2999.68836 | 22.380 | 0.110 | 2999.68700 | 2.875 | 0.460 | 89.558 | 88.892 | B |
| (90377) Sedna | 2999.69542 | 21.328 | 0.130 | 2999.69406 | 1.823 | 0.460 | 89.558 | 88.892 | V |
| (90377) Sedna | 2999.69832 | 21.353 | 0.135 | 2999.69696 | 1.848 | 0.460 | 89.558 | 88.892 | V |
| (90377) Sedna | 2999.70131 | 20.749 | 0.093 | 2999.69995 | 1.244 | 0.460 | 89.558 | 88.892 | R |
| (90377) Sedna | 2999.70436 | 19.889 | 0.095 | 2999.70300 | 0.384 | 0.460 | 89.558 | 88.892 | I |
| (90377) Sedna | 3000.67063 | 20.533 | 0.157 | 3000.66920 | 1.028 | 0.467 | 89.558 | 88.903 | R |
| (90377) Sedna | 3001.68258 | 20.634 | 0.051 | 3001.68109 | 1.129 | 0.474 | 89.557 | 88.915 | R |
| (90377) Sedna | 3002.67101 | 20.726 | 0.077 | 3002.66945 | 1.220 | 0.481 | 89.556 | 88.927 | R |
| (90377) Sedna | 3003.54110 | 20.677 | 0.066 | 3003.53948 | 1.171 | 0.487 | 89.555 | 88.937 | R |
| (90377) Sedna | 3003.59603 | 20.706 | 0.052 | 3003.59440 | 1.200 | 0.487 | 89.555 | 88.938 | R |
| (90377) Sedna | 3003.65036 | 20.611 | 0.050 | 3003.64873 | 1.105 | 0.488 | 89.555 | 88.939 | R |
| (90377) Sedna | 3003.67955 | 20.722 | 0.056 | 3003.67792 | 1.216 | 0.488 | 89.555 | 88.939 | R |
| (90377) Sedna | 3004.62275 | 20.708 | 0.075 | 3004.62105 | 1.202 | 0.494 | 89.554 | 88.951 | R |
| (90377) Sedna | 3004.65813 | 20.851 | 0.093 | 3004.65643 | 1.345 | 0.495 | 89.554 | 88.951 | R |
| (90377) Sedna | 3004.69149 | 20.626 | 0.097 | 3004.68978 | 1.120 | 0.495 | 89.554 | 88.952 | R |
| (90377) Sedna | 3009.66099 | 20.357 | 0.116 | 3009.65891 | 0.849 | 0.526 | 89.549 | 89.016 | R |
| (90377) Sedna | 3009.66099 | 20.392 | 0.120 | 3009.65891 | 0.884 | 0.526 | 89.549 | 89.016 | R |
| (90377) Sedna | 3012.58123 | 20.397 | 0.252 | 3012.57892 | 0.888 | 0.543 | 89.547 | 89.055 | R |
| (90377) Sedna | 3012.58123 | 20.585 | 0.254 | 3012.57892 | 1.076 | 0.543 | 89.547 | 89.055 | R |
| (90377) Sedna | 3012.61774 | 20.347 | 0.201 | 3012.61543 | 0.838 | 0.543 | 89.547 | 89.056 | R |
| (90377) Sedna | 3012.61774 | 20.529 | 0.216 | 3012.61543 | 1.020 | 0.543 | 89.547 | 89.056 | R |
| (90377) Sedna | 3012.64552 | 20.748 | 0.276 | 3012.64321 | 1.239 | 0.543 | 89.547 | 89.056 | R |
| (90377) Sedna | 3012.64552 | 20.879 | 0.280 | 3012.64321 | 1.370 | 0.543 | 89.547 | 89.056 | R |
| (90377) Sedna | 3015.56250 | 20.490 | 0.046 | 3015.55995 | 0.980 | 0.559 | 89.544 | 89.097 | R |
| (90377) Sedna | 3015.56250 | 20.548 | 0.117 | 3015.55995 | 1.038 | 0.559 | 89.544 | 89.097 | R |
| (90377) Sedna | 3015.60340 | 20.456 | 0.062 | 3015.60085 | 0.946 | 0.559 | 89.544 | 89.098 | R |
| (90377) Sedna | 3015.60340 | 20.519 | 0.126 | 3015.60085 | 1.009 | 0.559 | 89.544 | 89.098 | R |
| (90377) Sedna | 3015.63788 | 20.427 | 0.068 | 3015.63533 | 0.917 | 0.559 | 89.544 | 89.098 | R |
| (90377) Sedna | 3015.63788 | 20.482 | 0.127 | 3015.63533 | 0.972 | 0.559 | 89.544 | 89.098 | R |
| (90377) Sedna | 3018.54179 | 20.537 | 0.062 | 3018.53899 | 1.027 | 0.573 | 89.541 | 89.140 | R |
| (90377) Sedna | 3018.54179 | 20.592 | 0.120 | 3018.53899 | 1.082 | 0.573 | 89.541 | 89.140 | R |
| (90377) Sedna | 3018.58509 | 20.582 | 0.050 | 3018.58229 | 1.072 | 0.573 | 89.541 | 89.141 | R |
| (90377) Sedna | 3018.58509 | 20.644 | 0.120 | 3018.58229 | 1.134 | 0.573 | 89.541 | 89.141 | R |
| (90377) Sedna | 3018.61993 | 20.511 | 0.044 | 3018.61713 | 1.000 | 0.573 | 89.541 | 89.141 | R |
| (90377) Sedna | 3018.61993 | 20.567 | 0.126 | 3018.61713 | 1.056 | 0.573 | 89.541 | 89.141 | R |
| (90377) Sedna | 3021.55724 | 20.540 | 0.070 | 3021.55419 | 1.028 | 0.586 | 89.538 | 89.185 | R |
| (90377) Sedna | 3021.55724 | 20.657 | 0.075 | 3021.55419 | 1.145 | 0.586 | 89.538 | 89.185 | R |
| (90377) Sedna | 3021.59351 | 20.604 | 0.094 | 3021.59045 | 1.092 | 0.586 | 89.538 | 89.185 | R |
| (90377) Sedna | 3021.59351 | 20.722 | 0.093 | 3021.59045 | 1.210 | 0.586 | 89.538 | 89.185 | R |
| (90377) Sedna | 3021.64757 | 20.684 | 0.113 | 3021.64451 | 1.172 | 0.586 | 89.538 | 89.186 | R |
| (90377) Sedna | 3021.64757 | 20.805 | 0.111 | 3021.64451 | 1.293 | 0.586 | 89.538 | 89.186 | R |
| (90377) Sedna | 3023.54314 | 20.505 | 0.120 | 3023.53991 | 0.993 | 0.593 | 89.537 | 89.215 | R |
| (90377) Sedna | 3023.54314 | 20.505 | 0.120 | 3023.53991 | 0.993 | 0.593 | 89.537 | 89.215 | R |
| (90377) Sedna | 3023.59375 | 20.563 | 0.114 | 3023.59052 | 1.051 | 0.594 | 89.537 | 89.216 | R |
| (90377) Sedna | 3025.52923 | 20.380 | 0.155 | 3025.52583 | 0.867 | 0.600 | 89.535 | 89.245 | R |
| (90377) Sedna | 3025.58859 | 20.577 | 0.117 | 3025.58518 | 1.064 | 0.600 | 89.535 | 89.246 | R |
| (90377) Sedna | 3026.52960 | 20.572 | 0.153 | 3026.52611 | 1.059 | 0.604 | 89.534 | 89.261 | R |
| (90377) Sedna | 3026.55707 | 20.558 | 0.123 | 3026.55358 | 1.045 | 0.604 | 89.534 | 89.261 | R |
| (90377) Sedna | 3026.60160 | 20.666 | 0.053 | 3026.59810 | 1.153 | 0.604 | 89.534 | 89.262 | R |
| (90377) Sedna | 3027.59692 | 20.579 | 0.107 | 3027.59333 | 1.065 | 0.607 | 89.533 | 89.277 | R |
| (90377) Sedna | 3027.61992 | 20.592 | 0.120 | 3027.61633 | 1.078 | 0.607 | 89.533 | 89.278 | R |
| (90377) Sedna | 3029.61150 | 20.484 | 0.104 | 3029.60773 | 0.970 | 0.612 | 89.531 | 89.309 | R |
| (90377) Sedna | 3032.56053 | 20.630 | 0.052 | 3032.55649 | 1.115 | 0.619 | 89.528 | 89.355 | R |
| (90377) Sedna | 3032.58137 | 20.579 | 0.053 | 3032.57733 | 1.064 | 0.619 | 89.528 | 89.356 | R |
| (90377) Sedna | 3033.55789 | 20.621 | 0.087 | 3033.55376 | 1.105 | 0.621 | 89.528 | 89.371 | R |
| (90377) Sedna | 3033.57315 | 20.533 | 0.072 | 3033.56902 | 1.017 | 0.621 | 89.527 | 89.371 | R |
| (90377) Sedna | 3040.53464 | 20.768 | 0.116 | 3040.52986 | 1.250 | 0.630 | 89.521 | 89.483 | R |
| (90377) Sedna | 3040.53464 | 20.777 | 0.116 | 3040.52986 | 1.259 | 0.630 | 89.521 | 89.483 | R |
| (90377) Sedna | 3041.53090 | 20.688 | 0.121 | 3041.52603 | 1.169 | 0.630 | 89.520 | 89.499 | R |
| (90377) Sedna | 3041.53090 | 20.710 | 0.121 | 3041.52603 | 1.191 | 0.630 | 89.520 | 89.499 | R |
| (90377) Sedna | 3043.54005 | 20.760 | 0.092 | 3043.53499 | 1.241 | 0.630 | 89.518 | 89.531 | R |
| (90377) Sedna | 3047.53037 | 20.703 | 0.069 | 3047.52494 | 1.182 | 0.629 | 89.515 | 89.596 | R |
| (90377) Sedna | 3051.53284 | 20.572 | 0.118 | 3051.52704 | 1.050 | 0.625 | 89.511 | 89.660 | R |
| (90377) Sedna | 3054.52011 | 20.636 | 0.148 | 3054.51404 | 1.113 | 0.619 | 89.508 | 89.707 | R |
| (90377) Sedna | 3058.52355 | 20.641 | 0.087 | 3058.51712 | 1.116 | 0.610 | 89.505 | 89.770 | R |
| (90377) Sedna | 3060.52117 | 20.589 | 0.120 | 3060.51456 | 1.063 | 0.604 | 89.503 | 89.800 | R |
| (90377) Sedna | 3184.90517 | 20.636 | 0.084 | 3184.89712 | 1.107 | 0.491 | 89.389 | 90.050 | R |
| (90377) Sedna | 3190.88051 | 20.498 | 0.171 | 3190.87294 | 0.971 | 0.529 | 89.384 | 89.968 | R |
| (90377) Sedna | 3190.88051 | 20.498 | 0.171 | 3190.87294 | 0.971 | 0.529 | 89.384 | 89.968 | R |
| (90377) Sedna | 3192.90844 | 20.342 | 0.106 | 3192.90103 | 0.816 | 0.542 | 89.382 | 89.938 | R |
| (90377) Sedna | 3194.88242 | 20.657 | 0.127 | 3194.87518 | 1.132 | 0.553 | 89.380 | 89.909 | R |
| (90377) Sedna | 3202.84872 | 20.736 | 0.126 | 3202.84219 | 1.214 | 0.592 | 89.373 | 89.786 | R |
| (90377) Sedna | 3202.85703 | 20.909 | 0.103 | 3202.85050 | 1.387 | 0.592 | 89.373 | 89.786 | R |
| (90377) Sedna | 3204.84973 | 20.762 | 0.115 | 3204.84339 | 1.241 | 0.601 | 89.371 | 89.754 | R |
| (90377) Sedna | 3210.88164 | 20.682 | 0.099 | 3210.87587 | 1.163 | 0.622 | 89.366 | 89.655 | R |
| (90377) Sedna | 3212.90671 | 20.784 | 0.092 | 3212.90114 | 1.266 | 0.628 | 89.364 | 89.621 | R |
| (90377) Sedna | 3217.93400 | 20.368 | 0.139 | 3217.92892 | 0.852 | 0.640 | 89.359 | 89.535 | R |
| (90377) Sedna | 3218.85009 | 20.701 | 0.205 | 3218.84511 | 1.186 | 0.642 | 89.359 | 89.519 | R |
| (90377) Sedna | 3224.87923 | 20.871 | 0.163 | 3224.87485 | 1.358 | 0.649 | 89.353 | 89.415 | R |
| (90377) Sedna | 3224.91227 | 20.642 | 0.119 | 3224.90789 | 1.129 | 0.649 | 89.353 | 89.414 | R |
| (90377) Sedna | 3229.85973 | 20.610 | 0.055 | 3229.85585 | 1.100 | 0.651 | 89.349 | 89.327 | R |
| (90377) Sedna | 3230.79406 | 20.661 | 0.056 | 3230.79028 | 1.151 | 0.650 | 89.348 | 89.311 | R |
| (90377) Sedna | 3230.83936 | 20.830 | 0.061 | 3230.83558 | 1.320 | 0.650 | 89.348 | 89.310 | R |
| (90377) Sedna | 3231.84900 | 20.749 | 0.073 | 3231.84532 | 1.240 | 0.650 | 89.347 | 89.293 | R |
| (90377) Sedna | 3231.88446 | 20.696 | 0.073 | 3231.88079 | 1.187 | 0.650 | 89.347 | 89.292 | R |
| (90377) Sedna | 3232.79646 | 20.733 | 0.177 | 3232.79288 | 1.224 | 0.649 | 89.346 | 89.276 | R |
| (90377) Sedna | 3238.81246 | 20.669 | 0.074 | 3238.80948 | 1.163 | 0.642 | 89.340 | 89.172 | R |
| (90377) Sedna | 3238.85410 | 20.578 | 0.064 | 3238.85113 | 1.072 | 0.642 | 89.340 | 89.171 | R |
| (90377) Sedna | 3239.85190 | 20.707 | 0.057 | 3239.84903 | 1.201 | 0.640 | 89.339 | 89.154 | R |
| (90377) Sedna | 3239.89609 | 20.650 | 0.055 | 3239.89322 | 1.144 | 0.640 | 89.339 | 89.153 | R |
| (90377) Sedna | 3240.79675 | 20.651 | 0.064 | 3240.79397 | 1.146 | 0.638 | 89.339 | 89.137 | R |
| (90377) Sedna | 3240.84081 | 20.625 | 0.071 | 3240.83804 | 1.120 | 0.638 | 89.339 | 89.137 | R |
| (90377) Sedna | 3241.91180 | 20.659 | 0.054 | 3241.90913 | 1.154 | 0.636 | 89.338 | 89.118 | R |
| (90377) Sedna | 3242.79678 | 20.526 | 0.067 | 3242.79420 | 1.021 | 0.633 | 89.337 | 89.103 | R |
| (90377) Sedna | 3242.79678 | 20.601 | 0.067 | 3242.79420 | 1.096 | 0.633 | 89.337 | 89.103 | R |
| (90377) Sedna | 3243.79670 | 20.562 | 0.113 | 3243.79422 | 1.058 | 0.631 | 89.336 | 89.086 | R |
| (90377) Sedna | 3245.83226 | 20.471 | 0.098 | 3245.82998 | 0.968 | 0.625 | 89.334 | 89.052 | R |
| (90377) Sedna | 3246.77475 | 20.645 | 0.178 | 3246.77256 | 1.142 | 0.622 | 89.333 | 89.036 | R |
| (90377) Sedna | 3247.78812 | 20.745 | 0.235 | 3247.78602 | 1.243 | 0.619 | 89.332 | 89.019 | R |
| (90377) Sedna | 3248.78905 | 20.603 | 0.127 | 3248.78705 | 1.101 | 0.616 | 89.331 | 89.002 | R |
| (90377) Sedna | 3250.76009 | 20.293 | 0.126 | 3250.75828 | 0.792 | 0.609 | 89.330 | 88.970 | R |
| (90377) Sedna | 3250.79460 | 20.437 | 0.115 | 3250.79279 | 0.936 | 0.608 | 89.329 | 88.969 | R |
| (90377) Sedna | 3251.75349 | 20.812 | 0.242 | 3251.75177 | 1.311 | 0.605 | 89.329 | 88.954 | R |
| (90377) Sedna | 3251.78968 | 20.550 | 0.237 | 3251.78796 | 1.049 | 0.605 | 89.329 | 88.953 | R |
| (90377) Sedna | 3254.81083 | 20.562 | 0.165 | 3254.80940 | 1.062 | 0.592 | 89.326 | 88.905 | R |
| (90377) Sedna | 3254.84264 | 20.576 | 0.091 | 3254.84121 | 1.077 | 0.592 | 89.326 | 88.904 | R |
| (90377) Sedna | 3256.83958 | 20.464 | 0.074 | 3256.83833 | 0.965 | 0.582 | 89.324 | 88.873 | R |
| (90377) Sedna | 3256.89472 | 20.417 | 0.062 | 3256.89348 | 0.918 | 0.582 | 89.324 | 88.872 | R |
| (90377) Sedna | 3257.83125 | 20.572 | 0.054 | 3257.83009 | 1.074 | 0.577 | 89.323 | 88.857 | R |
| (90377) Sedna | 3257.86913 | 20.729 | 0.066 | 3257.86797 | 1.231 | 0.577 | 89.323 | 88.857 | R |
| (90377) Sedna | 3258.82338 | 20.717 | 0.054 | 3258.82231 | 1.219 | 0.572 | 89.322 | 88.842 | R |
| (90377) Sedna | 3260.77321 | 20.642 | 0.074 | 3260.77231 | 1.145 | 0.562 | 89.320 | 88.812 | R |
| (90377) Sedna | 3260.82378 | 20.792 | 0.102 | 3260.82288 | 1.295 | 0.562 | 89.320 | 88.811 | R |
| (90377) Sedna | 3262.80625 | 20.581 | 0.055 | 3262.80553 | 1.085 | 0.551 | 89.319 | 88.782 | R |
| (90377) Sedna | 3263.72745 | 20.629 | 0.048 | 3263.72680 | 1.133 | 0.546 | 89.318 | 88.768 | R |
| (90377) Sedna | 3263.81599 | 20.577 | 0.040 | 3263.81535 | 1.081 | 0.545 | 89.318 | 88.767 | R |
| (90377) Sedna | 3264.78850 | 20.689 | 0.046 | 3264.78794 | 1.193 | 0.539 | 89.317 | 88.753 | R |
| (90377) Sedna | 3265.81899 | 20.669 | 0.054 | 3265.81852 | 1.174 | 0.533 | 89.316 | 88.738 | R |
| (90377) Sedna | 3265.84149 | 20.548 | 0.042 | 3265.84102 | 1.053 | 0.533 | 89.316 | 88.737 | R |
| (90377) Sedna | 3266.68737 | 20.446 | 0.079 | 3266.68697 | 0.951 | 0.528 | 89.315 | 88.725 | R |
| (90377) Sedna | 3266.78043 | 20.593 | 0.066 | 3266.78004 | 1.098 | 0.527 | 89.315 | 88.724 | R |
| (90377) Sedna | 3267.74472 | 20.585 | 0.133 | 3267.74441 | 1.091 | 0.521 | 89.314 | 88.710 | R |
| (90377) Sedna | 3267.82486 | 20.693 | 0.069 | 3267.82455 | 1.199 | 0.520 | 89.314 | 88.709 | R |
| (90377) Sedna | 3269.78697 | 20.680 | 0.061 | 3269.78682 | 1.186 | 0.507 | 89.312 | 88.682 | R |
| (90377) Sedna | 3269.84267 | 20.635 | 0.054 | 3269.84252 | 1.141 | 0.507 | 89.312 | 88.681 | R |
| (90377) Sedna | 3270.80504 | 20.728 | 0.049 | 3270.80497 | 1.235 | 0.500 | 89.311 | 88.668 | R |
| (90377) Sedna | 3270.86085 | 20.551 | 0.045 | 3270.86078 | 1.058 | 0.500 | 89.311 | 88.668 | R |
| (90377) Sedna | 3271.81547 | 20.691 | 0.048 | 3271.81548 | 1.198 | 0.493 | 89.310 | 88.655 | R |
| (90377) Sedna | 3271.85614 | 20.629 | 0.045 | 3271.85615 | 1.136 | 0.493 | 89.310 | 88.654 | R |
| (90377) Sedna | 3273.76586 | 20.566 | 0.080 | 3273.76601 | 1.074 | 0.480 | 89.309 | 88.630 | R |
| (90377) Sedna | 3273.81025 | 20.538 | 0.070 | 3273.81041 | 1.046 | 0.479 | 89.309 | 88.629 | R |
| (90377) Sedna | 3274.74131 | 20.515 | 0.090 | 3274.74154 | 1.023 | 0.472 | 89.308 | 88.617 | R |
| (90377) Sedna | 3274.78021 | 20.529 | 0.071 | 3274.78044 | 1.037 | 0.472 | 89.308 | 88.617 | R |
| (90377) Sedna | 3275.80602 | 20.764 | 0.143 | 3275.80632 | 1.272 | 0.465 | 89.307 | 88.604 | R |
| (90377) Sedna | 3276.82478 | 20.643 | 0.139 | 3276.82516 | 1.152 | 0.457 | 89.306 | 88.591 | R |
| (90377) Sedna | 3276.87390 | 20.929 | 0.181 | 3276.87428 | 1.438 | 0.457 | 89.306 | 88.591 | R |
| (90377) Sedna | 3281.81626 | 20.257 | 0.124 | 3281.81697 | 0.767 | 0.418 | 89.301 | 88.533 | R |
| (90377) Sedna | 3283.76780 | 20.765 | 0.059 | 3283.76864 | 1.276 | 0.402 | 89.300 | 88.512 | R |
| (90377) Sedna | 3283.81118 | 20.710 | 0.059 | 3283.81202 | 1.221 | 0.401 | 89.299 | 88.511 | R |
| (90377) Sedna | 3284.74106 | 20.594 | 0.090 | 3284.74196 | 1.105 | 0.393 | 89.299 | 88.501 | R |
| (90377) Sedna | 3284.79245 | 20.511 | 0.095 | 3284.79335 | 1.022 | 0.393 | 89.299 | 88.501 | R |
| (90377) Sedna | 3289.74238 | 20.660 | 0.055 | 3289.74356 | 1.172 | 0.351 | 89.294 | 88.452 | R |
| (90377) Sedna | 3301.71454 | 20.616 | 0.072 | 3301.71625 | 1.131 | 0.244 | 89.283 | 88.360 | R |
| (90377) Sedna | 3301.75366 | 20.700 | 0.087 | 3301.75537 | 1.215 | 0.244 | 89.283 | 88.360 | R |
| (90377) Sedna | 3306.67109 | 20.603 | 0.117 | 3306.67295 | 1.119 | 0.202 | 89.279 | 88.334 | R |
| (90377) Sedna | 3314.66362 | 20.626 | 0.071 | 3314.66564 | 1.142 | 0.148 | 89.271 | 88.306 | R |
| (90377) Sedna | 3316.65488 | 20.466 | 0.058 | 3316.65693 | 0.983 | 0.140 | 89.270 | 88.302 | R |
| (90377) Sedna | 3568.90100 | 20.619 | 0.073 | 3568.89641 | 1.113 | 0.596 | 89.042 | 89.452 | R |
| (90377) Sedna | 3570.86283 | 20.626 | 0.168 | 3570.85842 | 1.121 | 0.604 | 89.040 | 89.420 | R |
| (90377) Sedna | 3573.88737 | 20.491 | 0.142 | 3573.88324 | 0.987 | 0.615 | 89.037 | 89.371 | R |
| (90377) Sedna | 3574.86711 | 20.702 | 0.163 | 3574.86308 | 1.199 | 0.619 | 89.036 | 89.355 | R |
| (90377) Sedna | 3575.88729 | 20.784 | 0.120 | 3575.88335 | 1.281 | 0.622 | 89.035 | 89.338 | R |
| (90377) Sedna | 3577.85530 | 20.828 | 0.271 | 3577.85155 | 1.326 | 0.628 | 89.034 | 89.305 | R |
| (90377) Sedna | 3585.84578 | 20.676 | 0.072 | 3585.84282 | 1.177 | 0.646 | 89.027 | 89.168 | R |
| (90377) Sedna | 3587.86438 | 20.765 | 0.248 | 3587.86163 | 1.267 | 0.649 | 89.025 | 89.133 | R |
| (90377) Sedna | 3588.86680 | 20.774 | 0.118 | 3588.86415 | 1.277 | 0.650 | 89.024 | 89.116 | R |
| (90377) Sedna | 3590.91261 | 20.654 | 0.052 | 3590.91016 | 1.158 | 0.652 | 89.022 | 89.080 | R |
| (90377) Sedna | 3591.84338 | 20.625 | 0.064 | 3591.84103 | 1.129 | 0.652 | 89.021 | 89.064 | R |
| (90377) Sedna | 3592.89015 | 20.655 | 0.057 | 3592.88790 | 1.159 | 0.653 | 89.020 | 89.046 | R |
| (90377) Sedna | 3593.86575 | 20.714 | 0.064 | 3593.86360 | 1.219 | 0.653 | 89.019 | 89.028 | R |
| (90377) Sedna | 3596.85621 | 20.806 | 0.113 | 3596.85436 | 1.312 | 0.653 | 89.017 | 88.976 | R |
| (90377) Sedna | 3599.87073 | 20.407 | 0.219 | 3599.86919 | 0.915 | 0.651 | 89.014 | 88.924 | R |
| (90377) Sedna | 3600.83703 | 20.819 | 0.202 | 3600.83558 | 1.327 | 0.650 | 89.013 | 88.907 | R |
| (90377) Sedna | 3603.88727 | 20.747 | 0.102 | 3603.88613 | 1.256 | 0.645 | 89.010 | 88.854 | R |
| (90377) Sedna | 3608.84012 | 20.789 | 0.141 | 3608.83947 | 1.301 | 0.635 | 89.006 | 88.769 | R |
| (90377) Sedna | 3614.77460 | 20.702 | 0.063 | 3614.77452 | 1.216 | 0.617 | 89.001 | 88.669 | R |
| (90377) Sedna | 3617.79431 | 20.728 | 0.077 | 3617.79452 | 1.243 | 0.606 | 88.998 | 88.620 | R |
| (90377) Sedna | 3618.81696 | 20.688 | 0.066 | 3618.81727 | 1.204 | 0.602 | 88.997 | 88.603 | R |
| (90377) Sedna | 3619.81067 | 20.804 | 0.068 | 3619.81107 | 1.320 | 0.598 | 88.996 | 88.587 | R |
| (90377) Sedna | 3620.72723 | 20.626 | 0.063 | 3620.72771 | 1.143 | 0.593 | 88.995 | 88.573 | R |
| (90377) Sedna | 3621.74292 | 20.545 | 0.061 | 3621.74350 | 1.062 | 0.589 | 88.994 | 88.557 | R |
| (90377) Sedna | 3622.73681 | 20.426 | 0.108 | 3622.73748 | 0.943 | 0.584 | 88.993 | 88.541 | R |
| (90377) Sedna | 3624.73882 | 20.519 | 0.113 | 3624.73966 | 1.037 | 0.574 | 88.992 | 88.510 | R |
| (90377) Sedna | 3625.79728 | 20.644 | 0.055 | 3625.79822 | 1.163 | 0.568 | 88.991 | 88.494 | R |
| (90377) Sedna | 3626.78062 | 20.682 | 0.144 | 3626.78164 | 1.201 | 0.563 | 88.990 | 88.479 | R |
| (90377) Sedna | 3628.80457 | 20.688 | 0.133 | 3628.80577 | 1.208 | 0.552 | 88.988 | 88.449 | R |
| (90377) Sedna | 3629.77667 | 20.352 | 0.086 | 3629.77795 | 0.872 | 0.546 | 88.987 | 88.434 | R |
| (90377) Sedna | 3630.72142 | 20.729 | 0.228 | 3630.72278 | 1.250 | 0.540 | 88.986 | 88.421 | R |
| (90377) Sedna | 3631.78522 | 20.794 | 0.181 | 3631.78667 | 1.315 | 0.534 | 88.985 | 88.405 | R |
| (90377) Sedna | 3637.80071 | 20.515 | 0.152 | 3637.80264 | 1.038 | 0.494 | 88.980 | 88.323 | R |
| (90377) Sedna | 3639.73292 | 20.698 | 0.049 | 3639.73499 | 1.222 | 0.480 | 88.978 | 88.297 | R |
| (90377) Sedna | 3643.70706 | 20.689 | 0.078 | 3643.70942 | 1.214 | 0.450 | 88.975 | 88.248 | R |
| (90377) Sedna | 3651.69419 | 20.586 | 0.058 | 3651.69706 | 1.114 | 0.385 | 88.967 | 88.159 | R |
| (90377) Sedna | 3653.63663 | 20.594 | 0.105 | 3653.63961 | 1.122 | 0.369 | 88.966 | 88.140 | R |
| (90377) Sedna | 3655.71600 | 20.564 | 0.062 | 3655.71909 | 1.093 | 0.350 | 88.964 | 88.120 | R |
| (90377) Sedna | 3657.77176 | 20.691 | 0.077 | 3657.77496 | 1.220 | 0.332 | 88.962 | 88.102 | R |
| (90377) Sedna | 3663.66216 | 20.271 | 0.210 | 3663.66563 | 0.801 | 0.280 | 88.957 | 88.056 | R |
| (90377) Sedna | 3665.71765 | 20.531 | 0.101 | 3665.72120 | 1.062 | 0.261 | 88.955 | 88.042 | R |
| (90377) Sedna | 3668.65608 | 20.527 | 0.050 | 3668.65973 | 1.058 | 0.236 | 88.952 | 88.024 | R |
| (90377) Sedna | 3670.67790 | 20.594 | 0.060 | 3670.68162 | 1.126 | 0.218 | 88.950 | 88.013 | R |
| (90377) Sedna | 3673.61945 | 20.650 | 0.065 | 3673.62325 | 1.182 | 0.194 | 88.948 | 87.999 | R |
| (90377) Sedna | 3675.66726 | 20.529 | 0.050 | 3675.67110 | 1.061 | 0.179 | 88.946 | 87.991 | R |
| (90377) Sedna | 3680.69337 | 20.581 | 0.071 | 3680.69730 | 1.114 | 0.148 | 88.941 | 87.976 | R |
| (90377) Sedna | 3682.60290 | 20.549 | 0.101 | 3682.60685 | 1.082 | 0.140 | 88.940 | 87.972 | R |
| (90377) Sedna | 3687.61875 | 20.315 | 0.079 | 3687.62273 | 0.848 | 0.133 | 88.935 | 87.967 | R |
| (90377) Sedna | 3693.59375 | 20.489 | 0.056 | 3693.59770 | 1.022 | 0.154 | 88.930 | 87.972 | R |
| (90377) Sedna | 3696.60825 | 20.497 | 0.058 | 3696.61217 | 1.030 | 0.173 | 88.927 | 87.978 | R |
| (90377) Sedna | 3700.64297 | 20.506 | 0.041 | 3700.64682 | 1.039 | 0.203 | 88.923 | 87.990 | R |
| (90377) Sedna | 3704.61252 | 20.574 | 0.045 | 3704.61627 | 1.106 | 0.237 | 88.920 | 88.007 | R |
| (90377) Sedna | 3707.61356 | 20.646 | 0.065 | 3707.61722 | 1.178 | 0.263 | 88.917 | 88.023 | R |
| (90377) Sedna | 3709.57576 | 20.540 | 0.086 | 3709.57935 | 1.072 | 0.281 | 88.915 | 88.035 | R |
| (90377) Sedna | 3711.59056 | 20.550 | 0.059 | 3711.59407 | 1.082 | 0.299 | 88.914 | 88.048 | R |
| (90377) Sedna | 3713.56215 | 20.591 | 0.065 | 3713.56558 | 1.122 | 0.317 | 88.912 | 88.062 | R |
| (90377) Sedna | 3715.62144 | 20.726 | 0.105 | 3715.62478 | 1.257 | 0.335 | 88.910 | 88.077 | R |
| (90377) Sedna | 3719.61505 | 20.293 | 0.119 | 3719.61820 | 0.823 | 0.369 | 88.906 | 88.110 | R |
| (90377) Sedna | 3725.60977 | 20.567 | 0.046 | 3725.61259 | 1.096 | 0.419 | 88.901 | 88.168 | R |
| (90377) Sedna | 3747.55046 | 20.631 | 0.127 | 3747.55170 | 1.154 | 0.565 | 88.881 | 88.441 | R |
| (90377) Sedna | 3749.55022 | 20.609 | 0.152 | 3749.55130 | 1.131 | 0.575 | 88.880 | 88.470 | R |
| (90377) Sedna | 3750.58404 | 20.721 | 0.124 | 3750.58503 | 1.243 | 0.579 | 88.879 | 88.485 | R |
| (90377) Sedna | 3751.55175 | 20.593 | 0.066 | 3751.55266 | 1.114 | 0.584 | 88.878 | 88.499 | R |
| (90377) Sedna | 3753.53740 | 20.811 | 0.136 | 3753.53814 | 1.332 | 0.592 | 88.876 | 88.529 | R |
| (90377) Sedna | 3755.55826 | 20.591 | 0.065 | 3755.55882 | 1.111 | 0.599 | 88.874 | 88.559 | R |
| (90377) Sedna | 3757.55720 | 20.381 | 0.060 | 3757.55758 | 0.900 | 0.606 | 88.873 | 88.590 | R |
| (90377) Sedna | 3761.58782 | 20.679 | 0.052 | 3761.58784 | 1.197 | 0.618 | 88.869 | 88.653 | R |
| (90377) Sedna | 3764.58921 | 20.643 | 0.073 | 3764.58895 | 1.160 | 0.625 | 88.866 | 88.700 | R |
| (90377) Sedna | 3765.59003 | 20.765 | 0.078 | 3765.58968 | 1.281 | 0.627 | 88.865 | 88.716 | R |
| | | | | # | | | | | |
| (90482) Orcus | 3057.66936 | 19.145 | 0.039 | 3057.66936 | 2.411 | 0.393 | 47.608 | 46.674 | V |
| (90482) Orcus | 3057.67071 | 19.182 | 0.039 | 3057.67071 | 2.448 | 0.393 | 47.608 | 46.674 | V |
| (90482) Orcus | 3057.67214 | 18.289 | 0.044 | 3057.67214 | 1.555 | 0.393 | 47.608 | 46.674 | I |
| (90482) Orcus | 3057.67349 | 18.288 | 0.044 | 3057.67349 | 1.554 | 0.393 | 47.608 | 46.674 | I |
| (90482) Orcus | 3061.62703 | 19.678 | 0.020 | 3061.62697 | 2.944 | 0.431 | 47.608 | 46.685 | B |
| (90482) Orcus | 3061.62981 | 19.018 | 0.041 | 3061.62975 | 2.284 | 0.431 | 47.608 | 46.685 | V |
| (90482) Orcus | 3061.63120 | 18.635 | 0.033 | 3061.63114 | 1.901 | 0.431 | 47.608 | 46.685 | R |
| (90482) Orcus | 3061.63264 | 18.243 | 0.046 | 3061.63258 | 1.509 | 0.431 | 47.608 | 46.685 | I |
| (90482) Orcus | 3061.67234 | 19.670 | 0.017 | 3061.67227 | 2.936 | 0.431 | 47.608 | 46.685 | B |
| (90482) Orcus | 3061.67513 | 19.020 | 0.039 | 3061.67506 | 2.286 | 0.431 | 47.608 | 46.685 | V |
| (90482) Orcus | 3061.67652 | 18.640 | 0.040 | 3061.67645 | 1.906 | 0.431 | 47.608 | 46.685 | R |
| (90482) Orcus | 3061.67796 | 18.259 | 0.046 | 3061.67789 | 1.525 | 0.431 | 47.608 | 46.685 | I |
| (90482) Orcus | 3065.70468 | 19.269 | 0.160 | 3065.70452 | 2.534 | 0.479 | 47.609 | 46.701 | V |
| (90482) Orcus | 3065.70603 | 19.080 | 0.104 | 3065.70587 | 2.345 | 0.479 | 47.609 | 46.701 | V |
| (90482) Orcus | 3065.70746 | 18.288 | 0.172 | 3065.70730 | 1.553 | 0.479 | 47.609 | 46.701 | I |
| (90482) Orcus | 3065.70881 | 18.154 | 0.133 | 3065.70865 | 1.419 | 0.479 | 47.609 | 46.701 | I |
| (90482) Orcus | 3069.60146 | 19.616 | 0.061 | 3069.60119 | 2.880 | 0.532 | 47.610 | 46.721 | B |
| (90482) Orcus | 3069.60425 | 19.066 | 0.102 | 3069.60398 | 2.330 | 0.532 | 47.610 | 46.721 | V |
| (90482) Orcus | 3069.60567 | 18.304 | 0.104 | 3069.60540 | 1.568 | 0.532 | 47.610 | 46.721 | I |
| (90482) Orcus | 3069.64677 | 19.713 | 0.066 | 3069.64649 | 2.977 | 0.532 | 47.610 | 46.722 | B |
| (90482) Orcus | 3069.65099 | 18.369 | 0.108 | 3069.65071 | 1.633 | 0.532 | 47.610 | 46.722 | I |
| (90482) Orcus | 3073.67904 | 19.750 | 0.043 | 3073.67862 | 3.013 | 0.590 | 47.610 | 46.747 | B |
| (90482) Orcus | 3073.68183 | 19.076 | 0.059 | 3073.68141 | 2.339 | 0.590 | 47.610 | 46.747 | V |
| (90482) Orcus | 3073.68325 | 18.342 | 0.053 | 3073.68283 | 1.605 | 0.590 | 47.610 | 46.747 | I |
| (90482) Orcus | 3073.72310 | 19.708 | 0.050 | 3073.72268 | 2.971 | 0.591 | 47.610 | 46.747 | B |
| (90482) Orcus | 3073.72588 | 19.160 | 0.085 | 3073.72546 | 2.423 | 0.591 | 47.610 | 46.747 | V |
| (90482) Orcus | 3073.72731 | 18.433 | 0.070 | 3073.72689 | 1.696 | 0.591 | 47.610 | 46.747 | I |
| (90482) Orcus | 3076.64779 | 19.786 | 0.024 | 3076.64725 | 3.048 | 0.634 | 47.611 | 46.768 | B |
| (90482) Orcus | 3076.65059 | 19.088 | 0.043 | 3076.65005 | 2.350 | 0.634 | 47.611 | 46.768 | V |
| (90482) Orcus | 3076.65202 | 18.397 | 0.047 | 3076.65148 | 1.659 | 0.634 | 47.611 | 46.768 | I |
| (90482) Orcus | 3076.70597 | 19.882 | 0.035 | 3076.70542 | 3.144 | 0.635 | 47.611 | 46.768 | B |
| (90482) Orcus | 3076.70875 | 19.316 | 0.062 | 3076.70820 | 2.578 | 0.635 | 47.611 | 46.768 | V |
| (90482) Orcus | 3076.71018 | 18.495 | 0.060 | 3076.70963 | 1.757 | 0.635 | 47.611 | 46.768 | I |
| (90482) Orcus | 3080.60463 | 19.770 | 0.017 | 3080.60390 | 3.030 | 0.693 | 47.612 | 46.800 | B |
| (90482) Orcus | 3080.60742 | 19.146 | 0.037 | 3080.60669 | 2.406 | 0.693 | 47.612 | 46.800 | V |
| (90482) Orcus | 3080.60885 | 18.506 | 0.046 | 3080.60812 | 1.766 | 0.693 | 47.612 | 46.800 | I |
| (90482) Orcus | 3080.64503 | 19.829 | 0.019 | 3080.64430 | 3.089 | 0.694 | 47.612 | 46.800 | B |
| (90482) Orcus | 3080.64782 | 19.186 | 0.042 | 3080.64709 | 2.446 | 0.694 | 47.612 | 46.800 | V |
| (90482) Orcus | 3080.64925 | 18.499 | 0.048 | 3080.64852 | 1.759 | 0.694 | 47.612 | 46.800 | I |
| (90482) Orcus | 3088.48572 | 19.689 | 0.064 | 3088.48456 | 2.946 | 0.808 | 47.613 | 46.874 | B |
| (90482) Orcus | 3088.48854 | 19.094 | 0.060 | 3088.48738 | 2.351 | 0.808 | 47.613 | 46.874 | V |
| (90482) Orcus | 3088.48996 | 18.337 | 0.067 | 3088.48880 | 1.594 | 0.808 | 47.613 | 46.874 | I |
| (90482) Orcus | 3088.52831 | 19.750 | 0.032 | 3088.52715 | 3.007 | 0.808 | 47.613 | 46.875 | B |
| (90482) Orcus | 3088.53110 | 19.174 | 0.037 | 3088.52994 | 2.431 | 0.808 | 47.613 | 46.875 | V |
| (90482) Orcus | 3088.53253 | 18.374 | 0.042 | 3088.53137 | 1.631 | 0.808 | 47.613 | 46.875 | I |
| (90482) Orcus | 3092.54502 | 19.768 | 0.022 | 3092.54361 | 3.023 | 0.864 | 47.613 | 46.918 | B |
| (90482) Orcus | 3092.54781 | 19.096 | 0.040 | 3092.54640 | 2.351 | 0.864 | 47.613 | 46.918 | V |
| (90482) Orcus | 3092.54923 | 18.436 | 0.049 | 3092.54782 | 1.691 | 0.864 | 47.613 | 46.918 | I |
| (90482) Orcus | 3092.59120 | 19.768 | 0.020 | 3092.58979 | 3.023 | 0.864 | 47.613 | 46.919 | B |
| (90482) Orcus | 3092.59399 | 19.161 | 0.040 | 3092.59258 | 2.416 | 0.864 | 47.613 | 46.919 | V |
| (90482) Orcus | 3092.59541 | 18.403 | 0.045 | 3092.59400 | 1.658 | 0.864 | 47.613 | 46.919 | I |
| (90482) Orcus | 3099.53128 | 19.779 | 0.141 | 3099.52939 | 3.030 | 0.953 | 47.615 | 47.001 | B |
| (90482) Orcus | 3099.53407 | 19.183 | 0.101 | 3099.53218 | 2.434 | 0.953 | 47.615 | 47.001 | V |
| (90482) Orcus | 3099.53549 | 18.121 | 0.113 | 3099.53360 | 1.372 | 0.953 | 47.615 | 47.001 | I |
| (90482) Orcus | 3103.63863 | 19.932 | 0.047 | 3103.63643 | 3.180 | 1.001 | 47.615 | 47.055 | B |
| (90482) Orcus | 3103.64142 | 19.270 | 0.073 | 3103.63922 | 2.518 | 1.001 | 47.615 | 47.055 | V |
| (90482) Orcus | 3103.64285 | 18.551 | 0.070 | 3103.64065 | 1.799 | 1.001 | 47.615 | 47.055 | I |
| (90482) Orcus | 3103.67271 | 19.923 | 0.059 | 3103.67051 | 3.171 | 1.001 | 47.615 | 47.055 | B |
| (90482) Orcus | 3103.67550 | 19.337 | 0.088 | 3103.67330 | 2.585 | 1.001 | 47.615 | 47.055 | V |
| (90482) Orcus | 3103.67693 | 18.543 | 0.082 | 3103.67473 | 1.791 | 1.001 | 47.615 | 47.055 | I |
| (90482) Orcus | 3107.57381 | 20.001 | 0.021 | 3107.57130 | 3.247 | 1.042 | 47.616 | 47.108 | B |
| (90482) Orcus | 3107.57661 | 19.334 | 0.042 | 3107.57410 | 2.580 | 1.042 | 47.616 | 47.108 | V |
| (90482) Orcus | 3107.57804 | 18.632 | 0.054 | 3107.57553 | 1.878 | 1.042 | 47.616 | 47.108 | I |
| (90482) Orcus | 3107.61660 | 19.981 | 0.020 | 3107.61409 | 3.227 | 1.043 | 47.616 | 47.109 | B |
| (90482) Orcus | 3107.61940 | 19.310 | 0.046 | 3107.61689 | 2.556 | 1.043 | 47.616 | 47.109 | V |
| (90482) Orcus | 3107.62083 | 18.521 | 0.057 | 3107.61832 | 1.767 | 1.043 | 47.616 | 47.109 | I |
| (90482) Orcus | 3111.50561 | 19.802 | 0.037 | 3111.50278 | 3.045 | 1.080 | 47.616 | 47.164 | B |
| (90482) Orcus | 3111.50834 | 19.197 | 0.049 | 3111.50551 | 2.440 | 1.080 | 47.616 | 47.164 | V |
| (90482) Orcus | 3111.50971 | 18.419 | 0.061 | 3111.50688 | 1.662 | 1.080 | 47.616 | 47.164 | I |
| (90482) Orcus | 3111.56506 | 19.851 | 0.037 | 3111.56222 | 3.094 | 1.080 | 47.616 | 47.165 | B |
| (90482) Orcus | 3111.56779 | 19.264 | 0.051 | 3111.56495 | 2.507 | 1.081 | 47.616 | 47.165 | V |
| (90482) Orcus | 3111.56916 | 18.451 | 0.062 | 3111.56632 | 1.694 | 1.081 | 47.616 | 47.165 | I |
| (90482) Orcus | 3121.52210 | 19.718 | 0.037 | 3121.51840 | 2.954 | 1.156 | 47.618 | 47.315 | B |
| (90482) Orcus | 3121.52620 | 18.424 | 0.128 | 3121.52250 | 1.660 | 1.156 | 47.618 | 47.315 | I |
| (90482) Orcus | 3121.57440 | 19.750 | 0.042 | 3121.57069 | 2.986 | 1.156 | 47.618 | 47.316 | B |
| (90482) Orcus | 3121.57713 | 19.124 | 0.106 | 3121.57342 | 2.360 | 1.156 | 47.618 | 47.316 | V |
| (90482) Orcus | 3121.57850 | 18.404 | 0.106 | 3121.57479 | 1.640 | 1.156 | 47.618 | 47.316 | I |
| (90482) Orcus | 3125.48453 | 19.736 | 0.043 | 3125.48047 | 2.969 | 1.178 | 47.619 | 47.377 | B |
| (90482) Orcus | 3125.48726 | 19.082 | 0.070 | 3125.48320 | 2.315 | 1.178 | 47.619 | 47.377 | V |
| (90482) Orcus | 3125.48864 | 18.292 | 0.071 | 3125.48458 | 1.525 | 1.178 | 47.619 | 47.377 | I |
| (90482) Orcus | 3125.53216 | 19.804 | 0.047 | 3125.52809 | 3.037 | 1.178 | 47.619 | 47.378 | B |
| (90482) Orcus | 3125.53490 | 19.191 | 0.079 | 3125.53083 | 2.424 | 1.178 | 47.619 | 47.378 | V |
| (90482) Orcus | 3125.53627 | 18.437 | 0.080 | 3125.53220 | 1.670 | 1.178 | 47.619 | 47.378 | I |
| (90482) Orcus | 3129.57127 | 19.755 | 0.064 | 3129.56683 | 2.985 | 1.195 | 47.619 | 47.443 | B |
| (90482) Orcus | 3129.57400 | 19.288 | 0.102 | 3129.56956 | 2.518 | 1.195 | 47.619 | 47.443 | V |
| (90482) Orcus | 3129.57537 | 18.420 | 0.069 | 3129.57093 | 1.650 | 1.195 | 47.619 | 47.443 | I |
| (90482) Orcus | 3129.61778 | 19.919 | 0.102 | 3129.61334 | 3.149 | 1.195 | 47.619 | 47.443 | B |
| (90482) Orcus | 3129.62051 | 19.292 | 0.151 | 3129.61607 | 2.522 | 1.195 | 47.619 | 47.443 | V |
| (90482) Orcus | 3129.62188 | 18.800 | 0.200 | 3129.61744 | 2.030 | 1.195 | 47.619 | 47.443 | I |
| (90482) Orcus | 3133.46242 | 19.852 | 0.025 | 3133.45762 | 3.079 | 1.207 | 47.620 | 47.505 | B |
| (90482) Orcus | 3133.46514 | 19.181 | 0.047 | 3133.46034 | 2.408 | 1.207 | 47.620 | 47.506 | V |
| (90482) Orcus | 3133.46653 | 18.493 | 0.063 | 3133.46173 | 1.720 | 1.207 | 47.620 | 47.506 | I |
| (90482) Orcus | 3133.50775 | 18.570 | 0.057 | 3133.50294 | 1.797 | 1.207 | 47.620 | 47.506 | I |
| (90482) Orcus | 3133.50775 | 19.197 | 0.040 | 3133.50294 | 2.424 | 1.207 | 47.620 | 47.506 | V |
| (90482) Orcus | 3133.50914 | 19.874 | 0.020 | 3133.50433 | 3.101 | 1.207 | 47.620 | 47.506 | B |
| (90482) Orcus | 3142.45300 | 19.772 | 0.043 | 3142.44735 | 2.993 | 1.215 | 47.621 | 47.652 | B |
| (90482) Orcus | 3142.45572 | 19.404 | 0.069 | 3142.45007 | 2.625 | 1.215 | 47.621 | 47.652 | V |
| (90482) Orcus | 3142.45710 | 18.503 | 0.071 | 3142.45145 | 1.724 | 1.215 | 47.621 | 47.652 | I |
| (90482) Orcus | 3142.48772 | 19.882 | 0.020 | 3142.48207 | 3.103 | 1.215 | 47.621 | 47.653 | B |
| (90482) Orcus | 3142.49046 | 19.188 | 0.040 | 3142.48481 | 2.409 | 1.215 | 47.621 | 47.653 | V |
| (90482) Orcus | 3142.49183 | 18.485 | 0.052 | 3142.48618 | 1.706 | 1.215 | 47.621 | 47.653 | I |
| (90482) Orcus | 3146.44584 | 19.765 | 0.269 | 3146.43981 | 2.983 | 1.211 | 47.622 | 47.717 | B |
| (90482) Orcus | 3146.44856 | 19.311 | 0.183 | 3146.44253 | 2.529 | 1.211 | 47.622 | 47.717 | V |
| (90482) Orcus | 3146.44992 | 18.733 | 0.224 | 3146.44389 | 1.951 | 1.211 | 47.622 | 47.717 | I |
| (90482) Orcus | 3146.48247 | 19.899 | 0.041 | 3146.47644 | 3.117 | 1.211 | 47.622 | 47.718 | B |
| (90482) Orcus | 3146.48519 | 19.306 | 0.066 | 3146.47916 | 2.524 | 1.211 | 47.622 | 47.718 | V |
| (90482) Orcus | 3146.48656 | 18.565 | 0.068 | 3146.48053 | 1.783 | 1.211 | 47.622 | 47.718 | I |
| (90482) Orcus | 3152.44815 | 20.002 | 0.084 | 3152.44156 | 3.215 | 1.195 | 47.623 | 47.814 | B |
| (90482) Orcus | 3152.45087 | 19.102 | 0.069 | 3152.44428 | 2.315 | 1.195 | 47.623 | 47.814 | V |
| (90482) Orcus | 3152.45223 | 18.496 | 0.074 | 3152.44564 | 1.709 | 1.195 | 47.623 | 47.814 | I |
| (90482) Orcus | 3152.49226 | 19.841 | 0.045 | 3152.48567 | 3.054 | 1.195 | 47.623 | 47.815 | B |
| (90482) Orcus | 3152.49499 | 19.196 | 0.092 | 3152.48840 | 2.409 | 1.195 | 47.623 | 47.815 | V |
| (90482) Orcus | 3152.49636 | 18.765 | 0.235 | 3152.48977 | 1.978 | 1.195 | 47.623 | 47.815 | I |
| (90482) Orcus | 3156.46892 | 19.655 | 0.047 | 3156.46197 | 2.865 | 1.178 | 47.624 | 47.878 | B |
| (90482) Orcus | 3156.47164 | 19.270 | 0.065 | 3156.46468 | 2.480 | 1.178 | 47.624 | 47.878 | V |
| (90482) Orcus | 3156.47301 | 18.448 | 0.066 | 3156.46605 | 1.658 | 1.178 | 47.624 | 47.878 | I |
| (90482) Orcus | 3156.51447 | 19.542 | 0.053 | 3156.50751 | 2.752 | 1.178 | 47.624 | 47.879 | B |
| (90482) Orcus | 3156.51720 | 19.259 | 0.080 | 3156.51024 | 2.469 | 1.178 | 47.624 | 47.879 | V |
| (90482) Orcus | 3156.51857 | 18.614 | 0.089 | 3156.51161 | 1.824 | 1.178 | 47.624 | 47.879 | I |
| (90482) Orcus | 3159.46065 | 19.242 | 0.080 | 3159.45343 | 2.450 | 1.163 | 47.624 | 47.925 | V |
| (90482) Orcus | 3159.46426 | 18.475 | 0.068 | 3159.45704 | 1.683 | 1.163 | 47.624 | 47.925 | I |
| (90482) Orcus | 3159.50952 | 19.386 | 0.124 | 3159.50229 | 2.594 | 1.162 | 47.624 | 47.926 | V |
| (90482) Orcus | 3159.51089 | 18.447 | 0.078 | 3159.50366 | 1.655 | 1.162 | 47.624 | 47.926 | I |
| (90482) Orcus | 3169.45449 | 19.466 | 0.055 | 3169.44640 | 2.667 | 1.092 | 47.626 | 48.075 | B |
| (90482) Orcus | 3169.45720 | 19.129 | 0.069 | 3169.44911 | 2.330 | 1.092 | 47.626 | 48.075 | V |
| (90482) Orcus | 3169.45858 | 18.564 | 0.128 | 3169.45049 | 1.765 | 1.092 | 47.626 | 48.075 | I |
| (90482) Orcus | 3169.50138 | 19.489 | 0.049 | 3169.49329 | 2.690 | 1.092 | 47.626 | 48.075 | B |
| (90482) Orcus | 3169.50411 | 19.066 | 0.063 | 3169.49602 | 2.267 | 1.092 | 47.626 | 48.075 | V |
| (90482) Orcus | 3169.50548 | 18.410 | 0.142 | 3169.49739 | 1.611 | 1.092 | 47.626 | 48.075 | I |
| (90482) Orcus | 3175.44653 | 19.862 | 0.116 | 3175.43795 | 3.059 | 1.038 | 47.627 | 48.159 | B |
| (90482) Orcus | 3175.44926 | 19.130 | 0.085 | 3175.44068 | 2.327 | 1.038 | 47.627 | 48.159 | V |
| (90482) Orcus | 3175.45063 | 18.864 | 0.245 | 3175.44205 | 2.061 | 1.038 | 47.627 | 48.159 | I |
| (90482) Orcus | 3175.47121 | 19.819 | 0.036 | 3175.46263 | 3.016 | 1.038 | 47.627 | 48.159 | B |
| (90482) Orcus | 3175.47393 | 19.214 | 0.052 | 3175.46535 | 2.411 | 1.038 | 47.627 | 48.159 | V |
| (90482) Orcus | 3175.47531 | 18.423 | 0.060 | 3175.46673 | 1.620 | 1.038 | 47.627 | 48.159 | I |
| (90482) Orcus | 3180.44673 | 19.801 | 0.104 | 3180.43777 | 2.995 | 0.986 | 47.627 | 48.225 | B |
| (90482) Orcus | 3180.44945 | 19.352 | 0.113 | 3180.44049 | 2.546 | 0.986 | 47.627 | 48.225 | V |
| (90482) Orcus | 3183.45657 | 19.932 | 0.057 | 3183.44739 | 3.125 | 0.952 | 47.628 | 48.262 | B |
| (90482) Orcus | 3183.45929 | 19.203 | 0.068 | 3183.45011 | 2.396 | 0.952 | 47.628 | 48.262 | V |
| (90482) Orcus | 3183.46066 | 18.461 | 0.080 | 3183.45148 | 1.654 | 0.952 | 47.628 | 48.263 | I |
| (90482) Orcus | 3189.45359 | 19.161 | 0.205 | 3189.44401 | 2.350 | 0.879 | 47.629 | 48.333 | V |
| (90482) Orcus | 3189.45496 | 18.520 | 0.188 | 3189.44538 | 1.709 | 0.879 | 47.629 | 48.333 | I |
| (90482) Orcus | 3193.45072 | 19.718 | 0.153 | 3193.44089 | 2.905 | 0.827 | 47.629 | 48.376 | B |
| | | | | # | | | | | |
| (50000) Quaoar | 2696.85404 | 19.452 | 0.112 | 2696.85404 | 3.072 | 1.302 | 43.413 | 43.490 | V |
| (50000) Quaoar | 2696.85540 | 19.451 | 0.107 | 2696.85540 | 3.071 | 1.302 | 43.413 | 43.490 | V |
| (50000) Quaoar | 2696.85819 | 20.165 | 0.069 | 2696.85819 | 3.785 | 1.302 | 43.413 | 43.490 | B |
| (50000) Quaoar | 2696.86089 | 18.124 | 0.097 | 2696.86089 | 1.744 | 1.302 | 43.413 | 43.490 | I |
| (50000) Quaoar | 2698.85722 | 19.455 | 0.112 | 2698.85742 | 3.077 | 1.306 | 43.413 | 43.456 | V |
| (50000) Quaoar | 2698.85857 | 19.377 | 0.106 | 2698.85877 | 2.999 | 1.306 | 43.413 | 43.456 | V |
| (50000) Quaoar | 2698.86137 | 20.337 | 0.074 | 2698.86157 | 3.959 | 1.306 | 43.413 | 43.456 | B |
| (50000) Quaoar | 2698.86406 | 18.241 | 0.105 | 2698.86426 | 1.863 | 1.306 | 43.413 | 43.456 | I |
| (50000) Quaoar | 2700.84523 | 19.274 | 0.087 | 2700.84563 | 2.897 | 1.308 | 43.413 | 43.422 | V |
| (50000) Quaoar | 2700.84659 | 19.141 | 0.074 | 2700.84699 | 2.764 | 1.308 | 43.413 | 43.422 | V |
| (50000) Quaoar | 2700.84938 | 20.084 | 0.050 | 2700.84978 | 3.707 | 1.308 | 43.413 | 43.422 | B |
| (50000) Quaoar | 2700.85208 | 17.986 | 0.074 | 2700.85248 | 1.609 | 1.308 | 43.413 | 43.422 | I |
| (50000) Quaoar | 2704.84205 | 19.270 | 0.081 | 2704.84285 | 2.897 | 1.309 | 43.413 | 43.353 | V |
| (50000) Quaoar | 2704.84341 | 19.249 | 0.085 | 2704.84421 | 2.876 | 1.309 | 43.413 | 43.353 | V |
| (50000) Quaoar | 2704.84620 | 20.144 | 0.051 | 2704.84700 | 3.771 | 1.309 | 43.413 | 43.353 | B |
| (50000) Quaoar | 2704.84890 | 18.089 | 0.082 | 2704.84970 | 1.716 | 1.309 | 43.413 | 43.353 | I |
| (50000) Quaoar | 2706.85972 | 19.241 | 0.081 | 2706.86072 | 2.870 | 1.307 | 43.412 | 43.318 | V |
| (50000) Quaoar | 2706.86108 | 19.276 | 0.083 | 2706.86208 | 2.905 | 1.307 | 43.412 | 43.318 | V |
| (50000) Quaoar | 2706.86388 | 20.322 | 0.060 | 2706.86488 | 3.951 | 1.307 | 43.412 | 43.318 | B |
| (50000) Quaoar | 2706.86657 | 18.283 | 0.099 | 2706.86757 | 1.912 | 1.307 | 43.412 | 43.318 | I |
| (50000) Quaoar | 2708.81668 | 19.249 | 0.083 | 2708.81787 | 2.879 | 1.303 | 43.412 | 43.284 | V |
| (50000) Quaoar | 2708.81804 | 19.266 | 0.081 | 2708.81923 | 2.896 | 1.303 | 43.412 | 43.284 | V |
| (50000) Quaoar | 2708.82083 | 20.261 | 0.057 | 2708.82202 | 3.891 | 1.303 | 43.412 | 43.284 | B |
| (50000) Quaoar | 2708.82354 | 17.997 | 0.079 | 2708.82473 | 1.627 | 1.303 | 43.412 | 43.284 | I |
| (50000) Quaoar | 2710.82968 | 19.324 | 0.073 | 2710.83107 | 2.956 | 1.298 | 43.412 | 43.250 | V |
| (50000) Quaoar | 2710.83104 | 19.486 | 0.083 | 2710.83243 | 3.118 | 1.298 | 43.412 | 43.250 | V |
| (50000) Quaoar | 2710.83384 | 20.205 | 0.045 | 2710.83523 | 3.837 | 1.298 | 43.412 | 43.250 | B |
| (50000) Quaoar | 2714.87749 | 19.364 | 0.078 | 2714.87927 | 2.999 | 1.282 | 43.412 | 43.182 | V |
| (50000) Quaoar | 2714.87884 | 19.254 | 0.066 | 2714.88063 | 2.889 | 1.282 | 43.412 | 43.182 | V |
| (50000) Quaoar | 2714.88164 | 20.155 | 0.042 | 2714.88342 | 3.790 | 1.282 | 43.412 | 43.181 | B |
| (50000) Quaoar | 2714.88433 | 17.948 | 0.058 | 2714.88611 | 1.583 | 1.282 | 43.412 | 43.181 | I |
| (50000) Quaoar | 2716.86227 | 19.360 | 0.159 | 2716.86425 | 2.997 | 1.273 | 43.412 | 43.148 | V |
| (50000) Quaoar | 2716.86363 | 19.273 | 0.144 | 2716.86561 | 2.910 | 1.273 | 43.412 | 43.148 | V |
| (50000) Quaoar | 2716.86642 | 20.504 | 0.202 | 2716.86840 | 4.141 | 1.273 | 43.412 | 43.148 | B |
| (50000) Quaoar | 2716.86912 | 18.005 | 0.079 | 2716.87110 | 1.642 | 1.273 | 43.412 | 43.148 | I |
| (50000) Quaoar | 2718.88142 | 19.210 | 0.132 | 2718.88359 | 2.849 | 1.261 | 43.411 | 43.115 | V |
| (50000) Quaoar | 2718.88278 | 19.124 | 0.121 | 2718.88495 | 2.763 | 1.261 | 43.411 | 43.115 | V |
| (50000) Quaoar | 2718.88558 | 20.138 | 0.132 | 2718.88775 | 3.777 | 1.261 | 43.411 | 43.115 | B |
| (50000) Quaoar | 2718.88828 | 17.883 | 0.079 | 2718.89045 | 1.522 | 1.261 | 43.411 | 43.115 | I |
| (50000) Quaoar | 2722.84788 | 19.585 | 0.266 | 2722.85042 | 3.227 | 1.234 | 43.411 | 43.050 | V |
| (50000) Quaoar | 2722.85203 | 20.259 | 0.142 | 2722.85457 | 3.901 | 1.234 | 43.411 | 43.050 | B |
| (50000) Quaoar | 2722.85473 | 18.248 | 0.171 | 2722.85727 | 1.890 | 1.234 | 43.411 | 43.050 | I |
| (50000) Quaoar | 2724.89572 | 19.617 | 0.099 | 2724.89845 | 3.261 | 1.218 | 43.411 | 43.017 | V |
| (50000) Quaoar | 2724.89708 | 19.453 | 0.089 | 2724.89981 | 3.097 | 1.218 | 43.411 | 43.017 | V |
| (50000) Quaoar | 2724.89988 | 20.477 | 0.076 | 2724.90261 | 4.121 | 1.218 | 43.411 | 43.017 | B |
| (50000) Quaoar | 2724.90258 | 18.327 | 0.078 | 2724.90531 | 1.971 | 1.218 | 43.411 | 43.017 | I |
| (50000) Quaoar | 2728.88956 | 19.233 | 0.073 | 2728.89266 | 2.880 | 1.182 | 43.410 | 42.955 | V |
| (50000) Quaoar | 2728.89092 | 19.353 | 0.077 | 2728.89402 | 3.000 | 1.182 | 43.410 | 42.955 | V |
| (50000) Quaoar | 2728.89372 | 20.423 | 0.064 | 2728.89682 | 4.070 | 1.182 | 43.410 | 42.954 | B |
| (50000) Quaoar | 2728.89642 | 18.317 | 0.089 | 2728.89952 | 1.964 | 1.182 | 43.410 | 42.954 | I |
| (50000) Quaoar | 2730.83995 | 19.343 | 0.085 | 2730.84322 | 2.992 | 1.163 | 43.410 | 42.925 | V |
| (50000) Quaoar | 2730.84131 | 19.397 | 0.090 | 2730.84458 | 3.046 | 1.163 | 43.410 | 42.925 | V |
| (50000) Quaoar | 2730.84410 | 20.229 | 0.060 | 2730.84737 | 3.878 | 1.163 | 43.410 | 42.925 | B |
| (50000) Quaoar | 2730.84680 | 18.050 | 0.083 | 2730.85007 | 1.699 | 1.162 | 43.410 | 42.925 | I |
| (50000) Quaoar | 2732.81711 | 19.211 | 0.091 | 2732.82055 | 2.861 | 1.141 | 43.410 | 42.895 | V |
| (50000) Quaoar | 2732.81847 | 19.206 | 0.093 | 2732.82191 | 2.856 | 1.141 | 43.410 | 42.895 | V |
| (50000) Quaoar | 2732.82127 | 20.108 | 0.066 | 2732.82471 | 3.758 | 1.141 | 43.410 | 42.895 | B |
| (50000) Quaoar | 2732.82398 | 18.081 | 0.094 | 2732.82742 | 1.731 | 1.141 | 43.410 | 42.895 | I |
| (50000) Quaoar | 2736.82823 | 19.378 | 0.075 | 2736.83201 | 3.031 | 1.094 | 43.410 | 42.837 | V |
| (50000) Quaoar | 2736.82959 | 19.285 | 0.063 | 2736.83337 | 2.938 | 1.094 | 43.410 | 42.837 | V |
| (50000) Quaoar | 2736.83239 | 20.227 | 0.042 | 2736.83617 | 3.880 | 1.094 | 43.410 | 42.837 | B |
| (50000) Quaoar | 2736.83506 | 18.095 | 0.066 | 2736.83884 | 1.748 | 1.094 | 43.410 | 42.837 | I |
| (50000) Quaoar | 2738.81598 | 19.310 | 0.078 | 2738.81992 | 2.964 | 1.069 | 43.410 | 42.809 | V |
| (50000) Quaoar | 2738.81734 | 19.147 | 0.070 | 2738.82128 | 2.801 | 1.069 | 43.410 | 42.809 | V |
| (50000) Quaoar | 2738.82013 | 20.093 | 0.039 | 2738.82407 | 3.747 | 1.069 | 43.410 | 42.809 | B |
| (50000) Quaoar | 2738.82282 | 18.015 | 0.066 | 2738.82676 | 1.669 | 1.069 | 43.410 | 42.809 | I |
| (50000) Quaoar | 2740.81194 | 19.341 | 0.075 | 2740.81604 | 2.997 | 1.042 | 43.409 | 42.781 | V |
| (50000) Quaoar | 2740.81329 | 19.352 | 0.076 | 2740.81739 | 3.008 | 1.042 | 43.409 | 42.781 | V |
| (50000) Quaoar | 2740.81609 | 20.312 | 0.047 | 2740.82019 | 3.968 | 1.042 | 43.409 | 42.781 | B |
| (50000) Quaoar | 2740.81878 | 18.051 | 0.068 | 2740.82288 | 1.707 | 1.042 | 43.409 | 42.781 | I |
| (50000) Quaoar | 2742.84166 | 19.278 | 0.068 | 2742.84591 | 2.935 | 1.014 | 43.409 | 42.754 | V |
| (50000) Quaoar | 2742.84301 | 19.074 | 0.062 | 2742.84726 | 2.731 | 1.014 | 43.409 | 42.754 | V |
| (50000) Quaoar | 2742.84581 | 20.125 | 0.039 | 2742.85006 | 3.782 | 1.014 | 43.409 | 42.754 | B |
| (50000) Quaoar | 2742.84851 | 18.003 | 0.064 | 2742.85276 | 1.660 | 1.014 | 43.409 | 42.754 | I |
| (50000) Quaoar | 2744.78946 | 19.155 | 0.286 | 2744.79386 | 2.813 | 0.986 | 43.409 | 42.729 | V |
| (50000) Quaoar | 2744.79495 | 17.777 | 0.210 | 2744.79935 | 1.435 | 0.986 | 43.409 | 42.729 | I |
| (50000) Quaoar | 2750.83303 | 19.163 | 0.125 | 2750.83785 | 2.825 | 0.891 | 43.408 | 42.656 | V |
| (50000) Quaoar | 2750.83439 | 19.391 | 0.150 | 2750.83921 | 3.053 | 0.891 | 43.408 | 42.656 | V |
| (50000) Quaoar | 2750.83719 | 20.048 | 0.102 | 2750.84201 | 3.710 | 0.891 | 43.408 | 42.656 | B |
| (50000) Quaoar | 2750.83989 | 17.963 | 0.109 | 2750.84471 | 1.625 | 0.891 | 43.408 | 42.656 | I |

| | | | | | | | | |
|---|---|---|---|---|---|---|---|---|
| (50000) Quaoar | 2752.81282 | 19.295 | 0.091 | 2752.81777 | 2.958 | 0.858 | 43.408 | 42.634 | V |
| (50000) Quaoar | 2752.81418 | 19.205 | 0.092 | 2752.81913 | 2.868 | 0.858 | 43.408 | 42.634 | V |
| (50000) Quaoar | 2752.81698 | 20.083 | 0.087 | 2752.82193 | 3.746 | 0.858 | 43.408 | 42.633 | B |
| (50000) Quaoar | 2752.81965 | 17.861 | 0.070 | 2752.82460 | 1.524 | 0.858 | 43.408 | 42.633 | I |
| (50000) Quaoar | 2754.75687 | 19.264 | 0.111 | 2754.76194 | 2.928 | 0.825 | 43.408 | 42.612 | V |
| (50000) Quaoar | 2754.75822 | 19.378 | 0.115 | 2754.76329 | 3.042 | 0.825 | 43.408 | 42.612 | V |
| (50000) Quaoar | 2754.76102 | 20.106 | 0.088 | 2754.76609 | 3.770 | 0.825 | 43.408 | 42.612 | B |
| (50000) Quaoar | 2754.76371 | 18.068 | 0.086 | 2754.76878 | 1.732 | 0.825 | 43.408 | 42.612 | I |
| (50000) Quaoar | 2756.83098 | 19.193 | 0.074 | 2756.83618 | 2.859 | 0.789 | 43.408 | 42.591 | V |
| (50000) Quaoar | 2756.83233 | 19.189 | 0.077 | 2756.83753 | 2.855 | 0.789 | 43.408 | 42.591 | V |
| (50000) Quaoar | 2756.83513 | 20.135 | 0.051 | 2756.84033 | 3.801 | 0.789 | 43.408 | 42.591 | B |
| (50000) Quaoar | 2756.83783 | 17.848 | 0.072 | 2756.84303 | 1.514 | 0.789 | 43.408 | 42.591 | I |
| (50000) Quaoar | 2758.83210 | 19.192 | 0.100 | 2758.83741 | 2.859 | 0.753 | 43.408 | 42.571 | V |
| (50000) Quaoar | 2758.83346 | 19.140 | 0.099 | 2758.83877 | 2.807 | 0.753 | 43.408 | 42.571 | V |
| (50000) Quaoar | 2758.83626 | 20.210 | 0.093 | 2758.84157 | 3.877 | 0.753 | 43.408 | 42.571 | B |
| (50000) Quaoar | 2758.83895 | 17.940 | 0.157 | 2758.84426 | 1.607 | 0.752 | 43.408 | 42.571 | I |
| (50000) Quaoar | 2760.79025 | 19.233 | 0.072 | 2760.79567 | 2.901 | 0.717 | 43.408 | 42.553 | V |
| (50000) Quaoar | 2760.79161 | 19.199 | 0.070 | 2760.79703 | 2.867 | 0.717 | 43.408 | 42.553 | V |
| (50000) Quaoar | 2760.79442 | 20.153 | 0.042 | 2760.79984 | 3.821 | 0.716 | 43.408 | 42.553 | B |
| (50000) Quaoar | 2760.79713 | 17.863 | 0.064 | 2760.80255 | 1.531 | 0.716 | 43.408 | 42.553 | I |
| (50000) Quaoar | 2764.68062 | 19.228 | 0.076 | 2764.68623 | 2.897 | 0.643 | 43.407 | 42.519 | V |
| (50000) Quaoar | 2764.68198 | 19.174 | 0.068 | 2764.68759 | 2.843 | 0.643 | 43.407 | 42.519 | V |
| (50000) Quaoar | 2764.68477 | 20.115 | 0.043 | 2764.69038 | 3.784 | 0.643 | 43.407 | 42.519 | B |
| (50000) Quaoar | 2764.68747 | 17.823 | 0.059 | 2764.69308 | 1.492 | 0.643 | 43.407 | 42.519 | I |
| (50000) Quaoar | 2770.81076 | 19.013 | 0.058 | 2770.81663 | 2.685 | 0.523 | 43.407 | 42.474 | V |
| (50000) Quaoar | 2770.81211 | 18.981 | 0.057 | 2770.81798 | 2.653 | 0.523 | 43.407 | 42.474 | V |
| (50000) Quaoar | 2770.81490 | 20.079 | 0.040 | 2770.82077 | 3.751 | 0.523 | 43.407 | 42.474 | B |
| (50000) Quaoar | 2770.81760 | 17.727 | 0.063 | 2770.82347 | 1.399 | 0.523 | 43.407 | 42.474 | I |
| (50000) Quaoar | 2773.71459 | 19.011 | 0.157 | 2773.72056 | 2.684 | 0.465 | 43.406 | 42.456 | V |
| (50000) Quaoar | 2773.71595 | 18.990 | 0.199 | 2773.72192 | 2.663 | 0.465 | 43.406 | 42.456 | V |
| (50000) Quaoar | 2773.71875 | 20.178 | 0.205 | 2773.72472 | 3.851 | 0.465 | 43.406 | 42.456 | B |
| (50000) Quaoar | 2773.72144 | 17.589 | 0.106 | 2773.72741 | 1.262 | 0.465 | 43.406 | 42.456 | I |
| (50000) Quaoar | 2775.59860 | 19.089 | 0.232 | 2775.60463 | 2.762 | 0.428 | 43.406 | 42.446 | V |
| (50000) Quaoar | 2775.60139 | 19.861 | 0.240 | 2775.60742 | 3.534 | 0.428 | 43.406 | 42.446 | B |
| (50000) Quaoar | 2775.60408 | 17.959 | 0.130 | 2775.61011 | 1.632 | 0.428 | 43.406 | 42.446 | I |
| (50000) Quaoar | 2782.66870 | 19.115 | 0.096 | 2782.67490 | 2.790 | 0.292 | 43.406 | 42.416 | V |
| (50000) Quaoar | 2782.67005 | 18.992 | 0.112 | 2782.67625 | 2.667 | 0.292 | 43.406 | 42.416 | V |
| (50000) Quaoar | 2782.67285 | 20.248 | 0.060 | 2782.67905 | 3.923 | 0.292 | 43.406 | 42.416 | B |
| (50000) Quaoar | 2782.67554 | 17.677 | 0.139 | 2782.68174 | 1.352 | 0.292 | 43.406 | 42.416 | I |
| (50000) Quaoar | 2786.76796 | 19.108 | 0.091 | 2786.77423 | 2.783 | 0.224 | 43.405 | 42.405 | V |
| (50000) Quaoar | 2786.76932 | 19.128 | 0.100 | 2786.77559 | 2.803 | 0.224 | 43.405 | 42.405 | V |
| (50000) Quaoar | 2786.77211 | 20.070 | 0.080 | 2786.77838 | 3.745 | 0.224 | 43.405 | 42.405 | B |
| (50000) Quaoar | 2786.77481 | 17.758 | 0.101 | 2786.78108 | 1.433 | 0.224 | 43.405 | 42.405 | I |
| (50000) Quaoar | 2788.62300 | 19.134 | 0.106 | 2788.62929 | 2.809 | 0.199 | 43.405 | 42.402 | V |
| (50000) Quaoar | 2788.62435 | 19.149 | 0.104 | 2788.63064 | 2.824 | 0.199 | 43.405 | 42.402 | V |
| (50000) Quaoar | 2788.62715 | 19.965 | 0.078 | 2788.63344 | 3.640 | 0.199 | 43.405 | 42.402 | B |
| (50000) Quaoar | 2788.62985 | 17.894 | 0.110 | 2788.63614 | 1.569 | 0.199 | 43.405 | 42.402 | I |
| (50000) Quaoar | 2790.58284 | 18.814 | 0.196 | 2790.58914 | 2.489 | 0.180 | 43.405 | 42.400 | V |
| (50000) Quaoar | 2790.58420 | 18.873 | 0.150 | 2790.59050 | 2.548 | 0.180 | 43.405 | 42.400 | V |
| (50000) Quaoar | 2790.58700 | 20.049 | 0.156 | 2790.59330 | 3.724 | 0.180 | 43.405 | 42.400 | B |
| (50000) Quaoar | 2790.58969 | 17.463 | 0.237 | 2790.59599 | 1.138 | 0.180 | 43.405 | 42.400 | I |
| (50000) Quaoar | 2792.65617 | 19.236 | 0.091 | 2792.66248 | 2.912 | 0.170 | 43.405 | 42.399 | V |
| (50000) Quaoar | 2792.65753 | 19.126 | 0.097 | 2792.66384 | 2.802 | 0.170 | 43.405 | 42.399 | V |
| (50000) Quaoar | 2792.66032 | 19.966 | 0.074 | 2792.66663 | 3.642 | 0.170 | 43.405 | 42.399 | B |
| (50000) Quaoar | 2792.66302 | 17.739 | 0.075 | 2792.66933 | 1.415 | 0.170 | 43.405 | 42.399 | I |
| (50000) Quaoar | 2794.68978 | 19.088 | 0.060 | 2794.69609 | 2.764 | 0.173 | 43.405 | 42.398 | V |
| (50000) Quaoar | 2794.69113 | 19.108 | 0.060 | 2794.69744 | 2.784 | 0.173 | 43.405 | 42.398 | V |
| (50000) Quaoar | 2794.69393 | 20.078 | 0.037 | 2794.70024 | 3.754 | 0.173 | 43.405 | 42.398 | B |
| (50000) Quaoar | 2794.69662 | 17.720 | 0.063 | 2794.70293 | 1.396 | 0.173 | 43.405 | 42.398 | I |
| (50000) Quaoar | 2811.57455 | 19.311 | 0.122 | 2811.58060 | 2.984 | 0.442 | 43.403 | 42.444 | V |
| (50000) Quaoar | 2811.57735 | 20.105 | 0.100 | 2811.58340 | 3.778 | 0.442 | 43.403 | 42.444 | B |
| (50000) Quaoar | 2811.58005 | 17.782 | 0.124 | 2811.58609 | 1.455 | 0.442 | 43.403 | 42.444 | I |
| (50000) Quaoar | 2813.49597 | 18.841 | 0.109 | 2813.50196 | 2.514 | 0.481 | 43.403 | 42.454 | V |
| (50000) Quaoar | 2813.49733 | 18.977 | 0.098 | 2813.50332 | 2.650 | 0.481 | 43.403 | 42.454 | V |
| (50000) Quaoar | 2813.50013 | 19.930 | 0.073 | 2813.50612 | 3.603 | 0.481 | 43.403 | 42.454 | B |
| (50000) Quaoar | 2813.50283 | 17.946 | 0.146 | 2813.50882 | 1.619 | 0.481 | 43.403 | 42.454 | I |
| (50000) Quaoar | 2818.74335 | 19.150 | 0.096 | 2818.74915 | 2.821 | 0.585 | 43.402 | 42.487 | V |
| (50000) Quaoar | 2818.74471 | 19.213 | 0.105 | 2818.75051 | 2.884 | 0.585 | 43.402 | 42.487 | V |
| (50000) Quaoar | 2818.74750 | 20.144 | 0.065 | 2818.75330 | 3.815 | 0.585 | 43.402 | 42.487 | B |
| (50000) Quaoar | 2818.75020 | 18.000 | 0.100 | 2818.75600 | 1.671 | 0.585 | 43.402 | 42.487 | I |
| (50000) Quaoar | 2820.53683 | 19.144 | 0.062 | 2820.54255 | 2.815 | 0.620 | 43.402 | 42.500 | V |
| (50000) Quaoar | 2820.53819 | 19.192 | 0.066 | 2820.54391 | 2.862 | 0.620 | 43.402 | 42.500 | V |
| (50000) Quaoar | 2820.54098 | 20.120 | 0.041 | 2820.54670 | 3.790 | 0.620 | 43.402 | 42.500 | B |
| (50000) Quaoar | 2820.54368 | 17.750 | 0.066 | 2820.54940 | 1.420 | 0.620 | 43.402 | 42.500 | I |
| (50000) Quaoar | 2830.65555 | 19.860 | 0.208 | 2830.66076 | 3.526 | 0.808 | 43.401 | 42.588 | B |
| (50000) Quaoar | 2830.65825 | 18.205 | 0.270 | 2830.66346 | 1.871 | 0.808 | 43.401 | 42.588 | I |
| (50000) Quaoar | 2832.63602 | 20.087 | 0.254 | 2832.64112 | 3.752 | 0.842 | 43.401 | 42.608 | B |
| (50000) Quaoar | 2832.63873 | 18.332 | 0.262 | 2832.64383 | 1.997 | 0.842 | 43.401 | 42.608 | I |
| (50000) Quaoar | 2834.51260 | 19.144 | 0.174 | 2834.51758 | 2.808 | 0.874 | 43.401 | 42.628 | V |
| (50000) Quaoar | 2834.51396 | 18.950 | 0.170 | 2834.51894 | 2.614 | 0.874 | 43.401 | 42.628 | V |
| (50000) Quaoar | 2834.51673 | 19.705 | 0.184 | 2834.52171 | 3.369 | 0.874 | 43.401 | 42.628 | B |
| (50000) Quaoar | 2834.51941 | 17.856 | 0.178 | 2834.52439 | 1.520 | 0.875 | 43.401 | 42.628 | I |
| (50000) Quaoar | 2837.57189 | 18.858 | 0.146 | 2837.57668 | 2.520 | 0.925 | 43.401 | 42.662 | V |
| (50000) Quaoar | 2837.57324 | 19.232 | 0.141 | 2837.57803 | 2.894 | 0.925 | 43.401 | 42.662 | V |
| (50000) Quaoar | 2837.57604 | 20.121 | 0.106 | 2837.58082 | 3.783 | 0.925 | 43.401 | 42.662 | B |
| (50000) Quaoar | 2837.57874 | 17.837 | 0.118 | 2837.58352 | 1.499 | 0.925 | 43.401 | 42.662 | I |
| | | | | # | | | | | |
| (28978) Ixion | 2697.83640 | 20.279 | 0.189 | 2697.83640 | 3.949 | 1.322 | 42.951 | 42.948 | V |
| (28978) Ixion | 2697.83811 | 20.575 | 0.228 | 2697.83811 | 4.245 | 1.322 | 42.951 | 42.948 | V |
| (28978) Ixion | 2697.84165 | 21.172 | 0.133 | 2697.84165 | 4.842 | 1.322 | 42.951 | 42.948 | B |
| (28978) Ixion | 2697.84500 | 18.995 | 0.191 | 2697.84500 | 2.665 | 1.322 | 42.951 | 42.948 | I |
| (28978) Ixion | 2699.86742 | 20.424 | 0.167 | 2699.86763 | 4.096 | 1.322 | 42.950 | 42.912 | V |
| (28978) Ixion | 2699.86913 | 20.402 | 0.174 | 2699.86934 | 4.074 | 1.322 | 42.950 | 42.912 | V |
| (28978) Ixion | 2699.87267 | 21.359 | 0.128 | 2699.87288 | 5.031 | 1.322 | 42.950 | 42.912 | B |
| (28978) Ixion | 2699.87602 | 19.116 | 0.174 | 2699.87623 | 2.788 | 1.322 | 42.950 | 42.912 | I |
| (28978) Ixion | 2701.84370 | 20.399 | 0.180 | 2701.84411 | 4.073 | 1.321 | 42.949 | 42.876 | V |
| (28978) Ixion | 2701.84541 | 20.286 | 0.163 | 2701.84582 | 3.960 | 1.321 | 42.949 | 42.876 | V |
| (28978) Ixion | 2701.84895 | 21.400 | 0.150 | 2701.84936 | 5.074 | 1.321 | 42.949 | 42.876 | B |
| (28978) Ixion | 2701.85229 | 19.417 | 0.255 | 2701.85270 | 3.091 | 1.321 | 42.949 | 42.876 | I |
| (28978) Ixion | 2703.84414 | 20.214 | 0.237 | 2703.84476 | 3.890 | 1.318 | 42.948 | 42.841 | V |
| (28978) Ixion | 2703.84939 | 21.252 | 0.188 | 2703.85001 | 4.928 | 1.318 | 42.948 | 42.841 | B |
| (28978) Ixion | 2705.83823 | 20.146 | 0.137 | 2705.83905 | 3.824 | 1.314 | 42.947 | 42.805 | V |
| (28978) Ixion | 2705.83993 | 20.456 | 0.176 | 2705.84075 | 4.134 | 1.314 | 42.947 | 42.805 | V |
| (28978) Ixion | 2705.84347 | 21.396 | 0.131 | 2705.84429 | 5.074 | 1.314 | 42.947 | 42.805 | B |
| (28978) Ixion | 2705.84682 | 19.141 | 0.194 | 2705.84764 | 2.819 | 1.314 | 42.947 | 42.805 | I |
| (28978) Ixion | 2707.83091 | 20.523 | 0.168 | 2707.83194 | 4.203 | 1.308 | 42.946 | 42.770 | V |
| (28978) Ixion | 2707.83263 | 20.167 | 0.125 | 2707.83366 | 3.847 | 1.308 | 42.946 | 42.770 | V |
| (28978) Ixion | 2707.83617 | 21.357 | 0.104 | 2707.83720 | 5.037 | 1.308 | 42.946 | 42.770 | B |
| (28978) Ixion | 2707.83949 | 19.027 | 0.159 | 2707.84052 | 2.707 | 1.308 | 42.946 | 42.770 | I |
| (28978) Ixion | 2709.83532 | 20.288 | 0.135 | 2709.83655 | 3.970 | 1.300 | 42.945 | 42.735 | V |
| (28978) Ixion | 2709.83702 | 20.438 | 0.157 | 2709.83825 | 4.120 | 1.300 | 42.945 | 42.735 | V |
| (28978) Ixion | 2709.84056 | 21.326 | 0.116 | 2709.84179 | 5.008 | 1.300 | 42.945 | 42.735 | B |
| (28978) Ixion | 2709.84390 | 19.013 | 0.161 | 2709.84513 | 2.695 | 1.300 | 42.945 | 42.735 | I |
| (28978) Ixion | 2711.82148 | 20.204 | 0.130 | 2711.82291 | 3.887 | 1.291 | 42.943 | 42.700 | V |
| (28978) Ixion | 2711.82318 | 20.099 | 0.124 | 2711.82461 | 3.782 | 1.291 | 42.943 | 42.700 | V |
| (28978) Ixion | 2711.82671 | 21.347 | 0.104 | 2711.82814 | 5.030 | 1.291 | 42.943 | 42.700 | B |
| (28978) Ixion | 2711.83006 | 19.187 | 0.165 | 2711.83149 | 2.870 | 1.291 | 42.943 | 42.700 | I |
| (28978) Ixion | 2713.86698 | 20.694 | 0.272 | 2713.86862 | 4.379 | 1.280 | 42.942 | 42.665 | V |
| (28978) Ixion | 2713.86869 | 20.125 | 0.167 | 2713.87033 | 3.810 | 1.280 | 42.942 | 42.665 | V |
| (28978) Ixion | 2713.87224 | 21.434 | 0.139 | 2713.87388 | 5.119 | 1.280 | 42.942 | 42.665 | B |
| (28978) Ixion | 2713.87558 | 19.211 | 0.193 | 2713.87722 | 2.896 | 1.280 | 42.942 | 42.665 | I |
| (28978) Ixion | 2715.83636 | 20.250 | 0.275 | 2715.83819 | 3.937 | 1.268 | 42.941 | 42.631 | V |
| (28978) Ixion | 2715.84161 | 20.839 | 0.198 | 2715.84344 | 4.526 | 1.268 | 42.941 | 42.631 | B |
| (28978) Ixion | 2715.84496 | 19.082 | 0.187 | 2715.84679 | 2.769 | 1.268 | 42.941 | 42.631 | I |
| (28978) Ixion | 2717.88091 | 19.712 | 0.230 | 2717.88294 | 3.401 | 1.254 | 42.940 | 42.596 | V |
| (28978) Ixion | 2717.88950 | 19.186 | 0.217 | 2717.89153 | 2.875 | 1.254 | 42.940 | 42.596 | I |
| (28978) Ixion | 2723.83756 | 20.269 | 0.152 | 2723.84016 | 3.963 | 1.204 | 42.937 | 42.498 | V |
| (28978) Ixion | 2723.83926 | 20.294 | 0.166 | 2723.84186 | 3.988 | 1.204 | 42.937 | 42.498 | V |
| (28978) Ixion | 2723.84615 | 19.444 | 0.243 | 2723.84875 | 3.138 | 1.204 | 42.937 | 42.497 | I |
| (28978) Ixion | 2725.81368 | 20.311 | 0.132 | 2725.81647 | 4.007 | 1.184 | 42.936 | 42.466 | V |
| (28978) Ixion | 2725.81539 | 20.478 | 0.154 | 2725.81818 | 4.174 | 1.184 | 42.936 | 42.466 | V |
| (28978) Ixion | 2725.81893 | 21.298 | 0.100 | 2725.82172 | 4.994 | 1.184 | 42.936 | 42.466 | B |
| (28978) Ixion | 2725.82227 | 19.050 | 0.143 | 2725.82506 | 2.746 | 1.184 | 42.936 | 42.466 | I |
| (28978) Ixion | 2727.87282 | 20.381 | 0.120 | 2727.87579 | 4.078 | 1.163 | 42.934 | 42.433 | V |
| (28978) Ixion | 2727.87453 | 20.330 | 0.111 | 2727.87750 | 4.027 | 1.163 | 42.934 | 42.433 | V |
| (28978) Ixion | 2727.87807 | 21.199 | 0.077 | 2727.88104 | 4.896 | 1.163 | 42.934 | 42.433 | B |
| (28978) Ixion | 2727.88142 | 19.345 | 0.171 | 2727.88439 | 3.042 | 1.162 | 42.934 | 42.433 | I |
| (28978) Ixion | 2729.80578 | 20.156 | 0.118 | 2729.80893 | 3.855 | 1.141 | 42.933 | 42.403 | V |
| (28978) Ixion | 2729.80748 | 20.038 | 0.113 | 2729.81063 | 3.737 | 1.141 | 42.933 | 42.403 | V |
| (28978) Ixion | 2729.81103 | 21.259 | 0.091 | 2729.81418 | 4.958 | 1.141 | 42.933 | 42.403 | B |
| (28978) Ixion | 2729.81437 | 19.324 | 0.170 | 2729.81752 | 3.023 | 1.141 | 42.933 | 42.403 | I |
| (28978) Ixion | 2731.82834 | 21.288 | 0.083 | 2731.83166 | 4.989 | 1.116 | 42.932 | 42.373 | B |
| (28978) Ixion | 2731.83170 | 18.873 | 0.118 | 2731.83502 | 2.574 | 1.116 | 42.932 | 42.373 | I |
| (28978) Ixion | 2733.79557 | 20.271 | 0.243 | 2733.79906 | 3.973 | 1.091 | 42.931 | 42.343 | V |
| (28978) Ixion | 2733.79728 | 20.566 | 0.264 | 2733.80077 | 4.268 | 1.091 | 42.931 | 42.343 | V |
| (28978) Ixion | 2733.80083 | 21.180 | 0.111 | 2733.80432 | 4.882 | 1.091 | 42.931 | 42.343 | B |
| (28978) Ixion | 2733.80418 | 19.142 | 0.174 | 2733.80767 | 2.844 | 1.091 | 42.931 | 42.343 | I |
| (28978) Ixion | 2735.79350 | 20.129 | 0.120 | 2735.79716 | 3.833 | 1.065 | 42.930 | 42.314 | V |
| (28978) Ixion | 2735.79521 | 20.143 | 0.123 | 2735.79887 | 3.847 | 1.065 | 42.930 | 42.314 | V |
| (28978) Ixion | 2735.79875 | 21.177 | 0.083 | 2735.80241 | 4.881 | 1.065 | 42.930 | 42.314 | B |
| (28978) Ixion | 2735.80210 | 19.139 | 0.159 | 2735.80576 | 2.843 | 1.064 | 42.930 | 42.314 | I |
| (28978) Ixion | 2737.77074 | 20.109 | 0.117 | 2737.77456 | 3.814 | 1.037 | 42.929 | 42.286 | V |
| (28978) Ixion | 2737.77245 | 20.276 | 0.131 | 2737.77627 | 3.981 | 1.037 | 42.929 | 42.286 | V |
| (28978) Ixion | 2737.77599 | 21.203 | 0.081 | 2737.77981 | 4.908 | 1.037 | 42.929 | 42.286 | B |
| (28978) Ixion | 2737.77933 | 19.226 | 0.175 | 2737.78315 | 2.931 | 1.037 | 42.929 | 42.286 | I |
| (28978) Ixion | 2739.75482 | 20.164 | 0.148 | 2739.75880 | 3.871 | 1.008 | 42.928 | 42.259 | V |
| (28978) Ixion | 2739.75653 | 20.128 | 0.132 | 2739.76051 | 3.835 | 1.008 | 42.928 | 42.259 | V |
| (28978) Ixion | 2739.76007 | 21.124 | 0.095 | 2739.76405 | 4.831 | 1.008 | 42.928 | 42.259 | B |
| (28978) Ixion | 2739.76341 | 19.463 | 0.218 | 2739.76739 | 3.170 | 1.008 | 42.928 | 42.259 | I |
| (28978) Ixion | 2741.74739 | 20.307 | 0.156 | 2741.75152 | 4.015 | 0.977 | 42.927 | 42.232 | V |
| (28978) Ixion | 2741.74909 | 20.125 | 0.131 | 2741.75322 | 3.833 | 0.977 | 42.927 | 42.232 | V |
| (28978) Ixion | 2741.75264 | 21.260 | 0.121 | 2741.75677 | 4.968 | 0.977 | 42.926 | 42.232 | B |
| (28978) Ixion | 2741.75598 | 19.557 | 0.250 | 2741.76011 | 3.265 | 0.977 | 42.926 | 42.232 | I |
| (28978) Ixion | 2743.69320 | 20.117 | 0.223 | 2743.69748 | 3.827 | 0.947 | 42.925 | 42.207 | V |
| (28978) Ixion | 2743.70009 | 19.201 | 0.250 | 2743.70437 | 2.911 | 0.946 | 42.925 | 42.207 | I |
| (28978) Ixion | 2751.78090 | 20.380 | 0.209 | 2751.78574 | 4.095 | 0.807 | 42.921 | 42.111 | V |
| (28978) Ixion | 2751.78261 | 20.063 | 0.176 | 2751.78745 | 3.778 | 0.807 | 42.921 | 42.111 | V |
| (28978) Ixion | 2751.78615 | 20.737 | 0.180 | 2751.79099 | 4.452 | 0.807 | 42.921 | 42.111 | B |
| (28978) Ixion | 2751.78950 | 19.510 | 0.264 | 2751.79434 | 3.225 | 0.807 | 42.921 | 42.111 | I |
| (28978) Ixion | 2753.87469 | 20.083 | 0.231 | 2753.87966 | 3.799 | 0.768 | 42.920 | 42.088 | V |
| (28978) Ixion | 2753.87822 | 20.935 | 0.239 | 2753.88319 | 4.651 | 0.768 | 42.920 | 42.088 | B |
| (28978) Ixion | 2753.88157 | 18.960 | 0.206 | 2753.88654 | 2.676 | 0.768 | 42.920 | 42.088 | I |
| (28978) Ixion | 2755.74978 | 20.418 | 0.192 | 2755.75486 | 4.135 | 0.733 | 42.919 | 42.069 | V |
| (28978) Ixion | 2755.75149 | 20.155 | 0.160 | 2755.75657 | 3.872 | 0.733 | 42.919 | 42.069 | V |
| (28978) Ixion | 2755.75503 | 21.184 | 0.109 | 2755.76011 | 4.901 | 0.732 | 42.919 | 42.069 | B |
| (28978) Ixion | 2755.75837 | 19.130 | 0.206 | 2755.76345 | 2.847 | 0.732 | 42.919 | 42.069 | I |
| (28978) Ixion | 2757.76061 | 20.552 | 0.168 | 2757.76580 | 4.270 | 0.693 | 42.917 | 42.049 | V |
| (28978) Ixion | 2757.76232 | 20.519 | 0.165 | 2757.76751 | 4.237 | 0.693 | 42.917 | 42.049 | V |
| (28978) Ixion | 2757.76586 | 21.307 | 0.086 | 2757.77105 | 5.025 | 0.693 | 42.917 | 42.049 | B |
| (28978) Ixion | 2757.76921 | 19.290 | 0.182 | 2757.77440 | 3.008 | 0.693 | 42.917 | 42.049 | I |
| (28978) Ixion | 2759.82072 | 20.211 | 0.137 | 2759.82602 | 3.930 | 0.653 | 42.916 | 42.030 | V |
| (28978) Ixion | 2759.82243 | 20.149 | 0.128 | 2759.82773 | 3.868 | 0.652 | 42.916 | 42.030 | V |
| (28978) Ixion | 2759.82597 | 21.145 | 0.089 | 2759.83127 | 4.864 | 0.652 | 42.916 | 42.030 | B |
| (28978) Ixion | 2759.82930 | 19.252 | 0.191 | 2759.83460 | 2.971 | 0.652 | 42.916 | 42.030 | I |
| (28978) Ixion | 2761.76778 | 19.927 | 0.223 | 2761.77318 | 3.647 | 0.613 | 42.915 | 42.013 | V |
| (28978) Ixion | 2761.76949 | 20.035 | 0.243 | 2761.77489 | 3.755 | 0.613 | 42.915 | 42.013 | V |
| (28978) Ixion | 2761.77303 | 20.610 | 0.188 | 2761.77843 | 4.330 | 0.613 | 42.915 | 42.013 | B |
| (28978) Ixion | 2761.77638 | 18.716 | 0.228 | 2761.78178 | 2.436 | 0.613 | 42.915 | 42.013 | I |
| (28978) Ixion | 2763.68498 | 19.645 | 0.173 | 2763.69047 | 3.366 | 0.573 | 42.914 | 41.998 | V |
| (28978) Ixion | 2763.68669 | 19.766 | 0.181 | 2763.69218 | 3.487 | 0.573 | 42.914 | 41.998 | V |
| (28978) Ixion | 2763.69022 | 20.608 | 0.188 | 2763.69571 | 4.329 | 0.573 | 42.914 | 41.998 | B |
| (28978) Ixion | 2763.69357 | 18.795 | 0.209 | 2763.69906 | 2.516 | 0.573 | 42.914 | 41.998 | I |
| (28978) Ixion | 2765.75524 | 19.921 | 0.171 | 2765.76082 | 3.643 | 0.530 | 42.913 | 41.982 | V |
| (28978) Ixion | 2765.75695 | 19.817 | 0.186 | 2765.76253 | 3.539 | 0.530 | 42.913 | 41.982 | V |
| (28978) Ixion | 2765.76049 | 20.830 | 0.170 | 2765.76607 | 4.552 | 0.530 | 42.913 | 41.982 | B |
| (28978) Ixion | 2765.76383 | 18.972 | 0.219 | 2765.76941 | 2.694 | 0.530 | 42.913 | 41.982 | I |
| (28978) Ixion | 2771.62173 | 21.233 | 0.246 | 2771.62753 | 4.957 | 0.403 | 42.910 | 41.943 | B |
| (28978) Ixion | 2771.62507 | 19.181 | 0.254 | 2771.63087 | 2.905 | 0.403 | 42.910 | 41.943 | I |
| (28978) Ixion | 2777.61899 | 20.453 | 0.258 | 2777.62496 | 4.179 | 0.270 | 42.906 | 41.914 | B |
| (28978) Ixion | 2785.54434 | 20.243 | 0.180 | 2785.55045 | 3.970 | 0.091 | 42.902 | 41.891 | V |
| (28978) Ixion | 2785.54605 | 20.061 | 0.169 | 2785.55216 | 3.788 | 0.091 | 42.902 | 41.891 | V |
| (28978) Ixion | 2785.54959 | 21.234 | 0.146 | 2785.55570 | 4.961 | 0.091 | 42.902 | 41.891 | B |
| (28978) Ixion | 2785.55293 | 19.195 | 0.253 | 2785.55904 | 2.922 | 0.091 | 42.902 | 41.891 | I |
| (28978) Ixion | 2787.55372 | 20.146 | 0.141 | 2787.55984 | 3.873 | 0.050 | 42.900 | 41.888 | V |
| (28978) Ixion | 2787.55543 | 20.094 | 0.125 | 2787.56155 | 3.821 | 0.049 | 42.900 | 41.888 | V |
| (28978) Ixion | 2787.55897 | 20.906 | 0.083 | 2787.56509 | 4.633 | 0.049 | 42.900 | 41.888 | B |
| (28978) Ixion | 2787.56231 | 18.917 | 0.157 | 2787.56843 | 2.644 | 0.049 | 42.900 | 41.888 | I |
| (28978) Ixion | 2789.58974 | 19.978 | 0.122 | 2789.59587 | 3.705 | 0.030 | 42.899 | 41.886 | V |
| (28978) Ixion | 2789.59145 | 20.029 | 0.136 | 2789.59758 | 3.756 | 0.030 | 42.899 | 41.886 | V |
| (28978) Ixion | 2789.59499 | 21.015 | 0.089 | 2789.60112 | 4.742 | 0.030 | 42.899 | 41.886 | B |
| (28978) Ixion | 2789.59832 | 18.970 | 0.161 | 2789.60445 | 2.697 | 0.030 | 42.899 | 41.886 | I |
| (28978) Ixion | 2791.55201 | 20.213 | 0.146 | 2791.55815 | 3.940 | 0.059 | 42.898 | 41.885 | V |
| (28978) Ixion | 2791.55372 | 19.906 | 0.129 | 2791.55986 | 3.633 | 0.060 | 42.898 | 41.885 | V |
| (28978) Ixion | 2791.55726 | 21.038 | 0.108 | 2791.56340 | 4.765 | 0.060 | 42.898 | 41.885 | B |
| (28978) Ixion | 2791.56061 | 19.209 | 0.202 | 2791.56675 | 2.936 | 0.060 | 42.898 | 41.885 | I |
| (28978) Ixion | 2793.67518 | 20.044 | 0.146 | 2793.68131 | 3.772 | 0.105 | 42.897 | 41.886 | V |
| (28978) Ixion | 2793.67689 | 20.235 | 0.137 | 2793.68302 | 3.963 | 0.105 | 42.897 | 41.886 | V |
| (28978) Ixion | 2793.68043 | 21.083 | 0.089 | 2793.68656 | 4.811 | 0.105 | 42.897 | 41.886 | B |
| (28978) Ixion | 2793.68378 | 18.845 | 0.155 | 2793.68991 | 2.573 | 0.105 | 42.897 | 41.886 | I |
| (28978) Ixion | 2819.51632 | 20.108 | 0.161 | 2819.52182 | 3.831 | 0.668 | 42.882 | 41.997 | V |
| (28978) Ixion | 2819.51803 | 20.694 | 0.212 | 2819.52353 | 4.417 | 0.668 | 42.882 | 41.997 | V |
| (28978) Ixion | 2819.52156 | 20.977 | 0.199 | 2819.52705 | 4.700 | 0.669 | 42.882 | 41.997 | B |
| (28978) Ixion | 2819.52488 | 19.402 | 0.273 | 2819.53037 | 3.125 | 0.669 | 42.882 | 41.997 | I |
| (28978) Ixion | 2821.54714 | 20.014 | 0.225 | 2821.55254 | 3.736 | 0.709 | 42.881 | 42.013 | V |
| (28978) Ixion | 2821.55068 | 21.434 | 0.287 | 2821.55608 | 5.156 | 0.709 | 42.881 | 42.013 | B |
| (28978) Ixion | 2837.50666 | 20.250 | 0.189 | 2837.51113 | 3.964 | 0.991 | 42.872 | 42.174 | V |
| (28978) Ixion | 2837.50837 | 20.109 | 0.195 | 2837.51284 | 3.823 | 0.991 | 42.872 | 42.174 | V |
| (28978) Ixion | 2837.51191 | 20.996 | 0.165 | 2837.51638 | 4.710 | 0.991 | 42.872 | 42.175 | B |
| (28978) Ixion | 2843.51916 | 20.439 | 0.145 | 2843.52320 | 4.149 | 1.080 | 42.869 | 42.249 | B |
| (28978) Ixion | 2846.53019 | 19.744 | 0.225 | 2846.53400 | 3.452 | 1.120 | 42.867 | 42.289 | V |
| (28978) Ixion | 2846.53374 | 20.844 | 0.161 | 2846.53755 | 4.552 | 1.120 | 42.867 | 42.289 | B |
| | | | | # | | | | | |
| (55636) 2002 TX300 | 3224.85445 | 20.344 | 0.068 | 3224.85445 | 4.250 | 1.245 | 40.935 | 40.434 | B |
| (55636) 2002 TX300 | 3224.85770 | 19.935 | 0.112 | 3224.85770 | 3.841 | 1.245 | 40.935 | 40.434 | V |
| (55636) 2002 TX300 | 3224.85941 | 19.050 | 0.111 | 3224.85941 | 2.956 | 1.245 | 40.935 | 40.434 | I |
| (55636) 2002 TX300 | 3249.71112 | 20.282 | 0.189 | 3249.71271 | 4.202 | 0.903 | 40.942 | 40.159 | B |
| (55636) 2002 TX300 | 3249.71437 | 19.582 | 0.210 | 3249.71596 | 3.502 | 0.903 | 40.942 | 40.159 | V |
| (55636) 2002 TX300 | 3249.71609 | 18.810 | 0.167 | 3249.71768 | 2.730 | 0.903 | 40.942 | 40.159 | I |
| (55636) 2002 TX300 | 3251.73481 | 20.366 | 0.100 | 3251.73650 | 4.287 | 0.871 | 40.943 | 40.142 | B |
| (55636) 2002 TX300 | 3251.73807 | 19.677 | 0.113 | 3251.73976 | 3.598 | 0.870 | 40.943 | 40.142 | V |
| (55636) 2002 TX300 | 3251.73979 | 18.974 | 0.109 | 3251.74148 | 2.895 | 0.870 | 40.943 | 40.142 | I |
| (55636) 2002 TX300 | 3254.76919 | 20.221 | 0.041 | 3254.77101 | 4.143 | 0.822 | 40.944 | 40.118 | B |
| (55636) 2002 TX300 | 3254.77245 | 19.469 | 0.067 | 3254.77427 | 3.391 | 0.822 | 40.944 | 40.118 | V |
| (55636) 2002 TX300 | 3254.77417 | 18.878 | 0.083 | 3254.77599 | 2.800 | 0.822 | 40.944 | 40.118 | I |
| (55636) 2002 TX300 | 3257.78003 | 20.401 | 0.052 | 3257.78197 | 4.324 | 0.774 | 40.945 | 40.097 | B |
| (55636) 2002 TX300 | 3257.78333 | 19.595 | 0.072 | 3257.78527 | 3.518 | 0.774 | 40.945 | 40.097 | V |
| (55636) 2002 TX300 | 3257.78508 | 19.015 | 0.097 | 3257.78702 | 2.938 | 0.774 | 40.945 | 40.097 | I |
| (55636) 2002 TX300 | 3259.74778 | 20.251 | 0.065 | 3259.74980 | 4.175 | 0.743 | 40.945 | 40.085 | B |
| (55636) 2002 TX300 | 3259.75104 | 19.332 | 0.096 | 3259.75306 | 3.256 | 0.743 | 40.945 | 40.085 | V |
| (55636) 2002 TX300 | 3259.75276 | 18.720 | 0.148 | 3259.75478 | 2.644 | 0.743 | 40.945 | 40.085 | I |

| | | | | | | | | | | |
|---|---|---|---|---|---|---|---|---|---|---|
| (55636) 2002 TX300 | 3261.74497 | 20.290 | 0.035 | 3261.74705 | 4.215 | 0.711 | 40.946 | 40.073 | B |
| (55636) 2002 TX300 | 3261.74823 | 19.633 | 0.063 | 3261.75031 | 3.558 | 0.711 | 40.946 | 40.073 | V |
| (55636) 2002 TX300 | 3261.74995 | 18.858 | 0.075 | 3261.75203 | 2.783 | 0.711 | 40.946 | 40.073 | I |
| (55636) 2002 TX300 | 3263.70981 | 20.246 | 0.024 | 3263.71195 | 4.171 | 0.681 | 40.947 | 40.062 | B |
| (55636) 2002 TX300 | 3263.71307 | 19.540 | 0.044 | 3263.71521 | 3.465 | 0.681 | 40.947 | 40.062 | V |
| (55636) 2002 TX300 | 3263.71480 | 18.840 | 0.056 | 3263.71694 | 2.765 | 0.681 | 40.947 | 40.062 | I |
| (55636) 2002 TX300 | 3265.74884 | 20.358 | 0.035 | 3265.75104 | 4.284 | 0.651 | 40.947 | 40.053 | B |
| (55636) 2002 TX300 | 3265.75213 | 19.545 | 0.061 | 3265.75433 | 3.471 | 0.651 | 40.947 | 40.053 | V |
| (55636) 2002 TX300 | 3265.75388 | 18.890 | 0.071 | 3265.75608 | 2.816 | 0.651 | 40.947 | 40.053 | I |
| (55636) 2002 TX300 | 3267.78162 | 19.407 | 0.148 | 3267.78387 | 3.333 | 0.623 | 40.948 | 40.044 | V |
| (55636) 2002 TX300 | 3267.78334 | 18.651 | 0.184 | 3267.78559 | 2.577 | 0.623 | 40.948 | 40.044 | I |
| (55636) 2002 TX300 | 3269.80140 | 20.299 | 0.058 | 3269.80370 | 4.226 | 0.596 | 40.949 | 40.036 | B |
| (55636) 2002 TX300 | 3269.80466 | 19.575 | 0.196 | 3269.80696 | 3.502 | 0.596 | 40.949 | 40.036 | V |
| (55636) 2002 TX300 | 3269.80638 | 19.028 | 0.295 | 3269.80868 | 2.955 | 0.596 | 40.949 | 40.036 | I |
| (55636) 2002 TX300 | 3271.72757 | 20.326 | 0.059 | 3271.72990 | 4.253 | 0.572 | 40.949 | 40.030 | B |
| (55636) 2002 TX300 | 3271.73084 | 19.492 | 0.067 | 3271.73317 | 3.419 | 0.572 | 40.949 | 40.030 | V |
| (55636) 2002 TX300 | 3271.73256 | 18.859 | 0.070 | 3271.73489 | 2.786 | 0.572 | 40.949 | 40.030 | I |
| (55636) 2002 TX300 | 3273.71659 | 20.141 | 0.115 | 3273.71895 | 4.068 | 0.550 | 40.950 | 40.025 | B |
| (55636) 2002 TX300 | 3273.71988 | 19.541 | 0.134 | 3273.72224 | 3.468 | 0.550 | 40.950 | 40.025 | V |
| (55636) 2002 TX300 | 3273.72164 | 18.823 | 0.088 | 3273.72400 | 2.750 | 0.550 | 40.950 | 40.025 | I |
| (55636) 2002 TX300 | 3281.71200 | 19.544 | 0.074 | 3281.71442 | 3.472 | 0.493 | 40.952 | 40.014 | V |
| (55636) 2002 TX300 | 3281.71374 | 18.831 | 0.070 | 3281.71616 | 2.759 | 0.493 | 40.952 | 40.014 | I |
| (55636) 2002 TX300 | 3283.70227 | 20.251 | 0.028 | 3283.70469 | 4.179 | 0.489 | 40.953 | 40.014 | B |
| (55636) 2002 TX300 | 3283.70555 | 19.576 | 0.052 | 3283.70797 | 3.504 | 0.489 | 40.953 | 40.014 | V |
| (55636) 2002 TX300 | 3283.70729 | 18.825 | 0.070 | 3283.70971 | 2.753 | 0.489 | 40.953 | 40.014 | I |
| (55636) 2002 TX300 | 3291.65234 | 20.323 | 0.040 | 3291.65469 | 4.250 | 0.512 | 40.955 | 40.026 | B |
| (55636) 2002 TX300 | 3291.65564 | 19.555 | 0.070 | 3291.65799 | 3.482 | 0.512 | 40.955 | 40.026 | V |
| (55636) 2002 TX300 | 3291.65739 | 18.720 | 0.083 | 3291.65974 | 2.647 | 0.512 | 40.955 | 40.026 | I |
| (55636) 2002 TX300 | 3293.64542 | 20.367 | 0.061 | 3293.64774 | 4.293 | 0.528 | 40.956 | 40.032 | B |
| (55636) 2002 TX300 | 3293.64871 | 19.559 | 0.104 | 3293.65103 | 3.485 | 0.528 | 40.956 | 40.032 | V |
| (55636) 2002 TX300 | 3293.65047 | 19.041 | 0.156 | 3293.65279 | 2.967 | 0.528 | 40.956 | 40.032 | I |
| (55636) 2002 TX300 | 3295.67248 | 20.295 | 0.030 | 3295.67476 | 4.221 | 0.547 | 40.956 | 40.039 | B |
| (55636) 2002 TX300 | 3295.67578 | 19.488 | 0.055 | 3295.67806 | 3.414 | 0.547 | 40.956 | 40.039 | V |
| (55636) 2002 TX300 | 3295.67754 | 18.911 | 0.072 | 3295.67982 | 2.837 | 0.547 | 40.956 | 40.039 | I |
| (55636) 2002 TX300 | 3297.67597 | 20.258 | 0.031 | 3297.67820 | 4.184 | 0.569 | 40.957 | 40.047 | B |
| (55636) 2002 TX300 | 3297.68103 | 18.868 | 0.233 | 3297.68326 | 2.794 | 0.569 | 40.957 | 40.047 | I |
| (55636) 2002 TX300 | 3299.67439 | 20.289 | 0.041 | 3299.67657 | 4.214 | 0.593 | 40.958 | 40.056 | B |
| (55636) 2002 TX300 | 3299.67769 | 19.584 | 0.064 | 3299.67987 | 3.509 | 0.593 | 40.958 | 40.056 | V |
| (55636) 2002 TX300 | 3299.67944 | 18.840 | 0.065 | 3299.68162 | 2.765 | 0.593 | 40.958 | 40.056 | I |
| (55636) 2002 TX300 | 3302.60379 | 20.121 | 0.146 | 3302.60588 | 4.045 | 0.632 | 40.959 | 40.071 | B |
| (55636) 2002 TX300 | 3302.60708 | 19.450 | 0.187 | 3302.60917 | 3.374 | 0.632 | 40.959 | 40.071 | V |
| (55636) 2002 TX300 | 3302.60884 | 18.964 | 0.166 | 3302.61093 | 2.888 | 0.632 | 40.959 | 40.071 | I |
| (55636) 2002 TX300 | 3307.62713 | 20.452 | 0.264 | 3307.62904 | 4.374 | 0.706 | 40.960 | 40.103 | B |
| (55636) 2002 TX300 | 3307.63044 | 19.426 | 0.210 | 3307.63235 | 3.348 | 0.706 | 40.960 | 40.103 | V |
| (55636) 2002 TX300 | 3307.63220 | 18.845 | 0.175 | 3307.63411 | 2.767 | 0.706 | 40.960 | 40.103 | I |
| (55636) 2002 TX300 | 3309.60679 | 19.433 | 0.076 | 3309.60862 | 3.354 | 0.736 | 40.961 | 40.118 | V |
| (55636) 2002 TX300 | 3309.60856 | 18.848 | 0.087 | 3309.61039 | 2.769 | 0.736 | 40.961 | 40.118 | I |
| (55636) 2002 TX300 | 3311.59818 | 20.279 | 0.027 | 3311.59992 | 4.200 | 0.767 | 40.961 | 40.133 | B |
| (55636) 2002 TX300 | 3311.60148 | 19.702 | 0.059 | 3311.60322 | 3.623 | 0.768 | 40.961 | 40.133 | V |
| (55636) 2002 TX300 | 3311.60324 | 18.923 | 0.067 | 3311.60498 | 2.844 | 0.768 | 40.961 | 40.133 | I |

| | | | | | | | | | | |
|---|---|---|---|---|---|---|---|---|---|---|
| (55636) 2002 TX300 | 3325.57787 | 20.286 | 0.031 | 3325.57882 | 4.199 | 0.985 | 40.966 | 40.269 | B |
| (55636) 2002 TX300 | 3325.58117 | 19.597 | 0.052 | 3325.58212 | 3.510 | 0.985 | 40.966 | 40.269 | V |
| (55636) 2002 TX300 | 3325.58293 | 18.955 | 0.079 | 3325.58388 | 2.868 | 0.985 | 40.966 | 40.269 | I |
| (55636) 2002 TX300 | 3329.58035 | 20.271 | 0.059 | 3329.58103 | 4.181 | 1.043 | 40.967 | 40.316 | B |
| (55636) 2002 TX300 | 3329.58365 | 19.493 | 0.077 | 3329.58433 | 3.403 | 1.043 | 40.967 | 40.316 | V |
| (55636) 2002 TX300 | 3329.58541 | 18.898 | 0.082 | 3329.58609 | 2.808 | 1.043 | 40.967 | 40.316 | I |
| (55636) 2002 TX300 | 3331.53421 | 20.279 | 0.152 | 3331.53475 | 4.188 | 1.071 | 40.967 | 40.340 | B |
| (55636) 2002 TX300 | 3331.53751 | 19.409 | 0.162 | 3331.53805 | 3.318 | 1.071 | 40.967 | 40.340 | V |
| (55636) 2002 TX300 | 3334.56090 | 20.413 | 0.253 | 3334.56122 | 4.320 | 1.111 | 40.968 | 40.378 | B |
| (55636) 2002 TX300 | 3334.56419 | 19.468 | 0.223 | 3334.56451 | 3.375 | 1.111 | 40.968 | 40.378 | V |
| (55636) 2002 TX300 | 3334.56595 | 18.787 | 0.169 | 3334.56627 | 2.694 | 1.111 | 40.968 | 40.378 | I |
| (55636) 2002 TX300 | 3338.55692 | 20.274 | 0.040 | 3338.55693 | 4.178 | 1.161 | 40.970 | 40.432 | B |
| (55636) 2002 TX300 | 3338.56022 | 19.361 | 0.061 | 3338.56023 | 3.265 | 1.161 | 40.970 | 40.432 | V |
| (55636) 2002 TX300 | 3338.56197 | 18.832 | 0.091 | 3338.56198 | 2.736 | 1.161 | 40.970 | 40.432 | I |
| (55636) 2002 TX300 | 3340.52714 | 20.245 | 0.054 | 3340.52699 | 4.148 | 1.183 | 40.970 | 40.459 | B |
| (55636) 2002 TX300 | 3340.53044 | 19.567 | 0.061 | 3340.53029 | 3.470 | 1.183 | 40.970 | 40.459 | V |
| (55636) 2002 TX300 | 3340.53220 | 18.911 | 0.074 | 3340.53205 | 2.814 | 1.183 | 40.970 | 40.459 | I |
| (55636) 2002 TX300 | 3344.53053 | 19.647 | 0.076 | 3344.53005 | 3.546 | 1.226 | 40.971 | 40.517 | V |
| (55636) 2002 TX300 | 3344.53228 | 18.867 | 0.078 | 3344.53180 | 2.766 | 1.226 | 40.971 | 40.517 | I |
| (55636) 2002 TX300 | 3349.54627 | 20.355 | 0.047 | 3349.54536 | 4.250 | 1.273 | 40.973 | 40.592 | B |
| (55636) 2002 TX300 | 3352.53220 | 20.205 | 0.079 | 3352.53102 | 4.098 | 1.297 | 40.974 | 40.638 | B |
| (55636) 2002 TX300 | 3352.53551 | 19.804 | 0.085 | 3352.53433 | 3.697 | 1.297 | 40.974 | 40.639 | V |
| (55636) 2002 TX300 | 3352.53726 | 18.867 | 0.169 | 3352.53608 | 2.760 | 1.297 | 40.974 | 40.639 | I |
| (55636) 2002 TX300 | 3354.53114 | 20.342 | 0.086 | 3354.52978 | 4.233 | 1.311 | 40.974 | 40.670 | B |
| (55636) 2002 TX300 | 3354.53444 | 19.817 | 0.082 | 3354.53308 | 3.708 | 1.311 | 40.974 | 40.670 | V |
| (55636) 2002 TX300 | 3354.53620 | 19.045 | 0.094 | 3354.53484 | 2.936 | 1.311 | 40.974 | 40.670 | I |
| (55636) 2002 TX300 | 3356.53380 | 20.569 | 0.139 | 3356.53225 | 4.458 | 1.324 | 40.975 | 40.702 | B |
| (55636) 2002 TX300 | 3356.53710 | 19.661 | 0.095 | 3356.53555 | 3.550 | 1.324 | 40.975 | 40.702 | V |
| (55636) 2002 TX300 | 3356.53885 | 18.935 | 0.096 | 3356.53730 | 2.824 | 1.324 | 40.975 | 40.702 | I |
| (55636) 2002 TX300 | 3358.52653 | 19.612 | 0.158 | 3358.52480 | 3.500 | 1.335 | 40.976 | 40.734 | V |
| (55636) 2002 TX300 | 3358.52828 | 18.914 | 0.155 | 3358.52655 | 2.802 | 1.335 | 40.976 | 40.734 | I |
| | | | | # | | | | | |
| (55565) 2002 AW197 | 2696.63646 | 20.052 | 0.101 | 2696.63646 | 3.349 | 0.497 | 47.264 | 46.363 | V |
| (55565) 2002 AW197 | 2696.63863 | 20.661 | 0.168 | 2696.63863 | 3.958 | 0.497 | 47.264 | 46.363 | V |
| (55565) 2002 AW197 | 2696.64357 | 21.174 | 0.083 | 2696.64357 | 4.471 | 0.497 | 47.264 | 46.363 | B |
| (55565) 2002 AW197 | 2696.64830 | 19.261 | 0.155 | 2696.64830 | 2.557 | 0.497 | 47.264 | 46.363 | I |
| (55565) 2002 AW197 | 2698.66134 | 21.099 | 0.096 | 2698.66126 | 4.395 | 0.534 | 47.263 | 46.376 | B |
| (55565) 2002 AW197 | 2698.66700 | 20.569 | 0.199 | 2698.66692 | 3.865 | 0.534 | 47.263 | 46.376 | V |
| (55565) 2002 AW197 | 2698.66916 | 20.079 | 0.133 | 2698.66908 | 3.375 | 0.534 | 47.263 | 46.376 | V |
| (55565) 2002 AW197 | 2698.67141 | 19.209 | 0.187 | 2698.67133 | 2.505 | 0.534 | 47.263 | 46.376 | I |
| (55565) 2002 AW197 | 2700.70830 | 20.364 | 0.152 | 2700.70814 | 3.659 | 0.572 | 47.263 | 46.391 | V |
| (55565) 2002 AW197 | 2700.71047 | 20.891 | 0.260 | 2700.71031 | 4.186 | 0.572 | 47.263 | 46.391 | V |
| (55565) 2002 AW197 | 2700.71541 | 21.170 | 0.093 | 2700.71525 | 4.465 | 0.572 | 47.263 | 46.391 | B |
| (55565) 2002 AW197 | 2700.72014 | 19.359 | 0.180 | 2700.71998 | 2.654 | 0.572 | 47.263 | 46.391 | I |
| (55565) 2002 AW197 | 2702.67722 | 20.535 | 0.211 | 2702.67697 | 3.830 | 0.607 | 47.262 | 46.406 | V |
| (55565) 2002 AW197 | 2702.67939 | 20.395 | 0.203 | 2702.67914 | 3.690 | 0.607 | 47.262 | 46.406 | V |
| (55565) 2002 AW197 | 2702.68433 | 21.540 | 0.174 | 2702.68408 | 4.835 | 0.607 | 47.262 | 46.406 | B |
| (55565) 2002 AW197 | 2702.69120 | 19.201 | 0.253 | 2702.69095 | 2.496 | 0.608 | 47.262 | 46.406 | I |
| (55565) 2002 AW197 | 2704.68183 | 20.555 | 0.206 | 2704.68148 | 3.849 | 0.643 | 47.261 | 46.423 | V |
| (55565) 2002 AW197 | 2704.68400 | 20.024 | 0.128 | 2704.68365 | 3.318 | 0.643 | 47.261 | 46.423 | V |
| (55565) 2002 AW197 | 2704.68893 | 21.420 | 0.144 | 2704.68858 | 4.714 | 0.643 | 47.261 | 46.423 | B |
| (55565) 2002 AW197 | 2704.69366 | 19.047 | 0.175 | 2704.69331 | 2.341 | 0.643 | 47.261 | 46.423 | I |
| (55565) 2002 AW197 | 2706.64926 | 20.540 | 0.160 | 2706.64881 | 3.833 | 0.677 | 47.261 | 46.440 | V |
| (55565) 2002 AW197 | 2706.65143 | 20.385 | 0.140 | 2706.65098 | 3.678 | 0.678 | 47.261 | 46.440 | V |
| (55565) 2002 AW197 | 2706.65636 | 21.295 | 0.097 | 2706.65591 | 4.588 | 0.678 | 47.261 | 46.440 | B |
| (55565) 2002 AW197 | 2706.66109 | 19.226 | 0.149 | 2706.66064 | 2.519 | 0.678 | 47.261 | 46.440 | I |
| (55565) 2002 AW197 | 2708.61425 | 20.416 | 0.137 | 2708.61370 | 3.708 | 0.711 | 47.260 | 46.458 | V |
| (55565) 2002 AW197 | 2708.61641 | 20.340 | 0.127 | 2708.61586 | 3.632 | 0.711 | 47.260 | 46.458 | V |
| (55565) 2002 AW197 | 2708.62135 | 21.296 | 0.090 | 2708.62080 | 4.588 | 0.711 | 47.260 | 46.458 | B |
| (55565) 2002 AW197 | 2708.62607 | 19.284 | 0.161 | 2708.62552 | 2.576 | 0.711 | 47.260 | 46.458 | I |
| (55565) 2002 AW197 | 2710.62509 | 20.347 | 0.147 | 2710.62443 | 3.638 | 0.745 | 47.259 | 46.478 | V |
| (55565) 2002 AW197 | 2710.62726 | 20.686 | 0.197 | 2710.62660 | 3.977 | 0.745 | 47.259 | 46.478 | V |
| (55565) 2002 AW197 | 2710.63219 | 21.255 | 0.125 | 2710.63153 | 4.546 | 0.745 | 47.259 | 46.478 | B |
| (55565) 2002 AW197 | 2710.63691 | 19.163 | 0.132 | 2710.63625 | 2.454 | 0.745 | 47.259 | 46.478 | I |
| (55565) 2002 AW197 | 2718.67207 | 18.963 | 0.171 | 2718.67090 | 2.250 | 0.871 | 47.257 | 46.566 | I |
| (55565) 2002 AW197 | 2724.69327 | 20.033 | 0.180 | 2724.69167 | 3.317 | 0.954 | 47.255 | 46.640 | V |
| (55565) 2002 AW197 | 2726.59242 | 20.297 | 0.261 | 2726.59068 | 3.580 | 0.979 | 47.254 | 46.665 | V |
| (55565) 2002 AW197 | 2726.59736 | 21.189 | 0.155 | 2726.59562 | 4.472 | 0.979 | 47.254 | 46.665 | B |
| (55565) 2002 AW197 | 2726.60209 | 19.638 | 0.185 | 2726.60034 | 2.921 | 0.979 | 47.254 | 46.665 | I |
| (55565) 2002 AW197 | 2728.60007 | 20.408 | 0.139 | 2728.59817 | 3.690 | 1.003 | 47.254 | 46.692 | V |
| (55565) 2002 AW197 | 2728.60224 | 20.407 | 0.134 | 2728.60034 | 3.689 | 1.003 | 47.254 | 46.692 | V |
| (55565) 2002 AW197 | 2728.60718 | 21.256 | 0.094 | 2728.60528 | 4.538 | 1.003 | 47.254 | 46.692 | B |
| (55565) 2002 AW197 | 2728.61192 | 19.343 | 0.222 | 2728.61002 | 2.625 | 1.003 | 47.254 | 46.692 | I |
| (55565) 2002 AW197 | 2730.58462 | 20.840 | 0.177 | 2730.58256 | 4.120 | 1.026 | 47.253 | 46.719 | V |
| (55565) 2002 AW197 | 2730.58678 | 20.603 | 0.140 | 2730.58472 | 3.883 | 1.026 | 47.253 | 46.720 | V |
| (55565) 2002 AW197 | 2730.59172 | 21.467 | 0.088 | 2730.58966 | 4.747 | 1.026 | 47.253 | 46.720 | B |
| (55565) 2002 AW197 | 2730.59646 | 19.260 | 0.132 | 2730.59440 | 2.540 | 1.026 | 47.253 | 46.720 | I |
| (55565) 2002 AW197 | 2732.58622 | 20.354 | 0.140 | 2732.58400 | 3.633 | 1.049 | 47.252 | 46.748 | V |
| (55565) 2002 AW197 | 2732.58839 | 20.304 | 0.140 | 2732.58617 | 3.583 | 1.049 | 47.252 | 46.748 | V |
| (55565) 2002 AW197 | 2732.59334 | 21.358 | 0.096 | 2732.59112 | 4.637 | 1.049 | 47.252 | 46.748 | B |
| (55565) 2002 AW197 | 2734.59869 | 20.953 | 0.261 | 2734.59630 | 4.231 | 1.069 | 47.252 | 46.777 | V |
| (55565) 2002 AW197 | 2734.60085 | 20.552 | 0.181 | 2734.59846 | 3.830 | 1.069 | 47.252 | 46.777 | V |
| (55565) 2002 AW197 | 2734.60579 | 21.153 | 0.219 | 2734.60340 | 4.431 | 1.069 | 47.252 | 46.777 | B |
| (55565) 2002 AW197 | 2734.61053 | 19.214 | 0.180 | 2734.60814 | 2.492 | 1.070 | 47.252 | 46.777 | I |
| (55565) 2002 AW197 | 2738.58422 | 20.551 | 0.167 | 2738.58149 | 3.826 | 1.107 | 47.250 | 46.836 | V |
| (55565) 2002 AW197 | 2738.58639 | 20.829 | 0.205 | 2738.58366 | 4.104 | 1.107 | 47.250 | 46.836 | V |
| (55565) 2002 AW197 | 2738.59133 | 21.051 | 0.233 | 2738.58860 | 4.326 | 1.107 | 47.250 | 46.836 | B |
| (55565) 2002 AW197 | 2738.59606 | 19.264 | 0.149 | 2738.59333 | 2.539 | 1.107 | 47.250 | 46.836 | I |
| (55565) 2002 AW197 | 2748.56978 | 19.877 | 0.289 | 2748.56615 | 3.145 | 1.179 | 47.247 | 46.991 | V |
| (55565) 2002 AW197 | 2750.59874 | 20.519 | 0.286 | 2750.59492 | 3.786 | 1.189 | 47.246 | 47.024 | V |
| (55565) 2002 AW197 | 2750.60367 | 21.223 | 0.188 | 2750.59985 | 4.490 | 1.189 | 47.246 | 47.024 | B |
| (55565) 2002 AW197 | 2750.60839 | 19.299 | 0.204 | 2750.60457 | 2.566 | 1.189 | 47.246 | 47.024 | I |
| (55565) 2002 AW197 | 2752.50506 | 20.295 | 0.143 | 2752.50106 | 3.560 | 1.198 | 47.246 | 47.055 | V |
| (55565) 2002 AW197 | 2752.50722 | 20.603 | 0.181 | 2752.50322 | 3.868 | 1.198 | 47.246 | 47.055 | V |
| (55565) 2002 AW197 | 2752.51214 | 21.315 | 0.081 | 2752.50814 | 4.580 | 1.198 | 47.246 | 47.055 | B |
| (55565) 2002 AW197 | 2752.51686 | 19.503 | 0.241 | 2752.51286 | 2.768 | 1.198 | 47.246 | 47.055 | I |
| (55565) 2002 AW197 | 2754.50548 | 20.418 | 0.135 | 2754.50130 | 3.682 | 1.205 | 47.245 | 47.088 | V |
| (55565) 2002 AW197 | 2754.50765 | 20.343 | 0.131 | 2754.50346 | 3.607 | 1.205 | 47.245 | 47.088 | V |
| (55565) 2002 AW197 | 2754.51257 | 21.500 | 0.098 | 2754.50838 | 4.764 | 1.205 | 47.245 | 47.088 | B |
| (55565) 2002 AW197 | 2754.51728 | 19.638 | 0.221 | 2754.51309 | 2.902 | 1.205 | 47.245 | 47.088 | I |

| | | | | | | | | | | |
|---|---|---|---|---|---|---|---|---|---|---|
| (55565) 2002 AW197 | 2760.48588 | 20.446 | 0.180 | 2760.48113 | 3.705 | 1.219 | 47.243 | 47.186 | V |
| (55565) 2002 AW197 | 2760.48805 | 20.467 | 0.150 | 2760.48330 | 3.726 | 1.219 | 47.243 | 47.186 | V |
| (55565) 2002 AW197 | 2760.49299 | 21.360 | 0.090 | 2760.48824 | 4.619 | 1.219 | 47.243 | 47.186 | B |
| (55565) 2002 AW197 | 2760.49773 | 19.261 | 0.203 | 2760.49298 | 2.520 | 1.219 | 47.243 | 47.186 | I |
| (55565) 2002 AW197 | 2762.49649 | 20.335 | 0.209 | 2762.49154 | 3.593 | 1.221 | 47.243 | 47.219 | V |
| (55565) 2002 AW197 | 2762.49866 | 20.498 | 0.255 | 2762.49371 | 3.756 | 1.221 | 47.243 | 47.219 | V |
| (55565) 2002 AW197 | 2762.50359 | 21.339 | 0.125 | 2762.49864 | 4.597 | 1.222 | 47.243 | 47.219 | B |
| (55565) 2002 AW197 | 2766.46231 | 20.313 | 0.206 | 2766.45699 | 3.568 | 1.221 | 47.241 | 47.285 | V |
| (55565) 2002 AW197 | 2766.46941 | 21.259 | 0.119 | 2766.46408 | 4.514 | 1.221 | 47.241 | 47.285 | B |
| (55565) 2002 AW197 | 2766.47414 | 18.922 | 0.232 | 2766.46881 | 2.177 | 1.221 | 47.241 | 47.285 | I |
| (55565) 2002 AW197 | 2786.45850 | 21.331 | 0.169 | 2786.45132 | 4.571 | 1.140 | 47.235 | 47.606 | B |
| (55565) 2002 AW197 | 2786.46323 | 19.431 | 0.256 | 2786.45605 | 2.671 | 1.140 | 47.235 | 47.606 | I |
| (55565) 2002 AW197 | 2790.49552 | 20.503 | 0.249 | 2790.48799 | 3.741 | 1.108 | 47.234 | 47.667 | V |
| (55565) 2002 AW197 | 2790.49769 | 20.588 | 0.274 | 2790.49016 | 3.826 | 1.108 | 47.234 | 47.667 | V |
| (55565) 2002 AW197 | 2794.47768 | 20.569 | 0.273 | 2794.46982 | 3.804 | 1.072 | 47.232 | 47.725 | V |
| (55565) 2002 AW197 | 2794.47984 | 20.519 | 0.251 | 2794.47197 | 3.754 | 1.072 | 47.232 | 47.725 | V |
| (55565) 2002 AW197 | 2794.48478 | 21.244 | 0.189 | 2794.47691 | 4.479 | 1.072 | 47.232 | 47.725 | B |
| (55565) 2002 AW197 | 2794.48951 | 19.210 | 0.252 | 2794.48164 | 2.445 | 1.072 | 47.232 | 47.725 | I |
| (55565) 2002 AW197 | 2973.75808 | 20.491 | 0.070 | 2973.75554 | 3.771 | 1.117 | 47.174 | 46.803 | V |
| (55565) 2002 AW197 | 2973.76040 | 20.506 | 0.067 | 2973.75786 | 3.786 | 1.117 | 47.174 | 46.803 | V |
| (55565) 2002 AW197 | 2973.76543 | 21.289 | 0.041 | 2973.76289 | 4.569 | 1.117 | 47.174 | 46.803 | B |
| (55565) 2002 AW197 | 2973.77048 | 19.322 | 0.066 | 2973.76794 | 2.602 | 1.117 | 47.174 | 46.803 | I |
| (55565) 2002 AW197 | 2975.76201 | 20.532 | 0.072 | 2975.75966 | 3.814 | 1.101 | 47.174 | 46.771 | V |
| (55565) 2002 AW197 | 2975.76433 | 20.434 | 0.064 | 2975.76198 | 3.716 | 1.101 | 47.174 | 46.770 | V |
| (55565) 2002 AW197 | 2975.76936 | 21.337 | 0.045 | 2975.76701 | 4.619 | 1.100 | 47.174 | 46.770 | B |
| (55565) 2002 AW197 | 2975.77442 | 19.408 | 0.068 | 2975.77207 | 2.690 | 1.100 | 47.174 | 46.770 | I |
| (55565) 2002 AW197 | 2980.72305 | 20.308 | 0.231 | 2980.72115 | 3.593 | 1.054 | 47.172 | 46.693 | V |
| (55565) 2002 AW197 | 2980.72536 | 20.239 | 0.182 | 2980.72346 | 3.524 | 1.054 | 47.172 | 46.693 | V |
| (55565) 2002 AW197 | 2980.73039 | 20.910 | 0.167 | 2980.72849 | 4.195 | 1.054 | 47.172 | 46.692 | B |
| (55565) 2002 AW197 | 2980.73544 | 19.181 | 0.091 | 2980.73354 | 2.466 | 1.054 | 47.172 | 46.692 | I |
| (55565) 2002 AW197 | 3001.77937 | 20.286 | 0.051 | 3001.77911 | 3.585 | 0.775 | 47.165 | 46.409 | V |
| (55565) 2002 AW197 | 3001.78169 | 20.330 | 0.053 | 3001.78143 | 3.629 | 0.775 | 47.165 | 46.409 | V |
| (55565) 2002 AW197 | 3001.82296 | 21.293 | 0.037 | 3001.82270 | 4.592 | 0.774 | 47.165 | 46.408 | B |
| (55565) 2002 AW197 | 3001.82802 | 19.351 | 0.055 | 3001.82776 | 2.650 | 0.774 | 47.165 | 46.408 | I |
| (55565) 2002 AW197 | 3022.77672 | 21.267 | 0.047 | 3022.77749 | 4.575 | 0.402 | 47.158 | 46.229 | B |
| (55565) 2002 AW197 | 3022.78176 | 20.357 | 0.066 | 3022.78253 | 3.665 | 0.401 | 47.158 | 46.229 | V |
| (55565) 2002 AW197 | 3022.78407 | 20.374 | 0.065 | 3022.78484 | 3.682 | 0.401 | 47.158 | 46.229 | V |
| (55565) 2002 AW197 | 3022.78653 | 19.210 | 0.084 | 3022.78730 | 2.518 | 0.401 | 47.158 | 46.229 | I |
| (55565) 2002 AW197 | 3024.71212 | 21.300 | 0.035 | 3024.71295 | 4.608 | 0.366 | 47.158 | 46.219 | B |
| (55565) 2002 AW197 | 3024.71717 | 20.308 | 0.049 | 3024.71800 | 3.616 | 0.366 | 47.158 | 46.219 | V |
| (55565) 2002 AW197 | 3024.71949 | 20.305 | 0.049 | 3024.72032 | 3.613 | 0.366 | 47.158 | 46.218 | V |
| (55565) 2002 AW197 | 3024.72194 | 19.245 | 0.059 | 3024.72277 | 2.553 | 0.366 | 47.158 | 46.218 | I |
| (55565) 2002 AW197 | 3039.71026 | 21.247 | 0.216 | 3039.71134 | 4.557 | 0.165 | 47.153 | 46.176 | B |
| (55565) 2002 AW197 | 3039.71636 | 20.357 | 0.188 | 3039.71744 | 3.667 | 0.165 | 47.153 | 46.176 | V |
| (55565) 2002 AW197 | 3039.71867 | 20.192 | 0.181 | 3039.71975 | 3.502 | 0.165 | 47.153 | 46.176 | V |
| (55565) 2002 AW197 | 3039.72119 | 19.226 | 0.121 | 3039.72227 | 2.536 | 0.165 | 47.153 | 46.176 | I |
| (55565) 2002 AW197 | 3044.71718 | 20.955 | 0.135 | 3044.71825 | 4.265 | 0.191 | 47.151 | 46.177 | B |
| (55565) 2002 AW197 | 3044.72222 | 20.128 | 0.136 | 3044.72329 | 3.438 | 0.191 | 47.151 | 46.177 | V |
| (55565) 2002 AW197 | 3044.72454 | 20.186 | 0.144 | 3044.72561 | 3.496 | 0.191 | 47.151 | 46.177 | V |
| (55565) 2002 AW197 | 3044.72701 | 19.158 | 0.086 | 3044.72808 | 2.468 | 0.191 | 47.151 | 46.177 | I |

| | | | | | | | | | | |
|---|---|---|---|---|---|---|---|---|---|---|
| (55565) | 2002 AW197 | 3046.71341 | 21.265 | 0.107 | 3046.71447 | 4.575 | 0.215 | 47.151 | 46.180 | B |
| (55565) | 2002 AW197 | 3046.71845 | 20.468 | 0.112 | 3046.71951 | 3.778 | 0.215 | 47.151 | 46.180 | V |
| (55565) | 2002 AW197 | 3046.72077 | 20.344 | 0.107 | 3046.72183 | 3.654 | 0.216 | 47.151 | 46.180 | V |
| (55565) | 2002 AW197 | 3046.72323 | 19.379 | 0.083 | 3046.72429 | 2.689 | 0.216 | 47.151 | 46.180 | I |
| (55565) | 2002 AW197 | 3050.69070 | 20.341 | 0.052 | 3050.69171 | 3.651 | 0.277 | 47.149 | 46.189 | V |
| (55565) | 2002 AW197 | 3050.69301 | 20.308 | 0.069 | 3050.69402 | 3.618 | 0.277 | 47.149 | 46.189 | V |
| (55565) | 2002 AW197 | 3050.69547 | 19.146 | 0.052 | 3050.69648 | 2.456 | 0.277 | 47.149 | 46.189 | I |
| (55565) | 2002 AW197 | 3050.69786 | 19.116 | 0.051 | 3050.69887 | 2.426 | 0.277 | 47.149 | 46.189 | I |
| (55565) | 2002 AW197 | 3064.66168 | 20.340 | 0.075 | 3064.66229 | 3.647 | 0.533 | 47.145 | 46.256 | V |
| (55565) | 2002 AW197 | 3064.66400 | 20.343 | 0.070 | 3064.66461 | 3.650 | 0.533 | 47.145 | 46.257 | V |
| (55565) | 2002 AW197 | 3064.66646 | 19.245 | 0.062 | 3064.66708 | 2.552 | 0.533 | 47.145 | 46.257 | I |
| (55565) | 2002 AW197 | 3064.66885 | 19.235 | 0.066 | 3064.66946 | 2.542 | 0.533 | 47.145 | 46.257 | I |
| | | | | | # | | | | | |
| (55637) | 2002 UX25 | 2845.89169 | 20.345 | 0.154 | 2845.89169 | 4.059 | 1.361 | 42.579 | 42.464 | V |
| (55637) | 2002 UX25 | 2845.89398 | 20.399 | 0.154 | 2845.89398 | 4.113 | 1.361 | 42.579 | 42.464 | V |
| (55637) | 2002 UX25 | 2845.89897 | 21.203 | 0.168 | 2845.89897 | 4.917 | 1.361 | 42.579 | 42.464 | B |
| (55637) | 2002 UX25 | 2845.90398 | 19.546 | 0.268 | 2845.90398 | 3.260 | 1.361 | 42.579 | 42.464 | I |
| (55637) | 2002 UX25 | 2854.80946 | 20.552 | 0.084 | 2854.81033 | 4.274 | 1.325 | 42.576 | 42.312 | V |
| (55637) | 2002 UX25 | 2854.81175 | 20.504 | 0.089 | 2854.81262 | 4.226 | 1.325 | 42.576 | 42.312 | V |
| (55637) | 2002 UX25 | 2854.81695 | 21.538 | 0.070 | 2854.81782 | 5.260 | 1.325 | 42.576 | 42.312 | B |
| (55637) | 2002 UX25 | 2854.82200 | 19.390 | 0.131 | 2854.82287 | 3.112 | 1.325 | 42.576 | 42.312 | I |
| (55637) | 2002 UX25 | 2856.82967 | 20.351 | 0.137 | 2856.83074 | 4.075 | 1.313 | 42.575 | 42.279 | V |
| (55637) | 2002 UX25 | 2856.83196 | 20.237 | 0.127 | 2856.83303 | 3.961 | 1.313 | 42.575 | 42.279 | V |
| (55637) | 2002 UX25 | 2856.83707 | 21.222 | 0.114 | 2856.83814 | 4.946 | 1.313 | 42.575 | 42.279 | B |
| (55637) | 2002 UX25 | 2856.84214 | 19.160 | 0.188 | 2856.84321 | 2.884 | 1.313 | 42.575 | 42.279 | I |
| (55637) | 2002 UX25 | 2858.82793 | 20.254 | 0.125 | 2858.82919 | 3.979 | 1.299 | 42.574 | 42.246 | V |
| (55637) | 2002 UX25 | 2858.83293 | 21.337 | 0.062 | 2858.83419 | 5.062 | 1.299 | 42.574 | 42.246 | B |
| (55637) | 2002 UX25 | 2858.83795 | 19.255 | 0.098 | 2858.83921 | 2.980 | 1.299 | 42.574 | 42.246 | I |
| (55637) | 2002 UX25 | 2859.87063 | 20.277 | 0.086 | 2859.87199 | 4.003 | 1.291 | 42.574 | 42.229 | V |
| (55637) | 2002 UX25 | 2859.87294 | 20.384 | 0.070 | 2859.87430 | 4.110 | 1.291 | 42.574 | 42.229 | V |
| (55637) | 2002 UX25 | 2859.87798 | 21.300 | 0.086 | 2859.87934 | 5.026 | 1.291 | 42.574 | 42.229 | B |
| (55637) | 2002 UX25 | 2859.88304 | 19.264 | 0.098 | 2859.88440 | 2.990 | 1.291 | 42.574 | 42.229 | I |
| (55637) | 2002 UX25 | 2860.89558 | 20.371 | 0.071 | 2860.89703 | 4.098 | 1.283 | 42.574 | 42.212 | V |
| (55637) | 2002 UX25 | 2860.90286 | 21.359 | 0.041 | 2860.90431 | 5.086 | 1.283 | 42.574 | 42.212 | B |
| (55637) | 2002 UX25 | 2860.90788 | 19.193 | 0.080 | 2860.90933 | 2.920 | 1.283 | 42.574 | 42.212 | I |
| (55637) | 2002 UX25 | 2862.85391 | 20.225 | 0.133 | 2862.85555 | 3.954 | 1.267 | 42.573 | 42.181 | V |
| (55637) | 2002 UX25 | 2862.85619 | 20.170 | 0.153 | 2862.85783 | 3.899 | 1.267 | 42.573 | 42.181 | V |
| (55637) | 2002 UX25 | 2862.86120 | 21.360 | 0.167 | 2862.86284 | 5.089 | 1.266 | 42.573 | 42.181 | B |
| (55637) | 2002 UX25 | 2862.86622 | 19.295 | 0.074 | 2862.86785 | 3.024 | 1.266 | 42.573 | 42.181 | I |
| (55637) | 2002 UX25 | 2863.84196 | 20.014 | 0.189 | 2863.84369 | 3.744 | 1.258 | 42.573 | 42.165 | V |
| (55637) | 2002 UX25 | 2863.84428 | 20.164 | 0.189 | 2863.84601 | 3.894 | 1.258 | 42.573 | 42.165 | V |
| (55637) | 2002 UX25 | 2863.84932 | 21.056 | 0.163 | 2863.85105 | 4.786 | 1.258 | 42.573 | 42.165 | B |
| (55637) | 2002 UX25 | 2863.85438 | 19.188 | 0.088 | 2863.85611 | 2.918 | 1.258 | 42.573 | 42.165 | I |
| (55637) | 2002 UX25 | 2865.85931 | 20.079 | 0.196 | 2865.86122 | 3.810 | 1.239 | 42.572 | 42.133 | V |
| (55637) | 2002 UX25 | 2866.81159 | 20.043 | 0.266 | 2866.81359 | 3.775 | 1.229 | 42.571 | 42.118 | V |
| (55637) | 2002 UX25 | 2866.81696 | 21.238 | 0.237 | 2866.81895 | 4.970 | 1.229 | 42.571 | 42.118 | B |
| (55637) | 2002 UX25 | 2866.82202 | 19.356 | 0.133 | 2866.82401 | 3.088 | 1.229 | 42.571 | 42.118 | I |
| (55637) | 2002 UX25 | 2873.81801 | 19.865 | 0.249 | 2873.82061 | 3.603 | 1.149 | 42.569 | 42.013 | V |
| (55637) | 2002 UX25 | 2873.82545 | 21.289 | 0.136 | 2873.82805 | 5.027 | 1.149 | 42.569 | 42.013 | B |
| (55637) | 2002 UX25 | 2875.82912 | 20.451 | 0.066 | 2875.83189 | 4.190 | 1.123 | 42.568 | 41.984 | V |

| | | | | | | | | | |
|---|---|---|---|---|---|---|---|---|---|
| (55637) 2002 UX25 | 2875.83120 | 20.379 | 0.064 | 2875.83397 | 4.118 | 1.123 | 42.568 | 41.984 | V |
| (55637) 2002 UX25 | 2875.83657 | 21.299 | 0.040 | 2875.83934 | 5.038 | 1.123 | 42.568 | 41.984 | B |
| (55637) 2002 UX25 | 2875.84163 | 19.342 | 0.070 | 2875.84440 | 3.081 | 1.123 | 42.568 | 41.984 | I |
| (55637) 2002 UX25 | 2876.75221 | 20.342 | 0.076 | 2876.75506 | 4.082 | 1.110 | 42.568 | 41.971 | V |
| (55637) 2002 UX25 | 2876.75452 | 20.319 | 0.086 | 2876.75737 | 4.059 | 1.110 | 42.568 | 41.971 | V |
| (55637) 2002 UX25 | 2876.75956 | 21.279 | 0.053 | 2876.76241 | 5.019 | 1.110 | 42.568 | 41.971 | B |
| (55637) 2002 UX25 | 2876.76462 | 19.351 | 0.085 | 2876.76747 | 3.091 | 1.110 | 42.568 | 41.971 | I |
| (55637) 2002 UX25 | 2886.79747 | 20.358 | 0.093 | 2886.80108 | 4.105 | 0.958 | 42.564 | 41.839 | V |
| (55637) 2002 UX25 | 2886.79978 | 20.384 | 0.088 | 2886.80339 | 4.131 | 0.957 | 42.564 | 41.839 | V |
| (55637) 2002 UX25 | 2886.80482 | 21.238 | 0.082 | 2886.80843 | 4.985 | 0.957 | 42.564 | 41.839 | B |
| (55637) 2002 UX25 | 2886.80988 | 19.390 | 0.087 | 2886.81349 | 3.137 | 0.957 | 42.564 | 41.839 | I |
| (55637) 2002 UX25 | 2890.72738 | 20.914 | 0.210 | 2890.73125 | 4.663 | 0.890 | 42.562 | 41.793 | B |
| (55637) 2002 UX25 | 2892.78011 | 20.286 | 0.216 | 2892.78411 | 4.037 | 0.852 | 42.562 | 41.771 | V |
| (55637) 2002 UX25 | 2892.78243 | 20.139 | 0.188 | 2892.78643 | 3.890 | 0.852 | 42.562 | 41.771 | V |
| (55637) 2002 UX25 | 2892.78747 | 21.045 | 0.210 | 2892.79147 | 4.796 | 0.852 | 42.562 | 41.771 | B |
| (55637) 2002 UX25 | 2892.79254 | 19.165 | 0.090 | 2892.79654 | 2.916 | 0.852 | 42.562 | 41.771 | I |
| (55637) 2002 UX25 | 2894.73951 | 19.209 | 0.199 | 2894.74363 | 2.961 | 0.816 | 42.561 | 41.750 | I |
| (55637) 2002 UX25 | 2897.76076 | 20.100 | 0.167 | 2897.76506 | 3.853 | 0.758 | 42.560 | 41.720 | V |
| (55637) 2002 UX25 | 2897.76308 | 20.154 | 0.191 | 2897.76738 | 3.907 | 0.758 | 42.560 | 41.720 | V |
| (55637) 2002 UX25 | 2897.76811 | 20.974 | 0.169 | 2897.77241 | 4.727 | 0.758 | 42.560 | 41.720 | B |
| (55637) 2002 UX25 | 2897.77317 | 19.358 | 0.175 | 2897.77747 | 3.111 | 0.758 | 42.560 | 41.720 | I |
| (55637) 2002 UX25 | 2901.75604 | 19.976 | 0.247 | 2901.76055 | 3.731 | 0.678 | 42.558 | 41.683 | V |
| (55637) 2002 UX25 | 2901.75835 | 20.083 | 0.249 | 2901.76286 | 3.838 | 0.678 | 42.558 | 41.683 | V |
| (55637) 2002 UX25 | 2901.76338 | 21.034 | 0.184 | 2901.76789 | 4.789 | 0.678 | 42.558 | 41.683 | B |
| (55637) 2002 UX25 | 2901.76844 | 18.874 | 0.247 | 2901.77295 | 2.629 | 0.677 | 42.558 | 41.683 | I |
| (55637) 2002 UX25 | 2903.75201 | 20.117 | 0.068 | 2903.75662 | 3.873 | 0.636 | 42.557 | 41.667 | V |
| (55637) 2002 UX25 | 2903.75433 | 20.244 | 0.056 | 2903.75894 | 4.000 | 0.636 | 42.557 | 41.667 | V |
| (55637) 2002 UX25 | 2903.75937 | 21.180 | 0.041 | 2903.76397 | 4.936 | 0.636 | 42.557 | 41.666 | B |
| (55637) 2002 UX25 | 2903.76443 | 19.096 | 0.173 | 2903.76903 | 2.852 | 0.636 | 42.557 | 41.666 | I |
| (55637) 2002 UX25 | 2905.80206 | 20.272 | 0.073 | 2905.80676 | 4.029 | 0.593 | 42.557 | 41.650 | V |
| (55637) 2002 UX25 | 2905.80438 | 20.245 | 0.075 | 2905.80908 | 4.002 | 0.593 | 42.557 | 41.650 | V |
| (55637) 2002 UX25 | 2905.80942 | 21.169 | 0.048 | 2905.81412 | 4.926 | 0.593 | 42.557 | 41.650 | B |
| (55637) 2002 UX25 | 2905.81448 | 19.253 | 0.077 | 2905.81918 | 3.010 | 0.593 | 42.557 | 41.650 | I |
| (55637) 2002 UX25 | 2907.75093 | 20.228 | 0.069 | 2907.75571 | 3.986 | 0.551 | 42.556 | 41.636 | V |
| (55637) 2002 UX25 | 2907.75324 | 20.220 | 0.062 | 2907.75802 | 3.978 | 0.551 | 42.556 | 41.636 | V |
| (55637) 2002 UX25 | 2907.76334 | 19.247 | 0.054 | 2907.76812 | 3.005 | 0.551 | 42.556 | 41.636 | I |
| (55637) 2002 UX25 | 2909.69621 | 20.104 | 0.123 | 2909.70107 | 3.863 | 0.509 | 42.555 | 41.623 | V |
| (55637) 2002 UX25 | 2909.69853 | 20.361 | 0.114 | 2909.70339 | 4.120 | 0.509 | 42.555 | 41.623 | V |
| (55637) 2002 UX25 | 2909.70356 | 21.201 | 0.277 | 2909.70842 | 4.960 | 0.509 | 42.555 | 41.623 | B |
| (55637) 2002 UX25 | 2911.70133 | 20.238 | 0.117 | 2911.70626 | 3.997 | 0.465 | 42.554 | 41.611 | V |
| (55637) 2002 UX25 | 2911.70869 | 20.920 | 0.109 | 2911.71362 | 4.679 | 0.464 | 42.554 | 41.611 | B |
| (55637) 2002 UX25 | 2911.71376 | 19.179 | 0.181 | 2911.71869 | 2.938 | 0.464 | 42.554 | 41.611 | I |
| (55637) 2002 UX25 | 2925.67138 | 19.921 | 0.131 | 2925.67662 | 3.683 | 0.143 | 42.549 | 41.556 | V |
| (55637) 2002 UX25 | 2925.67369 | 19.995 | 0.140 | 2925.67893 | 3.757 | 0.143 | 42.549 | 41.556 | V |
| (55637) 2002 UX25 | 2925.67872 | 21.207 | 0.158 | 2925.68396 | 4.969 | 0.143 | 42.549 | 41.556 | B |
| (55637) 2002 UX25 | 2925.68379 | 18.974 | 0.108 | 2925.68903 | 2.736 | 0.143 | 42.549 | 41.556 | I |
| (55637) 2002 UX25 | 2927.70558 | 19.958 | 0.098 | 2927.71084 | 3.721 | 0.095 | 42.548 | 41.553 | V |
| (55637) 2002 UX25 | 2927.70790 | 20.007 | 0.096 | 2927.71316 | 3.770 | 0.095 | 42.548 | 41.553 | V |
| (55637) 2002 UX25 | 2927.71293 | 21.045 | 0.088 | 2927.71819 | 4.808 | 0.095 | 42.548 | 41.553 | B |
| (55637) 2002 UX25 | 2927.71800 | 19.032 | 0.083 | 2927.72326 | 2.795 | 0.095 | 42.548 | 41.553 | I |
| (55637) 2002 UX25 | 2929.63498 | 20.266 | 0.079 | 2929.64025 | 4.029 | 0.049 | 42.547 | 41.551 | V |
| (55637) 2002 UX25 | 2929.63730 | 20.236 | 0.085 | 2929.64257 | 3.999 | 0.049 | 42.547 | 41.551 | V |
| (55637) 2002 UX25 | 2929.64234 | 21.242 | 0.050 | 2929.64761 | 5.005 | 0.049 | 42.547 | 41.551 | B |
| (55637) 2002 UX25 | 2929.64741 | 19.063 | 0.068 | 2929.65268 | 2.826 | 0.049 | 42.547 | 41.551 | I |
| (55637) 2002 UX25 | 2932.64440 | 20.074 | 0.047 | 2932.64967 | 3.837 | 0.023 | 42.546 | 41.551 | V |
| (55637) 2002 UX25 | 2932.64671 | 20.113 | 0.047 | 2932.65198 | 3.876 | 0.023 | 42.546 | 41.551 | V |
| (55637) 2002 UX25 | 2932.65174 | 21.052 | 0.030 | 2932.65701 | 4.815 | 0.023 | 42.546 | 41.551 | B |
| (55637) 2002 UX25 | 2932.65681 | 19.028 | 0.054 | 2932.66208 | 2.791 | 0.023 | 42.546 | 41.551 | I |
| (55637) 2002 UX25 | 2941.62687 | 20.245 | 0.055 | 2941.63206 | 4.007 | 0.234 | 42.543 | 41.565 | V |
| (55637) 2002 UX25 | 2941.62919 | 20.164 | 0.050 | 2941.63438 | 3.926 | 0.234 | 42.543 | 41.565 | V |
| (55637) 2002 UX25 | 2941.63422 | 21.206 | 0.035 | 2941.63941 | 4.968 | 0.235 | 42.543 | 41.565 | B |
| (55637) 2002 UX25 | 2941.63929 | 19.145 | 0.048 | 2941.64448 | 2.907 | 0.235 | 42.543 | 41.565 | I |
| (55637) 2002 UX25 | 2943.68814 | 20.108 | 0.060 | 2943.69329 | 3.870 | 0.282 | 42.542 | 41.572 | V |
| (55637) 2002 UX25 | 2943.68814 | 20.237 | 0.066 | 2943.69329 | 3.999 | 0.282 | 42.542 | 41.572 | V |
| (55637) 2002 UX25 | 2943.69560 | 21.107 | 0.051 | 2943.70075 | 4.869 | 0.282 | 42.542 | 41.572 | B |
| (55637) 2002 UX25 | 2943.70066 | 19.203 | 0.063 | 2943.70581 | 2.965 | 0.283 | 42.542 | 41.572 | I |
| (55637) 2002 UX25 | 2944.67872 | 20.133 | 0.070 | 2944.68385 | 3.895 | 0.305 | 42.542 | 41.576 | V |
| (55637) 2002 UX25 | 2944.68104 | 20.162 | 0.078 | 2944.68617 | 3.924 | 0.305 | 42.542 | 41.576 | V |
| (55637) 2002 UX25 | 2944.68608 | 21.141 | 0.060 | 2944.69121 | 4.903 | 0.305 | 42.542 | 41.576 | B |
| (55637) 2002 UX25 | 2944.69114 | 19.143 | 0.066 | 2944.69627 | 2.905 | 0.305 | 42.542 | 41.576 | I |
| (55637) 2002 UX25 | 2945.63328 | 20.221 | 0.111 | 2945.63839 | 3.983 | 0.327 | 42.541 | 41.580 | V |
| (55637) 2002 UX25 | 2945.63560 | 20.294 | 0.117 | 2945.64071 | 4.056 | 0.327 | 42.541 | 41.580 | V |
| (55637) 2002 UX25 | 2945.64064 | 20.802 | 0.081 | 2945.64575 | 4.564 | 0.327 | 42.541 | 41.580 | B |
| (55637) 2002 UX25 | 2945.64570 | 19.231 | 0.094 | 2945.65081 | 2.993 | 0.327 | 42.541 | 41.580 | I |
| (55637) 2002 UX25 | 2947.60159 | 20.047 | 0.110 | 2947.60665 | 3.808 | 0.372 | 42.541 | 41.588 | V |
| (55637) 2002 UX25 | 2947.60391 | 20.389 | 0.155 | 2947.60897 | 4.150 | 0.372 | 42.541 | 41.588 | V |
| (55637) 2002 UX25 | 2947.61400 | 19.140 | 0.086 | 2947.61905 | 2.901 | 0.372 | 42.541 | 41.588 | I |
| (55637) 2002 UX25 | 2952.63335 | 20.574 | 0.245 | 2952.63825 | 4.334 | 0.485 | 42.539 | 41.616 | V |
| (55637) 2002 UX25 | 2969.60661 | 20.333 | 0.066 | 2969.61067 | 4.086 | 0.831 | 42.532 | 41.761 | V |
| (55637) 2002 UX25 | 2969.60892 | 20.255 | 0.062 | 2969.61298 | 4.008 | 0.832 | 42.532 | 41.761 | V |
| (55637) 2002 UX25 | 2969.61901 | 19.202 | 0.104 | 2969.62307 | 2.955 | 0.832 | 42.532 | 41.761 | I |
| (55637) 2002 UX25 | 2973.59207 | 20.320 | 0.078 | 2973.59588 | 4.070 | 0.903 | 42.531 | 41.805 | V |
| (55637) 2002 UX25 | 2973.59439 | 20.342 | 0.119 | 2973.59820 | 4.092 | 0.903 | 42.531 | 41.805 | V |
| (55637) 2002 UX25 | 2973.59942 | 21.354 | 0.081 | 2973.60322 | 5.104 | 0.903 | 42.531 | 41.805 | B |
| (55637) 2002 UX25 | 2973.60448 | 19.184 | 0.118 | 2973.60828 | 2.934 | 0.903 | 42.531 | 41.805 | I |
| (55637) 2002 UX25 | 2975.55695 | 20.507 | 0.096 | 2975.56062 | 4.256 | 0.936 | 42.530 | 41.828 | V |
| (55637) 2002 UX25 | 2975.55927 | 20.300 | 0.084 | 2975.56294 | 4.049 | 0.936 | 42.530 | 41.828 | V |
| (55637) 2002 UX25 | 2975.56430 | 21.293 | 0.082 | 2975.56797 | 5.042 | 0.936 | 42.530 | 41.828 | B |
| (55637) 2002 UX25 | 2975.56936 | 19.175 | 0.063 | 2975.57303 | 2.924 | 0.936 | 42.530 | 41.828 | I |
| (55637) 2002 UX25 | 2996.59579 | 20.357 | 0.175 | 2996.59776 | 4.091 | 1.214 | 42.522 | 42.123 | V |
| (55637) 2002 UX25 | 2996.59811 | 20.160 | 0.153 | 2996.60008 | 3.894 | 1.214 | 42.522 | 42.123 | V |
| (55637) 2002 UX25 | 2996.60314 | 21.236 | 0.135 | 2996.60511 | 4.970 | 1.214 | 42.522 | 42.123 | B |
| (55637) 2002 UX25 | 2996.60819 | 19.172 | 0.159 | 2996.61016 | 2.906 | 1.214 | 42.522 | 42.123 | I |
| | | | | # | | | | | |
| (20000) Varuna | 3370.74537 | 21.288 | 0.151 | 3370.74537 | 4.978 | 0.131 | 43.248 | 42.270 | B |
| (20000) Varuna | 3370.75023 | 20.474 | 0.173 | 3370.75023 | 4.164 | 0.131 | 43.248 | 42.270 | V |
| (20000) Varuna | 3370.75233 | 19.129 | 0.088 | 3370.75233 | 2.819 | 0.131 | 43.248 | 42.270 | I |
| (20000) Varuna | 3372.64770 | 21.252 | 0.054 | 3372.64771 | 4.942 | 0.092 | 43.249 | 42.267 | B |
| (20000) Varuna | 3372.65256 | 19.995 | 0.068 | 3372.65257 | 3.685 | 0.092 | 43.249 | 42.267 | V |
| (20000) Varuna | 3372.65467 | 18.933 | 0.071 | 3372.65468 | 2.623 | 0.092 | 43.249 | 42.267 | I |
| (20000) Varuna | 3374.63803 | 21.045 | 0.060 | 3374.63805 | 4.735 | 0.061 | 43.249 | 42.267 | B |
| (20000) Varuna | 3374.64288 | 19.965 | 0.075 | 3374.64290 | 3.655 | 0.061 | 43.249 | 42.267 | V |
| (20000) Varuna | 3374.64498 | 18.957 | 0.089 | 3374.64500 | 2.647 | 0.061 | 43.249 | 42.267 | I |
| (20000) Varuna | 3376.67231 | 21.016 | 0.043 | 3376.67233 | 4.706 | 0.058 | 43.249 | 42.267 | B |
| (20000) Varuna | 3376.67717 | 20.247 | 0.084 | 3376.67719 | 3.937 | 0.058 | 43.249 | 42.267 | V |
| (20000) Varuna | 3376.67928 | 18.969 | 0.069 | 3376.67930 | 2.659 | 0.058 | 43.249 | 42.267 | I |
| (20000) Varuna | 3382.65440 | 21.277 | 0.058 | 3382.65437 | 4.967 | 0.170 | 43.250 | 42.275 | B |
| (20000) Varuna | 3382.65925 | 20.442 | 0.096 | 3382.65922 | 4.132 | 0.170 | 43.250 | 42.275 | V |
| (20000) Varuna | 3382.66135 | 19.126 | 0.134 | 3382.66132 | 2.816 | 0.170 | 43.250 | 42.275 | I |
| (20000) Varuna | 3384.65412 | 21.207 | 0.070 | 3384.65406 | 4.896 | 0.214 | 43.250 | 42.280 | B |
| (20000) Varuna | 3384.65897 | 20.331 | 0.099 | 3384.65891 | 4.020 | 0.215 | 43.250 | 42.281 | V |
| (20000) Varuna | 3384.66107 | 18.924 | 0.077 | 3384.66101 | 2.613 | 0.215 | 43.250 | 42.281 | I |
| (20000) Varuna | 3386.66026 | 21.099 | 0.040 | 3386.66016 | 4.788 | 0.260 | 43.250 | 42.287 | B |
| (20000) Varuna | 3386.66511 | 20.226 | 0.063 | 3386.66501 | 3.915 | 0.260 | 43.250 | 42.287 | V |
| (20000) Varuna | 3386.66722 | 19.029 | 0.063 | 3386.66712 | 2.718 | 0.260 | 43.250 | 42.287 | I |
| (20000) Varuna | 3399.62809 | 21.038 | 0.099 | 3399.62758 | 4.723 | 0.547 | 43.252 | 42.358 | B |
| (20000) Varuna | 3399.63294 | 20.025 | 0.105 | 3399.63243 | 3.710 | 0.547 | 43.252 | 42.358 | V |
| (20000) Varuna | 3399.63503 | 18.925 | 0.070 | 3399.63452 | 2.610 | 0.547 | 43.252 | 42.358 | I |
| (20000) Varuna | 3402.65069 | 21.313 | 0.072 | 3402.65004 | 4.997 | 0.611 | 43.253 | 42.382 | B |
| (20000) Varuna | 3402.65613 | 20.297 | 0.093 | 3402.65548 | 3.981 | 0.611 | 43.253 | 42.382 | V |
| (20000) Varuna | 3402.65821 | 19.050 | 0.065 | 3402.65756 | 2.734 | 0.611 | 43.253 | 42.382 | I |
| (20000) Varuna | 3404.63482 | 21.343 | 0.055 | 3404.63408 | 5.026 | 0.652 | 43.253 | 42.398 | B |
| (20000) Varuna | 3404.63966 | 20.330 | 0.084 | 3404.63892 | 4.013 | 0.652 | 43.253 | 42.399 | V |
| (20000) Varuna | 3404.64176 | 18.988 | 0.071 | 3404.64102 | 2.671 | 0.652 | 43.253 | 42.399 | I |
| (20000) Varuna | 3406.64153 | 20.963 | 0.038 | 3406.64068 | 4.645 | 0.692 | 43.253 | 42.417 | B |
| (20000) Varuna | 3406.64850 | 18.881 | 0.058 | 3406.64765 | 2.563 | 0.692 | 43.253 | 42.417 | I |
| (20000) Varuna | 3408.64232 | 21.058 | 0.045 | 3408.64136 | 4.739 | 0.732 | 43.253 | 42.436 | B |
| (20000) Varuna | 3408.64718 | 20.180 | 0.077 | 3408.64622 | 3.861 | 0.732 | 43.253 | 42.436 | V |
| (20000) Varuna | 3408.64929 | 19.061 | 0.078 | 3408.64833 | 2.742 | 0.732 | 43.253 | 42.436 | I |
| (20000) Varuna | 3410.62123 | 20.996 | 0.039 | 3410.62016 | 4.676 | 0.770 | 43.254 | 42.455 | B |
| (20000) Varuna | 3410.62608 | 20.078 | 0.066 | 3410.62501 | 3.758 | 0.770 | 43.254 | 42.455 | V |
| (20000) Varuna | 3410.62818 | 18.904 | 0.062 | 3410.62711 | 2.584 | 0.770 | 43.254 | 42.455 | I |
| (20000) Varuna | 3413.60831 | 21.457 | 0.061 | 3413.60705 | 5.136 | 0.826 | 43.254 | 42.487 | B |
| (20000) Varuna | 3413.61316 | 20.702 | 0.113 | 3413.61190 | 4.381 | 0.826 | 43.254 | 42.487 | V |
| (20000) Varuna | 3413.61527 | 19.146 | 0.081 | 3413.61401 | 2.825 | 0.826 | 43.254 | 42.487 | I |
| (20000) Varuna | 3415.59691 | 21.284 | 0.052 | 3415.59552 | 4.961 | 0.861 | 43.254 | 42.510 | B |
| (20000) Varuna | 3415.60177 | 20.479 | 0.084 | 3415.60038 | 4.156 | 0.861 | 43.254 | 42.510 | V |
| (20000) Varuna | 3415.60387 | 19.026 | 0.067 | 3415.60248 | 2.703 | 0.862 | 43.254 | 42.510 | I |
| (20000) Varuna | 3427.57984 | 21.090 | 0.153 | 3427.57757 | 4.760 | 1.053 | 43.256 | 42.662 | B |
| (20000) Varuna | 3427.58678 | 19.521 | 0.153 | 3427.58451 | 3.191 | 1.053 | 43.256 | 42.662 | I |
| (20000) Varuna | 3429.55001 | 21.418 | 0.095 | 3429.54758 | 5.086 | 1.080 | 43.256 | 42.690 | B |
| (20000) Varuna | 3429.55487 | 20.499 | 0.127 | 3429.55244 | 4.167 | 1.080 | 43.256 | 42.690 | V |
| (20000) Varuna | 3429.55697 | 19.252 | 0.120 | 3429.55454 | 2.920 | 1.080 | 43.256 | 42.690 | I |
| (20000) Varuna | 3431.55839 | 21.602 | 0.065 | 3431.55579 | 5.269 | 1.106 | 43.256 | 42.719 | B |
| (20000) Varuna | 3431.56324 | 20.618 | 0.096 | 3431.56064 | 4.285 | 1.106 | 43.256 | 42.719 | V |
| (20000) Varuna | 3431.56535 | 19.411 | 0.106 | 3431.56275 | 3.078 | 1.106 | 43.256 | 42.719 | I |
| (20000) Varuna | 3433.55058 | 21.375 | 0.063 | 3433.54781 | 5.040 | 1.131 | 43.257 | 42.748 | B |
| (20000) Varuna | 3433.55543 | 20.501 | 0.100 | 3433.55266 | 4.166 | 1.131 | 43.257 | 42.749 | V |
| (20000) Varuna | 3433.55754 | 19.077 | 0.077 | 3433.55477 | 2.742 | 1.131 | 43.257 | 42.749 | I |
| (20000) Varuna | 3435.54759 | 21.520 | 0.064 | 3435.54465 | 5.184 | 1.154 | 43.257 | 42.779 | B |
| (20000) Varuna | 3435.55244 | 20.595 | 0.101 | 3435.54950 | 4.259 | 1.154 | 43.257 | 42.779 | V |
| (20000) Varuna | 3435.55454 | 19.330 | 0.104 | 3435.55160 | 2.994 | 1.154 | 43.257 | 42.779 | I |
| (20000) Varuna | 3442.51511 | 21.604 | 0.124 | 3442.51154 | 5.262 | 1.224 | 43.258 | 42.888 | B |
| (20000) Varuna | 3442.51996 | 20.441 | 0.142 | 3442.51639 | 4.099 | 1.224 | 43.258 | 42.888 | V |
| (20000) Varuna | 3442.52206 | 19.538 | 0.183 | 3442.51849 | 3.196 | 1.224 | 43.258 | 42.888 | I |
| (20000) Varuna | 3444.52497 | 21.605 | 0.099 | 3444.52121 | 5.261 | 1.241 | 43.258 | 42.920 | B |
| (20000) Varuna | 3444.52982 | 20.548 | 0.137 | 3444.52606 | 4.204 | 1.241 | 43.258 | 42.921 | V |
| (20000) Varuna | 3450.52957 | 21.184 | 0.207 | 3450.52524 | 4.835 | 1.282 | 43.259 | 43.020 | B |
| (20000) Varuna | 3450.53442 | 19.994 | 0.203 | 3450.53009 | 3.645 | 1.282 | 43.259 | 43.020 | V |
| (20000) Varuna | 3450.53651 | 19.009 | 0.161 | 3450.53218 | 2.660 | 1.282 | 43.259 | 43.020 | I |
| (20000) Varuna | 3458.50093 | 21.498 | 0.105 | 3458.49581 | 5.142 | 1.315 | 43.260 | 43.156 | B |
| (20000) Varuna | 3458.50567 | 20.700 | 0.195 | 3458.50055 | 4.344 | 1.315 | 43.260 | 43.156 | V |
| (20000) Varuna | 3458.50767 | 19.625 | 0.155 | 3458.50255 | 3.269 | 1.315 | 43.260 | 43.156 | I |
| (20000) Varuna | 3460.49488 | 21.539 | 0.108 | 3460.48956 | 5.182 | 1.320 | 43.260 | 43.190 | B |
| (20000) Varuna | 3460.49966 | 20.839 | 0.177 | 3460.49434 | 4.482 | 1.320 | 43.260 | 43.190 | V |
| (20000) Varuna | 3460.50168 | 19.502 | 0.128 | 3460.49636 | 3.145 | 1.320 | 43.260 | 43.190 | I |
| (20000) Varuna | 3464.49919 | 21.354 | 0.090 | 3464.49348 | 4.993 | 1.324 | 43.261 | 43.259 | B |
| (20000) Varuna | 3464.50397 | 20.385 | 0.156 | 3464.49826 | 4.024 | 1.324 | 43.261 | 43.259 | V |
| (20000) Varuna | 3464.50599 | 19.169 | 0.117 | 3464.50028 | 2.808 | 1.324 | 43.261 | 43.259 | I |
| (20000) Varuna | 3478.46983 | 21.278 | 0.256 | 3478.46274 | 4.905 | 1.290 | 43.263 | 43.498 | B |
| (20000) Varuna | 3478.47463 | 20.559 | 0.237 | 3478.46754 | 4.186 | 1.290 | 43.263 | 43.498 | V |
| (20000) Varuna | 3478.47666 | 19.262 | 0.141 | 3478.46957 | 2.889 | 1.290 | 43.263 | 43.498 | I |
| | | | | # | | | | | |
| Nereid | 2729.88826 | 19.886 | 0.118 | 2729.88826 | 5.070 | 1.607 | 30.049 | 30.573 | B |
| Nereid | 2729.89056 | 18.960 | 0.267 | 2729.89056 | 4.144 | 1.607 | 30.049 | 30.573 | I |
| Nereid | 2731.90654 | 19.860 | 0.122 | 2731.90670 | 5.046 | 1.641 | 30.050 | 30.544 | B |
| Nereid | 2731.90884 | 18.792 | 0.190 | 2731.90900 | 3.978 | 1.641 | 30.050 | 30.544 | I |
| Nereid | 2749.92913 | 20.018 | 0.419 | 2749.93086 | 5.223 | 1.865 | 30.062 | 30.273 | V |
| Nereid | 2751.88279 | 19.601 | 0.197 | 2751.88470 | 4.808 | 1.879 | 30.063 | 30.242 | V |
| Nereid | 2751.88415 | 19.767 | 0.225 | 2751.88606 | 4.974 | 1.879 | 30.063 | 30.242 | V |
| Nereid | 2751.88660 | 20.382 | 0.166 | 2751.88851 | 5.589 | 1.879 | 30.063 | 30.242 | B |
| Nereid | 2751.88889 | 18.724 | 0.226 | 2751.89080 | 3.931 | 1.879 | 30.063 | 30.242 | I |
| Nereid | 2755.88941 | 19.965 | 0.136 | 2755.89168 | 5.176 | 1.902 | 30.067 | 30.179 | V |
| Nereid | 2755.89077 | 19.839 | 0.121 | 2755.89304 | 5.050 | 1.902 | 30.067 | 30.179 | V |
| Nereid | 2755.89321 | 20.441 | 0.102 | 2755.89548 | 5.652 | 1.902 | 30.067 | 30.179 | B |
| Nereid | 2755.89550 | 19.199 | 0.233 | 2755.89777 | 4.410 | 1.902 | 30.067 | 30.179 | I |
| Nereid | 2757.89371 | 19.977 | 0.159 | 2757.89617 | 5.190 | 1.910 | 30.070 | 30.148 | V |
| Nereid | 2757.89507 | 20.066 | 0.155 | 2757.89753 | 5.279 | 1.910 | 30.070 | 30.148 | V |
| Nereid | 2757.89752 | 20.513 | 0.072 | 2757.89998 | 5.726 | 1.910 | 30.070 | 30.147 | B |
| Nereid | 2759.91070 | 19.927 | 0.104 | 2759.91334 | 5.142 | 1.916 | 30.072 | 30.116 | V |
| Nereid | 2759.91450 | 20.383 | 0.056 | 2759.91714 | 5.598 | 1.916 | 30.072 | 30.116 | B |
| Nereid | 2759.91679 | 19.189 | 0.202 | 2759.91943 | 4.404 | 1.917 | 30.072 | 30.116 | I |
| Nereid | 2763.77658 | 19.845 | 0.121 | 2763.77957 | 5.064 | 1.922 | 30.078 | 30.056 | V |
| Nereid | 2763.77793 | 19.679 | 0.100 | 2763.78092 | 4.898 | 1.922 | 30.078 | 30.056 | V |
| Nereid | 2763.78038 | 20.489 | 0.076 | 2763.78337 | 5.708 | 1.922 | 30.078 | 30.056 | B |
| Nereid | 2763.78267 | 19.026 | 0.235 | 2763.78566 | 4.245 | 1.922 | 30.078 | 30.056 | I |
| Nereid | 2765.83174 | 19.682 | 0.116 | 2765.83491 | 4.903 | 1.922 | 30.081 | 30.024 | V |
| Nereid | 2765.83555 | 20.216 | 0.062 | 2765.83872 | 5.437 | 1.922 | 30.081 | 30.024 | B |
| Nereid | 2765.83784 | 18.839 | 0.190 | 2765.84101 | 4.060 | 1.922 | 30.081 | 30.024 | I |
| Nereid | 2771.79400 | 19.377 | 0.140 | 2771.79770 | 4.604 | 1.907 | 30.089 | 29.932 | V |
| Nereid | 2771.79535 | 19.252 | 0.117 | 2771.79905 | 4.479 | 1.907 | 30.089 | 29.932 | V |
| Nereid | 2771.79780 | 19.723 | 0.124 | 2771.80150 | 4.950 | 1.907 | 30.089 | 29.932 | B |
| Nereid | 2771.80009 | 19.081 | 0.373 | 2771.80379 | 4.308 | 1.907 | 30.089 | 29.932 | I |
| Nereid | 2778.85358 | 19.323 | 0.238 | 2778.85793 | 4.558 | 1.865 | 30.094 | 29.820 | V |
| Nereid | 2778.85494 | 19.423 | 0.281 | 2778.85929 | 4.658 | 1.865 | 30.094 | 29.820 | V |
| Nereid | 2778.85739 | 19.716 | 0.171 | 2778.86174 | 4.951 | 1.865 | 30.094 | 29.820 | B |
| Nereid | 2778.85968 | 19.057 | 0.380 | 2778.86403 | 4.292 | 1.865 | 30.094 | 29.819 | I |
| Nereid | 2787.86322 | 19.330 | 0.091 | 2787.86840 | 4.576 | 1.774 | 30.094 | 29.676 | V |
| Nereid | 2787.86457 | 19.444 | 0.081 | 2787.86975 | 4.690 | 1.774 | 30.094 | 29.676 | V |
| Nereid | 2787.86702 | 20.106 | 0.049 | 2787.87220 | 5.352 | 1.774 | 30.094 | 29.676 | B |
| Nereid | 2787.86931 | 18.853 | 0.159 | 2787.87449 | 4.099 | 1.774 | 30.094 | 29.676 | I |
| Nereid | 2789.85778 | 19.845 | 0.140 | 2789.86314 | 5.093 | 1.748 | 30.093 | 29.645 | V |
| Nereid | 2789.85914 | 19.404 | 0.100 | 2789.86450 | 4.652 | 1.748 | 30.093 | 29.645 | V |
| Nereid | 2789.86159 | 19.990 | 0.063 | 2789.86695 | 5.238 | 1.748 | 30.093 | 29.645 | B |
| Nereid | 2789.86388 | 18.632 | 0.185 | 2789.86924 | 3.880 | 1.748 | 30.093 | 29.645 | I |
| Nereid | 2791.85558 | 19.617 | 0.097 | 2791.86112 | 4.867 | 1.720 | 30.093 | 29.614 | V |
| Nereid | 2791.85694 | 19.650 | 0.098 | 2791.86248 | 4.900 | 1.719 | 30.093 | 29.614 | V |
| Nereid | 2791.85938 | 20.090 | 0.106 | 2791.86492 | 5.340 | 1.719 | 30.093 | 29.614 | B |
| Nereid | 2791.86167 | 18.631 | 0.136 | 2791.86721 | 3.881 | 1.719 | 30.093 | 29.614 | I |
| Nereid | 2793.89984 | 19.512 | 0.199 | 2793.90556 | 4.765 | 1.689 | 30.092 | 29.583 | V |
| Nereid | 2793.90120 | 19.491 | 0.162 | 2793.90692 | 4.744 | 1.689 | 30.092 | 29.583 | V |
| Nereid | 2793.90365 | 19.971 | 0.077 | 2793.90937 | 5.224 | 1.689 | 30.092 | 29.583 | B |
| Nereid | 2793.90594 | 18.507 | 0.237 | 2793.91166 | 3.760 | 1.689 | 30.092 | 29.583 | I |
| Nereid | 2804.90091 | 18.869 | 0.233 | 2804.90753 | 4.133 | 1.488 | 30.089 | 29.427 | V |
| Nereid | 2817.89530 | 19.391 | 0.096 | 2817.90282 | 4.667 | 1.183 | 30.084 | 29.270 | V |
| Nereid | 2817.89666 | 19.382 | 0.106 | 2817.90418 | 4.658 | 1.183 | 30.084 | 29.270 | V |
| Nereid | 2817.89911 | 20.113 | 0.069 | 2817.90663 | 5.389 | 1.183 | 30.084 | 29.270 | B |
| Nereid | 2817.90140 | 18.650 | 0.202 | 2817.90892 | 3.926 | 1.183 | 30.084 | 29.270 | I |
| Nereid | 2819.86645 | 19.465 | 0.086 | 2819.87409 | 4.743 | 1.131 | 30.083 | 29.250 | V |
| Nereid | 2819.86781 | 19.539 | 0.091 | 2819.87545 | 4.817 | 1.131 | 30.083 | 29.250 | V |
| Nereid | 2819.87025 | 20.160 | 0.052 | 2819.87789 | 5.438 | 1.131 | 30.083 | 29.250 | B |
| Nereid | 2819.87254 | 18.639 | 0.158 | 2819.88018 | 3.917 | 1.131 | 30.083 | 29.250 | I |
| Nereid | 2821.85577 | 19.221 | 0.198 | 2821.86352 | 4.500 | 1.078 | 30.082 | 29.230 | V |
| Nereid | 2821.85713 | 19.275 | 0.174 | 2821.86488 | 4.554 | 1.078 | 30.082 | 29.230 | V |
| Nereid | 2821.85957 | 19.893 | 0.112 | 2821.86732 | 5.172 | 1.077 | 30.082 | 29.230 | B |
| Nereid | 2821.86186 | 18.230 | 0.236 | 2821.86961 | 3.509 | 1.077 | 30.082 | 29.230 | I |
| Nereid | 2829.88483 | 19.247 | 0.092 | 2829.89299 | 4.532 | 0.848 | 30.079 | 29.160 | V |
| Nereid | 2829.88619 | 19.412 | 0.114 | 2829.89435 | 4.697 | 0.848 | 30.079 | 29.160 | V |
| Nereid | 2829.88863 | 19.944 | 0.070 | 2829.89679 | 5.229 | 0.848 | 30.079 | 29.160 | B |
| Nereid | 2829.89092 | 18.331 | 0.193 | 2829.89908 | 3.616 | 0.848 | 30.079 | 29.160 | I |
| Nereid | 2831.81892 | 19.405 | 0.150 | 2831.82716 | 4.691 | 0.790 | 30.078 | 29.146 | V |
| Nereid | 2831.82028 | 19.314 | 0.135 | 2831.82852 | 4.600 | 0.790 | 30.078 | 29.146 | V |
| Nereid | 2831.82273 | 20.161 | 0.126 | 2831.83097 | 5.447 | 0.790 | 30.078 | 29.146 | B |
| Nereid | 2831.82502 | 18.701 | 0.197 | 2831.83326 | 3.987 | 0.790 | 30.078 | 29.146 | I |
| Nereid | 2832.75756 | 19.432 | 0.168 | 2832.76584 | 4.718 | 0.762 | 30.078 | 29.139 | V |
| Nereid | 2832.75892 | 19.230 | 0.150 | 2832.76720 | 4.516 | 0.762 | 30.078 | 29.139 | V |
| Nereid | 2832.76137 | 19.989 | 0.142 | 2832.76965 | 5.275 | 0.762 | 30.078 | 29.139 | B |
| Nereid | 2832.76366 | 18.623 | 0.171 | 2832.77194 | 3.909 | 0.762 | 30.078 | 29.139 | I |
| Nereid | 2833.80230 | 19.083 | 0.293 | 2833.81062 | 4.370 | 0.730 | 30.077 | 29.132 | V |
| Nereid | 2833.80366 | 19.217 | 0.286 | 2833.81198 | 4.504 | 0.730 | 30.077 | 29.132 | V |
| Nereid | 2833.80610 | 20.025 | 0.336 | 2833.81442 | 5.312 | 0.730 | 30.077 | 29.132 | B |
| Nereid | 2833.80840 | 18.703 | 0.347 | 2833.81672 | 3.990 | 0.730 | 30.077 | 29.132 | I |
| Nereid | 2837.81653 | 18.790 | 0.250 | 2837.82499 | 4.079 | 0.605 | 30.076 | 29.108 | V |
| Nereid | 2837.81789 | 19.290 | 0.386 | 2837.82635 | 4.579 | 0.605 | 30.076 | 29.108 | V |
| Nereid | 2837.82033 | 19.351 | 0.185 | 2837.82879 | 4.640 | 0.605 | 30.076 | 29.108 | B |
| Nereid | 2837.82263 | 18.068 | 0.383 | 2837.83109 | 3.357 | 0.605 | 30.076 | 29.108 | I |
| Nereid | 2839.81048 | 19.248 | 0.118 | 2839.81900 | 4.538 | 0.542 | 30.075 | 29.097 | V |
| Nereid | 2839.81184 | 19.285 | 0.141 | 2839.82036 | 4.575 | 0.542 | 30.075 | 29.097 | V |
| Nereid | 2839.81429 | 20.011 | 0.117 | 2839.82281 | 5.301 | 0.542 | 30.075 | 29.097 | B |
| Nereid | 2839.81658 | 18.564 | 0.188 | 2839.82510 | 3.854 | 0.542 | 30.075 | 29.097 | I |
| Nereid | 2841.85964 | 19.086 | 0.118 | 2841.86822 | 4.376 | 0.477 | 30.074 | 29.088 | V |
| Nereid | 2841.86100 | 19.035 | 0.097 | 2841.86958 | 4.325 | 0.477 | 30.074 | 29.088 | V |
| Nereid | 2841.86344 | 19.750 | 0.075 | 2841.87202 | 5.040 | 0.477 | 30.074 | 29.088 | B |
| Nereid | 2841.86574 | 18.335 | 0.163 | 2841.87432 | 3.625 | 0.477 | 30.074 | 29.088 | I |
| Nereid | 2843.80130 | 19.337 | 0.083 | 2843.80992 | 4.628 | 0.414 | 30.074 | 29.080 | V |
| Nereid | 2843.80266 | 19.328 | 0.080 | 2843.81128 | 4.619 | 0.414 | 30.074 | 29.080 | V |
| Nereid | 2843.80511 | 19.993 | 0.055 | 2843.81373 | 5.284 | 0.414 | 30.074 | 29.080 | B |
| Nereid | 2843.80740 | 18.734 | 0.172 | 2843.81602 | 4.025 | 0.414 | 30.074 | 29.080 | I |
| Nereid | 2844.78792 | 19.112 | 0.069 | 2844.79656 | 4.403 | 0.382 | 30.073 | 29.076 | V |
| Nereid | 2844.78928 | 19.307 | 0.084 | 2844.79792 | 4.598 | 0.382 | 30.073 | 29.076 | V |
| Nereid | 2844.79173 | 19.954 | 0.047 | 2844.80037 | 5.245 | 0.382 | 30.073 | 29.076 | B |
| Nereid | 2844.79402 | 18.579 | 0.130 | 2844.80266 | 3.870 | 0.382 | 30.073 | 29.076 | I |
| Nereid | 2845.80261 | 19.247 | 0.074 | 2845.81127 | 4.539 | 0.349 | 30.073 | 29.073 | V |
| Nereid | 2845.80397 | 19.388 | 0.088 | 2845.81263 | 4.680 | 0.349 | 30.073 | 29.073 | V |
| Nereid | 2845.80641 | 19.765 | 0.045 | 2845.81507 | 5.057 | 0.349 | 30.073 | 29.072 | B |
| Nereid | 2845.80870 | 18.351 | 0.123 | 2845.81736 | 3.643 | 0.349 | 30.073 | 29.072 | I |
| Nereid | 2846.81332 | 19.271 | 0.161 | 2846.82200 | 4.563 | 0.316 | 30.072 | 29.069 | V |
| Nereid | 2846.81468 | 19.302 | 0.193 | 2846.82336 | 4.594 | 0.316 | 30.072 | 29.069 | V |
| Nereid | 2846.81713 | 19.837 | 0.106 | 2846.82581 | 5.129 | 0.316 | 30.072 | 29.069 | B |
| Nereid | 2846.81944 | 18.972 | 0.474 | 2846.82812 | 4.264 | 0.316 | 30.072 | 29.069 | I |
| Nereid | 2847.80421 | 19.261 | 0.111 | 2847.81291 | 4.553 | 0.284 | 30.072 | 29.067 | V |
| Nereid | 2847.80557 | 19.065 | 0.099 | 2847.81427 | 4.357 | 0.284 | 30.072 | 29.067 | V |
| Nereid | 2847.80803 | 19.753 | 0.066 | 2847.81673 | 5.045 | 0.283 | 30.072 | 29.067 | B |
| Nereid | 2847.81033 | 18.542 | 0.194 | 2847.81903 | 3.834 | 0.283 | 30.072 | 29.067 | I |
| Nereid | 2851.84076 | 19.172 | 0.043 | 2851.84951 | 4.465 | 0.150 | 30.070 | 29.058 | V |
| Nereid | 2851.84212 | 19.164 | 0.045 | 2851.85087 | 4.457 | 0.150 | 30.070 | 29.058 | V |
| Nereid | 2851.84458 | 19.738 | 0.043 | 2851.85333 | 5.031 | 0.150 | 30.070 | 29.058 | B |
| Nereid | 2851.84688 | 18.375 | 0.085 | 2851.85563 | 3.668 | 0.150 | 30.070 | 29.058 | I |
| Nereid | 2852.83791 | 19.114 | 0.043 | 2852.84667 | 4.407 | 0.117 | 30.070 | 29.057 | V |
| Nereid | 2852.83927 | 19.124 | 0.044 | 2852.84803 | 4.417 | 0.117 | 30.070 | 29.057 | V |
| Nereid | 2852.84172 | 19.798 | 0.027 | 2852.85048 | 5.091 | 0.117 | 30.070 | 29.057 | B |
| Nereid | 2852.84402 | 18.309 | 0.068 | 2852.85278 | 3.602 | 0.117 | 30.070 | 29.057 | I |
| Nereid | 2853.82273 | 19.079 | 0.038 | 2853.83149 | 4.372 | 0.085 | 30.070 | 29.056 | V |
| Nereid | 2853.82409 | 19.089 | 0.038 | 2853.83285 | 4.382 | 0.085 | 30.070 | 29.056 | V |
| Nereid | 2853.82654 | 19.801 | 0.022 | 2853.83530 | 5.094 | 0.085 | 30.070 | 29.056 | B |
| Nereid | 2853.82883 | 18.357 | 0.073 | 2853.83759 | 3.650 | 0.085 | 30.070 | 29.056 | I |
| Nereid | 2854.77191 | 19.029 | 0.035 | 2854.78068 | 4.322 | 0.053 | 30.069 | 29.055 | V |
| Nereid | 2854.77327 | 19.069 | 0.037 | 2854.78204 | 4.362 | 0.053 | 30.069 | 29.055 | V |
| Nereid | 2854.77803 | 18.292 | 0.060 | 2854.78680 | 3.585 | 0.053 | 30.069 | 29.055 | I |
| Nereid | 2855.53929 | 18.999 | 0.079 | 2855.54806 | 4.292 | 0.028 | 30.069 | 29.054 | V |

| | | | | | | | | |
|---|---|---|---|---|---|---|---|---|
| Nereid | 2855.54065 | 19.029 | 0.077 | 2855.54942 | 4.322 | 0.028 | 30.069 | 29.054 | V |
| Nereid | 2855.54311 | 19.738 | 0.057 | 2855.55188 | 5.031 | 0.028 | 30.069 | 29.054 | B |
| Nereid | 2855.54541 | 18.323 | 0.118 | 2855.55418 | 3.616 | 0.028 | 30.069 | 29.054 | I |
| Nereid | 2856.68175 | 18.995 | 0.063 | 2856.69052 | 4.288 | 0.010 | 30.069 | 29.054 | V |
| Nereid | 2856.68311 | 19.073 | 0.068 | 2856.69188 | 4.366 | 0.010 | 30.069 | 29.054 | V |
| Nereid | 2856.68557 | 19.682 | 0.047 | 2856.69434 | 4.975 | 0.010 | 30.069 | 29.054 | B |
| Nereid | 2856.68787 | 18.284 | 0.097 | 2856.69664 | 3.577 | 0.010 | 30.069 | 29.054 | I |
| Nereid | 2856.84461 | 19.085 | 0.088 | 2856.85338 | 4.378 | 0.016 | 30.069 | 29.054 | V |
| Nereid | 2856.84597 | 18.957 | 0.080 | 2856.85474 | 4.250 | 0.016 | 30.069 | 29.054 | V |
| Nereid | 2856.84842 | 19.754 | 0.048 | 2856.85719 | 5.047 | 0.016 | 30.069 | 29.054 | B |
| Nereid | 2856.85073 | 18.380 | 0.124 | 2856.85950 | 3.673 | 0.016 | 30.069 | 29.054 | I |
| Nereid | 2857.56725 | 19.019 | 0.049 | 2857.57602 | 4.312 | 0.040 | 30.068 | 29.054 | V |
| Nereid | 2857.56970 | 19.672 | 0.035 | 2857.57847 | 4.965 | 0.040 | 30.068 | 29.054 | B |
| Nereid | 2857.57200 | 18.340 | 0.064 | 2857.58077 | 3.633 | 0.040 | 30.068 | 29.054 | I |
| Nereid | 2857.60493 | 19.367 | 0.056 | 2857.61370 | 4.660 | 0.041 | 30.068 | 29.054 | V |
| Nereid | 2857.60629 | 19.343 | 0.045 | 2857.61506 | 4.636 | 0.041 | 30.068 | 29.054 | V |
| Nereid | 2857.60885 | 19.764 | 0.033 | 2857.61762 | 5.057 | 0.041 | 30.068 | 29.054 | B |
| Nereid | 2857.61115 | 18.542 | 0.073 | 2857.61992 | 3.835 | 0.041 | 30.068 | 29.054 | I |
| Nereid | 2857.64950 | 19.090 | 0.044 | 2857.65827 | 4.383 | 0.042 | 30.068 | 29.054 | V |
| Nereid | 2857.65086 | 19.157 | 0.049 | 2857.65963 | 4.450 | 0.042 | 30.068 | 29.054 | V |
| Nereid | 2857.65331 | 19.691 | 0.029 | 2857.66208 | 4.984 | 0.042 | 30.068 | 29.054 | B |
| Nereid | 2857.65561 | 18.448 | 0.066 | 2857.66438 | 3.741 | 0.043 | 30.068 | 29.054 | I |
| Nereid | 2857.84347 | 19.151 | 0.052 | 2857.85224 | 4.444 | 0.049 | 30.068 | 29.054 | V |
| Nereid | 2857.84483 | 19.092 | 0.053 | 2857.85360 | 4.385 | 0.049 | 30.068 | 29.054 | V |
| Nereid | 2857.84728 | 19.744 | 0.029 | 2857.85605 | 5.037 | 0.049 | 30.068 | 29.054 | B |
| Nereid | 2857.84959 | 18.372 | 0.085 | 2857.85836 | 3.665 | 0.049 | 30.068 | 29.054 | I |
| Nereid | 2858.54017 | 18.954 | 0.049 | 2858.54894 | 4.247 | 0.072 | 30.068 | 29.055 | V |
| Nereid | 2858.54154 | 19.102 | 0.053 | 2858.55031 | 4.395 | 0.072 | 30.068 | 29.055 | V |
| Nereid | 2858.54399 | 19.724 | 0.039 | 2858.55276 | 5.017 | 0.072 | 30.068 | 29.055 | B |
| Nereid | 2858.54629 | 18.391 | 0.071 | 2858.55506 | 3.684 | 0.072 | 30.068 | 29.055 | I |
| Nereid | 2858.67303 | 19.178 | 0.044 | 2858.68180 | 4.471 | 0.076 | 30.068 | 29.055 | V |
| Nereid | 2858.67439 | 19.128 | 0.042 | 2858.68316 | 4.421 | 0.076 | 30.068 | 29.055 | V |
| Nereid | 2858.67685 | 19.763 | 0.032 | 2858.68562 | 5.056 | 0.076 | 30.068 | 29.055 | B |
| Nereid | 2858.67914 | 18.555 | 0.063 | 2858.68791 | 3.848 | 0.077 | 30.068 | 29.055 | I |
| Nereid | 2858.81763 | 19.146 | 0.037 | 2858.82640 | 4.439 | 0.081 | 30.068 | 29.055 | V |
| Nereid | 2858.81899 | 19.167 | 0.038 | 2858.82776 | 4.460 | 0.081 | 30.068 | 29.055 | V |
| Nereid | 2858.82145 | 19.751 | 0.021 | 2858.83022 | 5.044 | 0.081 | 30.068 | 29.055 | B |
| Nereid | 2858.82375 | 18.513 | 0.076 | 2858.83252 | 3.806 | 0.081 | 30.068 | 29.055 | I |
| Nereid | 2859.80034 | 19.139 | 0.071 | 2859.80910 | 4.432 | 0.114 | 30.067 | 29.055 | V |
| Nereid | 2859.80170 | 19.094 | 0.066 | 2859.81046 | 4.387 | 0.114 | 30.067 | 29.055 | V |
| Nereid | 2859.80415 | 19.646 | 0.046 | 2859.81291 | 4.939 | 0.114 | 30.067 | 29.055 | B |
| Nereid | 2859.80645 | 18.366 | 0.093 | 2859.81521 | 3.659 | 0.114 | 30.067 | 29.055 | I |
| Nereid | 2860.78834 | 19.039 | 0.057 | 2860.79710 | 4.332 | 0.146 | 30.067 | 29.056 | V |
| Nereid | 2860.78970 | 19.119 | 0.066 | 2860.79846 | 4.412 | 0.146 | 30.067 | 29.056 | V |
| Nereid | 2860.79215 | 19.889 | 0.056 | 2860.80091 | 5.182 | 0.147 | 30.067 | 29.056 | B |
| Nereid | 2860.79445 | 18.361 | 0.071 | 2860.80321 | 3.654 | 0.147 | 30.067 | 29.056 | I |
| Nereid | 2862.77246 | 19.106 | 0.169 | 2862.78120 | 4.399 | 0.212 | 30.066 | 29.059 | V |
| Nereid | 2862.77382 | 19.042 | 0.160 | 2862.78256 | 4.335 | 0.212 | 30.066 | 29.059 | V |
| Nereid | 2862.77627 | 19.731 | 0.118 | 2862.78501 | 5.024 | 0.212 | 30.066 | 29.059 | B |
| Nereid | 2862.77857 | 18.488 | 0.220 | 2862.78731 | 3.781 | 0.212 | 30.066 | 29.059 | I |
| Nereid | 2865.78342 | 19.177 | 0.227 | 2865.79212 | 4.470 | 0.311 | 30.065 | 29.066 | V |
| Nereid | 2865.78481 | 18.911 | 0.201 | 2865.79351 | 4.204 | 0.311 | 30.065 | 29.066 | V |
| Nereid | 2865.78730 | 19.459 | 0.125 | 2865.79600 | 4.752 | 0.311 | 30.065 | 29.066 | B |
| Nereid | 2865.78964 | 18.663 | 0.336 | 2865.79834 | 3.956 | 0.311 | 30.065 | 29.066 | I |
| Nereid | 2866.74291 | 19.156 | 0.121 | 2866.75160 | 4.449 | 0.342 | 30.065 | 29.068 | V |
| Nereid | 2866.74430 | 19.339 | 0.116 | 2866.75299 | 4.632 | 0.342 | 30.065 | 29.068 | V |
| Nereid | 2866.74680 | 19.903 | 0.110 | 2866.75549 | 5.196 | 0.342 | 30.065 | 29.068 | B |
| Nereid | 2866.74913 | 18.336 | 0.154 | 2866.75782 | 3.629 | 0.343 | 30.065 | 29.068 | I |
| Nereid | 2867.72715 | 19.253 | 0.065 | 2867.73582 | 4.545 | 0.374 | 30.065 | 29.071 | V |
| Nereid | 2867.72851 | 19.205 | 0.060 | 2867.73718 | 4.497 | 0.374 | 30.065 | 29.071 | V |
| Nereid | 2867.73097 | 19.945 | 0.052 | 2867.73964 | 5.237 | 0.375 | 30.065 | 29.071 | B |
| Nereid | 2867.73327 | 18.584 | 0.079 | 2867.74194 | 3.876 | 0.375 | 30.065 | 29.071 | I |
| Nereid | 2868.70552 | 19.325 | 0.049 | 2868.71417 | 4.617 | 0.406 | 30.064 | 29.074 | V |
| Nereid | 2868.70688 | 19.391 | 0.050 | 2868.71553 | 4.683 | 0.406 | 30.064 | 29.074 | V |
| Nereid | 2868.70933 | 19.944 | 0.042 | 2868.71798 | 5.236 | 0.406 | 30.064 | 29.074 | B |
| Nereid | 2868.71163 | 18.635 | 0.074 | 2868.72028 | 3.927 | 0.406 | 30.064 | 29.074 | I |
| Nereid | 2869.71278 | 19.321 | 0.076 | 2869.72141 | 4.613 | 0.439 | 30.064 | 29.078 | V |
| Nereid | 2869.71414 | 19.285 | 0.057 | 2869.72277 | 4.577 | 0.439 | 30.064 | 29.078 | V |
| Nereid | 2869.71659 | 20.017 | 0.040 | 2869.72522 | 5.309 | 0.439 | 30.064 | 29.078 | B |
| Nereid | 2869.71889 | 18.406 | 0.107 | 2869.72752 | 3.698 | 0.439 | 30.064 | 29.078 | I |
| Nereid | 2871.73535 | 19.238 | 0.061 | 2871.74394 | 4.529 | 0.504 | 30.063 | 29.086 | V |
| Nereid | 2871.73670 | 19.277 | 0.085 | 2871.74529 | 4.568 | 0.504 | 30.063 | 29.086 | V |
| Nereid | 2871.73916 | 19.763 | 0.050 | 2871.74775 | 5.054 | 0.504 | 30.063 | 29.086 | B |
| Nereid | 2871.74147 | 18.616 | 0.144 | 2871.75006 | 3.907 | 0.504 | 30.063 | 29.086 | I |
| Nereid | 2872.69428 | 19.151 | 0.318 | 2872.70284 | 4.442 | 0.534 | 30.063 | 29.090 | V |
| Nereid | 2872.69567 | 19.501 | 0.331 | 2872.70423 | 4.792 | 0.535 | 30.063 | 29.090 | V |
| Nereid | 2872.69816 | 20.231 | 0.181 | 2872.70672 | 5.522 | 0.535 | 30.063 | 29.090 | B |
| Nereid | 3506.86357 | 19.570 | 0.023 | 3506.86734 | 4.799 | 1.910 | 30.077 | 29.920 | V |
| Nereid | 3508.87995 | 19.563 | 0.021 | 3508.88391 | 4.794 | 1.901 | 30.077 | 29.887 | V |
| Nereid | 3511.86633 | 19.526 | 0.018 | 3511.87058 | 4.761 | 1.883 | 30.077 | 29.837 | V |
| Nereid | 3515.86051 | 19.559 | 0.039 | 3515.86514 | 4.799 | 1.852 | 30.076 | 29.771 | V |
| Nereid | 3517.86826 | 19.362 | 0.046 | 3517.87308 | 4.604 | 1.833 | 30.075 | 29.738 | V |
| Nereid | 3521.85514 | 19.441 | 0.047 | 3521.86033 | 4.688 | 1.790 | 30.074 | 29.674 | V |
| Nereid | 3523.84658 | 19.425 | 0.155 | 3523.85195 | 4.674 | 1.765 | 30.074 | 29.642 | V |
| Nereid | 3525.83859 | 19.252 | 0.151 | 3525.84414 | 4.504 | 1.738 | 30.073 | 29.611 | V |
| Nereid | 3527.83547 | 19.182 | 0.142 | 3527.84120 | 4.436 | 1.709 | 30.073 | 29.581 | V |
| Nereid | 3529.82660 | 19.365 | 0.157 | 3529.83250 | 4.621 | 1.678 | 30.072 | 29.551 | V |
| Nereid | 3547.85135 | 19.525 | 0.045 | 3547.85865 | 4.800 | 1.314 | 30.066 | 29.308 | V |
| Nereid | 3547.85135 | 19.525 | 0.045 | 3547.85865 | 4.800 | 1.314 | 30.066 | 29.308 | V |
| Nereid | 3575.78693 | 19.309 | 0.035 | 3575.79558 | 4.602 | 0.515 | 30.056 | 29.074 | V |
| Nereid | 3586.59192 | 19.172 | 0.012 | 3586.60077 | 4.468 | 0.163 | 30.052 | 29.040 | V |
| Nereid | 3586.80013 | 19.178 | 0.011 | 3586.80898 | 4.474 | 0.156 | 30.052 | 29.040 | V |
| Nereid | 3587.84158 | 18.999 | 0.103 | 3587.85044 | 4.295 | 0.121 | 30.051 | 29.039 | V |
| Nereid | 3588.54083 | 18.752 | 0.108 | 3588.54969 | 4.048 | 0.098 | 30.051 | 29.038 | V |
| Nereid | 3588.82559 | 19.004 | 0.081 | 3588.83446 | 4.300 | 0.088 | 30.051 | 29.038 | V |
| Nereid | 3589.53756 | 19.116 | 0.016 | 3589.54643 | 4.412 | 0.065 | 30.051 | 29.037 | V |
| Nereid | 3589.85694 | 19.117 | 0.013 | 3589.86581 | 4.413 | 0.054 | 30.051 | 29.037 | V |
| Nereid | 3590.53516 | 19.049 | 0.011 | 3590.54403 | 4.345 | 0.032 | 30.050 | 29.037 | V |
| Nereid | 3590.85116 | 18.957 | 0.012 | 3590.86003 | 4.253 | 0.023 | 30.050 | 29.036 | V |
| Nereid | 3591.55444 | 19.066 | 0.017 | 3591.56331 | 4.362 | 0.006 | 30.050 | 29.036 | V |
| Nereid | 3591.82446 | 19.046 | 0.012 | 3591.83333 | 4.342 | 0.014 | 30.050 | 29.036 | V |
| Nereid | 3592.63274 | 19.130 | 0.010 | 3592.64161 | 4.426 | 0.039 | 30.050 | 29.036 | V |
| Nereid | 3592.83996 | 19.106 | 0.012 | 3592.84883 | 4.402 | 0.045 | 30.050 | 29.036 | V |
| Nereid | 3593.67382 | 19.167 | 0.010 | 3593.68269 | 4.463 | 0.073 | 30.049 | 29.037 | V |
| Nereid | 3593.82185 | 19.167 | 0.014 | 3593.83072 | 4.463 | 0.078 | 30.049 | 29.037 | V |
| Nereid | 3594.56794 | 19.189 | 0.014 | 3594.57681 | 4.485 | 0.103 | 30.049 | 29.037 | V |
| Nereid | 3594.80889 | 19.136 | 0.039 | 3594.81776 | 4.432 | 0.111 | 30.049 | 29.038 | V |
| Nereid | 3603.80141 | 19.270 | 0.052 | 3603.81016 | 4.565 | 0.407 | 30.046 | 29.057 | V |
| Nereid | 3607.74360 | 19.319 | 0.016 | 3607.75226 | 4.613 | 0.533 | 30.045 | 29.073 | V |
| Nereid | 3619.71711 | 19.373 | 0.014 | 3619.72534 | 4.661 | 0.901 | 30.041 | 29.148 | V |
| Nereid | 3621.67315 | 19.437 | 0.013 | 3621.68128 | 4.724 | 0.957 | 30.040 | 29.164 | V |
| Nereid | 3626.63859 | 19.184 | 0.138 | 3626.64646 | 4.468 | 1.096 | 30.039 | 29.209 | V |
| Nereid | 3629.69564 | 19.274 | 0.125 | 3629.70334 | 4.556 | 1.177 | 30.038 | 29.240 | V |
| Nereid | 3631.61633 | 19.407 | 0.046 | 3631.62391 | 4.687 | 1.226 | 30.037 | 29.261 | V |
| Nereid | 3635.64865 | 19.396 | 0.052 | 3635.65597 | 4.673 | 1.325 | 30.036 | 29.306 | V |
| Nereid | 3639.60743 | 19.489 | 0.012 | 3639.61447 | 4.762 | 1.415 | 30.035 | 29.354 | V |
| Nereid | 3642.66905 | 19.460 | 0.027 | 3642.67586 | 4.731 | 1.480 | 30.034 | 29.394 | V |
| Nereid | 3644.60862 | 19.447 | 0.074 | 3644.61528 | 4.716 | 1.519 | 30.034 | 29.419 | V |
| Nereid | 3650.60186 | 19.260 | 0.228 | 3650.60803 | 4.523 | 1.629 | 30.032 | 29.504 | V |
| Nereid | 3653.54624 | 19.548 | 0.238 | 3653.55216 | 4.808 | 1.676 | 30.031 | 29.547 | V |
| Nereid | 3655.61127 | 19.507 | 0.247 | 3655.61701 | 4.764 | 1.706 | 30.031 | 29.578 | V |
| Nereid | 3656.57414 | 19.526 | 0.267 | 3656.57980 | 4.782 | 1.720 | 30.031 | 29.593 | V |
| Nereid | 3661.59402 | 19.483 | 0.252 | 3661.59922 | 4.734 | 1.782 | 30.029 | 29.672 | V |
| Nereid | 3662.55658 | 19.437 | 0.070 | 3662.56169 | 4.686 | 1.792 | 30.029 | 29.687 | V |
| Nereid | 3665.56843 | 19.574 | 0.026 | 3665.57326 | 4.820 | 1.821 | 30.028 | 29.736 | V |
| Nereid | 3668.53941 | 19.481 | 0.019 | 3668.54396 | 4.723 | 1.845 | 30.028 | 29.785 | V |
| Nereid | 3670.52067 | 19.214 | 0.016 | 3670.52503 | 4.454 | 1.858 | 30.027 | 29.818 | V |
| | | | | # | | | | | |
| (119951) 2002 KX14 | 3067.76745 | 21.050 | 0.152 | 3067.76745 | 5.080 | 1.394 | 39.661 | 39.415 | V |
| (119951) 2002 KX14 | 3067.77027 | 19.927 | 0.108 | 3067.77027 | 3.957 | 1.394 | 39.661 | 39.415 | I |
| (119951) 2002 KX14 | 3067.81062 | 21.212 | 0.104 | 3067.81062 | 5.242 | 1.393 | 39.661 | 39.414 | V |
| (119951) 2002 KX14 | 3067.81343 | 19.774 | 0.078 | 3067.81343 | 3.804 | 1.393 | 39.661 | 39.414 | I |
| (119951) 2002 KX14 | 3071.78036 | 20.588 | 0.182 | 3071.78075 | 4.622 | 1.368 | 39.661 | 39.347 | V |
| (119951) 2002 KX14 | 3071.78317 | 19.635 | 0.098 | 3071.78356 | 3.669 | 1.368 | 39.661 | 39.347 | I |
| (119951) 2002 KX14 | 3071.82734 | 20.805 | 0.188 | 3071.82773 | 4.839 | 1.368 | 39.661 | 39.347 | V |
| (119951) 2002 KX14 | 3071.83015 | 19.673 | 0.101 | 3071.83054 | 3.707 | 1.368 | 39.661 | 39.347 | I |
| (119951) 2002 KX14 | 3077.75433 | 20.919 | 0.189 | 3077.75528 | 4.958 | 1.318 | 39.660 | 39.250 | V |
| (119951) 2002 KX14 | 3077.75715 | 20.127 | 0.177 | 3077.75810 | 4.166 | 1.318 | 39.660 | 39.250 | I |
| (119951) 2002 KX14 | 3077.80296 | 20.931 | 0.169 | 3077.80392 | 4.970 | 1.317 | 39.660 | 39.249 | V |
| (119951) 2002 KX14 | 3077.80578 | 19.991 | 0.142 | 3077.80674 | 4.030 | 1.317 | 39.660 | 39.249 | I |
| (119951) 2002 KX14 | 3081.75452 | 21.087 | 0.089 | 3081.75584 | 5.130 | 1.276 | 39.660 | 39.187 | V |
| (119951) 2002 KX14 | 3081.75734 | 19.877 | 0.080 | 3081.75866 | 3.920 | 1.276 | 39.660 | 39.187 | I |
| (119951) 2002 KX14 | 3081.79540 | 20.936 | 0.081 | 3081.79672 | 4.979 | 1.276 | 39.660 | 39.186 | V |
| (119951) 2002 KX14 | 3081.79821 | 19.989 | 0.097 | 3081.79953 | 4.032 | 1.276 | 39.660 | 39.186 | I |
| (119951) 2002 KX14 | 3089.77181 | 21.047 | 0.090 | 3089.77381 | 5.096 | 1.173 | 39.659 | 39.068 | V |
| (119951) 2002 KX14 | 3089.77463 | 19.746 | 0.074 | 3089.77663 | 3.795 | 1.173 | 39.659 | 39.068 | I |
| (119951) 2002 KX14 | 3089.81072 | 20.927 | 0.078 | 3089.81273 | 4.976 | 1.173 | 39.659 | 39.067 | V |
| (119951) 2002 KX14 | 3089.81354 | 19.946 | 0.085 | 3089.81555 | 3.995 | 1.173 | 39.659 | 39.067 | I |
| (119951) 2002 KX14 | 3097.66190 | 21.245 | 0.290 | 3097.66452 | 5.300 | 1.050 | 39.658 | 38.962 | V |
| (119951) 2002 KX14 | 3097.66471 | 19.940 | 0.143 | 3097.66733 | 3.995 | 1.050 | 39.658 | 38.962 | I |
| (119951) 2002 KX14 | 3097.70130 | 21.226 | 0.221 | 3097.70392 | 5.281 | 1.049 | 39.658 | 38.961 | V |
| (119951) 2002 KX14 | 3097.70412 | 19.792 | 0.111 | 3097.70674 | 3.847 | 1.049 | 39.658 | 38.961 | I |
| (119951) 2002 KX14 | 3101.70356 | 19.973 | 0.227 | 3101.70646 | 4.031 | 0.978 | 39.658 | 38.912 | I |
| (119951) 2002 KX14 | 3101.73573 | 20.139 | 0.236 | 3101.73864 | 4.197 | 0.978 | 39.658 | 38.912 | I |
| (119951) 2002 KX14 | 3105.71752 | 20.841 | 0.160 | 3105.72069 | 4.901 | 0.903 | 39.658 | 38.866 | V |
| (119951) 2002 KX14 | 3105.72034 | 19.391 | 0.081 | 3105.72351 | 3.451 | 0.903 | 39.658 | 38.866 | I |
| (119951) 2002 KX14 | 3105.76024 | 19.882 | 0.138 | 3105.76341 | 3.943 | 0.902 | 39.658 | 38.866 | I |
| (119951) 2002 KX14 | 3109.62033 | 20.929 | 0.116 | 3109.62373 | 4.992 | 0.825 | 39.657 | 38.826 | V |
| (119951) 2002 KX14 | 3109.62341 | 19.577 | 0.084 | 3109.62681 | 3.640 | 0.825 | 39.657 | 38.826 | I |
| (119951) 2002 KX14 | 3109.65897 | 21.070 | 0.106 | 3109.66238 | 5.133 | 0.824 | 39.657 | 38.825 | V |
| (119951) 2002 KX14 | 3109.66178 | 19.592 | 0.085 | 3109.66519 | 3.655 | 0.824 | 39.657 | 38.825 | I |
| (119951) 2002 KX14 | 3114.67664 | 20.803 | 0.076 | 3114.68032 | 4.868 | 0.718 | 39.657 | 38.778 | V |
| (119951) 2002 KX14 | 3114.67940 | 19.778 | 0.078 | 3114.68308 | 3.843 | 0.718 | 39.657 | 38.778 | I |
| (119951) 2002 KX14 | 3114.72142 | 20.934 | 0.085 | 3114.72510 | 4.999 | 0.717 | 39.657 | 38.778 | V |
| (119951) 2002 KX14 | 3114.72418 | 19.667 | 0.069 | 3114.72786 | 3.732 | 0.717 | 39.657 | 38.778 | I |
| (119951) 2002 KX14 | 3132.64762 | 20.782 | 0.262 | 3132.65195 | 4.854 | 0.299 | 39.655 | 38.666 | V |
| (119951) 2002 KX14 | 3132.65054 | 19.499 | 0.161 | 3132.65487 | 3.571 | 0.299 | 39.655 | 38.666 | I |
| (119951) 2002 KX14 | 3132.71380 | 20.483 | 0.169 | 3132.71813 | 4.555 | 0.297 | 39.655 | 38.666 | V |
| (119951) 2002 KX14 | 3132.71730 | 19.645 | 0.157 | 3132.72163 | 3.717 | 0.297 | 39.655 | 38.666 | I |
| (119951) 2002 KX14 | 3140.61616 | 20.705 | 0.062 | 3140.62060 | 4.778 | 0.101 | 39.654 | 38.645 | V |
| (119951) 2002 KX14 | 3140.61892 | 19.501 | 0.063 | 3140.62336 | 3.574 | 0.101 | 39.654 | 38.645 | I |
| (119951) 2002 KX14 | 3140.66358 | 20.883 | 0.078 | 3140.66802 | 4.956 | 0.100 | 39.654 | 38.645 | V |
| (119951) 2002 KX14 | 3140.66634 | 19.583 | 0.073 | 3140.67078 | 3.656 | 0.100 | 39.654 | 38.645 | I |
| (119951) 2002 KX14 | 3151.65431 | 20.758 | 0.074 | 3151.65874 | 4.831 | 0.176 | 39.653 | 38.648 | V |
| (119951) 2002 KX14 | 3160.54403 | 19.134 | 0.173 | 3160.54830 | 3.205 | 0.394 | 39.652 | 38.676 | I |
| (119951) 2002 KX14 | 3164.58859 | 20.988 | 0.063 | 3164.59274 | 5.058 | 0.491 | 39.652 | 38.696 | V |
| (119951) 2002 KX14 | 3164.59134 | 19.733 | 0.063 | 3164.59549 | 3.803 | 0.491 | 39.652 | 38.696 | I |
| (119951) 2002 KX14 | 3164.62510 | 21.005 | 0.074 | 3164.62925 | 5.075 | 0.492 | 39.652 | 38.696 | V |
| (119951) 2002 KX14 | 3164.62786 | 19.685 | 0.063 | 3164.63201 | 3.755 | 0.492 | 39.652 | 38.696 | I |
| (119951) 2002 KX14 | 3182.62538 | 20.850 | 0.096 | 3182.62871 | 4.913 | 0.886 | 39.650 | 38.837 | V |
| (119951) 2002 KX14 | 3182.62814 | 19.743 | 0.081 | 3182.63147 | 3.806 | 0.886 | 39.650 | 38.837 | I |
| (119951) 2002 KX14 | 3182.66648 | 21.110 | 0.134 | 3182.66981 | 5.172 | 0.887 | 39.650 | 38.838 | V |
| (119951) 2002 KX14 | 3182.66924 | 19.856 | 0.091 | 3182.67257 | 3.918 | 0.887 | 39.650 | 38.838 | I |
| (119951) 2002 KX14 | 3201.58532 | 21.029 | 0.089 | 3201.58733 | 5.079 | 1.209 | 39.648 | 39.066 | V |
| (119951) 2002 KX14 | 3201.58808 | 19.669 | 0.082 | 3201.59009 | 3.719 | 1.209 | 39.648 | 39.067 | I |
| (119951) 2002 KX14 | 3211.51723 | 21.183 | 0.270 | 3211.51841 | 5.225 | 1.330 | 39.647 | 39.211 | V |
| (119951) 2002 KX14 | 3211.52000 | 19.896 | 0.159 | 3211.52117 | 3.938 | 1.330 | 39.647 | 39.211 | I |
| | | | | # | | | | | |
| 2004 TY364 | 3292.71981 | 21.426 | 0.059 | 3292.71981 | 5.480 | 0.547 | 39.783 | 38.856 | B |
| 2004 TY364 | 3292.72397 | 20.422 | 0.097 | 3292.72397 | 4.476 | 0.547 | 39.783 | 38.856 | V |
| 2004 TY364 | 3292.72607 | 19.388 | 0.106 | 3292.72607 | 3.442 | 0.547 | 39.783 | 38.856 | I |
| 2004 TY364 | 3294.69541 | 21.408 | 0.047 | 3294.69541 | 5.462 | 0.552 | 39.783 | 38.856 | B |
| 2004 TY364 | 3294.69957 | 20.478 | 0.080 | 3294.69957 | 4.532 | 0.552 | 39.783 | 38.856 | V |
| 2004 TY364 | 3294.70168 | 19.291 | 0.065 | 3294.70168 | 3.345 | 0.552 | 39.783 | 38.856 | I |
| 2004 TY364 | 3296.70439 | 21.591 | 0.049 | 3296.70438 | 5.645 | 0.559 | 39.782 | 38.858 | B |
| 2004 TY364 | 3296.70855 | 20.639 | 0.082 | 3296.70854 | 4.693 | 0.559 | 39.782 | 38.858 | V |
| 2004 TY364 | 3296.71066 | 19.434 | 0.071 | 3296.71065 | 3.488 | 0.559 | 39.782 | 38.858 | I |
| 2004 TY364 | 3301.66835 | 21.331 | 0.118 | 3301.66828 | 5.385 | 0.593 | 39.779 | 38.868 | B |
| 2004 TY364 | 3301.67250 | 20.360 | 0.123 | 3301.67243 | 4.414 | 0.593 | 39.779 | 38.868 | V |
| 2004 TY364 | 3301.67461 | 19.512 | 0.097 | 3301.67454 | 3.566 | 0.593 | 39.779 | 38.868 | I |
| 2004 TY364 | 3306.65739 | 21.749 | 0.117 | 3306.65723 | 5.802 | 0.645 | 39.777 | 38.885 | B |
| 2004 TY364 | 3306.66155 | 20.514 | 0.177 | 3306.66139 | 4.567 | 0.645 | 39.777 | 38.885 | V |
| 2004 TY364 | 3306.66365 | 19.183 | 0.181 | 3306.66349 | 3.236 | 0.645 | 39.777 | 38.885 | I |
| 2004 TY364 | 3308.66153 | 21.367 | 0.152 | 3308.66132 | 5.420 | 0.670 | 39.776 | 38.893 | B |
| 2004 TY364 | 3308.66569 | 20.667 | 0.212 | 3308.66548 | 4.720 | 0.670 | 39.776 | 38.893 | V |
| 2004 TY364 | 3310.62597 | 21.270 | 0.061 | 3310.62571 | 5.322 | 0.696 | 39.775 | 38.902 | B |
| 2004 TY364 | 3310.63013 | 20.467 | 0.090 | 3310.62987 | 4.519 | 0.696 | 39.775 | 38.902 | V |
| 2004 TY364 | 3310.63223 | 19.328 | 0.092 | 3310.63197 | 3.380 | 0.696 | 39.775 | 38.902 | I |
| 2004 TY364 | 3324.61499 | 21.593 | 0.046 | 3324.61417 | 5.640 | 0.907 | 39.769 | 38.998 | B |
| 2004 TY364 | 3324.61916 | 20.620 | 0.079 | 3324.61834 | 4.667 | 0.907 | 39.769 | 38.998 | V |
| 2004 TY364 | 3324.62127 | 19.379 | 0.068 | 3324.62045 | 3.426 | 0.907 | 39.769 | 38.998 | I |
| 2004 TY364 | 3328.58582 | 21.596 | 0.094 | 3328.58479 | 5.641 | 0.967 | 39.767 | 39.034 | B |
| 2004 TY364 | 3328.58997 | 20.453 | 0.102 | 3328.58894 | 4.498 | 0.967 | 39.767 | 39.034 | V |
| 2004 TY364 | 3328.59208 | 19.664 | 0.111 | 3328.59105 | 3.709 | 0.967 | 39.767 | 39.034 | I |
| 2004 TY364 | 3330.61266 | 21.579 | 0.150 | 3330.61152 | 5.623 | 0.999 | 39.766 | 39.053 | B |
| 2004 TY364 | 3330.61682 | 20.684 | 0.167 | 3330.61568 | 4.728 | 0.999 | 39.766 | 39.053 | V |
| 2004 TY364 | 3330.61893 | 19.877 | 0.161 | 3330.61779 | 3.921 | 0.999 | 39.766 | 39.053 | I |
| 2004 TY364 | 3332.57859 | 21.281 | 0.240 | 3332.57733 | 5.324 | 1.028 | 39.765 | 39.074 | B |
| 2004 TY364 | 3332.58484 | 19.767 | 0.397 | 3332.58358 | 3.810 | 1.028 | 39.765 | 39.074 | I |
| 2004 TY364 | 3335.64076 | 21.539 | 0.236 | 3335.63931 | 5.580 | 1.072 | 39.764 | 39.107 | B |
| 2004 TY364 | 3335.64492 | 20.443 | 0.211 | 3335.64347 | 4.484 | 1.072 | 39.764 | 39.107 | V |
| 2004 TY364 | 3335.64703 | 19.415 | 0.134 | 3335.64558 | 3.456 | 1.072 | 39.764 | 39.107 | I |
| 2004 TY364 | 3337.60719 | 21.617 | 0.173 | 3337.60562 | 5.657 | 1.100 | 39.763 | 39.128 | B |
| 2004 TY364 | 3337.61134 | 20.558 | 0.153 | 3337.60977 | 4.598 | 1.100 | 39.763 | 39.128 | V |
| 2004 TY364 | 3337.61344 | 19.557 | 0.114 | 3337.61187 | 3.597 | 1.100 | 39.763 | 39.128 | I |
| 2004 TY364 | 3339.57067 | 21.598 | 0.104 | 3339.56897 | 5.637 | 1.127 | 39.762 | 39.151 | B |
| 2004 TY364 | 3339.57482 | 20.723 | 0.141 | 3339.57312 | 4.762 | 1.127 | 39.762 | 39.151 | V |
| 2004 TY364 | 3339.57692 | 19.589 | 0.164 | 3339.57522 | 3.628 | 1.127 | 39.762 | 39.151 | I |
| 2004 TY364 | 3343.54815 | 21.520 | 0.063 | 3343.54617 | 5.556 | 1.178 | 39.760 | 39.200 | B |
| 2004 TY364 | 3343.55230 | 20.444 | 0.094 | 3343.55032 | 4.480 | 1.178 | 39.760 | 39.200 | V |
| 2004 TY364 | 3343.55441 | 19.687 | 0.116 | 3343.55243 | 3.723 | 1.178 | 39.760 | 39.200 | I |
| 2004 TY364 | 3345.57723 | 21.553 | 0.052 | 3345.57510 | 5.588 | 1.203 | 39.759 | 39.225 | B |
| 2004 TY364 | 3345.58138 | 20.716 | 0.093 | 3345.57925 | 4.751 | 1.203 | 39.759 | 39.225 | V |
| 2004 TY364 | 3345.58348 | 19.629 | 0.099 | 3345.58135 | 3.664 | 1.203 | 39.759 | 39.225 | I |
| 2004 TY364 | 3348.56328 | 21.551 | 0.051 | 3348.56092 | 5.584 | 1.237 | 39.758 | 39.264 | B |
| 2004 TY364 | 3348.56743 | 20.592 | 0.697 | 3348.56507 | 4.625 | 1.237 | 39.758 | 39.264 | V |
| 2004 TY364 | 3348.56953 | 19.826 | 0.109 | 3348.56717 | 3.859 | 1.237 | 39.758 | 39.264 | I |
| 2004 TY364 | 3351.60069 | 21.656 | 0.064 | 3351.59809 | 5.687 | 1.269 | 39.756 | 39.305 | B |
| 2004 TY364 | 3351.60069 | 21.656 | 0.064 | 3351.59809 | 5.687 | 1.269 | 39.756 | 39.305 | B |
| 2004 TY364 | 3351.60485 | 20.656 | 0.094 | 3351.60225 | 4.687 | 1.269 | 39.756 | 39.305 | V |
| 2004 TY364 | 3351.60695 | 19.706 | 0.104 | 3351.60435 | 3.737 | 1.269 | 39.756 | 39.305 | I |
| 2004 TY364 | 3353.55063 | 21.611 | 0.056 | 3353.54788 | 5.640 | 1.288 | 39.755 | 39.333 | B |
| 2004 TY364 | 3353.55479 | 20.749 | 0.092 | 3353.55204 | 4.778 | 1.288 | 39.755 | 39.333 | V |
| 2004 TY364 | 3353.55689 | 19.746 | 0.097 | 3353.55414 | 3.775 | 1.289 | 39.755 | 39.333 | I |
| 2004 TY364 | 3355.58185 | 21.608 | 0.060 | 3355.57893 | 5.636 | 1.307 | 39.755 | 39.361 | B |
| 2004 TY364 | 3355.58600 | 20.840 | 0.100 | 3355.58308 | 4.868 | 1.307 | 39.755 | 39.361 | V |
| 2004 TY364 | 3355.58810 | 19.660 | 0.088 | 3355.58518 | 3.688 | 1.307 | 39.755 | 39.361 | I |
| 2004 TY364 | 3357.55629 | 21.613 | 0.097 | 3357.55321 | 5.639 | 1.323 | 39.754 | 39.389 | B |
| 2004 TY364 | 3357.56044 | 20.737 | 0.121 | 3357.55736 | 4.763 | 1.323 | 39.754 | 39.389 | V |
| 2004 TY364 | 3357.56254 | 19.713 | 0.120 | 3357.55946 | 3.739 | 1.324 | 39.754 | 39.389 | I |
| 2004 TY364 | 3359.59716 | 21.498 | 0.155 | 3359.59391 | 5.523 | 1.339 | 39.753 | 39.419 | B |
| 2004 TY364 | 3359.60130 | 20.289 | 0.151 | 3359.59805 | 4.314 | 1.339 | 39.753 | 39.419 | V |
| 2004 TY364 | 3359.60340 | 19.473 | 0.160 | 3359.60015 | 3.498 | 1.339 | 39.753 | 39.419 | I |
| 2004 TY364 | 3361.59772 | 21.352 | 0.140 | 3361.59430 | 5.375 | 1.354 | 39.752 | 39.448 | B |
| 2004 TY364 | 3361.60189 | 20.765 | 0.202 | 3361.59847 | 4.788 | 1.354 | 39.752 | 39.449 | V |
| 2004 TY364 | 3361.60400 | 19.734 | 0.144 | 3361.60058 | 3.757 | 1.354 | 39.752 | 39.449 | I |
| 2004 TY364 | 3363.55361 | 21.637 | 0.246 | 3363.55002 | 5.658 | 1.366 | 39.751 | 39.478 | B |
| 2004 TY364 | 3363.55776 | 20.620 | 0.206 | 3363.55417 | 4.641 | 1.366 | 39.751 | 39.478 | V |
| 2004 TY364 | 3363.55985 | 19.831 | 0.160 | 3363.55626 | 3.852 | 1.366 | 39.751 | 39.478 | I |
| 2004 TY364 | 3367.55898 | 21.568 | 0.121 | 3367.55504 | 5.586 | 1.387 | 39.749 | 39.539 | B |
| 2004 TY364 | 3367.56313 | 20.799 | 0.204 | 3367.55919 | 4.817 | 1.387 | 39.749 | 39.539 | V |
| 2004 TY364 | 3367.56523 | 19.928 | 0.200 | 3367.56129 | 3.946 | 1.387 | 39.749 | 39.539 | I |
| 2004 TY364 | 3369.58614 | 21.756 | 0.062 | 3369.58202 | 5.773 | 1.395 | 39.748 | 39.570 | B |
| 2004 TY364 | 3369.59029 | 21.053 | 0.110 | 3369.58616 | 5.070 | 1.395 | 39.748 | 39.570 | V |
| 2004 TY364 | 3369.59240 | 19.733 | 0.107 | 3369.58827 | 3.750 | 1.395 | 39.748 | 39.570 | I |
| 2004 TY364 | 3371.53294 | 21.785 | 0.305 | 3371.52864 | 5.800 | 1.402 | 39.747 | 39.600 | B |
| 2004 TY364 | 3371.53708 | 20.586 | 0.165 | 3371.53278 | 4.601 | 1.402 | 39.747 | 39.601 | V |
| 2004 TY364 | 3371.53918 | 19.866 | 0.160 | 3371.53488 | 3.881 | 1.402 | 39.747 | 39.601 | I |
| 2004 TY364 | 3373.59586 | 21.583 | 0.101 | 3373.59138 | 5.596 | 1.408 | 39.746 | 39.632 | B |
| 2004 TY364 | 3373.60002 | 20.666 | 0.163 | 3373.59554 | 4.679 | 1.408 | 39.746 | 39.633 | V |
| 2004 TY364 | 3373.60212 | 19.649 | 0.131 | 3373.59764 | 3.662 | 1.408 | 39.746 | 39.633 | I |
| 2004 TY364 | 3375.57002 | 21.627 | 0.069 | 3375.56536 | 5.639 | 1.412 | 39.745 | 39.663 | B |
| 2004 TY364 | 3375.57416 | 20.741 | 0.104 | 3375.56950 | 4.753 | 1.412 | 39.745 | 39.663 | V |
| 2004 TY364 | 3375.57626 | 19.717 | 0.119 | 3375.57160 | 3.729 | 1.412 | 39.745 | 39.663 | I |
| 2004 TY364 | 3383.56898 | 21.812 | 0.081 | 3383.56359 | 5.817 | 1.413 | 39.742 | 39.789 | B |
| 2004 TY364 | 3385.56128 | 21.892 | 0.139 | 3385.55571 | 5.895 | 1.410 | 39.741 | 39.820 | B |
| 2004 TY364 | 3387.55178 | 21.447 | 0.157 | 3387.54603 | 5.449 | 1.405 | 39.740 | 39.851 | B |
| 2004 TY364 | 3387.55594 | 20.689 | 0.223 | 3387.55019 | 4.691 | 1.405 | 39.740 | 39.851 | V |
| 2004 TY364 | 3387.55804 | 19.667 | 0.193 | 3387.55229 | 3.669 | 1.405 | 39.740 | 39.851 | I |
| 2004 TY364 | 3394.56885 | 21.288 | 0.227 | 3394.56248 | 5.284 | 1.376 | 39.737 | 39.959 | B |
| 2004 TY364 | 3394.57300 | 20.301 | 0.200 | 3394.56663 | 4.297 | 1.376 | 39.737 | 39.959 | V |
| 2004 TY364 | 3394.57511 | 19.728 | 0.189 | 3394.56874 | 3.724 | 1.376 | 39.737 | 39.959 | I |
| 2004 TY364 | 3395.53600 | 21.261 | 0.191 | 3395.52955 | 5.256 | 1.370 | 39.736 | 39.974 | B |
| 2004 TY364 | 3395.54016 | 20.669 | 0.224 | 3395.53371 | 4.664 | 1.370 | 39.736 | 39.974 | V |
| 2004 TY364 | 3395.54226 | 19.616 | 0.139 | 3395.53581 | 3.611 | 1.370 | 39.736 | 39.974 | I |
| | | | | # | | | | | |
| (26375) 1999 DE9 | 3403.76605 | 21.555 | 0.095 | 3403.76616 | 6.179 | 0.858 | 34.907 | 34.065 | B |
| (26375) 1999 DE9 | 3403.77126 | 20.514 | 0.088 | 3403.77137 | 5.138 | 0.858 | 34.907 | 34.065 | V |
| (26375) 1999 DE9 | 3403.77407 | 19.391 | 0.064 | 3403.77418 | 4.015 | 0.857 | 34.907 | 34.065 | I |
| (26375) 1999 DE9 | 3405.71294 | 21.620 | 0.051 | 3405.71314 | 6.245 | 0.809 | 34.909 | 34.049 | B |
| (26375) 1999 DE9 | 3405.71813 | 20.604 | 0.059 | 3405.71833 | 5.229 | 0.809 | 34.909 | 34.049 | V |
| (26375) 1999 DE9 | 3405.72093 | 19.500 | 0.061 | 3405.72113 | 4.125 | 0.809 | 34.909 | 34.049 | I |
| (26375) 1999 DE9 | 3407.72953 | 21.454 | 0.059 | 3407.72982 | 6.080 | 0.758 | 34.910 | 34.034 | B |
| (26375) 1999 DE9 | 3407.73474 | 20.544 | 0.076 | 3407.73503 | 5.170 | 0.758 | 34.910 | 34.034 | V |
| (26375) 1999 DE9 | 3407.73755 | 19.618 | 0.075 | 3407.73784 | 4.244 | 0.758 | 34.910 | 34.034 | I |
| (26375) 1999 DE9 | 3409.66372 | 21.447 | 0.099 | 3409.66409 | 6.074 | 0.709 | 34.911 | 34.020 | B |
| (26375) 1999 DE9 | 3409.66892 | 20.584 | 0.092 | 3409.66929 | 5.211 | 0.708 | 34.911 | 34.020 | V |
| (26375) 1999 DE9 | 3409.67172 | 19.594 | 0.084 | 3409.67209 | 4.221 | 0.708 | 34.911 | 34.020 | I |
| (26375) 1999 DE9 | 3411.77382 | 21.590 | 0.057 | 3411.77427 | 6.217 | 0.653 | 34.913 | 34.006 | B |
| (26375) 1999 DE9 | 3411.77903 | 20.685 | 0.075 | 3411.77948 | 5.312 | 0.653 | 34.913 | 34.006 | V |

| | | | | | | | | | | |
|---|---|---|---|---|---|---|---|---|---|---|
| (26375) | 1999 DE9 | 3411.78182 | 19.453 | 0.066 | 3411.78227 | 4.080 | 0.653 | 34.913 | 34.006 | I |
| (26375) | 1999 DE9 | 3413.74783 | 21.737 | 0.061 | 3413.74835 | 6.365 | 0.601 | 34.914 | 33.994 | B |
| (26375) | 1999 DE9 | 3413.75303 | 20.753 | 0.076 | 3413.75355 | 5.381 | 0.601 | 34.914 | 33.994 | V |
| (26375) | 1999 DE9 | 3413.75583 | 19.439 | 0.071 | 3413.75635 | 4.067 | 0.600 | 34.914 | 33.994 | I |
| (26375) | 1999 DE9 | 3415.79832 | 21.469 | 0.048 | 3415.79890 | 6.098 | 0.545 | 34.916 | 33.983 | B |
| (26375) | 1999 DE9 | 3415.80353 | 20.645 | 0.065 | 3415.80411 | 5.274 | 0.545 | 34.916 | 33.983 | V |
| (26375) | 1999 DE9 | 3415.80633 | 19.471 | 0.059 | 3415.80691 | 4.100 | 0.545 | 34.916 | 33.983 | I |
| (26375) | 1999 DE9 | 3431.74744 | 21.285 | 0.087 | 3431.74828 | 5.916 | 0.119 | 34.927 | 33.938 | B |
| (26375) | 1999 DE9 | 3431.75265 | 20.533 | 0.102 | 3431.75349 | 5.164 | 0.119 | 34.927 | 33.938 | V |
| (26375) | 1999 DE9 | 3431.75545 | 19.359 | 0.079 | 3431.75629 | 3.990 | 0.119 | 34.927 | 33.938 | I |
| (26375) | 1999 DE9 | 3433.68486 | 21.375 | 0.053 | 3433.68570 | 6.006 | 0.088 | 34.928 | 33.938 | B |
| (26375) | 1999 DE9 | 3433.69006 | 20.580 | 0.072 | 3433.69090 | 5.211 | 0.088 | 34.928 | 33.938 | V |
| (26375) | 1999 DE9 | 3433.69286 | 19.239 | 0.062 | 3433.69370 | 3.870 | 0.088 | 34.928 | 33.938 | I |
| (26375) | 1999 DE9 | 3435.72414 | 21.455 | 0.045 | 3435.72497 | 6.086 | 0.088 | 34.930 | 33.939 | B |
| (26375) | 1999 DE9 | 3435.72933 | 20.442 | 0.056 | 3435.73016 | 5.073 | 0.088 | 34.930 | 33.939 | V |
| (26375) | 1999 DE9 | 3435.73213 | 19.433 | 0.058 | 3435.73296 | 4.064 | 0.088 | 34.930 | 33.939 | I |
| (26375) | 1999 DE9 | 3438.69787 | 21.485 | 0.101 | 3438.69868 | 6.115 | 0.142 | 34.932 | 33.943 | B |
| (26375) | 1999 DE9 | 3438.70308 | 20.767 | 0.114 | 3438.70389 | 5.397 | 0.142 | 34.932 | 33.943 | V |
| (26375) | 1999 DE9 | 3438.70588 | 19.742 | 0.171 | 3438.70669 | 4.372 | 0.142 | 34.932 | 33.943 | I |
| (26375) | 1999 DE9 | 3442.65566 | 21.537 | 0.062 | 3442.65642 | 6.166 | 0.245 | 34.935 | 33.952 | B |
| (26375) | 1999 DE9 | 3442.66086 | 20.399 | 0.070 | 3442.66162 | 5.028 | 0.245 | 34.935 | 33.952 | V |
| (26375) | 1999 DE9 | 3442.66365 | 19.376 | 0.071 | 3442.66441 | 4.005 | 0.245 | 34.935 | 33.952 | I |
| (26375) | 1999 DE9 | 3444.68352 | 21.520 | 0.053 | 3444.68424 | 6.149 | 0.300 | 34.936 | 33.959 | B |
| (26375) | 1999 DE9 | 3444.68872 | 20.539 | 0.072 | 3444.68944 | 5.168 | 0.300 | 34.936 | 33.959 | V |
| (26375) | 1999 DE9 | 3444.69151 | 19.421 | 0.072 | 3444.69223 | 4.050 | 0.300 | 34.936 | 33.959 | I |
| (26375) | 1999 DE9 | 3446.60696 | 21.493 | 0.070 | 3446.60764 | 6.121 | 0.353 | 34.938 | 33.966 | B |
| (26375) | 1999 DE9 | 3446.61216 | 20.489 | 0.076 | 3446.61284 | 5.117 | 0.353 | 34.938 | 33.966 | V |
| (26375) | 1999 DE9 | 3446.61497 | 19.437 | 0.079 | 3446.61565 | 4.065 | 0.353 | 34.938 | 33.966 | I |
| (26375) | 1999 DE9 | 3448.64260 | 21.364 | 0.129 | 3448.64323 | 5.992 | 0.409 | 34.939 | 33.975 | B |
| (26375) | 1999 DE9 | 3448.64780 | 20.319 | 0.127 | 3448.64843 | 4.947 | 0.409 | 34.939 | 33.975 | V |
| (26375) | 1999 DE9 | 3448.65060 | 19.431 | 0.119 | 3448.65123 | 4.059 | 0.410 | 34.939 | 33.975 | I |
| (26375) | 1999 DE9 | 3450.69273 | 21.404 | 0.114 | 3450.69330 | 6.031 | 0.466 | 34.941 | 33.986 | B |
| (26375) | 1999 DE9 | 3450.69794 | 20.638 | 0.133 | 3450.69851 | 5.265 | 0.466 | 34.941 | 33.986 | V |
| (26375) | 1999 DE9 | 3450.70074 | 19.232 | 0.075 | 3450.70131 | 3.859 | 0.466 | 34.941 | 33.986 | I |
| (26375) | 1999 DE9 | 3459.57997 | 21.462 | 0.100 | 3459.58020 | 6.085 | 0.704 | 34.947 | 34.044 | B |
| (26375) | 1999 DE9 | 3459.58510 | 20.737 | 0.123 | 3459.58533 | 5.360 | 0.704 | 34.947 | 34.044 | V |
| (26375) | 1999 DE9 | 3459.58782 | 19.672 | 0.103 | 3459.58805 | 4.295 | 0.704 | 34.947 | 34.044 | I |
| (26375) | 1999 DE9 | 3461.65824 | 21.816 | 0.116 | 3461.65837 | 6.438 | 0.758 | 34.948 | 34.061 | B |
| (26375) | 1999 DE9 | 3461.66337 | 20.842 | 0.102 | 3461.66350 | 5.464 | 0.758 | 34.948 | 34.061 | V |
| (26375) | 1999 DE9 | 3461.66609 | 19.513 | 0.073 | 3461.66622 | 4.135 | 0.758 | 34.948 | 34.061 | I |
| (26375) | 1999 DE9 | 3469.52163 | 20.800 | 0.086 | 3469.52133 | 5.417 | 0.951 | 34.954 | 34.135 | V |
| (26375) | 1999 DE9 | 3469.52435 | 19.667 | 0.088 | 3469.52405 | 4.284 | 0.951 | 34.954 | 34.135 | I |
| (26375) | 1999 DE9 | 3473.57986 | 21.635 | 0.078 | 3473.57931 | 6.249 | 1.044 | 34.957 | 34.179 | B |
| (26375) | 1999 DE9 | 3473.58500 | 20.738 | 0.104 | 3473.58445 | 5.352 | 1.044 | 34.957 | 34.179 | V |
| (26375) | 1999 DE9 | 3473.58772 | 19.550 | 0.095 | 3473.58717 | 4.164 | 1.044 | 34.957 | 34.179 | I |
| (26375) | 1999 DE9 | 3475.56602 | 21.616 | 0.062 | 3475.56533 | 6.228 | 1.088 | 34.958 | 34.202 | B |
| (26375) | 1999 DE9 | 3475.57115 | 20.699 | 0.075 | 3475.57046 | 5.311 | 1.088 | 34.958 | 34.202 | V |
| (26375) | 1999 DE9 | 3475.57387 | 19.539 | 0.069 | 3475.57318 | 4.151 | 1.088 | 34.958 | 34.202 | I |
| (26375) | 1999 DE9 | 3477.52968 | 21.698 | 0.112 | 3477.52886 | 6.308 | 1.130 | 34.960 | 34.226 | B |
| (26375) | 1999 DE9 | 3477.53480 | 20.993 | 0.141 | 3477.53398 | 5.603 | 1.130 | 34.960 | 34.226 | V |

| | | | | | | | | | |
|---|---|---|---|---|---|---|---|---|---|
| (26375) 1999 DE9 | 3477.53753 | 19.519 | 0.092 | 3477.53671 | 4.129 | 1.130 | 34.960 | 34.226 | I |
| (26375) 1999 DE9 | 3501.52017 | 21.797 | 0.081 | 3501.51735 | 6.384 | 1.521 | 34.977 | 34.572 | B |
| (26375) 1999 DE9 | 3501.52531 | 20.768 | 0.114 | 3501.52249 | 5.355 | 1.521 | 34.977 | 34.572 | V |
| (26375) 1999 DE9 | 3501.52803 | 19.443 | 0.104 | 3501.52521 | 4.030 | 1.521 | 34.977 | 34.572 | I |
| (26375) 1999 DE9 | 3503.49507 | 20.846 | 0.210 | 3503.49206 | 5.431 | 1.543 | 34.978 | 34.604 | V |
| (26375) 1999 DE9 | 3503.49779 | 19.552 | 0.258 | 3503.49478 | 4.137 | 1.543 | 34.978 | 34.604 | I |
| (26375) 1999 DE9 | 3504.51564 | 21.786 | 0.112 | 3504.51254 | 6.370 | 1.553 | 34.979 | 34.621 | B |
| (26375) 1999 DE9 | 3504.52077 | 21.045 | 0.165 | 3504.51767 | 5.629 | 1.553 | 34.979 | 34.621 | V |
| (26375) 1999 DE9 | 3504.52349 | 19.692 | 0.127 | 3504.52039 | 4.276 | 1.553 | 34.979 | 34.621 | I |
| (26375) 1999 DE9 | 3509.55204 | 19.503 | 0.243 | 3509.54845 | 4.082 | 1.597 | 34.983 | 34.705 | I |
| (26375) 1999 DE9 | 3511.50648 | 21.512 | 0.224 | 3511.50270 | 6.089 | 1.611 | 34.984 | 34.739 | B |
| (26375) 1999 DE9 | 3511.51163 | 20.967 | 0.266 | 3511.50785 | 5.544 | 1.611 | 34.984 | 34.739 | V |
| (26375) 1999 DE9 | 3511.51435 | 19.617 | 0.141 | 3511.51057 | 4.194 | 1.611 | 34.984 | 34.739 | I |
| (26375) 1999 DE9 | 3520.56592 | 20.844 | 0.131 | 3520.56123 | 5.410 | 1.653 | 34.991 | 34.896 | V |
| (26375) 1999 DE9 | 3520.56864 | 19.841 | 0.150 | 3520.56395 | 4.407 | 1.653 | 34.991 | 34.896 | I |
| (26375) 1999 DE9 | 3521.50398 | 21.719 | 0.058 | 3521.49920 | 6.284 | 1.656 | 34.991 | 34.912 | B |
| (26375) 1999 DE9 | 3521.50913 | 20.897 | 0.098 | 3521.50434 | 5.462 | 1.656 | 34.991 | 34.912 | V |
| (26375) 1999 DE9 | 3521.51185 | 19.557 | 0.107 | 3521.50706 | 4.122 | 1.656 | 34.991 | 34.912 | I |
| (26375) 1999 DE9 | 3523.48278 | 21.672 | 0.061 | 3523.47779 | 6.235 | 1.659 | 34.993 | 34.947 | B |
| (26375) 1999 DE9 | 3523.48792 | 20.803 | 0.106 | 3523.48293 | 5.366 | 1.659 | 34.993 | 34.947 | V |
| (26375) 1999 DE9 | 3523.49065 | 19.682 | 0.164 | 3523.48566 | 4.245 | 1.659 | 34.993 | 34.947 | I |
| (26375) 1999 DE9 | 3527.45989 | 21.747 | 0.094 | 3527.45450 | 6.306 | 1.660 | 34.995 | 35.017 | B |
| (26375) 1999 DE9 | 3527.46504 | 20.899 | 0.106 | 3527.45965 | 5.458 | 1.660 | 34.995 | 35.017 | V |
| (26375) 1999 DE9 | 3527.46776 | 19.622 | 0.093 | 3527.46237 | 4.181 | 1.660 | 34.995 | 35.017 | I |
| | | | | # | | | | | |
| (47171) 1999 TC36 | 2845.88407 | 21.369 | 0.134 | 2845.88407 | 6.459 | 1.690 | 31.198 | 30.751 | B |
| (47171) 1999 TC36 | 2845.88706 | 19.038 | 0.260 | 2845.88706 | 4.128 | 1.690 | 31.198 | 30.750 | I |
| (47171) 1999 TC36 | 2846.89072 | 20.964 | 0.146 | 2846.89081 | 6.055 | 1.676 | 31.198 | 30.735 | B |
| (47171) 1999 TC36 | 2846.89517 | 20.039 | 0.206 | 2846.89526 | 5.130 | 1.676 | 31.198 | 30.735 | V |
| (47171) 1999 TC36 | 2847.88925 | 21.381 | 0.270 | 2847.88943 | 6.473 | 1.662 | 31.198 | 30.720 | B |
| (47171) 1999 TC36 | 2856.82281 | 21.367 | 0.105 | 2856.82373 | 6.469 | 1.517 | 31.195 | 30.592 | B |
| (47171) 1999 TC36 | 2856.82580 | 19.160 | 0.192 | 2856.82672 | 4.262 | 1.517 | 31.195 | 30.592 | I |
| (47171) 1999 TC36 | 2856.82725 | 20.792 | 0.198 | 2856.82817 | 5.894 | 1.517 | 31.195 | 30.592 | V |
| (47171) 1999 TC36 | 2857.85261 | 21.498 | 0.062 | 2857.85361 | 6.601 | 1.498 | 31.195 | 30.578 | B |
| (47171) 1999 TC36 | 2857.85561 | 19.104 | 0.108 | 2857.85661 | 4.207 | 1.498 | 31.195 | 30.578 | I |
| (47171) 1999 TC36 | 2857.85706 | 20.541 | 0.089 | 2857.85806 | 5.644 | 1.498 | 31.195 | 30.578 | V |
| (47171) 1999 TC36 | 2859.86353 | 21.545 | 0.070 | 2859.86468 | 6.649 | 1.460 | 31.194 | 30.551 | B |
| (47171) 1999 TC36 | 2859.86656 | 19.068 | 0.106 | 2859.86771 | 4.172 | 1.460 | 31.194 | 30.551 | I |
| (47171) 1999 TC36 | 2859.86805 | 20.471 | 0.091 | 2859.86920 | 5.575 | 1.459 | 31.194 | 30.551 | V |
| (47171) 1999 TC36 | 2860.88899 | 21.444 | 0.067 | 2860.89022 | 6.549 | 1.439 | 31.194 | 30.538 | B |
| (47171) 1999 TC36 | 2860.89199 | 19.176 | 0.110 | 2860.89322 | 4.281 | 1.439 | 31.194 | 30.538 | I |
| (47171) 1999 TC36 | 2860.89345 | 20.342 | 0.082 | 2860.89468 | 5.447 | 1.439 | 31.194 | 30.538 | V |
| (47171) 1999 TC36 | 2861.85712 | 20.942 | 0.219 | 2861.85842 | 6.048 | 1.420 | 31.194 | 30.525 | B |
| (47171) 1999 TC36 | 2861.86011 | 19.086 | 0.106 | 2861.86141 | 4.192 | 1.420 | 31.194 | 30.525 | I |
| (47171) 1999 TC36 | 2861.86156 | 20.281 | 0.118 | 2861.86286 | 5.387 | 1.420 | 31.194 | 30.525 | V |
| (47171) 1999 TC36 | 2862.84545 | 19.146 | 0.132 | 2862.84682 | 4.253 | 1.400 | 31.193 | 30.513 | I |
| (47171) 1999 TC36 | 2862.84690 | 20.204 | 0.138 | 2862.84827 | 5.311 | 1.400 | 31.193 | 30.513 | V |
| (47171) 1999 TC36 | 2863.82013 | 19.044 | 0.164 | 2863.82157 | 4.152 | 1.379 | 31.193 | 30.501 | I |
| (47171) 1999 TC36 | 2863.82162 | 20.159 | 0.176 | 2863.82306 | 5.267 | 1.379 | 31.193 | 30.501 | V |
| (47171) 1999 TC36 | 2864.84900 | 19.083 | 0.188 | 2864.85052 | 4.192 | 1.357 | 31.193 | 30.488 | I |
| (47171) 1999 TC36 | 2864.85048 | 20.327 | 0.246 | 2864.85200 | 5.436 | 1.357 | 31.193 | 30.488 | V |
| (47171) 1999 TC36 | 2865.85574 | 18.850 | 0.244 | 2865.85733 | 3.960 | 1.336 | 31.193 | 30.476 | I |
| (47171) 1999 TC36 | 2868.78180 | 20.760 | 0.285 | 2868.78358 | 5.872 | 1.270 | 31.192 | 30.442 | B |
| (47171) 1999 TC36 | 2868.78484 | 18.776 | 0.292 | 2868.78662 | 3.888 | 1.270 | 31.192 | 30.442 | I |
| (47171) 1999 TC36 | 2869.75921 | 20.589 | 0.213 | 2869.76106 | 5.702 | 1.248 | 31.191 | 30.431 | B |
| (47171) 1999 TC36 | 2869.76225 | 18.753 | 0.156 | 2869.76410 | 3.866 | 1.248 | 31.191 | 30.431 | I |
| (47171) 1999 TC36 | 2869.76374 | 20.229 | 0.174 | 2869.76559 | 5.342 | 1.248 | 31.191 | 30.431 | V |
| (47171) 1999 TC36 | 2874.81033 | 21.263 | 0.057 | 2874.81248 | 6.380 | 1.126 | 31.190 | 30.377 | B |
| (47171) 1999 TC36 | 2874.81337 | 18.988 | 0.109 | 2874.81552 | 4.105 | 1.126 | 31.190 | 30.377 | I |
| (47171) 1999 TC36 | 2874.81486 | 20.412 | 0.095 | 2874.81701 | 5.529 | 1.126 | 31.190 | 30.377 | V |
| (47171) 1999 TC36 | 2875.79882 | 21.454 | 0.056 | 2875.80103 | 6.572 | 1.102 | 31.190 | 30.368 | B |
| (47171) 1999 TC36 | 2875.80185 | 19.023 | 0.097 | 2875.80406 | 4.141 | 1.102 | 31.190 | 30.368 | I |
| (47171) 1999 TC36 | 2875.80334 | 20.383 | 0.075 | 2875.80555 | 5.501 | 1.102 | 31.190 | 30.368 | V |
| (47171) 1999 TC36 | 2876.74499 | 21.539 | 0.072 | 2876.74725 | 6.658 | 1.078 | 31.189 | 30.359 | B |
| (47171) 1999 TC36 | 2876.74803 | 18.898 | 0.099 | 2876.75029 | 4.017 | 1.078 | 31.189 | 30.359 | I |
| (47171) 1999 TC36 | 2876.74951 | 20.529 | 0.096 | 2876.75177 | 5.648 | 1.078 | 31.189 | 30.359 | V |
| (47171) 1999 TC36 | 2878.75879 | 21.566 | 0.094 | 2878.76116 | 6.686 | 1.026 | 31.189 | 30.340 | B |
| (47171) 1999 TC36 | 2878.76182 | 19.005 | 0.150 | 2878.76419 | 4.125 | 1.026 | 31.189 | 30.340 | I |
| (47171) 1999 TC36 | 2878.76331 | 20.373 | 0.105 | 2878.76568 | 5.493 | 1.026 | 31.189 | 30.340 | V |
| (47171) 1999 TC36 | 2892.75463 | 18.759 | 0.172 | 2892.75759 | 3.886 | 0.649 | 31.185 | 30.238 | I |
| (47171) 1999 TC36 | 2892.75613 | 20.183 | 0.257 | 2892.75909 | 5.310 | 0.649 | 31.185 | 30.238 | V |
| (47171) 1999 TC36 | 2895.73111 | 20.759 | 0.145 | 2895.73415 | 5.888 | 0.568 | 31.184 | 30.224 | B |
| (47171) 1999 TC36 | 2895.73414 | 18.594 | 0.186 | 2895.73718 | 3.723 | 0.568 | 31.184 | 30.224 | I |
| (47171) 1999 TC36 | 2895.73562 | 20.199 | 0.245 | 2895.73866 | 5.328 | 0.568 | 31.184 | 30.224 | V |
| (47171) 1999 TC36 | 2896.72539 | 21.483 | 0.239 | 2896.72846 | 6.612 | 0.542 | 31.184 | 30.219 | B |
| (47171) 1999 TC36 | 2896.72842 | 18.958 | 0.159 | 2896.73149 | 4.087 | 0.541 | 31.184 | 30.219 | I |
| (47171) 1999 TC36 | 2896.72990 | 19.804 | 0.125 | 2896.73297 | 4.933 | 0.541 | 31.184 | 30.219 | V |
| (47171) 1999 TC36 | 2901.72340 | 20.877 | 0.134 | 2901.72657 | 6.007 | 0.415 | 31.182 | 30.202 | B |
| (47171) 1999 TC36 | 2901.72792 | 19.970 | 0.152 | 2901.73109 | 5.100 | 0.415 | 31.182 | 30.202 | V |
| (47171) 1999 TC36 | 2903.74367 | 20.892 | 0.167 | 2903.74687 | 6.023 | 0.371 | 31.181 | 30.197 | B |
| (47171) 1999 TC36 | 2903.74820 | 19.856 | 0.219 | 2903.75140 | 4.987 | 0.370 | 31.181 | 30.197 | V |
| (47171) 1999 TC36 | 2907.71908 | 21.199 | 0.065 | 2907.72232 | 6.330 | 0.302 | 31.180 | 30.190 | B |
| (47171) 1999 TC36 | 2907.72211 | 18.956 | 0.094 | 2907.72535 | 4.087 | 0.302 | 31.180 | 30.190 | I |
| (47171) 1999 TC36 | 2907.72360 | 20.137 | 0.066 | 2907.72684 | 5.268 | 0.302 | 31.180 | 30.190 | V |
| (47171) 1999 TC36 | 2909.67507 | 21.245 | 0.094 | 2909.67831 | 6.376 | 0.284 | 31.180 | 30.189 | B |
| (47171) 1999 TC36 | 2909.67811 | 18.532 | 0.195 | 2909.68135 | 3.663 | 0.284 | 31.180 | 30.189 | I |
| (47171) 1999 TC36 | 2909.67959 | 20.112 | 0.112 | 2909.68283 | 5.243 | 0.284 | 31.180 | 30.189 | V |
| (47171) 1999 TC36 | 2915.69118 | 20.844 | 0.129 | 2915.69441 | 5.975 | 0.310 | 31.178 | 30.191 | B |
| (47171) 1999 TC36 | 2915.69421 | 18.770 | 0.141 | 2915.69744 | 3.901 | 0.310 | 31.178 | 30.191 | I |
| (47171) 1999 TC36 | 2915.69570 | 20.406 | 0.183 | 2915.69893 | 5.537 | 0.310 | 31.178 | 30.191 | V |
| (47171) 1999 TC36 | 2916.70305 | 21.127 | 0.109 | 2916.70627 | 6.258 | 0.326 | 31.178 | 30.193 | B |
| (47171) 1999 TC36 | 2916.70609 | 18.669 | 0.094 | 2916.70931 | 3.800 | 0.326 | 31.178 | 30.193 | I |
| (47171) 1999 TC36 | 2916.70757 | 20.518 | 0.129 | 2916.71079 | 5.649 | 0.326 | 31.178 | 30.193 | V |
| (47171) 1999 TC36 | 2928.64993 | 21.132 | 0.109 | 2928.65292 | 6.261 | 0.610 | 31.174 | 30.233 | B |
| (47171) 1999 TC36 | 2928.65296 | 18.845 | 0.095 | 2928.65595 | 3.974 | 0.610 | 31.174 | 30.233 | I |
| (47171) 1999 TC36 | 2928.65445 | 20.387 | 0.096 | 2928.65744 | 5.516 | 0.610 | 31.174 | 30.233 | V |
| (47171) 1999 TC36 | 2940.64073 | 18.872 | 0.101 | 2940.64325 | 3.995 | 0.936 | 31.171 | 30.315 | I |
| (47171) 1999 TC36 | 2940.64222 | 20.236 | 0.096 | 2940.64474 | 5.359 | 0.936 | 31.171 | 30.315 | V |
| (47171) 1999 TC36 | 2943.65343 | 18.913 | 0.096 | 2943.65579 | 4.034 | 1.014 | 31.170 | 30.342 | I |
| (47171) 1999 TC36 | 2943.65492 | 20.193 | 0.071 | 2943.65728 | 5.314 | 1.014 | 31.170 | 30.342 | V |
| (47171) 1999 TC36 | 2944.63653 | 21.311 | 0.124 | 2944.63884 | 6.432 | 1.039 | 31.170 | 30.351 | B |
| (47171) 1999 TC36 | 2944.63956 | 19.186 | 0.213 | 2944.64187 | 4.306 | 1.039 | 31.170 | 30.351 | I |
| (47171) 1999 TC36 | 2944.64106 | 20.446 | 0.164 | 2944.64337 | 5.566 | 1.039 | 31.170 | 30.351 | V |
| (47171) 1999 TC36 | 2945.59461 | 21.378 | 0.170 | 2945.59687 | 6.498 | 1.064 | 31.169 | 30.360 | B |
| (47171) 1999 TC36 | 2945.59765 | 19.172 | 0.202 | 2945.59991 | 4.292 | 1.064 | 31.169 | 30.360 | I |
| (47171) 1999 TC36 | 2945.59914 | 20.120 | 0.134 | 2945.60140 | 5.240 | 1.064 | 31.169 | 30.360 | V |
| (47171) 1999 TC36 | 2946.62921 | 21.456 | 0.184 | 2946.63141 | 6.575 | 1.090 | 31.169 | 30.370 | B |
| (47171) 1999 TC36 | 2946.63225 | 18.673 | 0.129 | 2946.63445 | 3.792 | 1.090 | 31.169 | 30.370 | I |
| (47171) 1999 TC36 | 2946.63374 | 20.260 | 0.155 | 2946.63594 | 5.379 | 1.090 | 31.169 | 30.370 | V |
| (47171) 1999 TC36 | 2947.59495 | 20.168 | 0.251 | 2947.59709 | 5.286 | 1.113 | 31.169 | 30.380 | B |
| (47171) 1999 TC36 | 2947.59948 | 20.242 | 0.281 | 2947.60162 | 5.360 | 1.114 | 31.169 | 30.380 | V |
| (47171) 1999 TC36 | 2951.61897 | 18.748 | 0.132 | 2951.62086 | 3.864 | 1.210 | 31.168 | 30.422 | I |
| (47171) 1999 TC36 | 2951.62047 | 20.061 | 0.183 | 2951.62236 | 5.177 | 1.210 | 31.168 | 30.423 | V |
| (47171) 1999 TC36 | 2953.58579 | 19.110 | 0.136 | 2953.58756 | 4.224 | 1.255 | 31.167 | 30.445 | I |
| (47171) 1999 TC36 | 2953.58728 | 20.296 | 0.163 | 2953.58905 | 5.410 | 1.255 | 31.167 | 30.445 | V |
| (47171) 1999 TC36 | 2953.85637 | 18.520 | 0.179 | 2953.85812 | 3.634 | 1.261 | 31.167 | 30.448 | I |
| (47171) 1999 TC36 | 2953.85666 | 19.840 | 0.209 | 2953.85841 | 4.954 | 1.261 | 31.167 | 30.448 | V |
| (47171) 1999 TC36 | 2957.62189 | 18.897 | 0.101 | 2957.62338 | 4.008 | 1.344 | 31.166 | 30.493 | I |
| (47171) 1999 TC36 | 2957.62338 | 20.266 | 0.100 | 2957.62487 | 5.377 | 1.344 | 31.166 | 30.493 | V |
| (47171) 1999 TC36 | 2963.56735 | 19.118 | 0.167 | 2963.56839 | 4.223 | 1.462 | 31.164 | 30.570 | I |
| (47171) 1999 TC36 | 2963.56882 | 20.011 | 0.090 | 2963.56986 | 5.116 | 1.462 | 31.164 | 30.570 | V |
| (47171) 1999 TC36 | 2974.61900 | 21.532 | 0.144 | 2974.61912 | 6.626 | 1.639 | 31.161 | 30.730 | B |
| (47171) 1999 TC36 | 2974.62203 | 19.324 | 0.178 | 2974.62215 | 4.418 | 1.639 | 31.161 | 30.730 | I |
| (47171) 1999 TC36 | 2974.62350 | 20.386 | 0.163 | 2974.62362 | 5.480 | 1.639 | 31.161 | 30.730 | V |
| (47171) 1999 TC36 | 2980.53097 | 21.336 | 0.192 | 2980.53055 | 6.424 | 1.709 | 31.159 | 30.822 | B |
| (47171) 1999 TC36 | 2980.53547 | 20.410 | 0.171 | 2980.53505 | 5.498 | 1.709 | 31.159 | 30.823 | V |
| | | | | # | | | | | |
| (55638) 2002 VE95 | 3224.90492 | 20.383 | 0.179 | 3224.90492 | 5.893 | 2.039 | 28.035 | 28.200 | V |
| (55638) 2002 VE95 | 3224.90698 | 19.213 | 0.127 | 3224.90698 | 4.723 | 2.039 | 28.035 | 28.200 | I |
| (55638) 2002 VE95 | 3238.86106 | 21.706 | 0.080 | 3238.86106 | 7.234 | 2.067 | 28.037 | 27.970 | B |
| (55638) 2002 VE95 | 3238.86516 | 20.593 | 0.113 | 3238.86649 | 6.121 | 2.067 | 28.037 | 27.970 | V |
| (55638) 2002 VE95 | 3238.86723 | 19.139 | 0.085 | 3238.86856 | 4.667 | 2.067 | 28.037 | 27.970 | I |
| (55638) 2002 VE95 | 3240.84778 | 21.810 | 0.100 | 3240.84797 | 7.340 | 2.062 | 28.037 | 27.937 | B |
| (55638) 2002 VE95 | 3240.85189 | 20.590 | 0.132 | 3240.85341 | 6.120 | 2.062 | 28.037 | 27.937 | V |
| (55638) 2002 VE95 | 3240.85396 | 18.894 | 0.099 | 3240.85548 | 4.424 | 2.062 | 28.037 | 27.937 | I |
| (55638) 2002 VE95 | 3242.81350 | 21.643 | 0.085 | 3242.81388 | 7.176 | 2.054 | 28.037 | 27.905 | B |
| (55638) 2002 VE95 | 3242.82458 | 21.701 | 0.075 | 3242.82496 | 7.234 | 2.054 | 28.037 | 27.905 | B |
| (55638) 2002 VE95 | 3242.82869 | 20.637 | 0.095 | 3242.83039 | 6.170 | 2.054 | 28.037 | 27.905 | V |
| (55638) 2002 VE95 | 3242.83076 | 18.984 | 0.079 | 3242.83246 | 4.517 | 2.054 | 28.037 | 27.905 | I |
| (55638) 2002 VE95 | 3245.84751 | 21.730 | 0.292 | 3245.84817 | 7.267 | 2.039 | 28.037 | 27.855 | B |
| (55638) 2002 VE95 | 3245.85162 | 20.306 | 0.176 | 3245.85361 | 5.843 | 2.039 | 28.037 | 27.855 | V |
| (55638) 2002 VE95 | 3245.85368 | 19.049 | 0.084 | 3245.85567 | 4.586 | 2.039 | 28.037 | 27.855 | I |
| (55638) 2002 VE95 | 3247.80674 | 20.193 | 0.255 | 3247.80891 | 5.732 | 2.026 | 28.038 | 27.823 | V |
| (55638) 2002 VE95 | 3247.80881 | 19.060 | 0.128 | 3247.81098 | 4.599 | 2.026 | 28.038 | 27.823 | I |
| (55638) 2002 VE95 | 3249.80158 | 21.060 | 0.212 | 3249.80261 | 6.602 | 2.011 | 28.038 | 27.791 | B |
| (55638) 2002 VE95 | 3249.80569 | 20.573 | 0.265 | 3249.80805 | 6.115 | 2.011 | 28.038 | 27.791 | V |
| (55638) 2002 VE95 | 3249.80776 | 18.977 | 0.106 | 3249.81012 | 4.519 | 2.011 | 28.038 | 27.791 | I |
| (55638) 2002 VE95 | 3251.79745 | 21.118 | 0.253 | 3251.79867 | 6.662 | 1.993 | 28.038 | 27.759 | B |
| (55638) 2002 VE95 | 3251.80157 | 20.191 | 0.257 | 3251.80411 | 5.735 | 1.993 | 28.038 | 27.759 | V |
| (55638) 2002 VE95 | 3251.80363 | 19.129 | 0.194 | 3251.80617 | 4.673 | 1.993 | 28.038 | 27.759 | I |

| | | | | | | | | | | |
|---|---|---|---|---|---|---|---|---|---|---|
| (55638) 2002 VE95 | 3257.83824 | 21.594 | 0.070 | 3257.84000 | 7.146 | 1.927 | 28.039 | 27.665 | B |
| (55638) 2002 VE95 | 3257.84239 | 20.347 | 0.092 | 3257.84548 | 5.899 | 1.927 | 28.039 | 27.665 | V |
| (55638) 2002 VE95 | 3257.84449 | 19.037 | 0.074 | 3257.84758 | 4.589 | 1.927 | 28.039 | 27.665 | I |
| (55638) 2002 VE95 | 3259.80268 | 21.609 | 0.133 | 3259.80462 | 7.163 | 1.901 | 28.039 | 27.635 | B |
| (55638) 2002 VE95 | 3259.80680 | 20.269 | 0.136 | 3259.81006 | 5.823 | 1.901 | 28.039 | 27.635 | V |
| (55638) 2002 VE95 | 3259.80887 | 18.866 | 0.105 | 3259.81213 | 4.420 | 1.901 | 28.039 | 27.635 | I |
| (55638) 2002 VE95 | 3261.82042 | 21.430 | 0.063 | 3261.82253 | 6.986 | 1.872 | 28.039 | 27.604 | B |
| (55638) 2002 VE95 | 3261.82454 | 20.602 | 0.102 | 3261.82798 | 6.158 | 1.872 | 28.039 | 27.604 | V |
| (55638) 2002 VE95 | 3261.82662 | 19.077 | 0.071 | 3261.83006 | 4.633 | 1.872 | 28.039 | 27.604 | I |
| (55638) 2002 VE95 | 3263.74395 | 21.497 | 0.059 | 3263.74623 | 7.056 | 1.842 | 28.039 | 27.576 | B |
| (55638) 2002 VE95 | 3263.74808 | 20.353 | 0.067 | 3263.75169 | 5.912 | 1.842 | 28.039 | 27.576 | V |
| (55638) 2002 VE95 | 3263.75015 | 18.986 | 0.063 | 3263.75376 | 4.545 | 1.842 | 28.039 | 27.576 | I |
| (55638) 2002 VE95 | 3265.83027 | 21.381 | 0.057 | 3265.83272 | 6.942 | 1.808 | 28.040 | 27.545 | B |
| (55638) 2002 VE95 | 3265.83442 | 20.456 | 0.090 | 3265.83820 | 6.017 | 1.808 | 28.040 | 27.545 | V |
| (55638) 2002 VE95 | 3265.83652 | 18.939 | 0.075 | 3265.84030 | 4.500 | 1.808 | 28.040 | 27.545 | I |
| (55638) 2002 VE95 | 3267.81334 | 21.327 | 0.092 | 3267.81596 | 6.890 | 1.773 | 28.040 | 27.517 | B |
| (55638) 2002 VE95 | 3267.81745 | 20.488 | 0.145 | 3267.82139 | 6.051 | 1.773 | 28.040 | 27.517 | V |
| (55638) 2002 VE95 | 3267.81952 | 18.881 | 0.090 | 3267.82346 | 4.444 | 1.773 | 28.040 | 27.517 | I |
| (55638) 2002 VE95 | 3269.82457 | 21.569 | 0.100 | 3269.82735 | 7.134 | 1.736 | 28.040 | 27.489 | B |
| (55638) 2002 VE95 | 3269.82868 | 20.461 | 0.117 | 3269.83279 | 6.026 | 1.736 | 28.040 | 27.489 | V |
| (55638) 2002 VE95 | 3269.83075 | 19.069 | 0.086 | 3269.83486 | 4.634 | 1.736 | 28.040 | 27.489 | I |
| (55638) 2002 VE95 | 3271.84501 | 21.516 | 0.049 | 3271.84795 | 7.083 | 1.696 | 28.040 | 27.461 | B |
| (55638) 2002 VE95 | 3271.84913 | 20.374 | 0.072 | 3271.85340 | 5.941 | 1.696 | 28.040 | 27.461 | V |
| (55638) 2002 VE95 | 3271.85119 | 19.115 | 0.059 | 3271.85546 | 4.682 | 1.696 | 28.040 | 27.461 | I |
| (55638) 2002 VE95 | 3273.77288 | 21.475 | 0.211 | 3273.77597 | 7.044 | 1.657 | 28.040 | 27.436 | B |
| (55638) 2002 VE95 | 3273.77704 | 20.256 | 0.139 | 3273.78145 | 5.825 | 1.657 | 28.040 | 27.436 | V |
| (55638) 2002 VE95 | 3273.77914 | 19.034 | 0.071 | 3273.78355 | 4.604 | 1.657 | 28.040 | 27.436 | I |
| (55638) 2002 VE95 | 3281.80470 | 21.067 | 0.173 | 3281.80836 | 6.644 | 1.474 | 28.041 | 27.336 | B |
| (55638) 2002 VE95 | 3281.80885 | 19.940 | 0.169 | 3281.81384 | 5.517 | 1.474 | 28.041 | 27.336 | V |
| (55638) 2002 VE95 | 3281.81095 | 18.759 | 0.116 | 3281.81594 | 4.336 | 1.473 | 28.041 | 27.336 | I |
| (55638) 2002 VE95 | 3283.81817 | 21.288 | 0.086 | 3283.82196 | 6.867 | 1.423 | 28.042 | 27.313 | B |
| (55638) 2002 VE95 | 3283.82231 | 20.443 | 0.107 | 3283.82743 | 6.022 | 1.423 | 28.042 | 27.313 | V |
| (55638) 2002 VE95 | 3283.82441 | 19.028 | 0.059 | 3283.82953 | 4.607 | 1.423 | 28.042 | 27.313 | I |
| (55638) 2002 VE95 | 3291.66977 | 21.571 | 0.119 | 3291.67403 | 7.156 | 1.213 | 28.042 | 27.232 | B |
| (55638) 2002 VE95 | 3291.67392 | 20.311 | 0.153 | 3291.67951 | 5.896 | 1.213 | 28.042 | 27.232 | V |
| (55638) 2002 VE95 | 3291.67602 | 18.933 | 0.113 | 3291.68161 | 4.518 | 1.213 | 28.042 | 27.232 | I |
| (55638) 2002 VE95 | 3293.71545 | 21.332 | 0.085 | 3293.71982 | 6.919 | 1.155 | 28.043 | 27.214 | B |
| (55638) 2002 VE95 | 3293.71960 | 20.213 | 0.113 | 3293.72530 | 5.800 | 1.155 | 28.043 | 27.214 | V |
| (55638) 2002 VE95 | 3293.72171 | 18.923 | 0.070 | 3293.72741 | 4.510 | 1.155 | 28.043 | 27.214 | I |
| (55638) 2002 VE95 | 3297.73231 | 21.325 | 0.045 | 3297.73687 | 6.915 | 1.037 | 28.043 | 27.180 | B |
| (55638) 2002 VE95 | 3297.73646 | 20.268 | 0.068 | 3297.74235 | 5.858 | 1.037 | 28.043 | 27.180 | V |
| (55638) 2002 VE95 | 3297.73856 | 18.913 | 0.053 | 3297.74445 | 4.503 | 1.037 | 28.043 | 27.180 | I |
| (55638) 2002 VE95 | 3299.68416 | 21.333 | 0.085 | 3299.68881 | 6.924 | 0.979 | 28.043 | 27.165 | B |
| (55638) 2002 VE95 | 3299.68832 | 20.292 | 0.089 | 3299.69429 | 5.883 | 0.979 | 28.043 | 27.165 | V |
| (55638) 2002 VE95 | 3299.69043 | 18.943 | 0.059 | 3299.69640 | 4.534 | 0.979 | 28.043 | 27.165 | I |
| (55638) 2002 VE95 | 3302.64297 | 21.283 | 0.266 | 3302.64774 | 6.875 | 0.890 | 28.044 | 27.145 | B |
| (55638) 2002 VE95 | 3302.64922 | 18.811 | 0.089 | 3302.65531 | 4.403 | 0.890 | 28.044 | 27.145 | I |
| (55638) 2002 VE95 | 3309.68153 | 21.311 | 0.213 | 3309.68652 | 6.906 | 0.677 | 28.045 | 27.106 | B |
| (55638) 2002 VE95 | 3309.68569 | 20.129 | 0.197 | 3309.69201 | 5.724 | 0.677 | 28.045 | 27.106 | V |
| (55638) 2002 VE95 | 3309.68779 | 18.728 | 0.117 | 3309.69411 | 4.323 | 0.677 | 28.045 | 27.106 | I |
| (55638) 2002 VE95 | 3311.66395 | 21.299 | 0.110 | 3311.66899 | 6.895 | 0.619 | 28.045 | 27.098 | B |
| (55638) 2002 VE95 | 3311.66811 | 20.271 | 0.106 | 3311.67448 | 5.867 | 0.619 | 28.045 | 27.098 | V |
| (55638) 2002 VE95 | 3311.67022 | 18.740 | 0.072 | 3311.67659 | 4.336 | 0.619 | 28.045 | 27.098 | I |
| (55638) 2002 VE95 | 3338.67801 | 21.352 | 0.165 | 3338.68303 | 6.948 | 0.564 | 28.048 | 27.101 | B |
| (55638) 2002 VE95 | 3338.68217 | 20.189 | 0.161 | 3338.68852 | 5.785 | 0.565 | 28.048 | 27.101 | V |
| (55638) 2002 VE95 | 3338.68428 | 18.859 | 0.132 | 3338.69063 | 4.455 | 0.565 | 28.048 | 27.101 | I |
| (55638) 2002 VE95 | 3340.61291 | 21.457 | 0.057 | 3340.61788 | 7.052 | 0.619 | 28.048 | 27.109 | B |
| (55638) 2002 VE95 | 3340.61706 | 20.204 | 0.085 | 3340.62336 | 5.799 | 0.620 | 28.048 | 27.109 | V |
| (55638) 2002 VE95 | 3340.61916 | 18.890 | 0.056 | 3340.62546 | 4.485 | 0.620 | 28.048 | 27.109 | I |
| (55638) 2002 VE95 | 3352.57893 | 21.327 | 0.060 | 3352.58345 | 6.916 | 0.978 | 28.050 | 27.187 | B |
| (55638) 2002 VE95 | 3352.58309 | 20.277 | 0.075 | 3352.58894 | 5.866 | 0.978 | 28.050 | 27.187 | V |
| (55638) 2002 VE95 | 3352.58520 | 18.899 | 0.057 | 3352.59105 | 4.488 | 0.978 | 28.050 | 27.187 | I |
| (55638) 2002 VE95 | 3354.58713 | 21.425 | 0.051 | 3354.59156 | 7.012 | 1.037 | 28.050 | 27.204 | B |
| (55638) 2002 VE95 | 3354.59128 | 20.287 | 0.066 | 3354.59703 | 5.874 | 1.038 | 28.050 | 27.204 | V |
| (55638) 2002 VE95 | 3354.59338 | 18.933 | 0.058 | 3354.59913 | 4.520 | 1.038 | 28.050 | 27.204 | I |
| (55638) 2002 VE95 | 3356.61728 | 21.366 | 0.058 | 3356.62160 | 6.952 | 1.097 | 28.050 | 27.222 | B |
| (55638) 2002 VE95 | 3356.62143 | 20.415 | 0.082 | 3356.62708 | 6.001 | 1.097 | 28.050 | 27.222 | V |
| (55638) 2002 VE95 | 3358.60921 | 20.576 | 0.113 | 3358.61475 | 6.160 | 1.154 | 28.050 | 27.241 | V |
| (55638) 2002 VE95 | 3360.55974 | 21.397 | 0.168 | 3360.56384 | 6.980 | 1.209 | 28.051 | 27.261 | B |
| (55638) 2002 VE95 | 3360.56388 | 20.183 | 0.133 | 3360.56930 | 5.766 | 1.209 | 28.051 | 27.261 | V |
| (55638) 2002 VE95 | 3360.56599 | 19.072 | 0.083 | 3360.57141 | 4.655 | 1.209 | 28.051 | 27.261 | I |
| (55638) 2002 VE95 | 3363.60903 | 19.195 | 0.208 | 3363.61427 | 4.775 | 1.292 | 28.051 | 27.293 | I |
| (55638) 2002 VE95 | 3366.56129 | 18.907 | 0.280 | 3366.56633 | 4.484 | 1.370 | 28.051 | 27.326 | I |
| (55638) 2002 VE95 | 3370.61213 | 21.457 | 0.055 | 3370.61556 | 7.030 | 1.471 | 28.052 | 27.375 | B |
| (55638) 2002 VE95 | 3370.61629 | 20.365 | 0.070 | 3370.62105 | 5.938 | 1.471 | 28.052 | 27.375 | V |
| (55638) 2002 VE95 | 3370.61840 | 19.059 | 0.060 | 3370.62316 | 4.632 | 1.471 | 28.052 | 27.376 | I |
| (55638) 2002 VE95 | 3372.59308 | 21.397 | 0.057 | 3372.59637 | 6.968 | 1.517 | 28.052 | 27.401 | B |
| (55638) 2002 VE95 | 3372.59723 | 20.533 | 0.092 | 3372.60185 | 6.104 | 1.517 | 28.052 | 27.401 | V |
| (55638) 2002 VE95 | 3372.59933 | 18.978 | 0.058 | 3372.60394 | 4.549 | 1.517 | 28.052 | 27.401 | I |
| (55638) 2002 VE95 | 3374.57558 | 21.400 | 0.070 | 3374.57872 | 6.969 | 1.562 | 28.052 | 27.427 | B |
| (55638) 2002 VE95 | 3374.57974 | 20.319 | 0.092 | 3374.58420 | 5.888 | 1.562 | 28.052 | 27.427 | V |
| (55638) 2002 VE95 | 3374.58184 | 18.937 | 0.072 | 3374.58630 | 4.506 | 1.563 | 28.052 | 27.427 | I |
| (55638) 2002 VE95 | 3376.58029 | 21.507 | 0.070 | 3376.58327 | 7.074 | 1.606 | 28.053 | 27.454 | B |
| (55638) 2002 VE95 | 3376.58445 | 20.551 | 0.092 | 3376.58876 | 6.118 | 1.606 | 28.053 | 27.454 | V |
| (55638) 2002 VE95 | 3376.58656 | 18.937 | 0.070 | 3376.59087 | 4.504 | 1.606 | 28.053 | 27.454 | I |
| (55638) 2002 VE95 | 3382.54885 | 21.342 | 0.094 | 3382.55134 | 6.902 | 1.724 | 28.053 | 27.539 | B |
| (55638) 2002 VE95 | 3382.55300 | 20.490 | 0.237 | 3382.55682 | 6.050 | 1.724 | 28.053 | 27.539 | V |
| (55638) 2002 VE95 | 3382.55511 | 19.242 | 0.186 | 3382.55893 | 4.802 | 1.724 | 28.053 | 27.539 | I |
| (55638) 2002 VE95 | 3394.58604 | 19.047 | 0.106 | 3394.58877 | 4.592 | 1.903 | 28.055 | 27.726 | I |
| (55638) 2002 VE95 | 3395.57787 | 20.105 | 0.154 | 3395.58051 | 5.649 | 1.915 | 28.055 | 27.743 | V |
| (55638) 2002 VE95 | 3395.57997 | 19.141 | 0.109 | 3395.58261 | 4.685 | 1.915 | 28.055 | 27.743 | I |
| (55638) 2002 VE95 | 3399.56274 | 21.504 | 0.066 | 3399.56367 | 7.043 | 1.953 | 28.055 | 27.809 | B |
| (55638) 2002 VE95 | 3399.56689 | 20.410 | 0.087 | 3399.56915 | 5.949 | 1.953 | 28.055 | 27.809 | V |
| (55638) 2002 VE95 | 3399.56900 | 19.098 | 0.067 | 3399.57126 | 4.637 | 1.953 | 28.055 | 27.809 | I |
| | | | | # | | | | | |
| (47932) 2000 GN171 | 3068.72357 | 21.276 | 0.207 | 3068.72357 | 6.789 | 1.464 | 28.526 | 27.681 | V |
| (47932) 2000 GN171 | 3068.72638 | 20.224 | 0.139 | 3068.72638 | 5.737 | 1.464 | 28.526 | 27.681 | I |
| (47932) 2000 GN171 | 3068.76636 | 21.001 | 0.149 | 3068.76636 | 6.514 | 1.463 | 28.526 | 27.681 | V |
| (47932) 2000 GN171 | 3068.76918 | 20.033 | 0.110 | 3068.76918 | 5.546 | 1.463 | 28.526 | 27.681 | I |
| (47932) 2000 GN171 | 3079.61034 | 20.992 | 0.095 | 3079.61063 | 6.509 | 1.091 | 28.524 | 27.630 | V |
| (47932) 2000 GN171 | 3079.61316 | 20.104 | 0.105 | 3079.61345 | 5.621 | 1.091 | 28.524 | 27.630 | I |
| (47932) 2000 GN171 | 3079.69659 | 21.067 | 0.085 | 3079.69688 | 6.584 | 1.088 | 28.524 | 27.630 | V |
| (47932) 2000 GN171 | 3079.69940 | 19.776 | 0.066 | 3079.69969 | 5.293 | 1.088 | 28.524 | 27.630 | I |
| (47932) 2000 GN171 | 3087.67983 | 20.927 | 0.072 | 3087.68034 | 6.447 | 0.814 | 28.522 | 27.593 | V |
| (47932) 2000 GN171 | 3087.68266 | 19.773 | 0.069 | 3087.68317 | 5.293 | 0.814 | 28.522 | 27.593 | I |
| (47932) 2000 GN171 | 3087.72192 | 20.788 | 0.064 | 3087.72243 | 6.308 | 0.812 | 28.522 | 27.593 | V |
| (47932) 2000 GN171 | 3087.72474 | 19.689 | 0.060 | 3087.72525 | 5.209 | 0.812 | 28.522 | 27.593 | I |
| (47932) 2000 GN171 | 3091.62193 | 21.336 | 0.143 | 3091.62254 | 6.858 | 0.679 | 28.522 | 27.575 | V |
| (47932) 2000 GN171 | 3091.62474 | 20.237 | 0.122 | 3091.62535 | 5.759 | 0.679 | 28.522 | 27.575 | I |
| (47932) 2000 GN171 | 3091.66477 | 20.955 | 0.076 | 3091.66538 | 6.477 | 0.677 | 28.522 | 27.575 | V |
| (47932) 2000 GN171 | 3091.66758 | 19.921 | 0.080 | 3091.66819 | 5.443 | 0.677 | 28.522 | 27.575 | I |
| (47932) 2000 GN171 | 3098.59334 | 19.819 | 0.134 | 3098.59414 | 5.343 | 0.439 | 28.520 | 27.543 | I |
| (47932) 2000 GN171 | 3098.64931 | 21.307 | 0.260 | 3098.65011 | 6.831 | 0.438 | 28.520 | 27.543 | V |
| (47932) 2000 GN171 | 3098.65212 | 19.764 | 0.105 | 3098.65292 | 5.288 | 0.437 | 28.520 | 27.543 | I |
| (47932) 2000 GN171 | 3102.67113 | 19.672 | 0.236 | 3102.67201 | 5.197 | 0.298 | 28.519 | 27.528 | I |
| (47932) 2000 GN171 | 3106.60128 | 20.832 | 0.083 | 3106.60222 | 6.358 | 0.159 | 28.518 | 27.519 | V |
| (47932) 2000 GN171 | 3106.60410 | 19.624 | 0.075 | 3106.60504 | 5.150 | 0.159 | 28.518 | 27.519 | I |
| (47932) 2000 GN171 | 3106.64479 | 20.744 | 0.080 | 3106.64573 | 6.270 | 0.158 | 28.518 | 27.519 | V |
| (47932) 2000 GN171 | 3106.64760 | 19.603 | 0.067 | 3106.64854 | 5.129 | 0.158 | 28.518 | 27.519 | I |
| (47932) 2000 GN171 | 3110.65355 | 20.601 | 0.072 | 3110.65451 | 6.128 | 0.020 | 28.518 | 27.514 | V |
| (47932) 2000 GN171 | 3110.65631 | 19.474 | 0.090 | 3110.65727 | 5.001 | 0.020 | 28.518 | 27.514 | I |
| (47932) 2000 GN171 | 3110.80916 | 20.785 | 0.098 | 3110.81012 | 6.312 | 0.019 | 28.518 | 27.514 | V |
| (47932) 2000 GN171 | 3110.81192 | 19.543 | 0.101 | 3110.81288 | 5.070 | 0.019 | 28.518 | 27.514 | I |
| (47932) 2000 GN171 | 3115.68178 | 21.037 | 0.097 | 3115.68274 | 6.564 | 0.162 | 28.517 | 27.515 | V |
| (47932) 2000 GN171 | 3115.68454 | 19.806 | 0.117 | 3115.68550 | 5.333 | 0.162 | 28.517 | 27.515 | I |
| (47932) 2000 GN171 | 3115.72778 | 21.304 | 0.142 | 3115.72874 | 6.831 | 0.164 | 28.517 | 27.515 | V |
| (47932) 2000 GN171 | 3115.73054 | 19.606 | 0.125 | 3115.73150 | 5.133 | 0.164 | 28.517 | 27.515 | I |
| (47932) 2000 GN171 | 3134.48533 | 21.385 | 0.148 | 3134.48589 | 6.906 | 0.805 | 28.513 | 27.584 | V |
| (47932) 2000 GN171 | 3134.48809 | 19.852 | 0.132 | 3134.48865 | 5.373 | 0.805 | 28.513 | 27.584 | I |
| (47932) 2000 GN171 | 3135.52116 | 21.231 | 0.202 | 3135.52168 | 6.752 | 0.838 | 28.512 | 27.591 | V |
| (47932) 2000 GN171 | 3139.52449 | 21.037 | 0.098 | 3139.52484 | 6.556 | 0.965 | 28.512 | 27.620 | V |
| (47932) 2000 GN171 | 3139.52725 | 19.807 | 0.110 | 3139.52760 | 5.326 | 0.965 | 28.512 | 27.620 | I |
| (47932) 2000 GN171 | 3139.56625 | 20.909 | 0.079 | 3139.56660 | 6.428 | 0.966 | 28.512 | 27.620 | V |
| (47932) 2000 GN171 | 3139.56902 | 19.607 | 0.075 | 3139.56937 | 5.126 | 0.966 | 28.512 | 27.620 | I |
| (47932) 2000 GN171 | 3144.48600 | 21.050 | 0.102 | 3144.48612 | 6.566 | 1.115 | 28.511 | 27.661 | V |
| (47932) 2000 GN171 | 3144.48876 | 20.069 | 0.111 | 3144.48888 | 5.585 | 1.115 | 28.511 | 27.661 | I |
| (47932) 2000 GN171 | 3144.59137 | 21.046 | 0.098 | 3144.59148 | 6.562 | 1.118 | 28.511 | 27.662 | V |
| (47932) 2000 GN171 | 3144.59413 | 19.862 | 0.081 | 3144.59424 | 5.378 | 1.118 | 28.511 | 27.662 | I |
| (47932) 2000 GN171 | 3149.60843 | 21.414 | 0.195 | 3149.60826 | 6.926 | 1.261 | 28.509 | 27.710 | V |
| (47932) 2000 GN171 | 3153.50751 | 21.248 | 0.205 | 3153.50711 | 6.757 | 1.366 | 28.509 | 27.751 | V |
| (47932) 2000 GN171 | 3153.51027 | 20.049 | 0.169 | 3153.50986 | 5.558 | 1.366 | 28.509 | 27.751 | I |
| (47932) 2000 GN171 | 3153.55268 | 20.289 | 0.245 | 3153.55227 | 5.798 | 1.367 | 28.509 | 27.752 | I |
| (47932) 2000 GN171 | 3162.58216 | 21.459 | 0.236 | 3162.58113 | 6.959 | 1.584 | 28.507 | 27.859 | V |
| (47932) 2000 GN171 | 3162.58492 | 20.271 | 0.147 | 3162.58389 | 5.771 | 1.584 | 28.507 | 27.859 | I |
| (47932) 2000 GN171 | 3168.65388 | 21.584 | 0.163 | 3168.65238 | 7.078 | 1.709 | 28.506 | 27.940 | V |
| (47932) 2000 GN171 | 3168.65664 | 20.516 | 0.153 | 3168.65514 | 6.010 | 1.709 | 28.506 | 27.940 | I |
| (47932) 2000 GN171 | 3168.69096 | 20.932 | 0.114 | 3168.68946 | 6.426 | 1.709 | 28.506 | 27.941 | V |
| (47932) 2000 GN171 | 3168.69372 | 20.184 | 0.166 | 3168.69222 | 5.678 | 1.709 | 28.506 | 27.941 | I |
| (47932) 2000 GN171 | 3174.48740 | 21.165 | 0.089 | 3174.48542 | 6.653 | 1.811 | 28.504 | 28.024 | V |
| (47932) 2000 GN171 | 3174.49017 | 19.962 | 0.114 | 3174.48819 | 5.450 | 1.811 | 28.504 | 28.024 | I |
| (47932) 2000 GN171 | 3174.54972 | 21.449 | 0.106 | 3174.54774 | 6.937 | 1.812 | 28.504 | 28.025 | V |
| (47932) 2000 GN171 | 3174.55249 | 20.549 | 0.197 | 3174.55051 | 6.037 | 1.812 | 28.504 | 28.025 | I |
| (47932) 2000 GN171 | 3178.50823 | 21.391 | 0.269 | 3178.50590 | 6.874 | 1.871 | 28.504 | 28.084 | V |
| (47932) 2000 GN171 | 3178.56058 | 20.241 | 0.203 | 3178.55825 | 5.724 | 1.872 | 28.504 | 28.085 | I |
| (47932) 2000 GN171 | 3185.61778 | 20.887 | 0.282 | 3185.61482 | 6.362 | 1.956 | 28.502 | 28.194 | V |
| (47932) 2000 GN171 | 3185.62054 | 19.967 | 0.233 | 3185.61757 | 5.442 | 1.956 | 28.502 | 28.194 | I |
| (47932) 2000 GN171 | 3194.50982 | 21.130 | 0.102 | 3194.50602 | 6.594 | 2.022 | 28.500 | 28.338 | V |
| (47932) 2000 GN171 | 3194.51257 | 20.726 | 0.282 | 3194.50877 | 6.190 | 2.022 | 28.500 | 28.338 | I |
| (47932) 2000 GN171 | 3203.46850 | 21.263 | 0.098 | 3203.46385 | 6.716 | 2.043 | 28.498 | 28.486 | V |
| (47932) 2000 GN171 | 3203.47125 | 20.083 | 0.094 | 3203.46660 | 5.536 | 2.043 | 28.498 | 28.486 | I |
| (47932) 2000 GN171 | 3207.47557 | 20.890 | 0.205 | 3207.47054 | 6.338 | 2.037 | 28.498 | 28.553 | V |
| (47932) 2000 GN171 | 3207.47833 | 20.534 | 0.270 | 3207.47329 | 5.982 | 2.037 | 28.498 | 28.553 | I |
| | | | | # | | | | | |
| (95626) 2002 GZ32 | 2696.77700 | 20.466 | 0.137 | 2696.77700 | 7.351 | 1.477 | 20.911 | 20.072 | V |
| (95626) 2002 GZ32 | 2696.77917 | 20.511 | 0.141 | 2696.77917 | 7.396 | 1.477 | 20.911 | 20.072 | V |
| (95626) 2002 GZ32 | 2698.80101 | 20.421 | 0.235 | 2698.80111 | 7.308 | 1.405 | 20.909 | 20.054 | V |
| (95626) 2002 GZ32 | 2698.80318 | 20.519 | 0.234 | 2698.80328 | 7.406 | 1.405 | 20.909 | 20.054 | V |
| (95626) 2002 GZ32 | 2698.80845 | 21.163 | 0.100 | 2698.80855 | 8.050 | 1.405 | 20.909 | 20.054 | B |
| (95626) 2002 GZ32 | 2698.81359 | 19.673 | 0.227 | 2698.81369 | 6.560 | 1.405 | 20.909 | 20.054 | I |
| (95626) 2002 GZ32 | 2700.80325 | 20.126 | 0.281 | 2700.80345 | 7.015 | 1.334 | 20.908 | 20.038 | V |
| (95626) 2002 GZ32 | 2700.80541 | 20.497 | 0.298 | 2700.80561 | 7.386 | 1.334 | 20.908 | 20.038 | V |
| (95626) 2002 GZ32 | 2700.81035 | 21.181 | 0.125 | 2700.81055 | 8.070 | 1.334 | 20.908 | 20.038 | B |
| (95626) 2002 GZ32 | 2702.75257 | 21.056 | 0.235 | 2702.75286 | 7.947 | 1.265 | 20.906 | 20.023 | B |
| (95626) 2002 GZ32 | 2704.79848 | 21.291 | 0.127 | 2704.79885 | 8.184 | 1.193 | 20.905 | 20.008 | B |
| (95626) 2002 GZ32 | 2704.80321 | 19.500 | 0.281 | 2704.80358 | 6.393 | 1.193 | 20.905 | 20.008 | I |
| (95626) 2002 GZ32 | 2706.77498 | 21.123 | 0.166 | 2706.77543 | 8.017 | 1.125 | 20.903 | 19.995 | B |
| (95626) 2002 GZ32 | 2708.72295 | 20.570 | 0.173 | 2708.72347 | 7.466 | 1.060 | 20.902 | 19.983 | V |
| (95626) 2002 GZ32 | 2708.72512 | 20.431 | 0.147 | 2708.72564 | 7.327 | 1.060 | 20.902 | 19.983 | V |
| (95626) 2002 GZ32 | 2708.73479 | 19.348 | 0.174 | 2708.73531 | 6.244 | 1.060 | 20.902 | 19.983 | I |
| (95626) 2002 GZ32 | 2710.76047 | 20.695 | 0.138 | 2710.76105 | 7.592 | 0.995 | 20.900 | 19.971 | V |
| (95626) 2002 GZ32 | 2710.76263 | 20.582 | 0.129 | 2710.76321 | 7.479 | 0.995 | 20.900 | 19.971 | V |
| (95626) 2002 GZ32 | 2710.76757 | 21.468 | 0.083 | 2710.76815 | 8.365 | 0.995 | 20.900 | 19.971 | B |
| (95626) 2002 GZ32 | 2710.77230 | 19.810 | 0.187 | 2710.77288 | 6.707 | 0.995 | 20.900 | 19.971 | I |
| (95626) 2002 GZ32 | 2712.79079 | 20.590 | 0.175 | 2712.79143 | 7.488 | 0.934 | 20.898 | 19.961 | V |
| (95626) 2002 GZ32 | 2712.79295 | 20.486 | 0.146 | 2712.79359 | 7.384 | 0.934 | 20.898 | 19.961 | V |
| (95626) 2002 GZ32 | 2712.79789 | 21.458 | 0.114 | 2712.79853 | 8.356 | 0.934 | 20.898 | 19.961 | B |
| (95626) 2002 GZ32 | 2712.80262 | 19.196 | 0.158 | 2712.80326 | 6.094 | 0.934 | 20.898 | 19.961 | I |
| (95626) 2002 GZ32 | 2714.79527 | 19.913 | 0.189 | 2714.79596 | 6.813 | 0.879 | 20.897 | 19.952 | V |
| (95626) 2002 GZ32 | 2714.80237 | 21.086 | 0.251 | 2714.80306 | 7.986 | 0.879 | 20.897 | 19.952 | B |
| (95626) 2002 GZ32 | 2714.80709 | 19.470 | 0.234 | 2714.80778 | 6.370 | 0.879 | 20.897 | 19.952 | I |
| (95626) 2002 GZ32 | 2724.73947 | 20.487 | 0.175 | 2724.74032 | 7.390 | 0.730 | 20.889 | 19.926 | V |
| (95626) 2002 GZ32 | 2724.74164 | 20.296 | 0.181 | 2724.74249 | 7.199 | 0.730 | 20.889 | 19.926 | V |
| (95626) 2002 GZ32 | 2724.74658 | 21.018 | 0.105 | 2724.74742 | 7.921 | 0.730 | 20.889 | 19.926 | B |
| (95626) 2002 GZ32 | 2724.75132 | 19.279 | 0.227 | 2724.75217 | 6.182 | 0.730 | 20.889 | 19.926 | I |
| (95626) 2002 GZ32 | 2728.79490 | 20.545 | 0.146 | 2728.79576 | 7.449 | 0.749 | 20.886 | 19.924 | V |
| (95626) 2002 GZ32 | 2728.79707 | 20.549 | 0.149 | 2728.79793 | 7.453 | 0.749 | 20.886 | 19.924 | V |
| (95626) 2002 GZ32 | 2728.80200 | 21.295 | 0.094 | 2728.80286 | 8.199 | 0.749 | 20.886 | 19.924 | B |
| (95626) 2002 GZ32 | 2728.80674 | 19.786 | 0.236 | 2728.80760 | 6.690 | 0.749 | 20.886 | 19.924 | I |
| (95626) 2002 GZ32 | 2730.71854 | 20.408 | 0.109 | 2730.71940 | 7.312 | 0.774 | 20.885 | 19.924 | V |
| (95626) 2002 GZ32 | 2730.72071 | 20.478 | 0.120 | 2730.72157 | 7.382 | 0.774 | 20.885 | 19.924 | V |
| (95626) 2002 GZ32 | 2730.72564 | 21.200 | 0.067 | 2730.72650 | 8.104 | 0.774 | 20.885 | 19.924 | B |
| (95626) 2002 GZ32 | 2730.73037 | 19.591 | 0.165 | 2730.73123 | 6.495 | 0.774 | 20.885 | 19.924 | I |
| (95626) 2002 GZ32 | 2732.70360 | 20.713 | 0.243 | 2732.70445 | 7.617 | 0.810 | 20.883 | 19.926 | V |
| (95626) 2002 GZ32 | 2732.70577 | 20.627 | 0.214 | 2732.70662 | 7.531 | 0.810 | 20.883 | 19.926 | V |
| (95626) 2002 GZ32 | 2732.71071 | 21.334 | 0.208 | 2732.71156 | 8.238 | 0.810 | 20.883 | 19.926 | B |
| (95626) 2002 GZ32 | 2732.71545 | 19.466 | 0.188 | 2732.71630 | 6.370 | 0.810 | 20.883 | 19.926 | I |
| (95626) 2002 GZ32 | 2736.75075 | 20.551 | 0.140 | 2736.75155 | 7.454 | 0.908 | 20.880 | 19.933 | V |
| (95626) 2002 GZ32 | 2736.75292 | 20.644 | 0.138 | 2736.75372 | 7.547 | 0.908 | 20.880 | 19.933 | V |
| (95626) 2002 GZ32 | 2736.75786 | 21.345 | 0.075 | 2736.75866 | 8.248 | 0.909 | 20.880 | 19.933 | B |
| (95626) 2002 GZ32 | 2736.76258 | 19.513 | 0.179 | 2736.76338 | 6.416 | 0.909 | 20.880 | 19.933 | I |
| (95626) 2002 GZ32 | 2738.67993 | 20.304 | 0.117 | 2738.68071 | 7.207 | 0.964 | 20.878 | 19.938 | V |
| (95626) 2002 GZ32 | 2738.68210 | 20.367 | 0.112 | 2738.68288 | 7.270 | 0.964 | 20.878 | 19.938 | V |
| (95626) 2002 GZ32 | 2738.68703 | 21.190 | 0.066 | 2738.68781 | 8.093 | 0.964 | 20.878 | 19.938 | B |
| (95626) 2002 GZ32 | 2738.69176 | 19.626 | 0.191 | 2738.69254 | 6.529 | 0.964 | 20.878 | 19.938 | I |
| (95626) 2002 GZ32 | 2740.69956 | 20.344 | 0.153 | 2740.70030 | 7.247 | 1.027 | 20.877 | 19.945 | V |
| (95626) 2002 GZ32 | 2740.70173 | 20.878 | 0.252 | 2740.70247 | 7.781 | 1.027 | 20.877 | 19.945 | V |
| (95626) 2002 GZ32 | 2740.70666 | 21.284 | 0.112 | 2740.70740 | 8.187 | 1.027 | 20.877 | 19.945 | B |
| (95626) 2002 GZ32 | 2742.68422 | 19.111 | 0.156 | 2742.68492 | 6.013 | 1.092 | 20.875 | 19.952 | I |
| (95626) 2002 GZ32 | 2750.73193 | 20.330 | 0.207 | 2750.73239 | 7.228 | 1.375 | 20.869 | 19.993 | V |
| (95626) 2002 GZ32 | 2750.73904 | 21.161 | 0.179 | 2750.73950 | 8.059 | 1.375 | 20.869 | 19.993 | B |
| (95626) 2002 GZ32 | 2750.74377 | 19.490 | 0.236 | 2750.74423 | 6.388 | 1.375 | 20.869 | 19.993 | I |
| (95626) 2002 GZ32 | 2752.69364 | 20.358 | 0.245 | 2752.69403 | 7.255 | 1.446 | 20.867 | 20.005 | V |
| (95626) 2002 GZ32 | 2752.69581 | 20.383 | 0.268 | 2752.69620 | 7.280 | 1.446 | 20.867 | 20.005 | V |
| (95626) 2002 GZ32 | 2752.70075 | 21.337 | 0.224 | 2752.70114 | 8.234 | 1.446 | 20.867 | 20.006 | B |
| (95626) 2002 GZ32 | 2756.70047 | 21.052 | 0.293 | 2756.70069 | 7.946 | 1.589 | 20.864 | 20.034 | V |
| (95626) 2002 GZ32 | 2756.70263 | 20.610 | 0.188 | 2756.70285 | 7.504 | 1.589 | 20.864 | 20.034 | V |
| (95626) 2002 GZ32 | 2760.69639 | 20.387 | 0.150 | 2760.69642 | 7.278 | 1.730 | 20.861 | 20.067 | V |
| (95626) 2002 GZ32 | 2760.69856 | 20.524 | 0.149 | 2760.69859 | 7.415 | 1.730 | 20.861 | 20.067 | V |
| (95626) 2002 GZ32 | 2760.70350 | 21.243 | 0.081 | 2760.70353 | 8.134 | 1.730 | 20.861 | 20.067 | B |
| (95626) 2002 GZ32 | 2760.70824 | 19.397 | 0.191 | 2760.70827 | 6.288 | 1.730 | 20.861 | 20.067 | I |
| (95626) 2002 GZ32 | 2762.57349 | 20.786 | 0.280 | 2762.57343 | 7.675 | 1.794 | 20.860 | 20.083 | V |
| (95626) 2002 GZ32 | 2762.57842 | 21.464 | 0.144 | 2762.57836 | 8.353 | 1.794 | 20.860 | 20.083 | B |
| (95626) 2002 GZ32 | 2764.56554 | 20.160 | 0.118 | 2764.56537 | 7.047 | 1.861 | 20.858 | 20.102 | V |
| (95626) 2002 GZ32 | 2764.56761 | 20.466 | 0.133 | 2764.56744 | 7.353 | 1.861 | 20.858 | 20.102 | V |
| (95626) 2002 GZ32 | 2764.57264 | 21.285 | 0.085 | 2764.57247 | 8.172 | 1.862 | 20.858 | 20.102 | B |
| (95626) 2002 GZ32 | 2764.57737 | 19.541 | 0.191 | 2764.57720 | 6.428 | 1.862 | 20.858 | 20.102 | I |
| (95626) 2002 GZ32 | 2777.55524 | 21.221 | 0.250 | 2777.55426 | 8.094 | 2.254 | 20.848 | 20.242 | B |
| (95626) 2002 GZ32 | 2777.55996 | 19.581 | 0.247 | 2777.55898 | 6.454 | 2.255 | 20.848 | 20.242 | I |
| (95626) 2002 GZ32 | 2782.55506 | 20.308 | 0.210 | 2782.55372 | 7.175 | 2.381 | 20.844 | 20.305 | V |
| (95626) 2002 GZ32 | 2782.56216 | 21.255 | 0.182 | 2782.56082 | 8.122 | 2.382 | 20.844 | 20.305 | B |
| (95626) 2002 GZ32 | 2793.58659 | 20.606 | 0.204 | 2793.58438 | 7.458 | 2.605 | 20.836 | 20.454 | V |
| (95626) 2002 GZ32 | 2793.59370 | 21.731 | 0.167 | 2793.59149 | 8.583 | 2.606 | 20.836 | 20.454 | B |
| (95626) 2002 GZ32 | 2794.52967 | 20.668 | 0.185 | 2794.52739 | 7.519 | 2.621 | 20.835 | 20.468 | V |
| (95626) 2002 GZ32 | 2794.53183 | 20.519 | 0.194 | 2794.52955 | 7.370 | 2.621 | 20.835 | 20.468 | V |
| (95626) 2002 GZ32 | 2794.53675 | 21.480 | 0.113 | 2794.53447 | 8.331 | 2.621 | 20.835 | 20.468 | B |
| (95626) 2002 GZ32 | 2818.45073 | 19.990 | 0.219 | 2818.44636 | 6.805 | 2.796 | 20.817 | 20.828 | V |
| (95626) 2002 GZ32 | 2818.45289 | 20.371 | 0.274 | 2818.44852 | 7.186 | 2.796 | 20.817 | 20.828 | V |
| (95626) 2002 GZ32 | 2818.45783 | 21.317 | 0.160 | 2818.45346 | 8.132 | 2.796 | 20.817 | 20.828 | B |
| (95626) 2002 GZ32 | 2820.48895 | 20.635 | 0.156 | 2820.48440 | 7.447 | 2.792 | 20.815 | 20.860 | V |
| (95626) 2002 GZ32 | 2820.49112 | 20.926 | 0.226 | 2820.48657 | 7.738 | 2.792 | 20.815 | 20.860 | V |
| (95626) 2002 GZ32 | 2820.49605 | 21.333 | 0.100 | 2820.49150 # | 8.145 | 2.792 | 20.815 | 20.860 | B |
| (42355 )2002 CR46 | 3402.69755 | 20.935 | 0.083 | 3402.69755 | 8.607 | 0.650 | 17.582 | 16.616 | B |
| (42355 )2002 CR46 | 3402.70066 | 20.129 | 0.076 | 3402.70066 | 7.801 | 0.650 | 17.582 | 16.616 | V |
| (42355 )2002 CR46 | 3402.70276 | 19.190 | 0.068 | 3402.70276 | 6.862 | 0.650 | 17.582 | 16.616 | I |
| (42355 )2002 CR46 | 3404.77591 | 20.567 | 0.080 | 3404.77595 | 8.240 | 0.530 | 17.582 | 16.609 | B |
| (42355 )2002 CR46 | 3404.78112 | 19.177 | 0.073 | 3404.78116 | 6.850 | 0.530 | 17.582 | 16.609 | I |
| (42355 )2002 CR46 | 3406.78207 | 20.834 | 0.052 | 3406.78215 | 8.508 | 0.414 | 17.581 | 16.603 | B |
| (42355 )2002 CR46 | 3406.78519 | 20.084 | 0.065 | 3406.78527 | 7.758 | 0.414 | 17.581 | 16.603 | V |
| (42355 )2002 CR46 | 3406.78729 | 19.097 | 0.069 | 3406.78737 | 6.771 | 0.414 | 17.581 | 16.603 | I |
| (42355 )2002 CR46 | 3408.79012 | 20.787 | 0.048 | 3408.79022 | 8.461 | 0.298 | 17.581 | 16.598 | B |
| (42355 )2002 CR46 | 3408.79324 | 20.076 | 0.064 | 3408.79334 | 7.750 | 0.298 | 17.581 | 16.598 | V |
| (42355 )2002 CR46 | 3408.79535 | 19.151 | 0.073 | 3408.79545 | 6.825 | 0.298 | 17.581 | 16.598 | I |
| (42355 )2002 CR46 | 3410.75299 | 20.757 | 0.046 | 3410.75311 | 8.432 | 0.188 | 17.581 | 16.595 | B |
| (42355 )2002 CR46 | 3410.75610 | 20.021 | 0.060 | 3410.75622 | 7.696 | 0.188 | 17.581 | 16.595 | V |
| (42355 )2002 CR46 | 3410.75819 | 19.082 | 0.064 | 3410.75831 | 6.757 | 0.188 | 17.581 | 16.595 | I |
| (42355 )2002 CR46 | 3432.69417 | 20.914 | 0.078 | 3432.69403 | 8.584 | 1.103 | 17.575 | 16.641 | B |
| (42355 )2002 CR46 | 3432.69728 | 20.021 | 0.072 | 3432.69714 | 7.691 | 1.103 | 17.575 | 16.641 | V |
| (42355 )2002 CR46 | 3432.69939 | 19.040 | 0.070 | 3432.69924 | 6.710 | 1.103 | 17.575 | 16.641 | I |
| (42355 )2002 CR46 | 3434.71475 | 21.080 | 0.064 | 3434.71454 | 8.748 | 1.214 | 17.575 | 16.653 | B |
| (42355 )2002 CR46 | 3434.71786 | 20.159 | 0.069 | 3434.71765 | 7.827 | 1.214 | 17.575 | 16.653 | V |
| (42355 )2002 CR46 | 3434.71995 | 19.188 | 0.074 | 3434.71974 | 6.856 | 1.214 | 17.575 | 16.653 | I |
| (42355 )2002 CR46 | 3445.69028 | 21.089 | 0.056 | 3445.68959 | 8.747 | 1.785 | 17.573 | 16.735 | B |
| (42355 )2002 CR46 | 3445.69339 | 20.320 | 0.074 | 3445.69270 | 7.978 | 1.785 | 17.573 | 16.735 | V |
| (42355 )2002 CR46 | 3445.69550 | 19.378 | 0.076 | 3445.69481 | 7.036 | 1.785 | 17.573 | 16.735 | I |
| (42355 )2002 CR46 | 3447.59513 | 21.104 | 0.111 | 3447.59434 | 8.760 | 1.877 | 17.572 | 16.752 | B |
| (42355 )2002 CR46 | 3447.59824 | 20.128 | 0.106 | 3447.59745 | 7.784 | 1.878 | 17.572 | 16.752 | V |
| (42355 )2002 CR46 | 3447.60035 | 19.289 | 0.106 | 3447.59956 | 6.945 | 1.878 | 17.572 | 16.752 | I |
| (42355 )2002 CR46 | 3449.66699 | 21.472 | 0.232 | 3449.66609 | 9.125 | 1.975 | 17.572 | 16.772 | B |
| (42355 )2002 CR46 | 3449.67009 | 20.643 | 0.229 | 3449.66919 | 8.296 | 1.975 | 17.572 | 16.772 | V |
| (42355 )2002 CR46 | 3453.62678 | 20.728 | 0.230 | 3453.62564 | 8.376 | 2.155 | 17.571 | 16.813 | B |
| (42355 )2002 CR46 | 3453.62984 | 20.226 | 0.294 | 3453.62870 | 7.874 | 2.155 | 17.571 | 16.813 | V |
| (42355 )2002 CR46 | 3453.63187 | 19.276 | 0.240 | 3453.63073 | 6.924 | 2.155 | 17.571 | 16.813 | I |
| (42355 )2002 CR46 | 3458.61858 | 21.032 | 0.186 | 3458.61711 | 8.673 | 2.365 | 17.570 | 16.870 | B |
| (42355 )2002 CR46 | 3458.62159 | 20.244 | 0.162 | 3458.62012 | 7.885 | 2.365 | 17.570 | 16.870 | V |
| (42355 )2002 CR46 | 3458.62359 | 19.472 | 0.136 | 3458.62212 | 7.113 | 2.365 | 17.570 | 16.870 | I |
| (42355 )2002 CR46 | 3460.55956 | 21.065 | 0.069 | 3460.55796 | 8.703 | 2.441 | 17.569 | 16.894 | B |
| (42355 )2002 CR46 | 3460.56259 | 20.328 | 0.090 | 3460.56099 | 7.966 | 2.441 | 17.569 | 16.894 | V |
| (42355 )2002 CR46 | 3460.56462 | 19.423 | 0.107 | 3460.56302 | 7.061 | 2.441 | 17.569 | 16.894 | I |
| (42355 )2002 CR46 | 3464.55503 | 21.142 | 0.087 | 3464.55313 | 8.773 | 2.590 | 17.569 | 16.944 | B |
| (42355 )2002 CR46 | 3464.55807 | 20.348 | 0.116 | 3464.55617 | 7.979 | 2.590 | 17.569 | 16.944 | V |
| (42355 )2002 CR46 | 3464.56010 | 19.438 | 0.114 | 3464.55820 | 7.069 | 2.590 | 17.569 | 16.944 | I |
| (42355 )2002 CR46 | 3466.52269 | 21.081 | 0.074 | 3466.52064 | 8.709 | 2.658 | 17.568 | 16.970 | B |
| (42355 )2002 CR46 | 3466.52573 | 20.359 | 0.101 | 3466.52368 | 7.987 | 2.658 | 17.568 | 16.970 | V |
| (42355 )2002 CR46 | 3466.52776 | 19.519 | 0.123 | 3466.52571 | 7.147 | 2.658 | 17.568 | 16.970 | I |
| (42355 )2002 CR46 | 3468.56894 | 21.099 | 0.064 | 3468.56673 | 8.723 | 2.726 | 17.568 | 16.998 | B |
| (42355 )2002 CR46 | 3468.57197 | 20.419 | 0.084 | 3468.56976 | 8.043 | 2.726 | 17.568 | 16.998 | V |
| (42355 )2002 CR46 | 3468.57399 | 19.290 | 0.085 | 3468.57178 | 6.914 | 2.726 | 17.568 | 16.998 | I |
| (42355 )2002 CR46 | 3470.56762 | 21.245 | 0.063 | 3470.56525 | 8.866 | 2.788 | 17.567 | 17.026 | B |
| (42355 )2002 CR46 | 3470.57066 | 20.439 | 0.083 | 3470.56829 | 8.060 | 2.788 | 17.567 | 17.026 | V |
| (42355 )2002 CR46 | 3470.57269 | 19.615 | 0.108 | 3470.57032 | 7.236 | 2.788 | 17.567 | 17.026 | I |
| (42355 )2002 CR46 | 3474.57005 | 21.344 | 0.121 | 3474.56735 | 8.958 | 2.903 | 17.566 | 17.083 | B |
| (42355 )2002 CR46 | 3474.57310 | 20.521 | 0.118 | 3474.57040 | 8.135 | 2.903 | 17.566 | 17.083 | V |
| (42355 )2002 CR46 | 3474.57513 | 19.666 | 0.159 | 3474.57243 | 7.280 | 2.903 | 17.566 | 17.083 | I |
| | | | | # | | | | | |
| (54598) Bienor | 2845.85807 | 21.736 | 0.115 | 2845.85807 | 9.036 | 2.145 | 18.996 | 18.256 | B |
| (54598) Bienor | 2845.86349 | 19.761 | 0.164 | 2845.86349 | 7.061 | 2.145 | 18.996 | 18.256 | I |
| (54598) Bienor | 2845.86649 | 20.623 | 0.137 | 2845.86649 | 7.923 | 2.145 | 18.996 | 18.256 | V |
| (54598) Bienor | 2846.87812 | 21.139 | 0.165 | 2846.87819 | 8.440 | 2.106 | 18.996 | 18.244 | B |
| (54598) Bienor | 2847.87541 | 21.184 | 0.138 | 2847.87555 | 8.487 | 2.068 | 18.995 | 18.232 | B |
| (54598) Bienor | 2847.88084 | 19.348 | 0.265 | 2847.88098 | 6.651 | 2.067 | 18.995 | 18.232 | I |
| (54598) Bienor | 2856.79604 | 21.581 | 0.082 | 2856.79673 | 8.895 | 1.696 | 18.991 | 18.137 | B |
| (54598) Bienor | 2856.80147 | 19.918 | 0.145 | 2856.80216 | 7.232 | 1.696 | 18.991 | 18.137 | I |
| (54598) Bienor | 2856.80448 | 20.772 | 0.118 | 2856.80517 | 8.086 | 1.696 | 18.991 | 18.137 | V |
| (54598) Bienor | 2857.83305 | 21.195 | 0.036 | 2857.83379 | 8.511 | 1.650 | 18.990 | 18.127 | B |
| (54598) Bienor | 2857.83848 | 19.578 | 0.075 | 2857.83922 | 6.894 | 1.650 | 18.990 | 18.127 | I |
| (54598) Bienor | 2857.84150 | 20.216 | 0.201 | 2857.84224 | 7.532 | 1.650 | 18.990 | 18.127 | V |
| (54598) Bienor | 2859.83702 | 21.577 | 0.085 | 2859.83787 | 8.895 | 1.559 | 18.989 | 18.109 | B |
| (54598) Bienor | 2859.84245 | 20.122 | 0.110 | 2859.84330 | 7.440 | 1.559 | 18.989 | 18.109 | I |
| (54598) Bienor | 2859.84546 | 20.576 | 0.229 | 2859.84631 | 7.894 | 1.559 | 18.989 | 18.109 | V |
| (54598) Bienor | 2860.86391 | 21.080 | 0.080 | 2860.86481 | 8.399 | 1.512 | 18.989 | 18.100 | B |
| (54598) Bienor | 2860.86933 | 19.402 | 0.085 | 2860.87023 | 6.721 | 1.512 | 18.989 | 18.100 | I |
| (54598) Bienor | 2860.87235 | 20.456 | 0.106 | 2860.87325 | 7.775 | 1.512 | 18.989 | 18.100 | V |
| (54598) Bienor | 2862.82118 | 19.571 | 0.092 | 2862.82217 | 6.892 | 1.421 | 18.988 | 18.084 | I |
| (54598) Bienor | 2862.82420 | 20.774 | 0.198 | 2862.82519 | 8.095 | 1.421 | 18.988 | 18.083 | V |
| (54598) Bienor | 2863.79662 | 19.793 | 0.156 | 2863.79766 | 7.115 | 1.375 | 18.987 | 18.076 | I |
| (54598) Bienor | 2863.79968 | 20.337 | 0.185 | 2863.80072 | 7.659 | 1.375 | 18.987 | 18.076 | V |
| (54598) Bienor | 2867.77607 | 21.103 | 0.089 | 2867.77728 | 8.429 | 1.183 | 18.985 | 18.047 | B |
| (54598) Bienor | 2867.78153 | 19.589 | 0.086 | 2867.78274 | 6.915 | 1.182 | 18.985 | 18.047 | I |
| (54598) Bienor | 2867.78459 | 20.329 | 0.111 | 2867.78580 | 7.655 | 1.182 | 18.985 | 18.047 | V |
| (54598) Bienor | 2868.74981 | 21.108 | 0.083 | 2868.75105 | 8.435 | 1.135 | 18.985 | 18.040 | B |
| (54598) Bienor | 2868.76172 | 19.610 | 0.113 | 2868.76296 | 6.937 | 1.134 | 18.985 | 18.040 | I |
| (54598) Bienor | 2868.76477 | 20.552 | 0.185 | 2868.76601 | 7.879 | 1.134 | 18.985 | 18.040 | V |
| (54598) Bienor | 2869.74702 | 21.402 | 0.101 | 2869.74830 | 8.730 | 1.085 | 18.984 | 18.034 | B |
| (54598) Bienor | 2869.75248 | 19.635 | 0.077 | 2869.75376 | 6.963 | 1.085 | 18.984 | 18.034 | I |
| (54598) Bienor | 2869.75554 | 20.667 | 0.101 | 2869.75682 | 7.995 | 1.085 | 18.984 | 18.034 | V |
| (54598) Bienor | 2870.76395 | 19.442 | 0.213 | 2870.76527 | 6.770 | 1.035 | 18.984 | 18.028 | I |
| (54598) Bienor | 2874.78265 | 21.190 | 0.049 | 2874.78409 | 8.521 | 0.832 | 18.982 | 18.006 | B |
| (54598) Bienor | 2874.78812 | 19.627 | 0.081 | 2874.78956 | 6.958 | 0.831 | 18.982 | 18.006 | I |
| (54598) Bienor | 2874.79117 | 20.489 | 0.073 | 2874.79261 | 7.820 | 0.831 | 18.982 | 18.006 | V |
| (54598) Bienor | 2878.74655 | 21.151 | 0.111 | 2878.74809 | 8.484 | 0.629 | 18.980 | 17.990 | B |
| (54598) Bienor | 2878.75197 | 19.461 | 0.156 | 2878.75351 | 6.794 | 0.629 | 18.980 | 17.990 | I |
| (54598) Bienor | 2878.75502 | 20.558 | 0.115 | 2878.75656 | 7.891 | 0.628 | 18.980 | 17.990 | V |
| (54598) Bienor | 2895.69872 | 20.839 | 0.085 | 2895.70036 | 8.176 | 0.327 | 18.972 | 17.971 | B |
| (54598) Bienor | 2895.70418 | 19.561 | 0.082 | 2895.70582 | 6.898 | 0.327 | 18.972 | 17.971 | I |
| (54598) Bienor | 2895.70724 | 20.332 | 0.106 | 2895.70888 | 7.669 | 0.327 | 18.972 | 17.971 | V |
| (54598) Bienor | 2897.70524 | 21.676 | 0.159 | 2897.70686 | 9.012 | 0.426 | 18.971 | 17.975 | B |
| (54598) Bienor | 2897.71071 | 19.431 | 0.080 | 2897.71233 | 6.767 | 0.426 | 18.971 | 17.975 | I |
| (54598) Bienor | 2897.71377 | 20.363 | 0.109 | 2897.71539 | 7.699 | 0.426 | 18.971 | 17.975 | V |
| (54598) Bienor | 2901.71046 | 21.061 | 0.099 | 2901.71202 | 8.396 | 0.630 | 18.969 | 17.985 | B |
| (54598) Bienor | 2901.71593 | 19.309 | 0.110 | 2901.71749 | 6.644 | 0.630 | 18.969 | 17.985 | I |
| (54598) Bienor | 2901.71899 | 20.186 | 0.101 | 2901.72055 | 7.521 | 0.630 | 18.969 | 17.985 | V |
| (54598) Bienor | 2903.72281 | 21.021 | 0.077 | 2903.72433 | 8.355 | 0.733 | 18.968 | 17.992 | B |
| (54598) Bienor | 2907.67444 | 21.128 | 0.028 | 2907.67586 | 8.461 | 0.934 | 18.966 | 18.010 | B |
| (54598) Bienor | 2907.67991 | 19.467 | 0.050 | 2907.68133 | 6.800 | 0.934 | 18.966 | 18.010 | I |
| (54598) Bienor | 2907.68297 | 20.310 | 0.048 | 2907.68439 | 7.643 | 0.935 | 18.966 | 18.010 | V |
| (54598) Bienor | 2911.68028 | 20.862 | 0.039 | 2911.68157 | 8.192 | 1.134 | 18.964 | 18.032 | B |
| (54598) Bienor | 2911.68624 | 19.385 | 0.060 | 2911.68753 | 6.715 | 1.135 | 18.964 | 18.032 | I |
| (54598) Bienor | 2911.68950 | 20.299 | 0.066 | 2911.69079 | 7.629 | 1.135 | 18.964 | 18.032 | V |
| (54598) Bienor | 2912.69604 | 20.343 | 0.281 | 2912.69730 | 7.672 | 1.184 | 18.963 | 18.038 | V |
| (54598) Bienor | 2915.66389 | 21.187 | 0.077 | 2915.66503 | 8.514 | 1.328 | 18.962 | 18.059 | B |
| (54598) Bienor | 2915.66935 | 19.546 | 0.080 | 2915.67049 | 6.873 | 1.328 | 18.962 | 18.059 | I |
| (54598) Bienor | 2915.67241 | 20.420 | 0.077 | 2915.67355 | 7.747 | 1.329 | 18.962 | 18.059 | V |
| (54598) Bienor | 2916.67266 | 21.027 | 0.098 | 2916.67376 | 8.353 | 1.376 | 18.961 | 18.066 | B |
| (54598) Bienor | 2916.67812 | 19.037 | 0.064 | 2916.67922 | 6.363 | 1.376 | 18.961 | 18.066 | I |
| (54598) Bienor | 2918.64149 | 19.664 | 0.141 | 2918.64250 | 6.989 | 1.469 | 18.960 | 18.081 | I |
| (54598) Bienor | 2918.64454 | 20.360 | 0.205 | 2918.64555 | 7.685 | 1.469 | 18.960 | 18.081 | V |
| (54598) Bienor | 2924.64943 | 21.027 | 0.133 | 2924.65013 | 8.346 | 1.739 | 18.957 | 18.134 | B |
| (54598) Bienor | 2924.65794 | 20.540 | 0.156 | 2924.65864 | 7.859 | 1.739 | 18.957 | 18.134 | V |
| (54598) Bienor | 2928.60990 | 19.329 | 0.124 | 2928.61037 | 6.643 | 1.906 | 18.955 | 18.174 | I |
| (54598) Bienor | 2928.61295 | 20.173 | 0.166 | 2928.61342 | 7.487 | 1.906 | 18.955 | 18.174 | V |
| (54598) Bienor | 2940.59591 | 20.738 | 0.101 | 2940.59556 | 8.036 | 2.350 | 18.949 | 18.316 | B |
| (54598) Bienor | 2940.60138 | 19.563 | 0.073 | 2940.60103 | 6.861 | 2.350 | 18.949 | 18.316 | I |
| (54598) Bienor | 2940.60443 | 20.353 | 0.163 | 2940.60408 | 7.651 | 2.350 | 18.949 | 18.316 | V |
| (54598) Bienor | 2944.60539 | 21.413 | 0.082 | 2944.60473 | 8.705 | 2.475 | 18.947 | 18.370 | B |
| (54598) Bienor | 2944.61085 | 19.737 | 0.101 | 2944.61019 | 7.029 | 2.475 | 18.947 | 18.370 | I |
| (54598) Bienor | 2944.61391 | 20.479 | 0.232 | 2944.61325 | 7.771 | 2.475 | 18.947 | 18.370 | V |
| (54598) Bienor | 2945.55460 | 21.382 | 0.132 | 2945.55387 | 8.672 | 2.503 | 18.947 | 18.383 | B |
| (54598) Bienor | 2945.56007 | 19.713 | 0.133 | 2945.55933 | 7.003 | 2.503 | 18.947 | 18.383 | I |
| (54598) Bienor | 2945.56312 | 20.402 | 0.135 | 2945.56238 | 7.692 | 2.503 | 18.947 | 18.383 | V |
| (54598) Bienor | 2946.59093 | 21.520 | 0.207 | 2946.59011 | 8.809 | 2.532 | 18.946 | 18.398 | B |
| (54598) Bienor | 2946.59640 | 19.797 | 0.210 | 2946.59558 | 7.086 | 2.532 | 18.946 | 18.398 | I |
| (54598) Bienor | 2946.59946 | 20.717 | 0.254 | 2946.59864 | 8.006 | 2.532 | 18.946 | 18.398 | V |
| (54598) Bienor | 2949.57597 | 21.231 | 0.160 | 2949.57490 | 8.515 | 2.612 | 18.945 | 18.440 | B |
| (54598) Bienor | 2949.58143 | 19.914 | 0.160 | 2949.58036 | 7.198 | 2.612 | 18.945 | 18.440 | I |
| (54598) Bienor | 2949.58449 | 20.397 | 0.163 | 2949.58342 | 7.681 | 2.612 | 18.945 | 18.441 | V |
| (54598) Bienor | 2950.58527 | 21.462 | 0.174 | 2950.58412 | 8.744 | 2.637 | 18.944 | 18.455 | B |
| (54598) Bienor | 2950.59072 | 19.856 | 0.140 | 2950.58957 | 7.138 | 2.638 | 18.944 | 18.455 | I |
| (54598) Bienor | 2950.59378 | 20.555 | 0.162 | 2950.59263 | 7.837 | 2.638 | 18.944 | 18.455 | V |
| (54598) Bienor | 2951.59745 | 21.548 | 0.246 | 2951.59621 | 8.828 | 2.662 | 18.944 | 18.470 | B |
| (54598) Bienor | 2951.60291 | 20.249 | 0.211 | 2951.60167 | 7.529 | 2.662 | 18.944 | 18.470 | I |
| (54598) Bienor | 2951.60597 | 20.585 | 0.215 | 2951.60473 | 7.865 | 2.662 | 18.944 | 18.470 | V |
| (54598) Bienor | 2952.56317 | 21.456 | 0.044 | 2952.56185 | 8.735 | 2.685 | 18.944 | 18.485 | B |
| (54598) Bienor | 2952.56864 | 19.918 | 0.092 | 2952.56732 | 7.197 | 2.685 | 18.944 | 18.485 | I |
| (54598) Bienor | 2952.57169 | 20.705 | 0.080 | 2952.57037 | 7.984 | 2.685 | 18.944 | 18.485 | V |
| (54598) Bienor | 2953.56360 | 21.156 | 0.101 | 2953.56219 | 8.433 | 2.707 | 18.943 | 18.500 | B |
| (54598) Bienor | 2953.56906 | 19.783 | 0.140 | 2953.56765 | 7.060 | 2.707 | 18.943 | 18.500 | I |
| (54598) Bienor | 2953.57211 | 20.557 | 0.158 | 2953.57070 | 7.834 | 2.707 | 18.943 | 18.500 | V |
| (54598) Bienor | 2953.85712 | 19.898 | 0.145 | 2953.85569 | 7.174 | 2.714 | 18.943 | 18.504 | I |
| (54598) Bienor | 2953.85724 | 20.627 | 0.158 | 2953.85581 | 7.903 | 2.714 | 18.943 | 18.504 | V |
| (54598) Bienor | 2953.85990 | 21.666 | 0.149 | 2953.85847 | 8.942 | 2.714 | 18.943 | 18.504 | B |
| (54598) Bienor | 2957.56864 | 21.353 | 0.042 | 2957.56688 | 8.623 | 2.789 | 18.941 | 18.561 | B |
| (54598) Bienor | 2957.57410 | 19.785 | 0.085 | 2957.57234 | 7.055 | 2.789 | 18.941 | 18.561 | I |
| (54598) Bienor | 2957.57716 | 20.540 | 0.074 | 2957.57540 | 7.810 | 2.789 | 18.941 | 18.561 | V |
| (54598) Bienor | 2963.54620 | 21.274 | 0.089 | 2963.54389 | 8.533 | 2.885 | 18.938 | 18.656 | B |
| (54598) Bienor | 2963.55165 | 19.753 | 0.155 | 2963.54934 | 7.012 | 2.885 | 18.938 | 18.656 | I |
| (54598) Bienor | 2963.55470 | 20.696 | 0.071 | 2963.55239 | 7.955 | 2.885 | 18.938 | 18.656 | V |
| (54598) Bienor | 3634.70550 | 20.513 | 0.103 | 3634.70940 | 7.943 | 0.453 | 18.573 | 17.580 | B |
| (54598) Bienor | 3634.70957 | 20.922 | 0.139 | 3634.71347 | 8.352 | 0.453 | 18.573 | 17.580 | B |
| (54598) Bienor | 3634.71298 | 20.182 | 0.108 | 3634.71688 | 7.612 | 0.453 | 18.573 | 17.580 | V |
| (54598) Bienor | 3634.71570 | 19.141 | 0.065 | 3634.71960 | 6.571 | 0.453 | 18.573 | 17.580 | I |
| (54598) Bienor | 3651.62553 | 21.124 | 0.069 | 3651.62904 | 8.547 | 1.262 | 18.563 | 17.648 | B |
| (54598) Bienor | 3651.62959 | 21.244 | 0.071 | 3651.63310 | 8.667 | 1.263 | 18.563 | 17.648 | B |
| (54598) Bienor | 3651.63300 | 20.320 | 0.063 | 3651.63651 | 7.743 | 1.263 | 18.563 | 17.648 | V |
| (54598) Bienor | 3651.63572 | 19.587 | 0.072 | 3651.63923 | 7.010 | 1.263 | 18.563 | 17.648 | I |
| (54598) Bienor | 3657.71184 | 20.651 | 0.231 | 3657.71509 | 8.069 | 1.552 | 18.560 | 17.693 | B |
| (54598) Bienor | 3657.71588 | 20.802 | 0.256 | 3657.71913 | 8.220 | 1.552 | 18.560 | 17.693 | B |
| (54598) Bienor | 3657.72201 | 19.675 | 0.223 | 3657.72526 | 7.093 | 1.552 | 18.560 | 17.693 | I |
| (54598) Bienor | 3668.56620 | 21.147 | 0.061 | 3668.56886 | 8.553 | 2.025 | 18.553 | 17.796 | B |
| (54598) Bienor | 3668.57025 | 21.195 | 0.065 | 3668.57291 | 8.601 | 2.025 | 18.553 | 17.796 | B |
| (54598) Bienor | 3668.57366 | 20.319 | 0.059 | 3668.57632 | 7.725 | 2.025 | 18.553 | 17.796 | V |
| (54598) Bienor | 3668.57638 | 19.601 | 0.077 | 3668.57904 | 7.007 | 2.025 | 18.553 | 17.796 | I |
| (54598) Bienor | 3670.54999 | 21.363 | 0.092 | 3670.55252 | 8.767 | 2.104 | 18.552 | 17.817 | B |
| (54598) Bienor | 3670.55404 | 21.311 | 0.088 | 3670.55657 | 8.715 | 2.104 | 18.552 | 17.818 | B |
| (54598) Bienor | 3670.55745 | 20.575 | 0.095 | 3670.55998 | 7.979 | 2.104 | 18.552 | 17.818 | V |
| (54598) Bienor | 3670.56018 | 19.518 | 0.098 | 3670.56271 | 6.922 | 2.104 | 18.552 | 17.818 | I |
| (54598) Bienor | 3673.54866 | 21.198 | 0.065 | 3673.55099 | 8.598 | 2.218 | 18.551 | 17.852 | B |
| (54598) Bienor | 3673.55271 | 21.228 | 0.063 | 3673.55504 | 8.628 | 2.218 | 18.551 | 17.852 | B |
| (54598) Bienor | 3673.55612 | 20.580 | 0.076 | 3673.55845 | 7.980 | 2.218 | 18.551 | 17.852 | V |
| (54598) Bienor | 3673.55885 | 19.630 | 0.089 | 3673.56118 | 7.030 | 2.218 | 18.551 | 17.852 | I |
| (54598) Bienor | 3675.61801 | 21.079 | 0.050 | 3675.62020 | 8.476 | 2.293 | 18.549 | 17.877 | B |
| (54598) Bienor | 3675.62207 | 21.097 | 0.050 | 3675.62426 | 8.494 | 2.293 | 18.549 | 17.877 | B |
| (54598) Bienor | 3677.58397 | 21.235 | 0.055 | 3677.58601 | 8.629 | 2.362 | 18.548 | 17.902 | B |
| (54598) Bienor | 3677.58802 | 21.287 | 0.057 | 3677.59006 | 8.681 | 2.362 | 18.548 | 17.902 | B |
| (54598) Bienor | 3677.59142 | 20.567 | 0.065 | 3677.59346 | 7.961 | 2.362 | 18.548 | 17.902 | V |
| (54598) Bienor | 3677.59416 | 19.534 | 0.078 | 3677.59620 | 6.928 | 2.362 | 18.548 | 17.902 | I |
| (54598) Bienor | 3680.58768 | 21.265 | 0.086 | 3680.58950 | 8.654 | 2.461 | 18.546 | 17.941 | B |
| (54598) Bienor | 3680.59172 | 21.142 | 0.097 | 3680.59354 | 8.531 | 2.461 | 18.546 | 17.941 | B |
| (54598) Bienor | 3680.59512 | 20.506 | 0.100 | 3680.59694 | 7.895 | 2.461 | 18.546 | 17.941 | V |
| (54598) Bienor | 3682.54968 | 21.030 | 0.095 | 3682.55135 | 8.416 | 2.522 | 18.545 | 17.967 | B |
| (54598) Bienor | 3682.55372 | 21.247 | 0.128 | 3682.55539 | 8.633 | 2.522 | 18.545 | 17.967 | B |
| (54598) Bienor | 3682.55712 | 20.403 | 0.094 | 3682.55879 | 7.789 | 2.522 | 18.545 | 17.967 | V |
| (54598) Bienor | 3682.55985 | 19.646 | 0.106 | 3682.56152 | 7.032 | 2.522 | 18.545 | 17.967 | I |
| (54598) Bienor | 3687.55280 | 21.116 | 0.242 | 3687.55406 | 8.494 | 2.663 | 18.542 | 18.037 | B |
| (54598) Bienor | 3687.55686 | 21.221 | 0.283 | 3687.55812 | 8.599 | 2.663 | 18.542 | 18.037 | B |
| (54598) Bienor | 3687.56298 | 19.740 | 0.256 | 3687.56424 | 7.118 | 2.663 | 18.542 | 18.038 | I |
| (54598) Bienor | 3689.55069 | 20.881 | 0.189 | 3689.55178 | 8.256 | 2.713 | 18.541 | 18.067 | B |
| (54598) Bienor | 3689.55474 | 20.806 | 0.185 | 3689.55583 | 8.181 | 2.713 | 18.541 | 18.067 | B |
| (54598) Bienor | 3689.55814 | 20.330 | 0.159 | 3689.55923 | 7.705 | 2.713 | 18.541 | 18.067 | V |
| (54598) Bienor | 3689.56087 | 19.330 | 0.099 | 3689.56196 | 6.705 | 2.714 | 18.541 | 18.067 | I |
| (54598) Bienor | 3693.52997 | 21.336 | 0.071 | 3693.53072 | 8.704 | 2.804 | 18.539 | 18.126 | B |
| (54598) Bienor | 3693.53402 | 21.230 | 0.064 | 3693.53477 | 8.598 | 2.804 | 18.539 | 18.126 | B |
| (54598) Bienor | 3693.53740 | 20.469 | 0.067 | 3693.53815 | 7.837 | 2.804 | 18.539 | 18.126 | V |
| (54598) Bienor | 3693.54013 | 19.588 | 0.095 | 3693.54088 | 6.956 | 2.804 | 18.539 | 18.126 | I |
| (54598) Bienor | 3700.60237 | 21.204 | 0.081 | 3700.60248 | 8.559 | 2.930 | 18.535 | 18.236 | B |
| (54598) Bienor | 3700.60641 | 21.209 | 0.079 | 3700.60652 | 8.564 | 2.930 | 18.535 | 18.236 | B |
| (54598) Bienor | 3700.60980 | 20.374 | 0.073 | 3700.60991 | 7.729 | 2.930 | 18.535 | 18.236 | V |
| (54598) Bienor | 3700.61252 | 19.564 | 0.107 | 3700.61263 | 6.919 | 2.930 | 18.535 | 18.236 | I |
| (54598) Bienor | 3702.54609 | 21.215 | 0.061 | 3702.54602 | 8.567 | 2.956 | 18.533 | 18.267 | B |
| (54598) Bienor | 3702.55017 | 21.086 | 0.056 | 3702.55010 | 8.438 | 2.956 | 18.533 | 18.268 | B |
| (54598) Bienor | 3702.55358 | 20.468 | 0.062 | 3702.55351 | 7.820 | 2.956 | 18.533 | 18.268 | V |
| (54598) Bienor | 3702.55633 | 19.681 | 0.085 | 3702.55626 | 7.033 | 2.956 | 18.533 | 18.268 | I |
| (54598) Bienor | 3704.56765 | 21.663 | 0.097 | 3704.56739 | 9.011 | 2.980 | 18.532 | 18.300 | B |
| (54598) Bienor | 3704.57172 | 21.530 | 0.084 | 3704.57146 | 8.878 | 2.980 | 18.532 | 18.300 | B |
| (54598) Bienor | 3704.57513 | 20.749 | 0.085 | 3704.57487 | 8.097 | 2.980 | 18.532 | 18.300 | V |
| (54598) Bienor | 3704.57788 | 19.734 | 0.111 | 3704.57762 | 7.082 | 2.980 | 18.532 | 18.300 | I |
| (54598) Bienor | 3707.56102 | 20.380 | 0.100 | 3707.56048 | 7.723 | 3.009 | 18.530 | 18.349 | V |
| (54598) Bienor | 3707.56377 | 19.569 | 0.114 | 3707.56323 | 6.912 | 3.009 | 18.530 | 18.349 | I |
| (54598) Bienor | 3709.54831 | 21.343 | 0.103 | 3709.54758 | 8.682 | 3.023 | 18.529 | 18.382 | B |
| (54598) Bienor | 3709.55241 | 21.309 | 0.130 | 3709.55168 | 8.648 | 3.024 | 18.529 | 18.382 | B |
| (54598) Bienor | 3709.55585 | 20.837 | 0.191 | 3709.55512 | 8.176 | 3.024 | 18.529 | 18.382 | V |
| (54598) Bienor | 3711.54664 | 21.251 | 0.151 | 3711.54572 | 8.586 | 3.034 | 18.528 | 18.415 | B |
| (54598) Bienor | 3711.55072 | 21.064 | 0.128 | 3711.54980 | 8.399 | 3.034 | 18.528 | 18.415 | B |
| (54598) Bienor | 3711.55413 | 20.136 | 0.097 | 3711.55321 | 7.471 | 3.034 | 18.528 | 18.415 | V |
| (54598) Bienor | 3711.55688 | 19.646 | 0.117 | 3711.55596 | 6.981 | 3.034 | 18.528 | 18.415 | I |
| (54598) Bienor | 3713.53174 | 20.419 | 0.259 | 3713.53063 | 7.750 | 3.041 | 18.527 | 18.448 | V |
| (54598) Bienor | 3715.57393 | 20.900 | 0.217 | 3715.57263 | 8.227 | 3.045 | 18.526 | 18.481 | B |
| (54598) Bienor | 3715.57734 | 20.409 | 0.170 | 3715.57604 | 7.736 | 3.045 | 18.526 | 18.481 | V |
| (54598) Bienor | 3715.58010 | 19.890 | 0.188 | 3715.57880 | 7.217 | 3.045 | 18.526 | 18.482 | I |
| (54598) Bienor | 3717.54132 | 20.756 | 0.259 | 3717.53983 | 8.080 | 3.045 | 18.524 | 18.514 | V |
| (54598) Bienor | 3717.54411 | 20.068 | 0.228 | 3717.54262 | 7.392 | 3.045 | 18.524 | 18.514 | I |
| (54598) Bienor | 3721.54515 | 21.546 | 0.158 | 3721.54328 | 8.862 | 3.034 | 18.522 | 18.580 | B |
| (54598) Bienor | 3721.54926 | 21.260 | 0.131 | 3721.54739 | 8.576 | 3.034 | 18.522 | 18.580 | B |
| (54598) Bienor | 3721.55270 | 20.533 | 0.117 | 3721.55082 | 7.849 | 3.034 | 18.522 | 18.581 | V |
| (54598) Bienor | 3721.55548 | 20.075 | 0.199 | 3721.55360 | 7.391 | 3.034 | 18.522 | 18.581 | I |
| | | | | # | | | | | |
| (73480) 2002 PN34 | 2733.90188 | 20.328 | 0.261 | 2733.90188 | 8.967 | 3.715 | 13.451 | 13.912 | V |
| (73480) 2002 PN34 | 2733.90713 | 20.963 | 0.136 | 2733.90713 | 9.602 | 3.716 | 13.451 | 13.912 | B |
| (73480) 2002 PN34 | 2733.91048 | 19.296 | 0.264 | 2733.91048 | 7.935 | 3.716 | 13.451 | 13.912 | I |
| (73480) 2002 PN34 | 2735.89257 | 20.273 | 0.162 | 2735.89273 | 8.917 | 3.783 | 13.452 | 13.884 | V |
| (73480) 2002 PN34 | 2735.89427 | 20.331 | 0.170 | 2735.89443 | 8.975 | 3.783 | 13.452 | 13.884 | V |
| (73480) 2002 PN34 | 2735.89782 | 21.057 | 0.095 | 2735.89798 | 9.701 | 3.783 | 13.452 | 13.884 | B |
| (73480) 2002 PN34 | 2735.90116 | 19.090 | 0.186 | 2735.90132 | 7.734 | 3.783 | 13.452 | 13.884 | I |
| (73480) 2002 PN34 | 2737.89428 | 20.339 | 0.177 | 2737.89461 | 8.987 | 3.847 | 13.453 | 13.855 | V |
| (73480) 2002 PN34 | 2737.89598 | 20.015 | 0.143 | 2737.89631 | 8.663 | 3.847 | 13.453 | 13.855 | V |
| (73480) 2002 PN34 | 2737.89952 | 21.101 | 0.101 | 2737.89985 | 9.749 | 3.847 | 13.453 | 13.855 | B |
| (73480) 2002 PN34 | 2737.90287 | 19.128 | 0.185 | 2737.90320 | 7.776 | 3.847 | 13.453 | 13.855 | I |
| (73480) 2002 PN34 | 2739.87920 | 20.416 | 0.212 | 2739.87970 | 9.068 | 3.906 | 13.453 | 13.826 | V |
| (73480) 2002 PN34 | 2739.88090 | 19.997 | 0.139 | 2739.88140 | 8.649 | 3.906 | 13.453 | 13.826 | V |
| (73480) 2002 PN34 | 2739.88445 | 21.177 | 0.118 | 2739.88495 | 9.829 | 3.906 | 13.453 | 13.826 | B |

| | | | | | | | | | | |
|---|---|---|---|---|---|---|---|---|---|---|
| (73480) 2002 PN34 | 2739.88779 | 19.105 | 0.205 | 2739.88829 | 7.757 | 3.906 | 13.453 | 13.826 | I |
| (73480) 2002 PN34 | 2741.89207 | 20.370 | 0.165 | 2741.89274 | 9.027 | 3.962 | 13.454 | 13.797 | V |
| (73480) 2002 PN34 | 2741.89378 | 20.183 | 0.138 | 2741.89445 | 8.840 | 3.962 | 13.454 | 13.797 | V |
| (73480) 2002 PN34 | 2741.89732 | 21.026 | 0.092 | 2741.89799 | 9.683 | 3.962 | 13.454 | 13.796 | B |
| (73480) 2002 PN34 | 2741.90066 | 19.453 | 0.262 | 2741.90133 | 8.110 | 3.963 | 13.454 | 13.796 | I |
| (73480) 2002 PN34 | 2751.89042 | 20.103 | 0.214 | 2751.89198 | 8.783 | 4.180 | 13.459 | 13.643 | V |
| (73480) 2002 PN34 | 2751.89212 | 20.142 | 0.230 | 2751.89368 | 8.822 | 4.180 | 13.459 | 13.643 | V |
| (73480) 2002 PN34 | 2751.89567 | 20.964 | 0.171 | 2751.89723 | 9.644 | 4.180 | 13.459 | 13.643 | B |
| (73480) 2002 PN34 | 2751.89902 | 19.167 | 0.267 | 2751.90058 | 7.847 | 4.180 | 13.459 | 13.643 | I |
| (73480) 2002 PN34 | 2757.90198 | 20.295 | 0.174 | 2757.90409 | 8.990 | 4.258 | 13.462 | 13.548 | V |
| (73480) 2002 PN34 | 2757.90369 | 20.142 | 0.126 | 2757.90580 | 8.837 | 4.258 | 13.462 | 13.548 | V |
| (73480) 2002 PN34 | 2757.90723 | 21.164 | 0.099 | 2757.90934 | 9.859 | 4.258 | 13.462 | 13.548 | B |
| (73480) 2002 PN34 | 2757.91058 | 19.467 | 0.291 | 2757.91269 | 8.162 | 4.258 | 13.462 | 13.548 | I |
| (73480) 2002 PN34 | 2759.91849 | 20.287 | 0.269 | 2759.92078 | 8.987 | 4.275 | 13.463 | 13.515 | V |
| (73480) 2002 PN34 | 2759.92020 | 20.309 | 0.277 | 2759.92249 | 9.009 | 4.275 | 13.463 | 13.515 | V |
| (73480) 2002 PN34 | 2759.92374 | 21.006 | 0.254 | 2759.92603 | 9.706 | 4.275 | 13.463 | 13.515 | B |
| (73480) 2002 PN34 | 2763.78436 | 19.896 | 0.179 | 2763.78701 | 8.606 | 4.295 | 13.465 | 13.453 | V |
| (73480) 2002 PN34 | 2763.78607 | 20.009 | 0.151 | 2763.78872 | 8.719 | 4.295 | 13.465 | 13.453 | V |
| (73480) 2002 PN34 | 2763.78962 | 20.882 | 0.093 | 2763.79227 | 9.592 | 4.295 | 13.465 | 13.453 | B |
| (73480) 2002 PN34 | 2763.79296 | 18.814 | 0.185 | 2763.79561 | 7.524 | 4.295 | 13.465 | 13.453 | I |
| (73480) 2002 PN34 | 2765.83966 | 20.174 | 0.132 | 2765.84250 | 8.889 | 4.298 | 13.466 | 13.420 | V |
| (73480) 2002 PN34 | 2765.84136 | 20.128 | 0.131 | 2765.84420 | 8.843 | 4.298 | 13.466 | 13.420 | V |
| (73480) 2002 PN34 | 2765.84490 | 20.843 | 0.073 | 2765.84774 | 9.558 | 4.298 | 13.466 | 13.420 | B |
| (73480) 2002 PN34 | 2765.84824 | 18.934 | 0.166 | 2765.85108 | 7.649 | 4.298 | 13.466 | 13.420 | I |
| (73480) 2002 PN34 | 2787.87286 | 19.992 | 0.100 | 2787.87770 | 8.762 | 4.018 | 13.477 | 13.074 | V |
| (73480) 2002 PN34 | 2787.87456 | 20.118 | 0.110 | 2787.87940 | 8.888 | 4.018 | 13.477 | 13.074 | V |
| (73480) 2002 PN34 | 2787.88145 | 19.139 | 0.163 | 2787.88629 | 7.909 | 4.018 | 13.477 | 13.074 | I |
| (73480) 2002 PN34 | 2789.86640 | 20.004 | 0.142 | 2789.87141 | 8.779 | 3.964 | 13.478 | 13.045 | V |
| (73480) 2002 PN34 | 2789.86811 | 20.050 | 0.146 | 2789.87312 | 8.825 | 3.964 | 13.478 | 13.045 | V |
| (73480) 2002 PN34 | 2789.87165 | 20.890 | 0.089 | 2789.87666 | 9.665 | 3.964 | 13.478 | 13.045 | B |
| (73480) 2002 PN34 | 2789.87499 | 19.081 | 0.195 | 2789.88000 | 7.856 | 3.964 | 13.478 | 13.045 | I |
| (73480) 2002 PN34 | 2791.81765 | 19.877 | 0.104 | 2791.82283 | 8.656 | 3.907 | 13.479 | 13.016 | V |
| (73480) 2002 PN34 | 2791.81936 | 19.920 | 0.110 | 2791.82454 | 8.699 | 3.907 | 13.479 | 13.016 | V |
| (73480) 2002 PN34 | 2791.82290 | 20.906 | 0.078 | 2791.82808 | 9.685 | 3.907 | 13.479 | 13.016 | B |
| (73480) 2002 PN34 | 2791.82625 | 19.179 | 0.185 | 2791.83143 | 7.958 | 3.907 | 13.479 | 13.016 | I |
| (73480) 2002 PN34 | 2793.89484 | 20.462 | 0.099 | 2793.90019 | 9.246 | 3.842 | 13.480 | 12.987 | B |
| (73480) 2002 PN34 | 2812.86608 | 20.179 | 0.216 | 2812.87282 | 9.002 | 3.026 | 13.491 | 12.745 | V |
| (73480) 2002 PN34 | 2812.86779 | 19.902 | 0.154 | 2812.87453 | 8.725 | 3.026 | 13.491 | 12.745 | V |
| (73480) 2002 PN34 | 2812.87133 | 20.624 | 0.102 | 2812.87807 | 9.447 | 3.026 | 13.491 | 12.745 | B |
| (73480) 2002 PN34 | 2812.87468 | 18.898 | 0.194 | 2812.88142 | 7.721 | 3.025 | 13.491 | 12.745 | I |
| (73480) 2002 PN34 | 2816.85300 | 19.903 | 0.103 | 2816.85998 | 8.733 | 2.809 | 13.493 | 12.703 | V |
| (73480) 2002 PN34 | 2816.85471 | 19.937 | 0.113 | 2816.86169 | 8.767 | 2.809 | 13.493 | 12.703 | V |
| (73480) 2002 PN34 | 2816.85825 | 20.831 | 0.069 | 2816.86523 | 9.661 | 2.809 | 13.493 | 12.703 | B |
| (73480) 2002 PN34 | 2816.86160 | 18.827 | 0.197 | 2816.86858 | 7.657 | 2.808 | 13.493 | 12.703 | I |
| (73480) 2002 PN34 | 2818.87715 | 19.947 | 0.108 | 2818.88425 | 8.780 | 2.693 | 13.494 | 12.683 | V |
| (73480) 2002 PN34 | 2818.87885 | 19.991 | 0.115 | 2818.88595 | 8.824 | 2.693 | 13.494 | 12.683 | V |
| (73480) 2002 PN34 | 2818.88239 | 20.798 | 0.067 | 2818.88949 | 9.631 | 2.693 | 13.494 | 12.683 | B |
| (73480) 2002 PN34 | 2818.88573 | 18.940 | 0.170 | 2818.89283 | 7.773 | 2.693 | 13.494 | 12.683 | I |
| (73480) 2002 PN34 | 2820.85760 | 19.971 | 0.112 | 2820.86481 | 8.807 | 2.577 | 13.495 | 12.664 | V |
| (73480) 2002 PN34 | 2820.85930 | 19.900 | 0.110 | 2820.86651 | 8.736 | 2.577 | 13.495 | 12.664 | V |

| | | | | | | | | | | |
|---|---|---|---|---|---|---|---|---|---|---|
| (73480) 2002 PN34 | 2820.86285 | 20.799 | 0.074 | 2820.87006 | 9.635 | 2.577 | 13.495 | 12.664 | B |
| (73480) 2002 PN34 | 2820.86619 | 18.871 | 0.172 | 2820.87340 | 7.707 | 2.577 | 13.495 | 12.664 | I |
| (73480) 2002 PN34 | 2831.82655 | 19.687 | 0.154 | 2831.83426 | 8.537 | 1.882 | 13.501 | 12.578 | V |
| (73480) 2002 PN34 | 2831.82825 | 19.956 | 0.194 | 2831.83596 | 8.806 | 1.882 | 13.501 | 12.578 | V |
| (73480) 2002 PN34 | 2831.83179 | 20.513 | 0.141 | 2831.83950 | 9.363 | 1.881 | 13.501 | 12.578 | B |
| (73480) 2002 PN34 | 2831.83513 | 18.612 | 0.131 | 2831.84284 | 7.462 | 1.881 | 13.501 | 12.578 | I |
| (73480) 2002 PN34 | 2832.77945 | 19.786 | 0.187 | 2832.78719 | 8.637 | 1.818 | 13.502 | 12.572 | V |
| (73480) 2002 PN34 | 2832.78116 | 20.212 | 0.271 | 2832.78890 | 9.063 | 1.818 | 13.502 | 12.572 | V |
| (73480) 2002 PN34 | 2832.78470 | 20.736 | 0.197 | 2832.79244 | 9.587 | 1.818 | 13.502 | 12.572 | B |
| (73480) 2002 PN34 | 2832.78805 | 19.017 | 0.205 | 2832.79579 | 7.868 | 1.817 | 13.502 | 12.572 | I |
| (73480) 2002 PN34 | 2840.83688 | 19.629 | 0.156 | 2840.84486 | 8.486 | 1.264 | 13.506 | 12.531 | V |
| (73480) 2002 PN34 | 2840.83858 | 19.783 | 0.155 | 2840.84656 | 8.640 | 1.264 | 13.506 | 12.531 | V |
| (73480) 2002 PN34 | 2840.84213 | 20.551 | 0.124 | 2840.85011 | 9.408 | 1.263 | 13.506 | 12.531 | B |
| (73480) 2002 PN34 | 2840.84547 | 18.584 | 0.145 | 2840.85345 | 7.441 | 1.263 | 13.506 | 12.531 | I |
| (73480) 2002 PN34 | 2843.80951 | 19.633 | 0.070 | 2843.81755 | 8.492 | 1.056 | 13.508 | 12.520 | V |
| (73480) 2002 PN34 | 2843.81237 | 19.781 | 0.090 | 2843.82041 | 8.640 | 1.056 | 13.508 | 12.520 | V |
| (73480) 2002 PN34 | 2843.81939 | 20.702 | 0.069 | 2843.82743 | 9.561 | 1.055 | 13.508 | 12.520 | B |
| (73480) 2002 PN34 | 2843.82644 | 18.873 | 0.100 | 2843.83448 | 7.732 | 1.055 | 13.508 | 12.520 | I |
| (73480) 2002 PN34 | 2844.79614 | 19.815 | 0.062 | 2844.80420 | 8.674 | 0.987 | 13.508 | 12.517 | V |
| (73480) 2002 PN34 | 2844.79900 | 19.530 | 0.049 | 2844.80706 | 8.389 | 0.987 | 13.508 | 12.517 | V |
| (73480) 2002 PN34 | 2844.80602 | 20.751 | 0.046 | 2844.81408 | 9.610 | 0.986 | 13.508 | 12.517 | B |
| (73480) 2002 PN34 | 2844.81306 | 18.749 | 0.084 | 2844.82112 | 7.608 | 0.986 | 13.508 | 12.517 | I |
| (73480) 2002 PN34 | 2846.82155 | 19.878 | 0.177 | 2846.82964 | 8.738 | 0.847 | 13.509 | 12.512 | V |
| (73480) 2002 PN34 | 2846.82442 | 20.012 | 0.215 | 2846.83251 | 8.872 | 0.847 | 13.509 | 12.512 | V |
| (73480) 2002 PN34 | 2847.81246 | 19.804 | 0.104 | 2847.82056 | 8.664 | 0.779 | 13.510 | 12.510 | V |
| (73480) 2002 PN34 | 2847.81532 | 19.891 | 0.130 | 2847.82342 | 8.751 | 0.779 | 13.510 | 12.510 | V |
| (73480) 2002 PN34 | 2847.82235 | 20.691 | 0.085 | 2847.83045 | 9.551 | 0.779 | 13.510 | 12.510 | B |
| (73480) 2002 PN34 | 2847.82941 | 19.134 | 0.188 | 2847.83751 | 7.994 | 0.778 | 13.510 | 12.510 | I |
| (73480) 2002 PN34 | 2852.81870 | 19.828 | 0.036 | 2852.82684 | 8.689 | 0.466 | 13.513 | 12.503 | V |
| (73480) 2002 PN34 | 2852.82156 | 19.872 | 0.044 | 2852.82970 | 8.733 | 0.466 | 13.513 | 12.503 | V |
| (73480) 2002 PN34 | 2852.82859 | 20.761 | 0.026 | 2852.83673 | 9.622 | 0.465 | 13.513 | 12.503 | B |
| (73480) 2002 PN34 | 2852.83563 | 18.841 | 0.043 | 2852.84377 | 7.702 | 0.465 | 13.513 | 12.503 | I |
| (73480) 2002 PN34 | 2853.78078 | 19.572 | 0.025 | 2853.78892 | 8.433 | 0.417 | 13.513 | 12.503 | V |
| (73480) 2002 PN34 | 2853.78365 | 19.696 | 0.041 | 2853.79179 | 8.557 | 0.417 | 13.513 | 12.503 | V |
| (73480) 2002 PN34 | 2853.79774 | 18.645 | 0.036 | 2853.80588 | 7.506 | 0.417 | 13.513 | 12.503 | I |
| (73480) 2002 PN34 | 2857.79803 | 19.820 | 0.035 | 2857.80616 | 8.681 | 0.330 | 13.516 | 12.504 | V |
| (73480) 2002 PN34 | 2857.80089 | 19.865 | 0.038 | 2857.80902 | 8.726 | 0.330 | 13.516 | 12.504 | V |
| (73480) 2002 PN34 | 2857.80792 | 20.795 | 0.028 | 2857.81605 | 9.656 | 0.330 | 13.516 | 12.504 | B |
| (73480) 2002 PN34 | 2857.81496 | 18.876 | 0.044 | 2857.82309 | 7.737 | 0.330 | 13.516 | 12.504 | I |
| (73480) 2002 PN34 | 2861.79506 | 19.795 | 0.089 | 2861.80316 | 8.654 | 0.475 | 13.518 | 12.510 | V |
| (73480) 2002 PN34 | 2861.79793 | 19.894 | 0.098 | 2861.80603 | 8.753 | 0.475 | 13.518 | 12.510 | V |
| (73480) 2002 PN34 | 2861.80496 | 20.692 | 0.100 | 2861.81306 | 9.551 | 0.475 | 13.518 | 12.510 | B |
| (73480) 2002 PN34 | 2861.81201 | 18.822 | 0.055 | 2861.82011 | 7.681 | 0.476 | 13.518 | 12.510 | I |
| (73480) 2002 PN34 | 2866.75419 | 19.879 | 0.153 | 2866.76221 | 8.735 | 0.787 | 13.521 | 12.524 | V |
| (73480) 2002 PN34 | 2866.76125 | 20.706 | 0.131 | 2866.76927 | 9.562 | 0.787 | 13.521 | 12.524 | B |
| (73480) 2002 PN34 | 2866.76834 | 19.047 | 0.088 | 2866.77636 | 7.903 | 0.787 | 13.521 | 12.524 | I |
| (73480) 2002 PN34 | 2867.73724 | 19.800 | 0.049 | 2867.74524 | 8.655 | 0.853 | 13.522 | 12.528 | V |
| (73480) 2002 PN34 | 2867.74011 | 19.933 | 0.061 | 2867.74811 | 8.788 | 0.854 | 13.522 | 12.528 | V |
| (73480) 2002 PN34 | 2867.74716 | 20.783 | 0.061 | 2867.75515 | 9.638 | 0.854 | 13.522 | 12.528 | B |
| (73480) 2002 PN34 | 2867.75425 | 18.944 | 0.042 | 2867.76224 | 7.799 | 0.855 | 13.522 | 12.528 | I |
| (73480) 2002 PN34 | 2875.74405 | 20.936 | 0.024 | 2875.75181 | 9.784 | 1.409 | 13.526 | 12.568 | B |
| (73480) 2002 PN34 | 2875.75114 | 19.036 | 0.043 | 2875.75890 | 7.884 | 1.409 | 13.526 | 12.568 | I |
| (73480) 2002 PN34 | 2876.70981 | 19.951 | 0.037 | 2876.71754 | 8.798 | 1.476 | 13.527 | 12.574 | V |
| (73480) 2002 PN34 | 2876.71268 | 20.028 | 0.038 | 2876.72041 | 8.875 | 1.476 | 13.527 | 12.574 | V |
| (73480) 2002 PN34 | 2876.71970 | 20.906 | 0.025 | 2876.72743 | 9.753 | 1.476 | 13.527 | 12.574 | B |
| (73480) 2002 PN34 | 2884.68911 | 20.111 | 0.035 | 2884.69649 | 8.947 | 2.011 | 13.532 | 12.634 | V |
| (73480) 2002 PN34 | 2884.69200 | 20.091 | 0.042 | 2884.69938 | 8.927 | 2.012 | 13.532 | 12.634 | V |
| (73480) 2002 PN34 | 2896.69101 | 20.025 | 0.076 | 2896.69770 | 8.839 | 2.741 | 13.539 | 12.755 | V |
| (73480) 2002 PN34 | 2896.69390 | 20.031 | 0.078 | 2896.70059 | 8.845 | 2.741 | 13.539 | 12.755 | V |
| (73480) 2002 PN34 | 2896.70097 | 20.829 | 0.067 | 2896.70765 | 9.643 | 2.742 | 13.539 | 12.755 | B |
| (73480) 2002 PN34 | 2896.70805 | 19.043 | 0.069 | 2896.71473 | 7.857 | 2.742 | 13.539 | 12.755 | I |
| (73480) 2002 PN34 | 2898.67926 | 20.376 | 0.079 | 2898.68581 | 9.186 | 2.851 | 13.540 | 12.778 | V |
| (73480) 2002 PN34 | 2898.68922 | 21.136 | 0.047 | 2898.69577 | 9.946 | 2.851 | 13.540 | 12.778 | B |
| (73480) 2002 PN34 | 2903.66058 | 19.917 | 0.145 | 2903.66677 | 8.715 | 3.109 | 13.543 | 12.840 | V |
| (73480) 2002 PN34 | 2903.66347 | 20.310 | 0.154 | 2903.66966 | 9.108 | 3.109 | 13.543 | 12.840 | V |
| (73480) 2002 PN34 | 2903.67053 | 20.773 | 0.181 | 2903.67672 | 9.571 | 3.110 | 13.543 | 12.840 | B |
| (73480) 2002 PN34 | 2905.65714 | 20.039 | 0.041 | 2905.66318 | 8.833 | 3.206 | 13.545 | 12.867 | V |
| (73480) 2002 PN34 | 2905.66003 | 20.078 | 0.045 | 2905.66607 | 8.872 | 3.206 | 13.545 | 12.867 | V |
| (73480) 2002 PN34 | 2905.66710 | 20.844 | 0.045 | 2905.67314 | 9.638 | 3.206 | 13.545 | 12.867 | B |
| (73480) 2002 PN34 | 2905.67419 | 19.095 | 0.046 | 2905.68023 | 7.889 | 3.207 | 13.545 | 12.867 | I |
| (73480) 2002 PN34 | 2907.65759 | 20.795 | 0.028 | 2907.66347 | 9.584 | 3.299 | 13.546 | 12.894 | B |
| (73480) 2002 PN34 | 2907.66467 | 19.077 | 0.054 | 2907.67055 | 7.866 | 3.299 | 13.546 | 12.894 | I |
| (73480) 2002 PN34 | 2915.62772 | 20.007 | 0.063 | 2915.63294 | 8.776 | 3.627 | 13.551 | 13.009 | V |
| (73480) 2002 PN34 | 2915.63062 | 19.988 | 0.067 | 2915.63584 | 8.757 | 3.627 | 13.551 | 13.009 | V |
| (73480) 2002 PN34 | 2915.63768 | 20.855 | 0.055 | 2915.64290 | 9.624 | 3.627 | 13.551 | 13.009 | B |
| (73480) 2002 PN34 | 2915.64477 | 19.072 | 0.069 | 2915.64999 | 7.841 | 3.627 | 13.551 | 13.009 | I |
| (73480) 2002 PN34 | 2921.61410 | 19.988 | 0.101 | 2921.61878 | 8.741 | 3.825 | 13.555 | 13.102 | V |
| (73480) 2002 PN34 | 2921.62117 | 20.944 | 0.105 | 2921.62585 | 9.697 | 3.825 | 13.555 | 13.102 | B |
| (73480) 2002 PN34 | 2921.62826 | 19.062 | 0.065 | 2921.63294 | 7.815 | 3.826 | 13.555 | 13.102 | I |
| (73480) 2002 PN34 | 2925.62447 | 20.085 | 0.088 | 2925.62878 | 8.827 | 3.935 | 13.557 | 13.167 | V |
| (73480) 2002 PN34 | 2925.62737 | 20.153 | 0.096 | 2925.63168 | 8.895 | 3.935 | 13.557 | 13.167 | V |
| (73480) 2002 PN34 | 2925.63443 | 20.880 | 0.082 | 2925.63873 | 9.622 | 3.935 | 13.557 | 13.167 | B |
| (73480) 2002 PN34 | 2925.64151 | 19.074 | 0.066 | 2925.64581 | 7.816 | 3.935 | 13.557 | 13.167 | I |
| (73480) 2002 PN34 | 2928.56754 | 19.866 | 0.265 | 2928.57157 | 8.599 | 4.003 | 13.559 | 13.215 | V |
| (73480) 2002 PN34 | 2928.57460 | 20.861 | 0.156 | 2928.57862 | 9.594 | 4.003 | 13.559 | 13.216 | B |
| (73480) 2002 PN34 | 2928.58168 | 18.909 | 0.190 | 2928.58570 | 7.642 | 4.003 | 13.559 | 13.216 | I |
| (73480) 2002 PN34 | 2929.55769 | 19.997 | 0.059 | 2929.56162 | 8.728 | 4.024 | 13.560 | 13.232 | V |
| (73480) 2002 PN34 | 2929.56058 | 20.093 | 0.063 | 2929.56451 | 8.824 | 4.024 | 13.560 | 13.232 | V |
| (73480) 2002 PN34 | 2929.56764 | 20.827 | 0.045 | 2929.57157 | 9.558 | 4.024 | 13.560 | 13.232 | B |
| (73480) 2002 PN34 | 2929.57473 | 18.972 | 0.061 | 2929.57866 | 7.703 | 4.024 | 13.560 | 13.232 | I |
| (73480) 2002 PN34 | 2932.56362 | 20.288 | 0.045 | 2932.56726 | 9.010 | 4.079 | 13.562 | 13.283 | V |
| (73480) 2002 PN34 | 2932.56651 | 20.238 | 0.045 | 2932.57015 | 8.960 | 4.079 | 13.562 | 13.283 | V |
| (73480) 2002 PN34 | 2932.57358 | 21.016 | 0.027 | 2932.57721 | 9.738 | 4.079 | 13.562 | 13.283 | B |
| (73480) 2002 PN34 | 2932.58067 | 19.382 | 0.054 | 2932.58430 | 8.104 | 4.079 | 13.562 | 13.283 | I |
| (73480) 2002 PN34 | 2936.59394 | 20.141 | 0.197 | 2936.59718 | 8.851 | 4.136 | 13.564 | 13.352 | V |
| (73480) 2002 PN34 | 2936.60101 | 20.901 | 0.083 | 2936.60425 | 9.611 | 4.136 | 13.564 | 13.352 | B |
| (73480) 2002 PN34 | 2936.60809 | 19.228 | 0.098 | 2936.61133 | 7.938 | 4.136 | 13.564 | 13.352 | I |
| (73480) 2002 PN34 | 2939.57858 | 20.099 | 0.044 | 2939.58152 | 8.801 | 4.165 | 13.566 | 13.404 | V |
| (73480) 2002 PN34 | 2939.58147 | 20.116 | 0.047 | 2939.58441 | 8.818 | 4.165 | 13.566 | 13.404 | V |
| (73480) 2002 PN34 | 2939.58854 | 20.984 | 0.032 | 2939.59148 | 9.686 | 4.165 | 13.566 | 13.404 | B |

| | | | | | | | | | | |
|---|---|---|---|---|---|---|---|---|---|---|
| (73480) 2002 PN34 | 2941.55540 | 20.154 | 0.057 | 2941.55814 | 8.850 | 4.178 | 13.568 | 13.438 | V |
| (73480) 2002 PN34 | 2941.55830 | 20.148 | 0.055 | 2941.56104 | 8.844 | 4.178 | 13.568 | 13.438 | V |
| (73480) 2002 PN34 | 2941.56536 | 21.104 | 0.036 | 2941.56810 | 9.800 | 4.178 | 13.568 | 13.438 | B |
| (73480) 2002 PN34 | 2941.57245 | 19.219 | 0.052 | 2941.57519 | 7.915 | 4.178 | 13.568 | 13.438 | I |
| (73480) 2002 PN34 | 2943.55308 | 20.152 | 0.082 | 2943.55562 | 8.842 | 4.187 | 13.569 | 13.473 | V |
| (73480) 2002 PN34 | 2943.55597 | 20.134 | 0.083 | 2943.55851 | 8.824 | 4.187 | 13.569 | 13.473 | V |
| (73480) 2002 PN34 | 2943.56304 | 21.012 | 0.063 | 2943.56558 | 9.702 | 4.187 | 13.569 | 13.473 | B |
| (73480) 2002 PN34 | 2943.57012 | 19.310 | 0.081 | 2943.57266 | 8.000 | 4.187 | 13.569 | 13.473 | I |
| (73480) 2002 PN34 | 2944.55421 | 19.890 | 0.209 | 2944.55665 | 8.577 | 4.189 | 13.570 | 13.490 | V |
| (73480) 2002 PN34 | 2944.55711 | 19.992 | 0.230 | 2944.55955 | 8.679 | 4.189 | 13.570 | 13.491 | V |
| (73480) 2002 PN34 | 2944.56417 | 20.657 | 0.121 | 2944.56661 | 9.344 | 4.189 | 13.570 | 13.491 | B |
| (73480) 2002 PN34 | 2944.57126 | 19.172 | 0.102 | 2944.57369 | 7.859 | 4.189 | 13.570 | 13.491 | I |
| (73480) 2002 PN34 | 2946.55426 | 20.300 | 0.145 | 2946.55649 | 8.981 | 4.190 | 13.571 | 13.526 | V |
| (73480) 2002 PN34 | 2946.55715 | 20.412 | 0.165 | 2946.55938 | 9.093 | 4.190 | 13.571 | 13.526 | V |
| (73480) 2002 PN34 | 2946.56421 | 20.882 | 0.102 | 2946.56644 | 9.563 | 4.190 | 13.571 | 13.526 | B |
| (73480) 2002 PN34 | 2946.57130 | 19.287 | 0.105 | 2946.57353 | 7.968 | 4.190 | 13.571 | 13.526 | I |
| (73480) 2002 PN34 | 2947.54319 | 20.150 | 0.187 | 2947.54532 | 8.828 | 4.189 | 13.572 | 13.543 | V |
| (73480) 2002 PN34 | 2947.54608 | 19.994 | 0.179 | 2947.54821 | 8.672 | 4.189 | 13.572 | 13.543 | V |
| (73480) 2002 PN34 | 2947.56023 | 18.994 | 0.124 | 2947.56236 | 7.672 | 4.189 | 13.572 | 13.543 | I |
| (73480) 2002 PN34 | 2949.54138 | 20.252 | 0.102 | 2949.54331 | 8.924 | 4.183 | 13.573 | 13.578 | V |
| (73480) 2002 PN34 | 2949.54427 | 20.137 | 0.094 | 2949.54620 | 8.809 | 4.183 | 13.573 | 13.578 | V |
| (73480) 2002 PN34 | 2949.55134 | 20.917 | 0.082 | 2949.55327 | 9.589 | 4.183 | 13.573 | 13.578 | B |
| (73480) 2002 PN34 | 2949.55842 | 19.154 | 0.072 | 2949.56035 | 7.826 | 4.183 | 13.573 | 13.578 | I |
| (73480) 2002 PN34 | 2950.53524 | 20.135 | 0.086 | 2950.53707 | 8.805 | 4.179 | 13.574 | 13.595 | V |
| (73480) 2002 PN34 | 2950.53813 | 20.368 | 0.106 | 2950.53996 | 9.038 | 4.179 | 13.574 | 13.595 | V |
| (73480) 2002 PN34 | 2950.54519 | 21.197 | 0.099 | 2950.54702 | 9.867 | 4.179 | 13.574 | 13.595 | B |
| (73480) 2002 PN34 | 2950.55227 | 19.435 | 0.077 | 2950.55410 | 8.105 | 4.179 | 13.574 | 13.596 | I |
| (73480) 2002 PN34 | 2951.53773 | 20.295 | 0.131 | 2951.53946 | 8.962 | 4.173 | 13.574 | 13.613 | V |
| (73480) 2002 PN34 | 2951.54063 | 20.254 | 0.129 | 2951.54236 | 8.921 | 4.173 | 13.574 | 13.613 | V |
| (73480) 2002 PN34 | 2951.54777 | 21.158 | 0.133 | 2951.54950 | 9.825 | 4.173 | 13.574 | 13.613 | B |
| (73480) 2002 PN34 | 2951.55508 | 19.344 | 0.092 | 2951.55681 | 8.011 | 4.173 | 13.574 | 13.613 | I |
| (73480) 2002 PN34 | 2953.53636 | 20.182 | 0.092 | 2953.53789 | 8.843 | 4.158 | 13.576 | 13.648 | V |
| (73480) 2002 PN34 | 2953.53925 | 20.068 | 0.084 | 2953.54078 | 8.729 | 4.158 | 13.576 | 13.648 | V |
| (73480) 2002 PN34 | 2953.55339 | 19.231 | 0.084 | 2953.55492 | 7.892 | 4.157 | 13.576 | 13.648 | I |
| (73480) 2002 PN34 | 2955.53619 | 20.289 | 0.089 | 2955.53752 | 8.944 | 4.138 | 13.577 | 13.683 | V |
| (73480) 2002 PN34 | 2955.53909 | 20.470 | 0.125 | 2955.54042 | 9.125 | 4.138 | 13.577 | 13.683 | V |
| (73480) 2002 PN34 | 2955.54615 | 21.164 | 0.062 | 2955.54748 | 9.819 | 4.137 | 13.577 | 13.683 | B |
| (73480) 2002 PN34 | 2955.55324 | 19.283 | 0.106 | 2955.55457 | 7.938 | 4.137 | 13.577 | 13.683 | I |
| (73480) 2002 PN34 | 2957.52918 | 20.224 | 0.050 | 2957.53031 | 8.873 | 4.113 | 13.578 | 13.717 | V |
| (73480) 2002 PN34 | 2957.53207 | 20.140 | 0.045 | 2957.53320 | 8.789 | 4.113 | 13.578 | 13.717 | V |
| (73480) 2002 PN34 | 2957.53912 | 21.058 | 0.031 | 2957.54025 | 9.707 | 4.113 | 13.578 | 13.717 | B |
| (73480) 2002 PN34 | 2957.54619 | 19.161 | 0.058 | 2957.54732 | 7.810 | 4.113 | 13.578 | 13.717 | I |
| | | | | # | | | | | |
| (29981) 1999TD10 | 2866.83916 | 19.802 | 0.167 | 2866.83916 | 8.438 | 4.139 | 13.801 | 13.578 | I |
| (29981) 1999TD10 | 2874.82041 | 20.863 | 0.115 | 2874.82105 | 9.514 | 3.989 | 13.818 | 13.467 | V |
| (29981) 1999TD10 | 2874.82307 | 20.802 | 0.120 | 2874.82371 | 9.453 | 3.989 | 13.818 | 13.467 | V |
| (29981) 1999TD10 | 2874.82899 | 21.524 | 0.064 | 2874.82963 | 10.176 | 3.989 | 13.818 | 13.467 | B |
| (29981) 1999TD10 | 2874.83491 | 19.896 | 0.131 | 2874.83555 | 8.548 | 3.989 | 13.818 | 13.467 | I |
| (29981) 1999TD10 | 2875.85338 | 21.502 | 0.051 | 2875.85410 | 10.155 | 3.964 | 13.820 | 13.453 | B |
| (29981) 1999TD10 | 2875.85930 | 19.721 | 0.098 | 2875.86002 | 8.374 | 3.964 | 13.820 | 13.453 | I |
| (29981) 1999TD10 | 2877.78970 | 21.186 | 0.049 | 2877.79057 | 9.843 | 3.914 | 13.824 | 13.427 | B |
| (29981) 1999TD10 | 2877.79562 | 19.512 | 0.067 | 2877.79649 | 8.169 | 3.914 | 13.824 | 13.427 | I |
| (29981) 1999TD10 | 2883.77943 | 20.902 | 0.078 | 2883.78076 | 9.570 | 3.733 | 13.836 | 13.349 | V |
| (29981) 1999TD10 | 2883.78210 | 20.804 | 0.071 | 2883.78343 | 9.472 | 3.733 | 13.836 | 13.349 | V |
| (29981) 1999TD10 | 2883.78827 | 21.716 | 0.050 | 2883.78960 | 10.384 | 3.733 | 13.836 | 13.349 | B |
| (29981) 1999TD10 | 2883.79431 | 19.855 | 0.075 | 2883.79564 | 8.523 | 3.733 | 13.836 | 13.349 | I |
| (29981) 1999TD10 | 2885.78177 | 20.997 | 0.143 | 2885.78324 | 9.668 | 3.663 | 13.840 | 13.324 | V |
| (29981) 1999TD10 | 2885.78444 | 20.900 | 0.133 | 2885.78591 | 9.571 | 3.663 | 13.840 | 13.324 | V |
| (29981) 1999TD10 | 2885.79035 | 21.633 | 0.072 | 2885.79182 | 10.304 | 3.663 | 13.840 | 13.324 | B |
| (29981) 1999TD10 | 2885.79628 | 19.670 | 0.110 | 2885.79775 | 8.341 | 3.663 | 13.840 | 13.324 | I |
| (29981) 1999TD10 | 2891.77774 | 20.764 | 0.268 | 2891.77962 | 9.445 | 3.429 | 13.852 | 13.253 | V |
| (29981) 1999TD10 | 2891.78040 | 20.890 | 0.285 | 2891.78228 | 9.571 | 3.429 | 13.852 | 13.253 | V |
| (29981) 1999TD10 | 2891.78632 | 21.209 | 0.287 | 2891.78820 | 9.890 | 3.429 | 13.852 | 13.252 | B |
| (29981) 1999TD10 | 2891.79225 | 19.734 | 0.150 | 2891.79413 | 8.415 | 3.428 | 13.852 | 13.252 | I |
| (29981) 1999TD10 | 2893.78846 | 19.611 | 0.161 | 2893.79047 | 8.295 | 3.342 | 13.856 | 13.230 | I |
| (29981) 1999TD10 | 2895.73819 | 20.308 | 0.187 | 2895.74033 | 8.995 | 3.253 | 13.860 | 13.209 | V |
| (29981) 1999TD10 | 2895.74085 | 20.491 | 0.193 | 2895.74299 | 9.178 | 3.253 | 13.860 | 13.209 | V |
| (29981) 1999TD10 | 2895.74675 | 20.877 | 0.137 | 2895.74889 | 9.564 | 3.253 | 13.860 | 13.208 | B |
| (29981) 1999TD10 | 2895.75268 | 19.312 | 0.126 | 2895.75482 | 7.999 | 3.253 | 13.860 | 13.208 | I |
| (29981) 1999TD10 | 2908.72360 | 20.385 | 0.077 | 2908.72645 | 9.088 | 2.571 | 13.887 | 13.086 | V |
| (29981) 1999TD10 | 2908.72627 | 20.331 | 0.083 | 2908.72912 | 9.034 | 2.571 | 13.887 | 13.086 | V |
| (29981) 1999TD10 | 2908.73218 | 21.159 | 0.198 | 2908.73503 | 9.862 | 2.571 | 13.887 | 13.086 | B |
| (29981) 1999TD10 | 2908.73811 | 19.497 | 0.077 | 2908.74096 | 8.200 | 2.570 | 13.887 | 13.086 | I |
| (29981) 1999TD10 | 2910.79692 | 20.631 | 0.118 | 2910.79986 | 9.336 | 2.449 | 13.891 | 13.070 | V |
| (29981) 1999TD10 | 2910.80283 | 20.982 | 0.219 | 2910.80577 | 9.687 | 2.448 | 13.891 | 13.070 | B |
| (29981) 1999TD10 | 2910.80875 | 18.997 | 0.174 | 2910.81169 | 7.702 | 2.448 | 13.891 | 13.070 | I |
| (29981) 1999TD10 | 2922.65735 | 19.680 | 0.220 | 2922.66070 | 8.393 | 1.689 | 13.915 | 12.998 | V |
| (29981) 1999TD10 | 2922.66002 | 19.544 | 0.186 | 2922.66337 | 8.257 | 1.689 | 13.915 | 12.998 | V |
| (29981) 1999TD10 | 2930.70408 | 20.190 | 0.048 | 2930.70759 | 8.905 | 1.134 | 13.932 | 12.971 | V |
| (29981) 1999TD10 | 2930.70674 | 20.149 | 0.050 | 2930.71025 | 8.864 | 1.134 | 13.932 | 12.971 | V |
| (29981) 1999TD10 | 2930.71266 | 21.017 | 0.031 | 2930.71617 | 9.732 | 1.133 | 13.932 | 12.971 | B |
| (29981) 1999TD10 | 2930.71859 | 19.062 | 0.049 | 2930.72210 | 7.777 | 1.133 | 13.932 | 12.971 | I |
| (29981) 1999TD10 | 2935.68781 | 20.276 | 0.055 | 2935.69136 | 8.991 | 0.785 | 13.942 | 12.964 | V |
| (29981) 1999TD10 | 2935.69639 | 21.298 | 0.044 | 2935.69994 | 10.013 | 0.785 | 13.942 | 12.964 | B |
| (29981) 1999TD10 | 2935.70232 | 19.322 | 0.057 | 2935.70587 | 8.037 | 0.784 | 13.942 | 12.964 | I |
| (29981) 1999TD10 | 2937.64880 | 20.470 | 0.082 | 2937.65235 | 9.184 | 0.650 | 13.947 | 12.964 | V |
| (29981) 1999TD10 | 2937.65147 | 20.336 | 0.067 | 2937.65502 | 9.050 | 0.650 | 13.947 | 12.964 | V |
| (29981) 1999TD10 | 2937.65738 | 21.345 | 0.050 | 2937.66093 | 10.059 | 0.650 | 13.947 | 12.964 | B |
| (29981) 1999TD10 | 2937.66331 | 19.526 | 0.086 | 2937.66686 | 8.240 | 0.650 | 13.947 | 12.964 | I |
| (29981) 1999TD10 | 2942.68266 | 18.991 | 0.050 | 2942.68619 | 7.703 | 0.342 | 13.957 | 12.967 | I |
| (29981) 1999TD10 | 2944.70834 | 19.389 | 0.079 | 2944.71185 | 8.100 | 0.270 | 13.961 | 12.971 | I |
| (29981) 1999TD10 | 2945.66355 | 20.665 | 0.144 | 2945.66705 | 9.375 | 0.260 | 13.963 | 12.973 | V |
| (29981) 1999TD10 | 2945.66621 | 20.219 | 0.096 | 2945.66971 | 8.929 | 0.260 | 13.963 | 12.973 | V |
| (29981) 1999TD10 | 2945.67213 | 21.275 | 0.094 | 2945.67563 | 9.985 | 0.260 | 13.963 | 12.973 | B |
| (29981) 1999TD10 | 2945.67805 | 19.408 | 0.086 | 2945.68155 | 8.118 | 0.260 | 13.963 | 12.973 | I |
| (29981) 1999TD10 | 2946.67610 | 20.202 | 0.078 | 2946.67958 | 8.911 | 0.271 | 13.965 | 12.975 | V |
| (29981) 1999TD10 | 2946.67876 | 20.313 | 0.082 | 2946.68224 | 9.022 | 0.271 | 13.965 | 12.975 | V |
| (29981) 1999TD10 | 2946.68467 | 21.196 | 0.070 | 2946.68815 | 9.905 | 0.271 | 13.965 | 12.975 | B |
| (29981) 1999TD10 | 2956.66188 | 20.097 | 0.082 | 2956.66512 | 8.796 | 0.852 | 13.986 | 13.018 | V |
| (29981) 1999TD10 | 2956.66454 | 20.182 | 0.104 | 2956.66778 | 8.881 | 0.852 | 13.986 | 13.018 | V |
| (29981) 1999TD10 | 2956.67046 | 20.991 | 0.075 | 2956.67370 | 9.690 | 0.853 | 13.986 | 13.018 | B |
| (29981) 1999TD10 | 2956.67638 | 19.099 | 0.066 | 2956.67962 | 7.798 | 0.853 | 13.986 | 13.018 | I |
| (29981) 1999TD10 | 2972.59304 | 20.726 | 0.082 | 2972.59553 | 9.398 | 1.930 | 14.020 | 13.147 | V |
| (29981) 1999TD10 | 2972.59570 | 20.672 | 0.089 | 2972.59819 | 9.344 | 1.930 | 14.020 | 13.147 | V |
| (29981) 1999TD10 | 2972.60161 | 21.521 | 0.060 | 2972.60410 | 10.193 | 1.930 | 14.020 | 13.147 | B |
| (29981) 1999TD10 | 2972.60752 | 19.778 | 0.086 | 2972.61001 | 8.450 | 1.931 | 14.020 | 13.147 | I |
| (29981) 1999TD10 | 2974.66089 | 20.102 | 0.111 | 2974.66326 | 8.770 | 2.060 | 14.024 | 13.169 | V |
| (29981) 1999TD10 | 2974.66355 | 20.290 | 0.068 | 2974.66592 | 8.958 | 2.061 | 14.024 | 13.169 | V |
| (29981) 1999TD10 | 2974.66946 | 21.209 | 0.074 | 2974.67183 | 9.877 | 2.061 | 14.024 | 13.169 | B |
| (29981) 1999TD10 | 2974.67537 | 19.390 | 0.066 | 2974.67773 | 8.058 | 2.061 | 14.024 | 13.169 | I |
| (29981) 1999TD10 | 2976.55816 | 20.331 | 0.090 | 2976.56040 | 8.995 | 2.178 | 14.028 | 13.190 | V |
| (29981) 1999TD10 | 2976.56083 | 20.321 | 0.083 | 2976.56307 | 8.985 | 2.178 | 14.028 | 13.190 | V |
| (29981) 1999TD10 | 2976.56673 | 21.000 | 0.070 | 2976.56897 | 9.664 | 2.178 | 14.028 | 13.190 | B |
| (29981) 1999TD10 | 2976.57265 | 19.336 | 0.067 | 2976.57489 | 8.000 | 2.178 | 14.028 | 13.190 | I |
| | | | | # | | | | | |
| (8405) Asbolus | 2845.84195 | 18.145 | 0.048 | 2845.84195 | 9.941 | 3.820 | 7.086 | 6.171 | B |
| (8405) Asbolus | 2845.84304 | 16.524 | 0.035 | 2845.84304 | 8.320 | 3.820 | 7.086 | 6.171 | I |
| (8405) Asbolus | 2845.84411 | 17.513 | 0.036 | 2845.84411 | 9.309 | 3.820 | 7.086 | 6.171 | V |
| (8405) Asbolus | 2847.84039 | 18.054 | 0.104 | 2847.84044 | 9.852 | 3.596 | 7.089 | 6.162 | B |
| (8405) Asbolus | 2847.84141 | 16.419 | 0.094 | 2847.84146 | 8.217 | 3.596 | 7.089 | 6.162 | I |
| (8405) Asbolus | 2847.84244 | 17.341 | 0.159 | 2847.84249 | 9.139 | 3.596 | 7.089 | 6.162 | V |
| (8405) Asbolus | 2851.85783 | 18.005 | 0.036 | 2851.85797 | 9.807 | 3.155 | 7.095 | 6.147 | B |
| (8405) Asbolus | 2851.85885 | 16.368 | 0.029 | 2851.85899 | 8.170 | 3.155 | 7.095 | 6.147 | I |
| (8405) Asbolus | 2851.85988 | 17.391 | 0.029 | 2851.86002 | 9.193 | 3.155 | 7.095 | 6.147 | V |
| (8405) Asbolus | 2855.80134 | 17.856 | 0.061 | 2855.80154 | 9.660 | 2.749 | 7.100 | 6.136 | B |
| (8405) Asbolus | 2855.80235 | 16.358 | 0.050 | 2855.80255 | 8.162 | 2.749 | 7.100 | 6.136 | I |
| (8405) Asbolus | 2855.80339 | 17.429 | 0.100 | 2855.80359 | 9.233 | 2.749 | 7.100 | 6.136 | V |
| (8405) Asbolus | 2856.78916 | 18.114 | 0.046 | 2856.78937 | 9.919 | 2.654 | 7.101 | 6.134 | B |
| (8405) Asbolus | 2856.79034 | 16.551 | 0.034 | 2856.79055 | 8.356 | 2.654 | 7.101 | 6.134 | I |
| (8405) Asbolus | 2856.79139 | 17.386 | 0.038 | 2856.79160 | 9.191 | 2.654 | 7.101 | 6.134 | V |
| (8405) Asbolus | 2857.82597 | 18.249 | 0.029 | 2857.82620 | 10.054 | 2.559 | 7.103 | 6.132 | B |
| (8405) Asbolus | 2857.82712 | 17.528 | 0.023 | 2857.82735 | 9.333 | 2.559 | 7.103 | 6.132 | V |
| (8405) Asbolus | 2857.82827 | 16.578 | 0.021 | 2857.82850 | 8.383 | 2.558 | 7.103 | 6.132 | I |
| (8405) Asbolus | 2859.82861 | 18.154 | 0.038 | 2859.82885 | 9.959 | 2.388 | 7.106 | 6.129 | B |
| (8405) Asbolus | 2859.82971 | 16.420 | 0.024 | 2859.82995 | 8.225 | 2.387 | 7.106 | 6.129 | I |
| (8405) Asbolus | 2859.83113 | 17.336 | 0.025 | 2859.83137 | 9.141 | 2.387 | 7.106 | 6.129 | V |
| (8405) Asbolus | 2862.80823 | 18.082 | 0.053 | 2862.80848 | 9.886 | 2.176 | 7.110 | 6.128 | B |
| (8405) Asbolus | 2862.80925 | 16.499 | 0.036 | 2862.80950 | 8.303 | 2.176 | 7.110 | 6.128 | I |
| (8405) Asbolus | 2862.81029 | 17.354 | 0.037 | 2862.81054 | 9.158 | 2.176 | 7.110 | 6.128 | V |
| (8405) Asbolus | 2863.78406 | 18.108 | 0.091 | 2863.78431 | 9.912 | 2.120 | 7.111 | 6.128 | B |
| (8405) Asbolus | 2863.78511 | 16.490 | 0.132 | 2863.78536 | 8.294 | 2.120 | 7.111 | 6.128 | I |
| (8405) Asbolus | 2863.78619 | 17.466 | 0.147 | 2863.78644 | 9.270 | 2.120 | 7.111 | 6.128 | V |
| (8405) Asbolus | 2865.83373 | 16.294 | 0.036 | 2865.83398 | 8.097 | 2.030 | 7.114 | 6.128 | I |
| (8405) Asbolus | 2866.77860 | 18.215 | 0.114 | 2866.77884 | 10.017 | 2.002 | 7.115 | 6.129 | B |
| (8405) Asbolus | 2866.77963 | 16.540 | 0.037 | 2866.77987 | 8.342 | 2.002 | 7.115 | 6.129 | I |
| (8405) Asbolus | 2866.78069 | 17.522 | 0.042 | 2866.78093 | 9.324 | 2.002 | 7.115 | 6.129 | V |
| (8405) Asbolus | 2868.71287 | 17.924 | 0.062 | 2868.71310 | 9.724 | 1.972 | 7.118 | 6.131 | B |
| (8405) Asbolus | 2868.71389 | 16.388 | 0.031 | 2868.71412 | 8.188 | 1.972 | 7.118 | 6.131 | I |
| (8405) Asbolus | 2868.71498 | 17.311 | 0.038 | 2868.71521 | 9.111 | 1.972 | 7.118 | 6.131 | V |
| (8405) Asbolus | 2869.74013 | 18.147 | 0.052 | 2869.74035 | 9.946 | 1.973 | 7.120 | 6.133 | B |
| (8405) Asbolus | 2869.74121 | 16.479 | 0.024 | 2869.74143 | 8.278 | 1.973 | 7.120 | 6.133 | I |
| (8405) Asbolus | 2869.74236 | 17.417 | 0.026 | 2869.74258 | 9.216 | 1.973 | 7.120 | 6.133 | V |
| (8405) Asbolus | 2870.75163 | 17.711 | 0.070 | 2870.75184 | 9.509 | 1.984 | 7.121 | 6.135 | B |
| (8405) Asbolus | 2870.75272 | 16.437 | 0.020 | 2870.75293 | 8.235 | 1.984 | 7.121 | 6.135 | I |
| (8405) Asbolus | 2870.75384 | 17.194 | 0.043 | 2870.75405 | 8.992 | 1.984 | 7.121 | 6.135 | V |
| (8405) Asbolus | 2872.70165 | 18.114 | 0.136 | 2872.70183 | 9.910 | 2.034 | 7.124 | 6.140 | B |
| (8405) Asbolus | 2872.70269 | 17.373 | 0.075 | 2872.70287 | 9.169 | 2.034 | 7.124 | 6.140 | V |
| (8405) Asbolus | 2872.70378 | 16.284 | 0.124 | 2872.70396 | 8.080 | 2.034 | 7.124 | 6.140 | I |
| (8405) Asbolus | 2873.79545 | 18.044 | 0.074 | 2873.79561 | 9.838 | 2.079 | 7.125 | 6.143 | B |
| (8405) Asbolus | 2873.79650 | 16.387 | 0.039 | 2873.79666 | 8.181 | 2.079 | 7.125 | 6.143 | I |
| (8405) Asbolus | 2873.79757 | 17.377 | 0.037 | 2873.79773 | 9.171 | 2.079 | 7.125 | 6.143 | V |
| (8405) Asbolus | 2874.77554 | 17.488 | 0.097 | 2874.77569 | 9.281 | 2.128 | 7.127 | 6.146 | B |
| (8405) Asbolus | 2874.77658 | 16.242 | 0.091 | 2874.77673 | 8.035 | 2.128 | 7.127 | 6.146 | I |
| (8405) Asbolus | 2874.77765 | 17.259 | 0.080 | 2874.77780 | 9.052 | 2.128 | 7.127 | 6.146 | V |
| (8405) Asbolus | 2875.66604 | 18.156 | 0.028 | 2875.66617 | 9.947 | 2.178 | 7.128 | 6.149 | B |
| (8405) Asbolus | 2875.66709 | 16.435 | 0.030 | 2875.66722 | 8.226 | 2.178 | 7.128 | 6.149 | I |
| (8405) Asbolus | 2875.66816 | 17.415 | 0.022 | 2875.66829 | 9.206 | 2.179 | 7.128 | 6.149 | V |
| (8405) Asbolus | 2876.73865 | 18.244 | 0.030 | 2876.73876 | 10.033 | 2.248 | 7.130 | 6.153 | B |
| (8405) Asbolus | 2876.73970 | 16.490 | 0.020 | 2876.73981 | 8.279 | 2.248 | 7.130 | 6.153 | I |
| (8405) Asbolus | 2876.74077 | 17.447 | 0.023 | 2876.74088 | 9.236 | 2.248 | 7.130 | 6.153 | V |
| (8405) Asbolus | 2878.64483 | 18.016 | 0.084 | 2878.64489 | 9.802 | 2.389 | 7.132 | 6.160 | B |
| (8405) Asbolus | 2878.64588 | 16.396 | 0.118 | 2878.64594 | 8.182 | 2.389 | 7.132 | 6.160 | I |
| (8405) Asbolus | 2886.73687 | 18.050 | 0.059 | 2886.73667 | 9.817 | 3.159 | 7.144 | 6.205 | B |
| (8405) Asbolus | 2886.73803 | 17.499 | 0.028 | 2886.73783 | 9.266 | 3.159 | 7.144 | 6.205 | V |
| (8405) Asbolus | 2886.73909 | 16.505 | 0.040 | 2886.73889 | 8.272 | 3.160 | 7.144 | 6.205 | I |
| (8405) Asbolus | 2892.74612 | 18.243 | 0.083 | 2892.74567 | 9.991 | 3.808 | 7.153 | 6.250 | B |
| (8405) Asbolus | 2892.74718 | 16.604 | 0.048 | 2892.74673 | 8.352 | 3.808 | 7.153 | 6.250 | I |
| (8405) Asbolus | 2892.74826 | 17.648 | 0.069 | 2892.74781 | 9.396 | 3.808 | 7.153 | 6.250 | V |
| (8405) Asbolus | 2893.69578 | 17.543 | 0.137 | 2893.69528 | 9.288 | 3.912 | 7.154 | 6.258 | V |
| (8405) Asbolus | 2894.69900 | 18.259 | 0.048 | 2894.69845 | 10.001 | 4.022 | 7.156 | 6.266 | B |
| (8405) Asbolus | 2894.70015 | 16.580 | 0.026 | 2894.69960 | 8.322 | 4.022 | 7.156 | 6.266 | I |
| (8405) Asbolus | 2894.70136 | 17.560 | 0.040 | 2894.70081 | 9.302 | 4.022 | 7.156 | 6.266 | V |
| (8405) Asbolus | 2895.66891 | 18.308 | 0.038 | 2895.66831 | 10.046 | 4.127 | 7.157 | 6.275 | B |
| (8405) Asbolus | 2895.67031 | 16.674 | 0.021 | 2895.66971 | 8.412 | 4.128 | 7.157 | 6.275 | I |
| (8405) Asbolus | 2895.67143 | 17.590 | 0.027 | 2895.67083 | 9.328 | 4.128 | 7.157 | 6.275 | V |
| (8405) Asbolus | 2896.71966 | 18.288 | 0.069 | 2896.71900 | 10.022 | 4.241 | 7.159 | 6.285 | B |
| (8405) Asbolus | 2896.72079 | 16.681 | 0.024 | 2896.72013 | 8.415 | 4.242 | 7.159 | 6.285 | I |
| (8405) Asbolus | 2896.72217 | 17.650 | 0.031 | 2896.72151 | 9.384 | 4.242 | 7.159 | 6.285 | V |
| (8405) Asbolus | 2902.71224 | 18.417 | 0.138 | 2902.71124 | 10.128 | 4.878 | 7.168 | 6.344 | B |
| (8405) Asbolus | 2902.71329 | 16.621 | 0.161 | 2902.71229 | 8.332 | 4.878 | 7.168 | 6.344 | I |
| (8405) Asbolus | 2902.71436 | 17.817 | 0.115 | 2902.71336 | 9.528 | 4.878 | 7.168 | 6.344 | V |
| (8405) Asbolus | 2903.71542 | 18.303 | 0.059 | 2903.71436 | 10.010 | 4.981 | 7.170 | 6.355 | B |
| (8405) Asbolus | 2903.71647 | 16.658 | 0.039 | 2903.71541 | 8.365 | 4.981 | 7.170 | 6.355 | I |
| (8405) Asbolus | 2903.71754 | 17.649 | 0.047 | 2903.71648 | 9.356 | 4.981 | 7.170 | 6.355 | V |
| (8405) Asbolus | 2907.66673 | 18.406 | 0.030 | 2907.66541 | 10.096 | 5.375 | 7.176 | 6.400 | B |
| (8405) Asbolus | 2907.66779 | 16.790 | 0.048 | 2907.66647 | 8.480 | 5.375 | 7.176 | 6.400 | I |
| (8405) Asbolus | 2907.66886 | 17.656 | 0.023 | 2907.66754 | 9.346 | 5.375 | 7.176 | 6.400 | V |
| (8405) Asbolus | 2908.65481 | 18.561 | 0.034 | 2908.65342 | 10.246 | 5.470 | 7.177 | 6.412 | B |
| (8405) Asbolus | 2908.65598 | 16.735 | 0.024 | 2908.65459 | 8.420 | 5.471 | 7.177 | 6.412 | I |
| (8405) Asbolus | 2908.65708 | 17.725 | 0.028 | 2908.65569 | 9.410 | 5.471 | 7.177 | 6.412 | V |
| (8405) Asbolus | 2911.67151 | 18.523 | 0.039 | 2911.66990 | 10.194 | 5.752 | 7.182 | 6.450 | B |
| (8405) Asbolus | 2911.67328 | 17.758 | 0.029 | 2911.67167 | 9.429 | 5.752 | 7.182 | 6.450 | V |
| (8405) Asbolus | 2911.67479 | 16.742 | 0.028 | 2911.67318 | 8.413 | 5.752 | 7.182 | 6.450 | I |
| (8405) Asbolus | 2912.66831 | 18.063 | 0.114 | 2912.66663 | 9.729 | 5.842 | 7.183 | 6.462 | B |
| (8405) Asbolus | 2912.66945 | 16.597 | 0.076 | 2912.66777 | 8.263 | 5.842 | 7.183 | 6.462 | I |
| (8405) Asbolus | 2915.65684 | 18.293 | 0.059 | 2915.65493 | 9.945 | 6.101 | 7.188 | 6.502 | B |
| (8405) Asbolus | 2915.65789 | 16.786 | 0.039 | 2915.65598 | 8.438 | 6.101 | 7.188 | 6.502 | I |
| (8405) Asbolus | 2915.65896 | 17.733 | 0.040 | 2915.65705 | 9.385 | 6.101 | 7.188 | 6.502 | V |
| (8405) Asbolus | 2916.66544 | 17.751 | 0.162 | 2916.66345 | 9.398 | 6.184 | 7.190 | 6.515 | B |
| (8405) Asbolus | 2916.66649 | 16.733 | 0.113 | 2916.66450 | 8.380 | 6.184 | 7.190 | 6.516 | I |
| (8405) Asbolus | 2916.66757 | 17.787 | 0.070 | 2916.66558 | 9.434 | 6.185 | 7.190 | 6.516 | V |
| (8405) Asbolus | 2924.62702 | 17.809 | 0.137 | 2924.62437 | 9.414 | 6.778 | 7.202 | 6.630 | B |
| (8405) Asbolus | 2924.62813 | 16.811 | 0.057 | 2924.62548 | 8.416 | 6.778 | 7.202 | 6.630 | I |
| (8405) Asbolus | 2924.62946 | 17.720 | 0.218 | 2924.62681 | 9.325 | 6.779 | 7.202 | 6.630 | V |
| (8405) Asbolus | 2928.59508 | 18.811 | 0.056 | 2928.59208 | 10.394 | 7.028 | 7.209 | 6.691 | B |
| (8405) Asbolus | 2928.59620 | 16.953 | 0.034 | 2928.59320 | 8.536 | 7.028 | 7.209 | 6.691 | I |
| (8405) Asbolus | 2928.59731 | 17.905 | 0.066 | 2928.59431 | 9.488 | 7.028 | 7.209 | 6.691 | V |
| (8405) Asbolus | 2942.58633 | 18.761 | 0.056 | 2942.58200 | 10.264 | 7.641 | 7.231 | 6.920 | B |
| (8405) Asbolus | 2942.58749 | 17.024 | 0.036 | 2942.58316 | 8.527 | 7.641 | 7.231 | 6.920 | I |
| (8405) Asbolus | 2942.58902 | 17.959 | 0.033 | 2942.58469 | 9.462 | 7.641 | 7.231 | 6.920 | V |
| (8405) Asbolus | 2943.62128 | 18.670 | 0.047 | 2943.61685 | 10.167 | 7.670 | 7.233 | 6.938 | B |
| (8405) Asbolus | 2943.62234 | 17.039 | 0.032 | 2943.61791 | 8.536 | 7.670 | 7.233 | 6.938 | I |
| (8405) Asbolus | 2943.62341 | 18.014 | 0.036 | 2943.61898 | 9.511 | 7.670 | 7.233 | 6.938 | V |
| (8405) Asbolus | 2944.59835 | 18.755 | 0.080 | 2944.59382 | 10.246 | 7.695 | 7.235 | 6.955 | B |
| (8405) Asbolus | 2944.59940 | 17.106 | 0.056 | 2944.59487 | 8.597 | 7.695 | 7.235 | 6.955 | I |
| (8405) Asbolus | 2944.60047 | 18.043 | 0.055 | 2944.59594 | 9.534 | 7.695 | 7.235 | 6.955 | V |
| (8405) Asbolus | 2950.57824 | 18.848 | 0.070 | 2950.57311 | 10.304 | 7.803 | 7.245 | 7.059 | B |
| (8405) Asbolus | 2950.57929 | 17.235 | 0.034 | 2950.57416 | 8.691 | 7.803 | 7.245 | 7.059 | I |
| (8405) Asbolus | 2950.58040 | 17.768 | 0.040 | 2950.57527 | 9.224 | 7.803 | 7.245 | 7.059 | V |
| (8405) Asbolus | 2951.56644 | 18.636 | 0.079 | 2951.56121 | 10.086 | 7.813 | 7.246 | 7.076 | B |
| (8405) Asbolus | 2951.56774 | 17.100 | 0.036 | 2951.56251 | 8.550 | 7.813 | 7.246 | 7.076 | I |
| (8405) Asbolus | 2951.56892 | 17.943 | 0.048 | 2951.56369 | 9.393 | 7.814 | 7.246 | 7.076 | V |
| | | | | # | | | | | |
| (32532) Thereus | 2854.79593 | 19.468 | 0.037 | 2854.79593 | 9.868 | 5.002 | 9.440 | 8.810 | V |
| (32532) Thereus | 2854.79764 | 19.377 | 0.035 | 2854.79764 | 9.777 | 5.002 | 9.440 | 8.810 | V |
| (32532) Thereus | 2854.80240 | 20.217 | 0.017 | 2854.80240 | 10.617 | 5.002 | 9.440 | 8.810 | B |
| (32532) Thereus | 2854.80715 | 18.355 | 0.037 | 2854.80715 | 8.755 | 5.001 | 9.440 | 8.810 | I |
| (32532) Thereus | 2858.77855 | 19.274 | 0.128 | 2858.77882 | 9.685 | 4.743 | 9.444 | 8.764 | V |
| (32532) Thereus | 2858.78026 | 19.290 | 0.146 | 2858.78053 | 9.701 | 4.743 | 9.444 | 8.764 | V |
| (32532) Thereus | 2858.78497 | 20.175 | 0.023 | 2858.78524 | 10.586 | 4.742 | 9.444 | 8.764 | B |
| (32532) Thereus | 2858.78970 | 18.474 | 0.153 | 2858.78997 | 8.885 | 4.742 | 9.444 | 8.764 | I |
| (32532) Thereus | 2859.84761 | 19.268 | 0.126 | 2859.84795 | 9.681 | 4.669 | 9.445 | 8.752 | V |
| (32532) Thereus | 2859.85419 | 20.138 | 0.021 | 2859.85453 | 10.551 | 4.669 | 9.445 | 8.752 | B |
| (32532) Thereus | 2859.85896 | 18.435 | 0.147 | 2859.85930 | 8.848 | 4.669 | 9.445 | 8.752 | I |
| (32532) Thereus | 2860.87447 | 19.450 | 0.054 | 2860.87487 | 9.866 | 4.597 | 9.446 | 8.741 | V |
| (32532) Thereus | 2860.87618 | 19.355 | 0.046 | 2860.87658 | 9.771 | 4.597 | 9.446 | 8.741 | V |
| (32532) Thereus | 2860.88089 | 20.169 | 0.028 | 2860.88129 | 10.585 | 4.597 | 9.446 | 8.741 | B |
| (32532) Thereus | 2860.88563 | 18.410 | 0.042 | 2860.88603 | 8.826 | 4.597 | 9.446 | 8.741 | I |
| (32532) Thereus | 2862.82759 | 19.269 | 0.065 | 2862.82811 | 9.690 | 4.456 | 9.448 | 8.720 | V |
| (32532) Thereus | 2862.82929 | 19.142 | 0.058 | 2862.82981 | 9.563 | 4.456 | 9.448 | 8.720 | V |
| (32532) Thereus | 2862.83400 | 19.974 | 0.047 | 2862.83452 | 10.395 | 4.456 | 9.448 | 8.720 | B |
| (32532) Thereus | 2862.83874 | 18.327 | 0.042 | 2862.83926 | 8.748 | 4.456 | 9.448 | 8.720 | I |
| (32532) Thereus | 2863.80234 | 19.424 | 0.099 | 2863.80292 | 9.847 | 4.384 | 9.449 | 8.710 | V |
| (32532) Thereus | 2863.80893 | 20.201 | 0.076 | 2863.80951 | 10.624 | 4.384 | 9.449 | 8.710 | B |
| (32532) Thereus | 2863.81370 | 18.535 | 0.062 | 2863.81428 | 8.958 | 4.383 | 9.449 | 8.710 | I |
| (32532) Thereus | 2865.83633 | 18.907 | 0.095 | 2865.83703 | 9.335 | 4.230 | 9.451 | 8.689 | V |
| (32532) Thereus | 2865.83807 | 18.969 | 0.128 | 2865.83877 | 9.397 | 4.230 | 9.451 | 8.689 | V |
| (32532) Thereus | 2865.84281 | 20.025 | 0.123 | 2865.84351 | 10.453 | 4.229 | 9.451 | 8.689 | B |
| (32532) Thereus | 2865.84759 | 18.424 | 0.166 | 2865.84829 | 8.852 | 4.229 | 9.451 | 8.689 | I |
| (32532) Thereus | 2866.80255 | 19.659 | 0.103 | 2866.80330 | 10.089 | 4.155 | 9.452 | 8.680 | B |
| (32532) Thereus | 2866.80732 | 18.219 | 0.088 | 2866.80807 | 8.649 | 4.154 | 9.452 | 8.680 | I |
| (32532) Thereus | 2874.79465 | 19.177 | 0.055 | 2874.79580 | 9.623 | 3.492 | 9.460 | 8.610 | V |
| (32532) Thereus | 2874.80123 | 19.727 | 0.155 | 2874.80238 | 10.173 | 3.492 | 9.460 | 8.610 | B |
| (32532) Thereus | 2874.80600 | 18.186 | 0.123 | 2874.80715 | 8.632 | 3.491 | 9.460 | 8.610 | I |
| (32532) Thereus | 2875.77644 | 19.315 | 0.034 | 2875.77764 | 9.762 | 3.406 | 9.461 | 8.603 | V |
| (32532) Thereus | 2875.77818 | 19.301 | 0.034 | 2875.77938 | 9.748 | 3.406 | 9.461 | 8.603 | V |
| (32532) Thereus | 2875.78292 | 20.193 | 0.018 | 2875.78412 | 10.640 | 3.406 | 9.461 | 8.603 | B |
| (32532) Thereus | 2877.71629 | 19.348 | 0.036 | 2877.71757 | 9.798 | 3.235 | 9.463 | 8.589 | V |
| (32532) Thereus | 2877.71800 | 19.320 | 0.100 | 2877.71928 | 9.770 | 3.234 | 9.463 | 8.589 | V |
| (32532) Thereus | 2877.72271 | 20.166 | 0.018 | 2877.72399 | 10.616 | 3.234 | 9.463 | 8.589 | B |
| (32532) Thereus | 2877.72747 | 18.379 | 0.038 | 2877.72875 | 8.829 | 3.234 | 9.463 | 8.589 | I |
| (32532) Thereus | 2879.74448 | 19.321 | 0.031 | 2879.74584 | 9.774 | 3.052 | 9.465 | 8.575 | V |
| (32532) Thereus | 2879.74622 | 19.302 | 0.031 | 2879.74758 | 9.755 | 3.052 | 9.465 | 8.575 | V |
| (32532) Thereus | 2879.75096 | 20.270 | 0.016 | 2879.75232 | 10.723 | 3.051 | 9.465 | 8.575 | B |
| (32532) Thereus | 2879.75574 | 18.494 | 0.034 | 2879.75710 | 8.947 | 3.051 | 9.465 | 8.575 | I |
| (32532) Thereus | 2885.69135 | 19.248 | 0.046 | 2885.69290 | 9.709 | 2.505 | 9.470 | 8.541 | V |
| (32532) Thereus | 2885.69309 | 19.332 | 0.049 | 2885.69464 | 9.793 | 2.505 | 9.470 | 8.541 | V |
| (32532) Thereus | 2885.69784 | 20.234 | 0.027 | 2885.69939 | 10.695 | 2.504 | 9.470 | 8.541 | B |
| (32532) Thereus | 2888.72021 | 18.906 | 0.039 | 2888.72184 | 9.369 | 2.225 | 9.473 | 8.527 | V |
| (32532) Thereus | 2888.72195 | 18.975 | 0.044 | 2888.72358 | 9.438 | 2.225 | 9.473 | 8.527 | V |
| (32532) Thereus | 2888.72670 | 19.863 | 0.030 | 2888.72833 | 10.326 | 2.224 | 9.473 | 8.527 | B |
| (32532) Thereus | 2888.73148 | 18.026 | 0.031 | 2888.73311 | 8.489 | 2.224 | 9.473 | 8.527 | I |
| (32532) Thereus | 2891.72697 | 19.037 | 0.081 | 2891.72867 | 9.503 | 1.951 | 9.476 | 8.516 | V |
| (32532) Thereus | 2891.72871 | 18.901 | 0.075 | 2891.73041 | 9.367 | 1.951 | 9.476 | 8.516 | V |
| (32532) Thereus | 2891.73346 | 20.085 | 0.080 | 2891.73516 | 10.551 | 1.950 | 9.476 | 8.516 | B |
| (32532) Thereus | 2891.73823 | 18.207 | 0.054 | 2891.73993 | 8.673 | 1.950 | 9.476 | 8.516 | I |
| (32532) Thereus | 2895.71642 | 19.073 | 0.069 | 2895.71818 | 9.540 | 1.605 | 9.480 | 8.506 | V |
| (32532) Thereus | 2895.71815 | 19.047 | 0.069 | 2895.71991 | 9.514 | 1.604 | 9.480 | 8.506 | V |
| (32532) Thereus | 2895.72289 | 19.897 | 0.053 | 2895.72465 | 10.364 | 1.604 | 9.480 | 8.506 | B |
| (32532) Thereus | 2895.72766 | 18.188 | 0.057 | 2895.72942 | 8.655 | 1.604 | 9.480 | 8.506 | I |
| (32532) Thereus | 2897.72748 | 19.039 | 0.065 | 2897.72926 | 9.507 | 1.445 | 9.482 | 8.502 | V |
| (32532) Thereus | 2897.72922 | 18.975 | 0.054 | 2897.73100 | 9.443 | 1.444 | 9.482 | 8.502 | V |
| (32532) Thereus | 2897.73396 | 19.832 | 0.046 | 2897.73574 | 10.300 | 1.444 | 9.482 | 8.502 | B |
| (32532) Thereus | 2897.73874 | 18.045 | 0.040 | 2897.74052 | 8.513 | 1.444 | 9.482 | 8.502 | I |
| (32532) Thereus | 2906.71708 | 19.045 | 0.034 | 2906.71887 | 9.511 | 1.031 | 9.491 | 8.501 | V |
| (32532) Thereus | 2906.71882 | 19.032 | 0.035 | 2906.72061 | 9.498 | 1.031 | 9.491 | 8.501 | V |
| (32532) Thereus | 2906.72357 | 19.899 | 0.014 | 2906.72536 | 10.365 | 1.031 | 9.491 | 8.501 | B |
| (32532) Thereus | 2906.72834 | 18.139 | 0.035 | 2906.73012 | 8.605 | 1.031 | 9.491 | 8.501 | I |
| (32532) Thereus | 2908.65897 | 18.971 | 0.028 | 2908.66074 | 9.436 | 1.048 | 9.493 | 8.504 | V |
| (32532) Thereus | 2908.66071 | 18.996 | 0.031 | 2908.66248 | 9.461 | 1.048 | 9.493 | 8.504 | V |
| (32532) Thereus | 2908.66545 | 19.537 | 0.014 | 2908.66722 | 10.002 | 1.048 | 9.493 | 8.504 | B |
| (32532) Thereus | 2908.67023 | 17.998 | 0.027 | 2908.67200 | 8.463 | 1.048 | 9.493 | 8.504 | I |
| (32532) Thereus | 2930.61433 | 19.018 | 0.030 | 2930.61545 | 9.450 | 2.710 | 9.514 | 8.615 | V |
| (32532) Thereus | 2930.61606 | 19.151 | 0.033 | 2930.61718 | 9.583 | 2.710 | 9.514 | 8.615 | V |
| (32532) Thereus | 2930.62080 | 19.939 | 0.017 | 2930.62192 | 10.371 | 2.710 | 9.514 | 8.615 | B |
| (32532) Thereus | 2930.62557 | 18.172 | 0.037 | 2930.62669 | 8.604 | 2.711 | 9.514 | 8.615 | I |
| (32532) Thereus | 2935.60485 | 19.172 | 0.029 | 2935.60572 | 9.591 | 3.156 | 9.519 | 8.660 | V |
| (32532) Thereus | 2935.60658 | 19.151 | 0.030 | 2935.60745 | 9.570 | 3.157 | 9.519 | 8.660 | V |
| (32532) Thereus | 2935.61133 | 19.991 | 0.014 | 2935.61220 | 10.410 | 3.157 | 9.519 | 8.660 | B |
| (32532) Thereus | 2935.61610 | 18.155 | 0.031 | 2935.61697 | 8.574 | 3.157 | 9.519 | 8.660 | I |
| (32532) Thereus | 2937.58351 | 19.190 | 0.034 | 2937.58426 | 9.604 | 3.329 | 9.521 | 8.679 | V |
| (32532) Thereus | 2937.58524 | 19.155 | 0.036 | 2937.58599 | 9.569 | 3.329 | 9.521 | 8.679 | V |
| (32532) Thereus | 2937.58999 | 20.039 | 0.017 | 2937.59074 | 10.453 | 3.329 | 9.521 | 8.679 | B |
| (32532) Thereus | 2937.59476 | 18.414 | 0.038 | 2937.59551 | 8.828 | 3.330 | 9.521 | 8.679 | I |
| (32532) Thereus | 2942.63799 | 19.345 | 0.044 | 2942.63843 | 9.744 | 3.752 | 9.526 | 8.733 | V |
| (32532) Thereus | 2942.63973 | 19.336 | 0.041 | 2942.64017 | 9.735 | 3.752 | 9.526 | 8.733 | V |
| (32532) Thereus | 2942.64448 | 20.192 | 0.022 | 2942.64492 | 10.591 | 3.752 | 9.526 | 8.734 | B |
| (32532) Thereus | 2942.64926 | 18.404 | 0.042 | 2942.64970 | 8.803 | 3.753 | 9.526 | 8.734 | I |
| (32532) Thereus | 2944.61658 | 19.013 | 0.142 | 2944.61689 | 9.406 | 3.910 | 9.528 | 8.756 | V |
| (32532) Thereus | 2944.61831 | 19.182 | 0.112 | 2944.61862 | 9.575 | 3.910 | 9.528 | 8.756 | V |
| (32532) Thereus | 2944.62307 | 20.039 | 0.045 | 2944.62338 | 10.432 | 3.910 | 9.528 | 8.756 | B |
| (32532) Thereus | 2944.62784 | 18.200 | 0.060 | 2944.62815 | 8.593 | 3.910 | 9.528 | 8.757 | I |
| (32532) Thereus | 2945.56619 | 19.176 | 0.059 | 2945.56643 | 9.566 | 3.984 | 9.529 | 8.768 | V |
| (32532) Thereus | 2945.56792 | 19.305 | 0.065 | 2945.56816 | 9.695 | 3.984 | 9.529 | 8.768 | V |
| (32532) Thereus | 2945.57267 | 19.989 | 0.041 | 2945.57291 | 10.379 | 3.984 | 9.529 | 8.768 | B |
| (32532) Thereus | 2945.57745 | 18.510 | 0.067 | 2945.57769 | 8.900 | 3.985 | 9.529 | 8.768 | I |
| (32532) Thereus | 2946.60597 | 19.170 | 0.076 | 2946.60614 | 9.557 | 4.064 | 9.530 | 8.780 | V |
| (32532) Thereus | 2946.60770 | 19.052 | 0.066 | 2946.60787 | 9.439 | 4.064 | 9.530 | 8.780 | V |
| (32532) Thereus | 2946.61245 | 20.109 | 0.068 | 2946.61262 | 10.496 | 4.064 | 9.530 | 8.780 | B |
| (32532) Thereus | 2946.61722 | 18.304 | 0.056 | 2946.61739 | 8.691 | 4.064 | 9.530 | 8.780 | I |
| (32532) Thereus | 2950.59680 | 19.260 | 0.076 | 2950.59668 | 9.634 | 4.357 | 9.534 | 8.831 | V |
| (32532) Thereus | 2950.59853 | 19.274 | 0.075 | 2950.59841 | 9.648 | 4.357 | 9.534 | 8.831 | V |
| (32532) Thereus | 2950.60328 | 19.983 | 0.148 | 2950.60316 | 10.356 | 4.357 | 9.534 | 8.831 | B |
| (32532) Thereus | 2950.60805 | 18.112 | 0.064 | 2950.60793 | 8.485 | 4.357 | 9.534 | 8.831 | I |
| (32532) Thereus | 2952.57454 | 19.438 | 0.039 | 2952.57427 | 9.805 | 4.494 | 9.536 | 8.857 | V |
| (32532) Thereus | 2952.57628 | 19.464 | 0.046 | 2952.57601 | 9.831 | 4.494 | 9.536 | 8.857 | V |
| (32532) Thereus | 2952.58103 | 20.193 | 0.027 | 2952.58075 | 10.560 | 4.494 | 9.536 | 8.858 | B |
| (32532) Thereus | 2952.58581 | 18.468 | 0.052 | 2952.58553 | 8.835 | 4.495 | 9.536 | 8.858 | I |
| (32532) Thereus | 2958.54420 | 19.263 | 0.062 | 2958.54344 | 9.608 | 4.873 | 9.542 | 8.942 | V |
| (32532) Thereus | 2958.54593 | 19.315 | 0.074 | 2958.54517 | 9.660 | 4.873 | 9.542 | 8.942 | V |
| (32532) Thereus | 2958.55068 | 20.026 | 0.038 | 2958.54992 | 10.371 | 4.873 | 9.542 | 8.942 | B |
| (32532) Thereus | 2958.55545 | 18.177 | 0.069 | 2958.55469 | 8.522 | 4.873 | 9.542 | 8.942 | I |
| (32532) Thereus | 2972.53913 | 19.687 | 0.049 | 2972.53710 | 9.976 | 5.535 | 9.556 | 9.162 | V |
| (32532) Thereus | 2972.54087 | 19.625 | 0.046 | 2972.53884 | 9.914 | 5.535 | 9.556 | 9.162 | V |
| (32532) Thereus | 2972.54561 | 20.400 | 0.034 | 2972.54358 | 10.689 | 5.535 | 9.556 | 9.162 | B |
| (32532) Thereus | 2972.55037 | 18.598 | 0.044 | 2972.54834 | 8.887 | 5.536 | 9.556 | 9.162 | I |
| (32532) Thereus | 2974.58960 | 19.598 | 0.095 | 2974.58737 | 9.878 | 5.604 | 9.558 | 9.196 | V |
| (32532) Thereus | 2974.59134 | 19.485 | 0.070 | 2974.58911 | 9.765 | 5.604 | 9.558 | 9.196 | V |
| (32532) Thereus | 2974.59608 | 20.217 | 0.051 | 2974.59385 | 10.497 | 5.604 | 9.558 | 9.196 | B |
| (32532) Thereus | 2974.60085 | 18.673 | 0.063 | 2974.59862 | 8.953 | 5.604 | 9.558 | 9.196 | I |
| (32532) Thereus | 3638.67727 | 20.119 | 0.038 | 3638.67431 | 10.217 | 2.220 | 10.252 | 9.323 | B |
| (32532) Thereus | 3638.68132 | 20.246 | 0.039 | 3638.67836 | 10.344 | 2.220 | 10.252 | 9.323 | B |
| (32532) Thereus | 3638.68438 | 19.384 | 0.050 | 3638.68142 | 9.482 | 2.219 | 10.252 | 9.323 | V |
| (32532) Thereus | 3638.68643 | 18.466 | 0.052 | 3638.68347 | 8.564 | 2.219 | 10.252 | 9.323 | I |
| (32532) Thereus | 3640.69001 | 18.739 | 0.052 | 3640.68710 | 8.839 | 2.028 | 10.254 | 9.313 | I |
| (32532) Thereus | 3640.69001 | 19.608 | 0.043 | 3640.68710 | 9.708 | 2.028 | 10.254 | 9.313 | V |
| (32532) Thereus | 3640.69106 | 20.458 | 0.037 | 3640.68815 | 10.558 | 2.027 | 10.254 | 9.313 | B |
| (32532) Thereus | 3640.69106 | 20.532 | 0.039 | 3640.68815 | 10.632 | 2.027 | 10.254 | 9.313 | B |
| (32532) Thereus | 3652.62533 | 20.210 | 0.043 | 3652.62262 | 10.315 | 0.836 | 10.267 | 9.278 | B |
| (32532) Thereus | 3652.62938 | 20.226 | 0.042 | 3652.62667 | 10.331 | 0.835 | 10.267 | 9.278 | B |
| (32532) Thereus | 3652.63243 | 19.403 | 0.048 | 3652.62973 | 9.508 | 0.835 | 10.267 | 9.278 | V |
| (32532) Thereus | 3652.63446 | 18.503 | 0.041 | 3652.63175 | 8.608 | 0.835 | 10.267 | 9.278 | I |
| (32532) Thereus | 3654.71304 | 20.182 | 0.039 | 3654.71034 | 10.287 | 0.622 | 10.269 | 9.276 | B |
| (32532) Thereus | 3654.71709 | 20.177 | 0.038 | 3654.71439 | 10.282 | 0.621 | 10.269 | 9.276 | B |
| (32532) Thereus | 3654.72016 | 19.434 | 0.044 | 3654.71747 | 9.539 | 0.621 | 10.269 | 9.276 | V |
| (32532) Thereus | 3654.72219 | 18.472 | 0.042 | 3654.71949 | 8.577 | 0.621 | 10.269 | 9.276 | I |
| (32532) Thereus | 3658.61420 | 20.237 | 0.174 | 3658.61150 | 10.341 | 0.224 | 10.273 | 9.277 | B |
| (32532) Thereus | 3658.61825 | 19.997 | 0.142 | 3658.61555 | 10.101 | 0.224 | 10.273 | 9.277 | B |
| (32532) Thereus | 3658.62130 | 19.169 | 0.108 | 3658.61860 | 9.273 | 0.224 | 10.273 | 9.277 | V |
| (32532) Thereus | 3658.62333 | 18.186 | 0.085 | 3658.62063 | 8.290 | 0.223 | 10.273 | 9.277 | I |
| (32532) Thereus | 3664.63355 | 20.247 | 0.056 | 3664.63079 | 10.348 | 0.411 | 10.280 | 9.287 | B |
| (32532) Thereus | 3664.63761 | 20.226 | 0.058 | 3664.63485 | 10.327 | 0.411 | 10.280 | 9.287 | B |
| (32532) Thereus | 3664.64066 | 19.685 | 0.071 | 3664.63790 | 9.786 | 0.412 | 10.280 | 9.287 | V |
| (32532) Thereus | 3664.64269 | 18.729 | 0.058 | 3664.63993 | 8.830 | 0.412 | 10.280 | 9.287 | I |
| (32532) Thereus | 3669.65740 | 20.254 | 0.143 | 3669.65455 | 10.350 | 0.923 | 10.285 | 9.304 | B |
| (32532) Thereus | 3669.66453 | 19.569 | 0.075 | 3669.66168 | 9.665 | 0.924 | 10.285 | 9.304 | V |
| (32532) Thereus | 3669.66656 | 18.734 | 0.099 | 3669.66371 | 8.830 | 0.924 | 10.285 | 9.304 | I |
| (32532) Thereus | 3674.58265 | 20.287 | 0.031 | 3674.57966 | 10.376 | 1.419 | 10.290 | 9.328 | B |
| (32532) Thereus | 3674.58670 | 20.344 | 0.032 | 3674.58371 | 10.433 | 1.419 | 10.290 | 9.328 | B |
| (32532) Thereus | 3674.58974 | 19.542 | 0.042 | 3674.58675 | 9.631 | 1.419 | 10.290 | 9.328 | V |
| (32532) Thereus | 3674.59177 | 18.672 | 0.048 | 3674.58878 | 8.761 | 1.420 | 10.290 | 9.328 | I |
| (32532) Thereus | 3676.57100 | 20.122 | 0.027 | 3676.56794 | 10.208 | 1.615 | 10.293 | 9.340 | B |
| (32532) Thereus | 3676.57506 | 20.123 | 0.026 | 3676.57200 | 10.209 | 1.616 | 10.293 | 9.340 | B |
| (32532) Thereus | 3676.57812 | 19.450 | 0.038 | 3676.57506 | 9.536 | 1.616 | 10.293 | 9.340 | V |
| (32532) Thereus | 3676.58015 | 18.470 | 0.041 | 3676.57709 | 8.556 | 1.616 | 10.293 | 9.340 | I |
| (32532) Thereus | 3678.60821 | 20.235 | 0.030 | 3678.60507 | 10.317 | 1.814 | 10.295 | 9.353 | B |
| (32532) Thereus | 3678.61226 | 20.265 | 0.030 | 3678.60912 | 10.347 | 1.814 | 10.295 | 9.353 | B |
| (32532) Thereus | 3678.61533 | 19.423 | 0.040 | 3678.61219 | 9.505 | 1.815 | 10.295 | 9.353 | V |
| (32532) Thereus | 3678.61736 | 18.544 | 0.045 | 3678.61422 | 8.626 | 1.815 | 10.295 | 9.353 | I |
| (32532) Thereus | 3681.60638 | 20.254 | 0.047 | 3681.60311 | 10.330 | 2.101 | 10.298 | 9.375 | B |
| (32532) Thereus | 3681.61043 | 20.181 | 0.047 | 3681.60716 | 10.257 | 2.101 | 10.298 | 9.375 | B |
| (32532) Thereus | 3681.61349 | 19.342 | 0.059 | 3681.61022 | 9.418 | 2.101 | 10.298 | 9.375 | V |
| (32532) Thereus | 3681.61552 | 18.529 | 0.061 | 3681.61225 | 8.605 | 2.102 | 10.298 | 9.375 | I |
| (32532) Thereus | 3686.54247 | 20.640 | 0.168 | 3686.53896 | 10.706 | 2.557 | 10.303 | 9.417 | B |
| (32532) Thereus | 3686.54653 | 20.526 | 0.146 | 3686.54302 | 10.592 | 2.557 | 10.303 | 9.417 | B |
| (32532) Thereus | 3686.54958 | 19.743 | 0.149 | 3686.54607 | 9.809 | 2.557 | 10.303 | 9.417 | V |
| (32532) Thereus | 3690.62500 | 20.414 | 0.178 | 3690.62127 | 10.470 | 2.916 | 10.307 | 9.456 | B |
| (32532) Thereus | 3690.62906 | 20.140 | 0.135 | 3690.62533 | 10.196 | 2.916 | 10.307 | 9.456 | B |
| (32532) Thereus | 3690.63211 | 19.593 | 0.142 | 3690.62838 | 9.649 | 2.916 | 10.308 | 9.456 | V |
| (32532) Thereus | 3690.63415 | 18.443 | 0.070 | 3690.63041 | 8.499 | 2.916 | 10.308 | 9.456 | I |
| (32532) Thereus | 3695.55354 | 20.608 | 0.039 | 3695.54950 | 10.650 | 3.325 | 10.313 | 9.510 | B |
| (32532) Thereus | 3695.55759 | 20.630 | 0.040 | 3695.55355 | 10.672 | 3.325 | 10.313 | 9.510 | B |
| (32532) Thereus | 3695.56064 | 19.841 | 0.052 | 3695.55660 | 9.883 | 3.325 | 10.313 | 9.510 | V |
| (32532) Thereus | 3703.60048 | 20.562 | 0.039 | 3703.59586 | 10.580 | 3.927 | 10.321 | 9.610 | B |
| (32532) Thereus | 3703.60454 | 20.521 | 0.036 | 3703.59992 | 10.539 | 3.927 | 10.321 | 9.610 | B |
| (32532) Thereus | 3703.60759 | 19.756 | 0.047 | 3703.60297 | 9.774 | 3.927 | 10.321 | 9.610 | V |
| (32532) Thereus | 3703.60964 | 18.848 | 0.054 | 3703.60502 | 8.866 | 3.927 | 10.321 | 9.610 | I |
| (32532) Thereus | 3708.55528 | 20.375 | 0.038 | 3708.55026 | 10.376 | 4.253 | 10.327 | 9.679 | B |
| (32532) Thereus | 3708.55936 | 20.438 | 0.042 | 3708.55434 | 10.439 | 4.253 | 10.327 | 9.679 | B |
| (32532) Thereus | 3708.56243 | 19.668 | 0.057 | 3708.55741 | 9.669 | 4.253 | 10.327 | 9.679 | V |
| (32532) Thereus | 3708.56449 | 18.731 | 0.056 | 3708.55947 | 8.732 | 4.253 | 10.327 | 9.679 | I |
| (32532) Thereus | 3710.57468 | 20.504 | 0.050 | 3710.56949 | 10.498 | 4.375 | 10.329 | 9.708 | B |
| (32532) Thereus | 3710.57775 | 19.770 | 0.053 | 3710.57256 | 9.764 | 4.375 | 10.329 | 9.708 | V |
| (32532) Thereus | 3710.57980 | 18.713 | 0.045 | 3710.57461 | 8.707 | 4.375 | 10.329 | 9.708 | I |
| (32532) Thereus | 3714.54238 | 20.765 | 0.164 | 3714.53685 | 10.745 | 4.597 | 10.333 | 9.768 | B |
| (32532) Thereus | 3714.54646 | 20.569 | 0.148 | 3714.54093 | 10.549 | 4.597 | 10.333 | 9.768 | B |
| (32532) Thereus | 3714.54953 | 19.948 | 0.167 | 3714.54400 | 9.928 | 4.597 | 10.333 | 9.768 | V |
| (32532) Thereus | 3714.55160 | 19.064 | 0.157 | 3714.54606 | 9.044 | 4.597 | 10.333 | 9.768 | I |
| (32532) Thereus | 3720.56582 | 20.695 | 0.137 | 3720.55973 | 10.652 | 4.885 | 10.339 | 9.864 | B |
| (32532) Thereus | 3720.56992 | 20.646 | 0.143 | 3720.56383 | 10.603 | 4.885 | 10.339 | 9.864 | B |
| (32532) Thereus | 3720.57302 | 19.812 | 0.105 | 3720.56693 | 9.769 | 4.886 | 10.339 | 9.864 | V |
| (32532) Thereus | 3720.57512 | 18.858 | 0.083 | 3720.56903 | 8.815 | 4.886 | 10.339 | 9.864 | I |
| (32532) Thereus | 3722.56166 | 20.644 | 0.053 | 3722.55538 | 10.594 | 4.968 | 10.342 | 9.897 | B |
| (32532) Thereus | 3722.56576 | 20.631 | 0.059 | 3722.55948 | 10.581 | 4.968 | 10.342 | 9.897 | B |
| (32532) Thereus | 3722.56887 | 19.850 | 0.068 | 3722.56259 | 9.800 | 4.968 | 10.342 | 9.897 | V |
| (32532) Thereus | 3722.57096 | 18.988 | 0.097 | 3722.56468 | 8.938 | 4.968 | 10.342 | 9.897 | I |
| (32532) Thereus | 3724.54050 | 20.529 | 0.059 | 3724.53403 | 10.471 | 5.043 | 10.344 | 9.930 | B |
| (32532) Thereus | 3724.54461 | 20.519 | 0.046 | 3724.53814 | 10.461 | 5.043 | 10.344 | 9.930 | B |
| (32532) Thereus | 3724.54771 | 19.730 | 0.064 | 3724.54124 | 9.672 | 5.044 | 10.344 | 9.930 | V |
| (32532) Thereus | 3724.54981 | 18.780 | 0.055 | 3724.54334 | 8.722 | 5.044 | 10.344 | 9.930 | I |
| (32532) Thereus | 3726.53634 | 20.621 | 0.098 | 3726.52968 | 10.555 | 5.113 | 10.346 | 9.963 | B |
| (32532) Thereus | 3726.54044 | 20.537 | 0.077 | 3726.53378 | 10.471 | 5.113 | 10.346 | 9.963 | B |
| (32532) Thereus | 3726.54355 | 19.892 | 0.063 | 3726.53689 | 9.826 | 5.113 | 10.346 | 9.964 | V |
| (32532) Thereus | 3726.54564 | 18.966 | 0.071 | 3726.53898 | 8.900 | 5.113 | 10.346 | 9.964 | I |
| (32532) Thereus | 3728.53376 | 20.646 | 0.112 | 3728.52690 | 10.572 | 5.176 | 10.348 | 9.998 | B |
| (32532) Thereus | 3728.54097 | 19.923 | 0.064 | 3728.53411 | 9.849 | 5.176 | 10.348 | 9.998 | V |
| (32532) Thereus | 3728.54306 | 19.031 | 0.073 | 3728.53620 | 8.957 | 5.176 | 10.348 | 9.998 | I |

TABLE 4

Measured Absolute Magnitudes and Phase Coeficients for TNOs, Centaurs, and Nereid

| Target | $B_0$ (mag) | $B_0$ Err. | $V_0$ (mag) | $V_0$ Err. | $R_0$ (mag) | $R_0$ Err. | $I_0$ (mag) | $I_0$ Err. | B' (mag deg$^{-1}$) | B' Err. | V' (mag deg$^{-1}$) | V' Err. | R' (mag deg$^{-1}$) | R' Err. | I' (mag deg$^{-1}$) | I' Err. |
|---|---|---|---|---|---|---|---|---|---|---|---|---|---|---|---|---|
| TNOs: | | | | | | | | | | | | | | | | |
| 2003 UB313 | -0.314 | 0.018 | -1.128 | 0.017 | ... | ... | -1.893 | 0.020 | -0.002 | 0.043 | 0.145 | 0.040 | ... | ... | 0.115 | 0.046 |
| 2005 FY9 | 0.869 | 0.027 | 0.089 | 0.025 | ... | ... | -0.754 | 0.025 | 0.136 | 0.031 | 0.057 | 0.030 | ... | ... | 0.100 | 0.030 |
| 2003 EL61* | 1.081 | 0.016 | 0.444 | 0.021 | … | … | -0.259 | 0.028 | 0.085 | 0.020 | 0.091 | 0.025 | … | … | 0.132 | 0.033 |
| (90377) Sedna† | 2.985 | 0.051 | 1.829 | 0.048 | 1.052 | 0.021 | 0.367 | 0.047 | ... | ... | ... | ... | 0.134 | 0.044 | ... | ... |
| (90482) Orcus* | 2.908 | 0.026 | 2.326 | 0.030 | ... | ... | 1.503 | 0.032 | 0.179 | 0.032 | 0.115 | 0.033 | ... | ... | 0.197 | 0.036 |
| (50000) Quaoar* | 3.776 | 0.026 | 2.722 | 0.027 | ... | ... | 1.306 | 0.036 | 0.081 | 0.028 | 0.172 | 0.029 | ... | ... | 0.290 | 0.038 |
| (28978) Ixion | 4.733 | 0.043 | 3.812 | 0.044 | ... | ... | 2.713 | 0.074 | 0.186 | 0.045 | 0.100 | 0.045 | ... | ... | 0.082 | 0.074 |
| (55636) 2002 TX300 | 4.226 | 0.046 | 3.407 | 0.089 | ... | ... | 2.736 | 0.048 | -0.003 | 0.050 | 0.090 | 0.099 | ... | ... | 0.079 | 0.052 |
| (55565) 2002 AW197 | 4.572 | 0.039 | 3.565 | 0.032 | ... | ... | 2.439 | 0.036 | 0.047 | 0.046 | 0.131 | 0.042 | ... | ... | 0.122 | 0.052 |
| (55637) 2002 UX25* | 4.874 | 0.022 | 3.868 | 0.022 | ... | ... | 2.826 | 0.028 | 0.151 | 0.028 | 0.164 | 0.028 | ... | ... | 0.130 | 0.036 |
| (20000) Varuna* | 4.732 | 0.025 | 3.764 | 0.035 | ... | ... | 2.569 | 0.033 | 0.262 | 0.033 | 0.274 | 0.048 | ... | ... | 0.195 | 0.045 |
| Nereid | 5.025 | 0.016 | 4.470 | 0.015 | ... | ... | 3.687 | 0.023 | 0.310 | 0.019 | 0.181 | 0.015 | ... | ... | 0.205 | 0.037 |
| (119951) 2002 KX14 | ... | ... | 4.861 | 0.040 | ... | ... | 3.570 | 0.040 | ... | ... | 0.161 | 0.046 | ... | ... | 0.191 | 0.043 |
| (120348) 2004 TY364 | 5.500 | 0.052 | 4.395 | 0.075 | ... | ... | 3.151 | 0.069 | 0.136 | 0.047 | 0.237 | 0.069 | ... | ... | 0.413 | 0.064 |
| (38628) Huya† | 6.038 | 0.014 | 5.048 | 0.021 | 4.418 | 0.005 | 4.068 | 0.110 | ... | ... | 0.155 | 0.041 | 0.125 | 0.009 | | |
| (26375) 1999 DE9 | 6.094 | 0.025 | 5.098 | 0.030 | ... | ... | 3.963 | 0.029 | 0.154 | 0.028 | 0.213 | 0.036 | ... | ... | 0.136 | 0.035 |
| (47171) 1999 TC36 | 6.195 | 0.051 | 5.255 | 0.054 | ... | ... | 3.758 | 0.063 | 0.244 | 0.044 | 0.120 | 0.048 | ... | ... | 0.239 | 0.057 |
| (55638) 2002 VE95 | 6.891 | 0.049 | 5.748 | 0.061 | ... | ... | 4.338 | 0.045 | 0.112 | 0.032 | 0.121 | 0.040 | ... | ... | 0.107 | 0.030 |
| (47932) 2000 GN171* | ... | ... | 6.368 | 0.035 | ... | ... | 5.084 | 0.035 | ... | ... | 0.143 | 0.031 | ... | ... | 0.281 | 0.033 |
| Centaurs: | | | | | | | | | | | | | | | | |
| (95626) 2002 GZ32 | 8.134 | 0.052 | 7.390 | 0.060 | ... | ... | 6.345 | 0.149 | 0.042 | 0.035 | -0.026 | 0.041 | ... | ... | -0.004 | 0.124 |
| (42355) 2002 CR46 | 8.449 | 0.030 | 7.676 | 0.037 | ... | ... | 6.714 | 0.037 | 0.139 | 0.017 | 0.125 | 0.022 | ... | ... | 0.130 | 0.023 |
| (54598) Bienor* | 8.317 | 0.027 | 7.605 | 0.036 | ... | ... | 6.642 | 0.033 | 0.118 | 0.013 | 0.082 | 0.017 | ... | ... | 0.132 | 0.017 |
| (73480) 2002 PN34 | 9.636 | 0.020 | 8.616 | 0.023 | ... | ... | 7.626 | 0.023 | 0.015 | 0.007 | 0.057 | 0.007 | ... | ... | 0.060 | 0.008 |
| (29981) 1999TD10* | 9.837 | 0.028 | 8.793 | 0.030 | ... | ... | 7.776 | 0.031 | 0.048 | 0.013 | 0.153 | 0.015 | ... | ... | 0.129 | 0.014 |
| (8405) Asbolus* | 9.762 | 0.031 | 9.087 | 0.020 | ... | ... | 8.130 | 0.019 | 0.065 | 0.006 | 0.055 | 0.005 | ... | ... | 0.052 | 0.004 |
| (32532) Thereus* | 10.252 | 0.018 | 9.427 | 0.017 | ... | ... | 8.510 | 0.019 | 0.070 | 0.005 | 0.067 | 0.005 | ... | ... | 0.053 | 0.006 |

* the phase curve was determined after subtraction of a rotation curve (see Table 1 for measured rotation periods).

† $B_0$, $V_0$, and $I_0$ are derived assuming B', V', and I' match the measured value for R'.

TABLE 5

$\chi^2$, N, and P values for linear fits to solar phase curves ($\chi^2$ fits for which P < 0.01 are highlighted in bold font)

| Target | B | | | V | | | R | | | I | | |
|---|---|---|---|---|---|---|---|---|---|---|---|---|
| | $\chi^2$ | N | P | $\chi^2$ | N | P | $\chi^2$ | N | P | $\chi^2$ | N | P |
| TNOs: | | | | | | | | | | | | |
| 2003 UB313 | 19.6 | 10 | 0.012 | 10.2 | 10 | 0.252 | ... | ... | | 20.2 | 10 | 0.010 |
| 2005 FY9 | 15.8 | 8 | 0.015 | 16.3 | 10 | 0.038 | ... | ... | | 4.9 | 10 | 0.773 |
| 2003 EL61 | 11.0 | 12 | 0.358 | 23.5 | 13 | 0.015 | ... | ... | | 19.1 | 13 | 0.060 |
| (90377) Sedna | | | | | | | 19.3 | 12 | 0.037 | | | |
| (90482) Orcus | 17.4 | 9 | 0.015 | 16.6 | 14 | 0.167 | ... | ... | | 22.4 | 14 | 0.033 |
| (50000) Quaoar | 19.4 | 12 | 0.036 | 8.8 | 11 | 0.452 | ... | ... | | 11.4 | 12 | 0.327 |
| (28978) Ixion | 25.3 | 13 | **0.008** | 5.8 | 10 | 0.668 | ... | ... | | 10.9 | 12 | 0.363 |
| (55636) 2002 TX300 | 7.6 | 15 | 0.870 | 17.8 | 14 | 0.121 | ... | ... | | 5.0 | 15 | 0.976 |
| (55565) 2002 AW197 | 11.6 | 11 | 0.234 | 11.7 | 11 | 0.231 | ... | ... | | 12.7 | 12 | 0.240 |
| (55637) 2002 UX25 | 20.9 | 13 | 0.035 | 15.8 | 13 | 0.148 | ... | ... | | 10.6 | 13 | 0.476 |
| (20000) Varuna | 5.0 | 11 | 0.835 | 20.0 | 11 | 0.018 | ... | ... | | 7.0 | 11 | 0.642 |
| Nereid | 11.4 | 7 | 0.044 | 25.23 | 10 | **0.001** | ... | ... | | 8.1 | 9 | 0.326 |
| (119951) 2002 KX14 | | | | 14.7 | 10 | 0.065 | ... | ... | | 17.5 | 11 | 0.041 |
| (120348) 2004 TY364 | 12.2 | 14 | 0.431 | 9.4 | 14 | 0.667 | ... | ... | | 14.6 | 15 | 0.330 |
| | | | | | | | | | | | | |
| (26375) 1999 DE9 | 9.6 | 10 | 0.294 | 11.6 | 10 | 0.172 | ... | ... | | 7.2 | 10 | 0.515 |
| (47171) 1999 TC36 | 17.6 | 10 | 0.025 | 27.7 | 14 | **0.006** | ... | ... | | 10.7 | 12 | 0.379 |
| (55638) 2002 VE95 | 12.7 | 11 | 0.175 | 7.6 | 10 | 0.472 | ... | ... | | 8.1 | 11 | 0.522 |
| (47932) 2000 GN171 | | | | 10.7 | 10 | 0.217 | ... | ... | | 10.6 | 11 | 0.301 |
| Centaurs: | | | | | | | | | | | | |
| (95626) 2002 GZ32 | 7.8 | 9 | 0.350 | 6.4 | 9 | 0.489 | ... | ... | | 5.2 | 8 | 0.516 |
| (42355) 2002 CR46 | 14.8 | 9 | 0.038 | 3.0 | 9 | 0.886 | ... | ... | | 9.2 | 9 | 0.238 |
| (54598) Bienor | 29.2 | 11 | **0.001** | 27.3 | 12 | **0.002** | ... | ... | | 16.8 | 11 | 0.052 |
| (73480) 2002 PN34 | 17.2 | 10 | 0.028 | 27.1 | 9 | **0.000** | ... | ... | | 24.5 | 11 | **0.004** |
| (29981) 1999 TD10 | 15.7 | 8 | 0.015 | 5.4 | 9 | 0.612 | ... | ... | | 11.9 | 8 | 0.064 |
| (8405) Asbolus | 7.8 | 9 | 0.353 | 7.1 | 11 | 0.627 | ... | ... | | 17.2 | 11 | 0.046 |

| | | | | | | | | | | | |
|---|---|---|---|---|---|---|---|---|---|---|---|
| (32532) Thereus | 8.4 | 8 | 0.211 | 11.7 | 12 | 0.302 | ... | ... | 8.9 | 11 | 0.450 |